\documentclass{aa}  

\usepackage{graphicx}
\usepackage{epstopdf}
\usepackage{txfonts}
\usepackage{xspace}
\usepackage{natbib}
\usepackage{color}
\usepackage{textcomp}
\newcommand{\Msun}{M$_{\odot}$\xspace}
\newcommand{\Lsun}{L$_{\odot}$\xspace}

\newcommand{\Msunyr}{M$_{\odot}$\,yr$^{-1}$\xspace}

\newcommand{\vterminal}{$\varv_{\infty}$\xspace}
\newcommand\Mdotgas{$\dot{M}_{\rm{gas}}$}
\newcommand\Mdotdustpd{$\dot{M}_{\rm{pd,d}}$}
\newcommand\Mdotdust{$\dot{M}_{\rm{d}}$}
\newcommand{\Tstar}{$T_{\rm{st}}$}
\newcommand{\Lstar}{$L_{\rm{st}}$}

\newcommand{\kms}{km\,s$^{-1}$\xspace}
\newcommand{\micron}{$\mu$m}
\newcommand\Jyarcsec{Jy/\arcsec$^{2}$}
\newcommand{\Pblue}{70\,$\mu$m}
\newcommand{\Pred}{160\,$\mu$m}
\newcommand{\Fblue}{$F_{\rm{70\,\mu m}}$}
\newcommand{\Fred}{$F_{\rm{160\,\mu m}}$}

\newcommand{\gcm}{g\,cm$^{-3}$}
\newcommand{\percm}{cm$^{-3}$}

\newcommand\Mdust{$M_{\rm{d}}$}

\newcommand\thetalim{$\theta_{\rm{lim}}$}
\newcommand\ria{$r_{\rm{IA}}$}
\newcommand\dRia{$\Delta r_{\rm{IA}}$}
\newcommand\Rso{$R_{\rm{0}}$}
\newcommand\Rp{$R_{\rm{p}}$}
\newcommand\tia{$t_{\rm{IA}}$}
\newcommand\tcross{$t_{\rm{cross}}$}

\newcommand\rhoism{$\rho_{\rm{ISM}}$}
\newcommand\rhoiapeak{$\rho_{\rm{IA}}^{peak}$}

\newcommand\Mia{$M_{\rm{IA}}$}
\newcommand\Via{$V_{\rm{IA}} $}
\newcommand\Mism{$M_{\rm{ISM}}$}
\newcommand\Fratio{$R_{\rm{FIR}}$}

\newcommand{\Tdust}{$T_{\rm{d}}$}
\newcommand{\agrain}{$a_{\rm{gr}}$\xspace}

\newcommand\Tmod{$T_{\rm{d}}^{\rm{mod}}$}
\newcommand\Mmod{$M_{\rm{d}}^{\rm{mod}}$}
\newcommand\Tdir{$T_{\rm{d}}^{\rm{dir}}$}
\newcommand\Mdir{$M_{\rm{d}}^{\rm{dir}}$}

\xspace

\xspace
\xspace

\DeclareMathSymbol{\omicron}{\mathord}{letters}{"6F}

\begin{document}

   \title{Dust in AGB wind-ISM interaction regions}
  \author{M.~Maercker
          \inst{1}
          \and
          T.~Khouri \inst{1}
          \and
          M.~Mecina \inst{2}
           \and
          E.~De Beck \inst{1}
          }

     \institute{Department of Space, Earth and Environment, Chalmers University of Technology, Onsala Space Observatory, 43992 Onsala, Sweden\\
   \email{maercker@chalmers.se}
         \and
             Department of Astrophysics, University of Vienna, T\"urkenschanzstr. 17, 1180 Vienna, Austria\\
             }

  \date{Accepted by A\&A in April 2022}
 
  \abstract
  {}
   {In this paper we aim to constrain the dust mass and grain sizes in the interaction regions between the stellar winds and the interstellar medium (ISM) around asymptotic giant branch stars. By describing the dust in these regions, we aim to shed light on the role of low mass evolved stars in the origin of dust in galaxies.} 
   {We use images in the far-infrared at \Pblue~and \Pred~to derive dust temperatures and dust masses in the wind-ISM interaction regions around a sample of carbon-rich and oxygen-rich asymptotic giant branch (AGB) stars. The dust temperature and mass are determined in two ways. First directly from the data using the ratio of the measured fluxes and assuming opacities for dust with a constant grain size of 0.1\,\micron. We then perform 3-dimensional dust-radiative transfer models spatially constrained by the observations to consistently calculate the temperature and mass. For the radiative transfer models each model contains one constant grain size, which is varied between 0.01\,\micron~to 5.0\,\micron.}
   {We find that the observed dust mass in the wind-ISM interaction regions is consistent with mass accumulated from the stellar winds. For the carbon-rich sources adding the spatial constraints in the radiative transfer models results in preferentially larger grain sizes ($\approx$2\,\micron). For the oxygen-rich sources the spatial constraints result in too high temperatures in the models, making it impossible to fit the observed far-infrared ratio irrespective of the grain size used, indicating a more complex interplay of grain properties and the stellar radiation field. }
   {The results have implications for how likely it is for the grains to survive the transition into the ISM, and the properties of dust particles that later act as seeds for grain growth in the ISM. However, the results for the oxygen-rich sources show that the derivation of dust properties is not straight forward, requiring more complex modelling}

   \keywords{stars: AGB and post-AGB -- stars: circumstellar matter -- stars: carbon -- stars: mass-loss -- stars: late-type
               }

   \maketitle
%

\section{Introduction}
\label{s:intro}

Dust grains are solid state particles with sizes ranging from a few nm to a few $\mu$m, a large variety of chemical compositions, and with complex surface morphologies. The formation and evolution of dust in astrophysical sources is strongly affected by the density and temperature of the surrounding gas, shocks and gas compression, the radiation fields, and the chemical abundances. 
The bulk of the dust observed in galaxies is likely a consequence of grain growth in the interstellar medium (ISM) on seed particles released into the ISM~\citep[e.g.,][and references therein]{ginolfietal2018}. The efficiency of the ISM grain growth is still heavily debated and depends on the chemical evolution of the galaxies and the properties of the seed particles~\citep{ferraraetal2016,zhukovskaetal2016,ginolfietal2018}. There is little doubt that evolved stars provide the seeds for grain growth in the universe. The seeds are the result of reprocessing of grains that were originally formed in the winds and ejecta of evolved stars. High-mass stars (M$_{\rm{star}}\gtrsim8-10$\,\Msun) end their evolution in supernova (SN) explosions. As the ejecta from the SN cools behind the shock moving through the pre-SN wind, dust particles form which are released into the ISM where they can act as seed particles. However, the reverse shock from the SN likely destroys a large fraction of the formed dust~\citep[possibly more than 90\% of the original dust mass;][]{bocchioetal2016}. In contrast, the majority of all stars that have died in the Milky Way evolved along the asymptotic giant branch (AGB) and ended their lives in a less violent manner. Through their mass-loss, AGB stars are major contributors of dust grains~\citep[e.g.,][and references therein]{ferrarottico2006,hofnerco2018}. The dust grains are released to the ISM where the stellar wind collides with the surrounding interstellar matter, providing seed particles for ISM grain growth. In models of the Large Magellanic Cloud (LMC), total dust production rates fit the observed dust mass if destruction through a SN reverse shock is included~\citep{schneideretal2014}. In this case the dust from AGB stars may account for up to 70\% of the total dust budget observed at the present time. At earlier times the contribution from SN explosions dominates the dust production, owing to the longer evolutionary timescales for low-mass stars. At which point dust production is dominated by AGB stars is very model dependent, and may happen already for galaxies at redshift 4-5~\citep{michalowskietal2010}.\\
It is possible that up to 90\% of the dust produced in \emph{stars} comes from low-mass stars~\citep{draine2009}, and the detection of pre-solar grains from low-mass stars in meteorites shows that (at least) some of this dust survives and can act as seeds for grain growth in the ISM~\citep{hecketal2020}. In fact, models of the origin of dust in the Milky~Way can not explain the observed dust masses from stellar sources alone, strongly indicating that ISM grain growth must take place~\citep{inoue2011,ginolfietal2018}.\\

The dust grains from AGB stars form in the extended atmosphere of the stars, where radiation pressure on the grains initiates the stellar wind~\citep[e.g.,][]{woitke2006,khourietal2016a,khourietal2020}. The wind and resulting circumstellar envelope (CSE) may be asymmetric, are likely inhomogeneous, and experience continued grain growth and processing in denser regions out to a few stellar radii. Eventually the dust will be released into the ISM, when the stellar wind collides with the surrounding medium in the wind-ISM interaction region~\citep{hofnerco2018}. The dust grains may contain complex molecules, and shield them from destruction due to the interstellar radiation field and passage into the ISM. The mass, effectiveness of shielding, and the properties of seeds released into the ISM, depend on the grain sizes, the porosity, and the structure of these stellar grains~\citep{draine2009}. These properties in turn depend on the physical conditions in the dust formation and destruction regions. Attempts to determine the properties of dust grains have been made by looking at distinct features~\citep[e.g.,][]{devriesetal2010} and continuum emission~\citep[e.g.,][]{groenewegenetal2009,ladjaletal2010,maerckeretal2018b}.

The properties of these original dust grains formed in the winds of evolved stars are directly connected to the properties of the seeds in the ISM, and the efficiency of the ISM-growth depends directly on the properties of the stellar grains that are released into the ISM. Specifically, properties such as the size and structure of the seed grains are critical, as they determine the likelihood for the grains to grow~\citep{draine2009}. Theoretical models of grain growth in the ISM combine critical dust properties in a generic "sticking coefficient", constituting one of the largest uncertainties in grain growth models~\citep{zhukovskaetal2016}, emphasising the importance of determining these properties directly. 
In this paper we study a sample of wind-ISM interaction regions observed at far-infrared wavelengths as a first step to study the properties of the dust grains released into the ISM by AGB stars and their significance for the origin of dust in the ISM. We focus our analysis on the interaction regions between the wind and the ISM to empirically study the properties of the dust in these regions, instead of the formation of the dust in the inner wind. A similar study using the same data was done by~\cite{coxetal2012}, albeit with a different approach and a focus on the overall geometry of the interaction regions rather than the dust properties. Section~\ref{s:data} describes the archival data we use, Section~\ref{s:methods} we use to determine dust masses and temperatures. The results are presented in Sect.~\ref{s:results}, including a comparison to the analysis in~\cite{coxetal2012}.. Finally, Sect.~\ref{s:discussion} contains a discussion putting the results into the context of dust return to the ISM, an investigation of what constraints can be put on the grain sizes, and difficulties in reproducing the observations and determining the grain properties. Our conclusions on the dust in AGB wind-ISM interaction regions are then summarised in Sect.~\ref{s:conclusions}.

\section{Data}
\label{s:data}

The wind-ISM interaction regions around a sample of AGB stars and supergiants were observed as part of the MESS (Mass-loss of Evolved Stars) programme~\citep{groenewegenetal2011} with PACS~\citep[Photodetector Array Camera and Spectrometer][]{poglitschetal2010} aboard \emph{Herschel Space Observatory}~\citep{pilbrattetal2010}. The interaction regions were observed in scan maps in the blue (at \Pblue) and red (at \Pred) filters. The emission in the infrared and the spatial resolution make it possible to determine the dust masses and dust temperatures in the wind-ISM interaction regions (Sects.~\ref{s:tempandmass} and~\ref{s:radmc3d}). The data were fully reduced by the MESS team and presented in~\citet[][hereafter C2012]{coxetal2012}. We use the highest-level of the JScanam maps of the reduced data from the archive (data products are provided at varying quality levels). At \Pblue~and \Pred~the pixel scales are 1\farcs6 and 3\farcs2, respectively. The absolute flux calibration uncertainty of the observations is 15\%. For more details on the data properties, see C2012 and the Herschel/PACS documentation\footnote{https://irsa.ipac.caltech.edu/data/Herschel/docs/nhsc/pacs/pacs.html}.

C2012 classified the observed structures based on their geometrical appearance into "fermata", "rings", "eyes", and "irregular". Here we investigate the subsample of sources classified as "fermata" (a central source surrounded by a bow-shape, e.g.,  Fig.~\ref{f:uuaur} or~\ref{f:rhya}) and "rings" (e.g., Fig.~\ref{f:rtcap}). All the data included in this study is presented in Appendices~\ref{a:models} and~\ref{b:images}. The origin of the structures for these sources is likely dominated by the interaction of a spherical wind with the ISM, while the "eyes" and "irregular" shapes may have additional shaping mechanisms (e.g. interactions with companions). The sources included in this study, with some basic parameters, are listed in Table~\ref{t:tempandmasses}. 

\section{Methods}
\label{s:methods}

We use two methods to determine the temperatures and dust masses in the wind-ISM interaction regions: direct estimates from the data, and radiative transfer models. In the first the temperature is directly obtained from the ratio of the observed fluxes and assumed opacity values for the grains (Sect.~\ref{s:tempandmass}), while in the second method we explicitly model the wind-ISM interaction regions using 3-dimensional dust-radiative transfer to consistently derive the temperatures (Sect.~\ref{s:radmc3d}).

\subsection{Direct estimates of temperatures and masses from the FIR images}
\label{s:tempandmass}

In the far-infrared (FIR), dust grains emit thermal radiation as a black body with temperature \Tdust~, modified by the dust opacity.

\begin{equation}
\label{e:flux}
F_{\lambda}= {{M_d B(\lambda,T_{\rm{d}})\kappa_{\lambda}} \over {D^2}},
\end{equation}

\noindent
where $M_d$ is the dust mass, $B(\lambda,T_d)$ is the blackbody flux at the dust temperature $T_d$ at wavelength $\lambda$, $\kappa_{\lambda}$ is the absorption coefficient at wavelength $\lambda$, and $D$ is the distance. The ratio of observed fluxes (\Fratio$=$\Fblue/\Fred) in the far-infrared hence effectively constrain the (average) dust temperature and total dust mass, assuming optically thin dust emission and specific grain properties (that is, $\kappa_{\lambda}$ for a particular grain). The PACS observations can therefore be used to directly determine the dust temperature and mass in the observed wind-ISM interaction regions. 

We measure the total flux towards the wind-ISM interaction regions in the \Pblue~and \Pred~filters after subtracting a local background outside the wind-ISM interaction regions, and summing the flux above the rms value (determined in the background regions). The star and present-day mass-loss dominate the flux along the line-of-sight towards the central star, and these regions are masked out. The error in the individual measurements of the total flux is $\sim$1\%. In some cases the rms cut-off is not sufficient to exclude emission outside of the wind-ISM interaction structure, and does not exclude the contribution from background sources. In these cases the flux is determined by hand within a polygon surrounding the main structure. By comparing the change in the measured flux using different polygon areas, we find that while the manual selection generally does not affect the fluxes measured in the \Pblue~images (by less than 10\%), this can have a much stronger effect on the measured \Pred~fluxes. Adding the absolute calibration uncertainty of 15\%, we estimate the total uncertainties in the measured flux values to be $\approx$18\% in the \Pblue~images and $\approx$25\% in the \Pred~images. The measured \Pblue~and \Pred~fluxes, and the resulting \Fratio~are shown in Table~\ref{t:tempandmasses}.

In all cases we assume optically thin dust emission from spherical grains with a constant radius of \agrain=0.1\micron. This is comparable to what is typically assumed in AGB CSEs~\citep[e.g.,][]{schoieretal2005,ramstedtetal2009,hofner2008,mattssonco2011,norrisetal2012}, {and allows to directly compare to C2012. However, see Sect.~\ref{s:grainsizes} for s discussion on grain sizes.} For oxygen-rich sources (that is, M-type and S-type stars) we use the absorption and scattering coefficients for astronomical silicates~\citep{suh1999} with a grain density of 3.3\,g/cm$^3$, giving a $\kappa_{\lambda}$ of 71.08\,cm$^2$/g and 12.71\,cm$^2$/g at \Pblue~and \Pred, respectively, calculated using Mie theory. For carbon-rich sources we use the absorption and scattering coefficients for amorphous carbon grains~\citep{suh2000} with a grain density of 1.8\,g/cm$^3$, giving  a $\kappa_{\lambda}$ of 86\,cm$^2$/g and 26\,cm$^2$/g at \Pblue~and \Pred, respectively. The measured fluxes and derived masses and temperatures (\Mdir~and \Tdir, respectively) are presented in Table~\ref{t:tempandmasses}. The uncertainties in the flux measurement lead to uncertainties of $\approx$60\% in \Mdir~and $\approx$15\% in \Tdir. Note that this method does not take any spatial constraints into consideration. The results merely provide the temperature for which the input dust grains will produce the observed ratios in the FIR. Different dust grains will produce the same FIR ratio, albeit at different temperatures. The \emph{actual} temperature of the grains in the observed wind-ISM interaction regions depends on the stellar radiation field (the spectral energy distribution (SED) of the star), the effect of the present-day wind on the SED, the distance between the wind-ISM interaction region and the star, and on the grain properties. Using the spatial information from the PACS observations, and modelling the observed structures and radiation fields explicitly, hence provides additional constraints on the dust properties in the observed regions.

\begin{table*}[t]
\tiny
  \centering
  \caption{Estimated masses and temperatures in the wind-ISM interaction regions based on the measured fluxes in the PACS \Pblue~and \Pred~images and assuming \agrain\,=\,0.1\,\micron. The observed far infrared-ratio \Fratio is given by $F_{\rm{70\,\mu m}}^{im}$/$F_{\rm{160\,\mu m}}^{im}$. {All corresponding PACS images and radiative transfer models are shown in Figs.~\ref{f:aqand} to~\ref{f:alphaori}. }}
\begin{tabular}{lccccccc}
\hline\hline
Source & Mask  & $F_{\rm{70\,\mu m}}^{im}$ & $F_{\rm{160\,\mu m}}^{im}$ & \Fratio & $\Delta$\Fratio& \Mdust$^{im}$ & \Tdir \\
      & [\arcsec] & [Jy]  & [Jy]  & && [$10^{-5}$\,\Msun] & [K] \\
\hline 
\multicolumn{2}{l}{\textbf{Carbon dust}} &       &       & & &       &  \\
AQ And & 10 & 3.2    & 1.8   & 1.8 & 0.5 & 23.3  & 38 \\
U Ant  & 20 & 22.0   & 6.2   & 3.6 & 1.1 & 3.9   & 52 \\
UU Aur & 40 & 6.0    & 3.3   & 1.8 & 0.6 & 6.5   & 39 \\
U Cam*  & 30 & 4.3    & 4.5   & 1.0 & 0.3 & 23.7  & 32 \\
RT Cap & 40 & 1.0    & 1.4   & 0.8 & 0.2 & 4.0   & 30 \\
S Cep*  & 20 & 2.5    & 1.0   & 2.5 & 0.8 & 2.1   & 44 \\
Y CVn*  & 30 & 2.2    & 11.8  & 0.2 & 0.1 & 126.2 & 22 \\
TT Cyg & 10 & 2.6    & 1.3   & 1.9 & 0.6 & 6.6   & 40 \\
U Hya  & 30 & 28.9   & 14.7  & 2.0 & 0.6 & 10.2  & 40 \\
CW Leo* & 0  & 3459.6 & 598.6 & 5.8 & 1.8 & 42.8  & 70 \\
W Ori  & 45 & 1.2    & 0.3   & 2.0 & 0.6 & 0.3   & 55 \\
W Pic  & 20 & 1.2    & 0.3   & 1.4 & 0.4 & 0.6   & 55 \\
TX Psc* & 10 & 5.8    & 1.9   & 3.1 & 1.0 & 1.5   & 48 \\
R Scl  & 10 & 28.1   & 5.8   & 4.8 & 1.5 & 4.6   & 62 \\
S Sct  & 20 & 17.3   & 14.9  & 1.2 & 0.4 & 62.8  & 33 \\
X TrA*  & 20 & 7.6    & 7.4   & 1.0 & 0.3 & 13.3  & 32\\
      &       &&&       &       &       &  \\
\multicolumn{2}{l}{\textbf{Silicate dust}} &       &  & &     &       &  \\
$\theta$ Aps & 35  & 6.6  & 1.4  & 4.8 & 1.5 & 0.4  & 46 \\
W Aql        & 30  & 9.6  & 3.3  & 2.9 & 0.9 & 7.6  & 43 \\
EP Aqr       & 20  & 8.2  & 2.9  & 2.9 & 0.9 & 1.9  & 38 \\
R Cas        & 45  & 19.9 & 6.3  & 3.2 & 1.0 & 3.6  & 39 \\
S Cas*        & 20  & 2.6  & 2.2  & 1.2 & 0.4 & 35.4 & 29 \\
$\mu$ Cep    & 45  & 46.8 & 6.2  & 7.5 & 2.3 & 13.0 & 59 \\
o Cet*        & 30  & 72.4 & 18.4 & 3.9 & 1.2 & 4.2  & 43 \\
$\chi$ Cyg*   & 20  & 7.6  & 2.8  & 2.7 & 0.8 & 2.5  & 37 \\
X Her*        & 20  & 10.4 & 3.4  & 3.1 & 1.0 & 2.2  & 39 \\
R Hya        & 55  & 11.3 & 5.5  & 2.1 & 0.6 & 3.8  & 34 \\
W Hya        & 50  & 32.1 & 8.6  & 3.7 & 1.2 & 2.7  & 42 \\
R Leo        & 50  & 20.7 & 3.3  & 6.4 & 2.0 & 0.3  & 53 \\
T Mic*        & 20  & 7.4  & 2.4  & 3.1 & 0.9 & 3.7  & 39 \\
$\alpha$ Ori* & 200 & 77.9 & 31.3 & 2.5 & 0.8 & 23.3 & 36 \\
X Pav        & 30  & 9.3  & 4.5  & 2.1 & 0.6 & 16.5 & 34 \\
V1943 Sgr    & 30  & 6.4  & 3.3  & 2.0 & 0.6 & 7.3  & 33 \\
NML Tau      & 55  & 10.3 & 4.8  & 2.1 & 0.7 & 15.0 & 34 \\
RT Vir       & 50  & 4.9  & 3.5  & 1.4 & 0.4 & 4.9  & 30\\
\hline
\multicolumn{5}{l}{*\emph{Sources not modelled with radiative transfer (Sect.~\ref{s:radmc3d}).}}\\
\end{tabular}%
  \label{t:tempandmasses}%
\end{table*}%

\subsection{Radiative transfer models with RADMC-3D}
\label{s:radmc3d}

In addition to deriving a first estimate of the masses and temperatures using the total PACS fluxes, we therefore also model the \emph{main} structures in the wind-ISM interaction regions. The wind-ISM interaction regions do not have a simple spherical symmetry making the use of 3-dimensional models necessary. We calculate models using v2.0 of the 3D radiative transfer code RADMC-3D~\citep{dullemond2012} RADMC-3D calculates the full frequency-dependent dust-radiative transfer using the Monte Carlo method, and produces images and spectra. Models can be calculated for any arbitrary 3D density distribution, allowing us to model the wind-ISM interaction regions explicitly, and to constrain the sizes and densities in the regions. 

We focus here on the main observable structure of the bow-shock. The bow-shock is formed when the stellar wind collides with the surrounding ISM, and is affected by the velocity of the wind relative to the ISM and the density of the ISM~\citep{wilkin1996,villaveretal2012,coxetal2012}. It is assumed that the stellar wind expands freely within the bubble created by the bow-shock, and the position of the star and present-day wind with relation to the bow-shock depend on the angle of the star's space motion with respect to the local ISM.

\begin{figure}
\centering
\includegraphics[width=8cm]{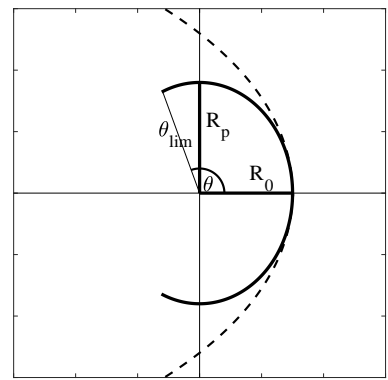}
\caption{The parameters of the ellipse shape describing the wind-ISM interaction regions: $R_{\rm{0}}$ (stand-off distance), $R_{\rm{p}}$ (the distance perpendicular to $R_{\rm{0}}$), and $\theta$ (the angle to the apex of the bow shock). The star is assumed to move towards the right relative to the ISM. The black dashed line shows the shape of the wilkinoid for the same stand-off distance (see text for details).}
\label{f:ellipseparams}
\end{figure}

In principle, the 3D surface of the bow-shock can be described by a hollow paraboloid~\citep[a \emph{wilkinoid}, see e.g.,][]{wilkin1996,coxetal2012}. This surface is derived assuming a ram pressure balance between the stellar wind and the flow of the ISM and the conservation of momentum flux across the shell. The geometry is described by the stand-off distance between the star and the wind-ISM interaction region. Using this description for the bow-shock surface has the advantage that it depends on physical parameters (the present-day mass-loss rate, the stellar wind velocity, the density of the surrounding ISM, and the stellar space velocity), which in principle can be constrained by fitting the observed structures. However, we find that the observed structures are generally better described by \emph{ellipsoids} rather than \emph{wilkinoids}, in particular at large angles relative to the direction of the apex of the bow shock (Sect.~\ref{s:coxetal} and Fig.~\ref{f:ellipseparams}).

In order to estimate the masses and densities in the wind-ISM interaction regions, we therefore approximate the 3D structure of the interaction regions with ellipsoidal shells that follow a Gaussian density distribution in the radial direction,

\begin{equation}
\rho_{\rm{IA}}(r)\propto exp({{-(r-r_{\rm{IA}})^2} \over {2\sigma_{\rm{IA}}^2}})\times [-cos({\theta \over 2})^A],
\end{equation}

\noindent
where $\rho_{\rm{IA}}$ is the density in the wind-ISM interaction region, \ria~is the distance from the star where the wind-ISM interaction region has its peak density, $\sigma_{\rm{IA}}$ is the width of the Gaussian distribution, $\theta$ is the angle to the apex of the bow shock, and $A$ is an exponent that allows the density to taper off behind the star. The density is scaled so that the integral over the entire volume of the wind-ISM interaction region gives the total mass \Mdust. The distance from the star is given by the ellipsoid

\begin{equation}
r_{\rm{IA}}= {{R_0 \times R_{\rm{p}}} \over \sqrt{{R_0^2 cos(\theta) + R_{\rm{p}}^2 sin(\theta)}}},
\end{equation}

\noindent
where $R_{\rm{0}}$ is the distance between the apex of the bow-shock and the star (the stand-off distance), and $R_{\rm{p}}$ is the distance perpendicular to $R_{\rm{0}}$ (Fig.~\ref{f:ellipseparams}). Depending on the geometry, $R_{\rm{0}}$ can either be the major or minor axis of the ellipsoid (with $R_{\rm{p}}$ then being the respective other axis). We construct "open" ellipsoids by setting the density to 0 at angles $\theta>\theta_{\rm{lim}}$. For most sources \thetalim$=100$\textdegree, with the exception of R~Cas ($\theta_{\rm{lim}}=140^{o}$) and W~Hya (a closed ellipsoid). We use a mild tapering off with $A=1 - 2$. Only for $\theta$~Aps the shell is more strongly tapered with $A=4$. Note that the parameters \Rso, \Rp, and \thetalim~are \emph{physical} parameters. Their observed projection will change depending on the inclination angle of the wind-ISM interaction region. 

In addition to the wind-ISM interaction region, we also model the stellar radiation field and the present-day (dusty) mass-loss \Mdotdustpd.  As the focus of this work is the modelling of the wind-ISM interaction regions, we use a simplified approach in modelling the stellar radiation field and present-day dust mass-loss to include a reasonable incident radiation field. We investigate the effect of changes in the incident radiation field in Sect.~\ref{s:grainsizes}.
The central star is included assuming black body radiation with temperature \Tstar~and luminosity \Lstar. The density of the present-day mass-loss follows the standard 1/$r^2$ density profile, assuming the same dust expansion velocity as the gas. This is a simplified approach, as in reality the dust will have a drift velocity relative to the gas. {Observationally this drift velocity can not be measured, as it is generally not possibly to measure the velocity of the dust. Theoretical models however predict significant drift-velocities, especially for low mass-loss-rate objects~\citep[e.g.,][]{sandinco2003,sandinco2020}.}

The modelled {dust mass-loss rates \Mdotdustpd~are hence lower limits}. However, the resulting density profile will still produce the required radiation field. The resulting SEDs are constrained by data downloaded from the Vizier online photometry viewer\footnote{http://vizier.u-strasbg.fr} and by varying the \Mdotdustpd\, until a satisfactory fit to the SED is found. The present-day dusty mass-loss is not strongly constrained {~\citep[at best to a factor of a few,][]{ramstedtetal2008}}, and for our purposes a reasonable fit to the data {by eye} is sufficient. The present-day wind expansion velocities \vterminal~and gas-mass-loss-rates \Mdotgas~are from C2012 (see references therein). The final \Tstar, \Lstar, and \Mdotdustpd~are presented in Table~\ref{t:basicparams}.

The model grids are 400\arcsec$\times$400\arcsec$\times$400\arcsec~with a step size of 2\arcsec, and are converted to cm using the distances provided in Table~\ref{t:basicparams}. We tested models with a finer grid, which did not affect the results significantly for the wind-ISM interaction regions. In the central pixel the grid is refined to a 6$\times$6$\times$6 sub-grid to properly sample the increasing density of the present-day wind. The number of photons in the Monte Carlo simulation is set to 500000. We assume isotropic scattering by the dust grains. The details in the ray-tracing are described in the RADMC-3D manual.

Although RADMC-3D allows to model an arbitrary 3-dimensional structure, there are some limitations to what is feasible in this study. The description of the wind-ISM interaction region with an ellipsoid and a Gaussian density profile can not reproduce deviations from this smooth geometry, such as arcs and/or clumpy structures, or extended tails behind the star. Hence, we generally expect the models to not be able to reproduce all the emission that is observed in the images. Instead we focus on the \emph{main, overall} structures of the wind-ISM region, and rather than fitting the total model flux to the observed flux, we try to reproduce the surface brightness and overall spatial distribution of the emission in the images. This fitting is done by eye by varying the radii, widths, angles, and dust masses in the wind-ISM interaction regions (see the Figs.~\ref{f:aqand} to~\ref{f:rtvir} to see the observed and modelled structures). In general, the radii (\Rso~and \Rp) are constrained to within $\sim$10\% of the width of \dRia, with \dRia being constrained to within $\approx$15\%. Although we are only considering sources classified as "fermata" or "rings", the structures in some sources are either too weak, and/or too complex to be feasibly modelled. These sources are indicated with a * in Table~\ref{t:tempandmasses}. We initially model all sources assuming the same carbon and silicate grains as in Sect.~\ref{s:tempandmass} with \agrain$=0.1$\,\micron. Generally we find it difficult to fit the emission in \emph{both} the \Pblue~and \Pred~filters (i.e. to reproduce the ratio between the two filters, \Fratio, see Sect~\ref{s:grainsizes}), and we therefore focus on reproducing the flux at 70\,\micron. Based on our experience while manually fitting the interaction regions, we estimate the uncertainty in the dust mass to be on the order of $\approx\pm$20\% (similar to the uncertainty in the flux measurement). The uncertainty in the derived \Rso~is estimated to be $\approx\pm$10\%. This may however be affected by the degree of clumpiness in the interaction region (making it more difficult to fit the surface brightness) and/or projection effects.

The results of the radiative transfer modelling of the wind-ISM interaction regions are given in Table~\ref{t:models}. The density \rhoiapeak~refers to the peak density of the dust in the model in the wind-ISM interaction region. The model images are compared to the observations in Appendix~\ref{a:models}.

\begin{table*}[htbp]
\tiny
  \centering
  \caption{Basic stellar and present-day wind parameters used as input in the modelling of the wind-ISM interaction regions. All models assume \agrain~=~0.1\,\micron.}
    \begin{tabular}{lccccccc}
\hline\hline
    Source & {D} & {\Tstar} & \Lstar & {\Mdotdustpd} & {\Mdotgas} & {g/d} & {\vterminal} \\
          & {[pc]} & {[K]} & {[\Lsun]} & {[$10^{-10}$\,\Msunyr]} & {[$10^{-8}$\,\Msunyr]} &       & {[\kms]} \\
\hline

\textbf{C-rich} & & & & & & & \\
    AQ And & 825   & 2200  & 5200  & 1     & 65    & 6500  & 17.0 \\
    U Ant & 268   & 2500  & 5000  & 0.3   & 2     & 667   & 4.0 \\
    UU Aur & 341   & 2500  & 8000  & 2.5   & 27    & 1080  & 11.0 \\
    RT Cap & 291   & 2500  & 2200  & 0.32  & 3.2   & 1000  & 8.0 \\
    TT Cyg & 562   & 2825  & 2700  & 0.2   & 3.2   & 1600  & 4.0 \\
    U Hya & 208   & 3000  & 3750  & 0.63  & 4.9   & 778   & 8.5 \\
    W Ori & 377   & 2200  & 7600  & 1.5   & 23    & 1533  & 11.0 \\
    W Pic & 512   & 2200  & 4000  & 0.5   & 30    & 6000  & 7.0 \\
    R Scl & 360   & 2325  & 5000  & 6     & 30    & 500   & 10.5 \\
    S Sct & 386   & 2500  & 4000  & 0.08  & 2     & 2500  & 4.0 \\
    & & & & & & & \\
\textbf{O-rich and S-type} & & & & & & & \\
    $\theta$ Aps & 113   & 2500  & 3500  & 0.5   & 11    & 2200  & 4.5 \\
    W Aql & 340   & 1350  & 6000  & 100   & 1300  & 1300  & 20.0 \\
    EP Aqr & 135   & 2500  & 3000  & 3.1   & 31    & 1000  & 11.5 \\
    R Cas & 127   & 1900  & 3000  & 12    & 120   & 1000  & 13.5 \\
    $\mu$ Cep & 390   & 2000  & 35000 & 100   & 200   & 200   & 35.0 \\
    R Hya & 118   & 2500  & 6500  & 1.5   & 16    & 1067  & 12.5 \\
    W Hya & 104   & 2200  & 9600  & 4     & 7.8   & 195   & 8.5 \\
    R Leo & 71    & 2000  & 1800  & 0.92  & 9.2   & 1000  & 9.0 \\
    X Pav & 270   & 2000  & 8000  & 10    & 52    & 520   & 11.0 \\
    V1943 Sgr & 197   & 2300  & 6000  & 0.43  & 13    & 3023  & 5.4 \\
    NML Tau & 245   & 1250  & 7000  & 175   & 320   & 183   & 18.5 \\
    RT Vir & 136   & 2250  & 2000  & 3     & 50    & 1667  & 7.8 \\
\hline
    \end{tabular}%
  \label{t:basicparams}%
\end{table*}%

\begin{table*}[htbp]
\tiny
\centering
\caption{Results from the radiative transfer modelling with RADMC-3D assuming \agrain\,=\,0.1\,\micron. For R~Scl and S~Sct \Mmod~is the \emph{total} (detached shell $+$ wind-ISM interaction region) dust mass. The value in parentheses is the dust mass in the wind-ISM interaction region only. \dRia is the FWHM of the wind-ISM interaction region. The other parameters are explained in the text.}
\begin{tabular}{lcccccccccc}
\hline\hline
Source & \Mmod & \Rso  & \Rp   & \dRia & \Tmod & \rhoiapeak & $F_{\rm{70\,\mu m}}^{mod}$ & $F_{\rm{160\,\mu m}}^{mod}$ & \tcross & \tia \\
      & [$10^{-5}$\,\Msun] & [\arcsec] & [\arcsec] & [\arcsec] & [K]   & [$10^{-24}$\gcm] & [Jy]  & [Jy]  & [10$^3$ yr] & [10$^3$ yr] \\
\hline 
\textbf{C-rich}&       &       &       &       &       &       &       &       &       &  \\
AQ And & 25.0  & 54    & 54    & 7     & 38    & 1.0   & 3.1   & 1.8   & 12    & 2500 \\
U Ant & 3.0   & 43    & 43    & 4     & 62    & 9.4   & 28.9  & 5.9   & 14    & 1000 \\
UU Aur & 2.0   & 90    & 100   & 8     & 47    & 0.7   & 4.8   & 1.6   & 13    & 80 \\
RT Cap & 1.8   & 97    & 97    & 8     & 39    & 0.4   & 2.0   & 1.1   & 17    & 563 \\
TT Cyg & 3.6   & 35    & 35    & 2     & 47    & 3.7   & 2.9   & 1.0   & 23    & 1800 \\
U Hya & 6.0   & 120   & 120   & 10    & 47    & 2.1   & 30.3  & 10.3  & 14    & 952 \\
W Ori & 1.2   & 98    & 98    & 15    & 43    & 0.1   & 1.5   & 0.6   & 16    & 80 \\
W Pic & 1.2   & 52    & 46    & 5     & 48    & 0.8   & 1.1   & 0.4   & 18    & 240 \\
R Scl & 4.3 (1.3) & 52    & 61    & 10    & 65    & 0.6   & 29.4  & 5.6   & 8     & 72 \\
S Sct & 18 (9.0) & 132   & 132   & 10    & 42    & 2.6   & 13.4  & 5.9   & 61    & 22500 \\
&       &       &       &       &       &       &       &       &       &  \\
\textbf{O-rich and S-type}&       &       &       &       &       &       &       &       &       &  \\
$\theta$ Aps & 0.2   & 75    & 67    & 7     & 52    & 5.6   & 5.3   & 0.9   & 9     & 40 \\
W Aql & 5.0   & 49    & 56    & 7     & 37    & 6.8   & 5.5   & 2.1   & 4     & 5 \\
EP Aqr & 0.4   & 37    & 50    & 7     & 56    & 14.4  & 11.7  & 1.7   & 2     & 14 \\
R Cas & 4.0   & 138   & 142   & 14    & 41    & 6.3   & 18.5  & 5.3   & 6     & 33 \\
$\mu$ Cep & 30.0  & 125   & 100   & 14    & 44    & 3.3   & 25.6  & 6.2   & 7     & 30 \\
R Hya & 0.8   & 94    & 125   & 15    & 48    & 1.7   & 11.9  & 2.3   & 4     & 50 \\
W Hya & 0.5   & 90    & 75    & 10    & 56    & 4.7   & 26.2  & 3.8   & 5     & 13 \\
R Leo & 0.3   & 110   & 118   & 13    & 45    & 5.2   & 9.5   & 2.1   & 4     & 33 \\
X Pav & 3.0   & 55    & 75    & 5     & 44    & 6.1   & 7.1   & 1.7   & 6     & 30 \\
V1943 Sgr & 1.0   & 65    & 80    & 10    & 47    & 2.1   & 4.8   & 1.0   & 11    & 233 \\
NML Tau & 8.0   & 95    & 95    & 10    & 35    & 5.5   & 12.0  & 5.1   & 6     & 5 \\
RT Vir & 0.6   & 85    & 120   & 10    & 39    & 1.9   & 3.2   & 1.0   & 7     & 20 \\
\hline &       &       &       &       &       &       &       &       &       &  \\
\end{tabular}%

\label{t:models}%
\end{table*}%

\section{Results}
\label{s:results}

\subsection{Direct estimates vs. radiative transfer modelling}
\label{s:fluxmodelcomp}

\begin{figure}
\centering
\includegraphics[width=10cm]{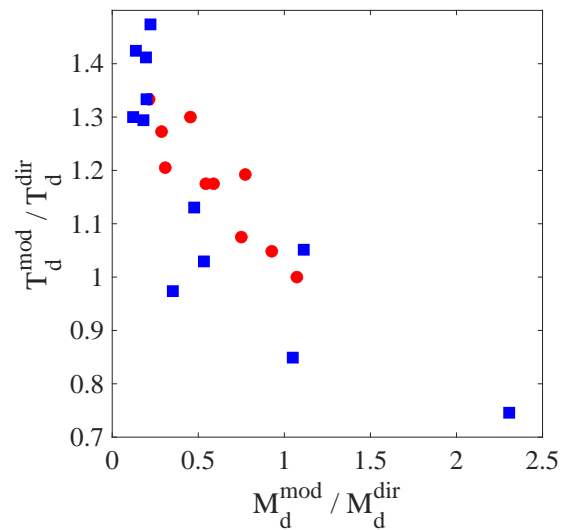}
\caption{Ratios between dust temperatures and dust masses obtained through two different methods using 0.1\,\micron~grains: full radiative transfer (\Tmod~and \Mmod) and derived directly from the PACS images (\Tdir~and \Mdir) for C-rich sources (red dots) and O-rich and S-type AGB stars (blue squares).}
\label{f:MmodvsMim}
\end{figure}

Figure~\ref{f:MmodvsMim} shows the ratio between the dust masses and temperatures derived consistently in the radiative transfer modelling (\Mmod~and \Tmod, respectively; Sect.~\ref{s:radmc3d}) to the dust masses and temperatures estimated directly from the flux-ratios in the images (\Mdir~and \Tdir, respectively; Sect.~\ref{s:tempandmass}). Both the radiative transfer models and the direct estimates  assume 0.1\,\micron~sized carbon or silicate grains. For the models the total dust mass is systematically less than the dust mass determined directly. On average, the ratio between \Mmod~and \Mdir~is 0.6, both for the C-rich and the O-rich sources. This difference is caused by the temperatures in the radiative transfer models being higher than those inferred directly from the flux density ratios. The consequence is that the dust mass has to be reduced in the model in order to still fit the emission at 70\,\micron. As mentioned in Sect.~\ref{s:tempandmass}, determining the temperatures and dust masses from the ratios in the PACS images directly does not take spatial constraints into account. As a consequence, when including the spatial constraints in the models and calculating the temperatures solving the radiative transfer, the grains become too warm, and the models do not reproduce the observed \Fratio. This discrepancy likely reflects issues in the assumed dust properties and/or radiation fields (Sect.~\ref{s:grainsizes}).

\subsection{Comparison to Cox et al. 2012}
\label{s:coxetal}

The data that is used here was first presented by~\cite{coxetal2012}. They provided a classification for the different wind-ISM interaction regions {(see Sect.~\ref{s:data})} and determine the dust masses in the wind-ISM interaction similar to our direct method. \cite{coxetal2012} further discuss in detail the effect of the space motion and possible binarity on the observed structures, find an estimate of the density of the surrounding ISM, and run hydrodynamical models to investigate the morphology of the bow-shock and the formation of instabilities. While we make use of a number of the results in C2012 (e.g., the classification scheme, the reduced data, and the derived position angles and inclinations), there are some notable differences in the method and focus:
\begin{itemize}
\item C2012 measure the flux in apertures that follow the shape of the bow shock, within a certain radial range and angle, while we measure the entire flux in the image that is attributed to the wind-ISM interaction. As such, we generally measure more flux than C2012.
\item We use the same formalism to derive the dust mass based on the flux in the images (Eq.~\ref{e:flux}). However, C2012 assume a fixed temperature of 30\,K and determine separate dust masses based on the \Pblue\,and\Pred\,fluxes, respectively. We use the ratio of the measured fluxes to constrain the temperature and then determine one dust mass.
\item C2012 use the same opacities for all sources~\citep{lico2001}, while we separate between carbon dust and silicate dust.
\item C2012 measure the stand-off distance (corrected for inclination) and use the description of wilkinoids to estimate the density of the surrounding \rhoism.
\item C2012 calculate hydrodynamical models of the wind-ISM interaction regions to study the morphology and instabilities. However, they do not perform radiative transfer models of the dust to compare with the observed images.
\item C2012 derive estimates of the total dust mass in the wind-ISM interaction regions. However, they do not derive densities in the wind-ISM interaction regions.
\end{itemize}

In comparison to C2012, we focus on modelling the (typical) dust masses and dust temperatures, explicitly taking the shape and size of the wind-ISM interaction regions into account in 3D-radiative transfer models, and investigate the constraints that can be set on the dust properties.

Comparing our dust masses derived using Eq.~\ref{e:flux} to the dust masses from C2012 (note that C2012 presented a corrigendum to their original results), our masses are systematically lower at only $\approx$30\% of their masses measured at 70\micron, despite our generally larger fluxes. However, using our measured fluxes and fixing the temperature to T=30\,K, our masses using Eq.~\ref{e:flux} increase by a factor 3.9 to an average of 17\% larger than C2012. Additionally, using the opacities that C2012 use, the masses based on our measured fluxes increase by an additional factor of 1.25 to an average of 47\% larger compared to C2012. From Table~\ref{t:tempandmasses} it is clear that the temperature in the dust emitting regions deviates significantly from 30\,K (between 29\,K to 62\,K, with an average of 40\,K) for several objects when assuming 0.1\,\micron-sized grains. Hence, although we in principle probe more of the mass in the wind-ISM interaction region by measuring the flux in the entire image (as opposed to using a limited aperture), by deriving the temperature instead of assuming one temperature, the total estimated mass decreases significantly compared to C2012. It is also worth noting that C2012 derive significantly \emph{different} masses in the \Pblue~and \Pred~images, indicating the same discrepancy in reproducing the observed \Fratio~(Sects.~\ref{s:fluxmodelcomp} and~\ref{s:grainsizes}).

The other parameter we can directly compare is the stand-off distance between the star and the bow-shock. In C2012 this corresponds to the radius that defines the shape of the wilkinoid, while here it is the major or minor axis of the ellipsoid that is used to model the wind-ISM interaction region. C2012 give two radii, one that is predicted based on the stellar space-velocity, the stellar mass-loss rate, the gas velocity of the stellar wind, and the density in the surrounding ISM, and a radius measured in the observations (de-projected for the inclination angle). Their observed stand-off distances for the objects studied here are a factor of $\approx$2 smaller than what is predicted from the stellar space-velocity. The radii derived in our models are consistent with the observed distances in C2012 (and the distances measured in the images). In order to investigate the formation and evolution of the interaction regions, C2012 present hydrodynamical models of the wind-ISM interaction. They find that the predicted stand-off distance is consistent with the models. However, the shape of the interaction region deviates significantly from what is predicted by the wilkinoid in most models. Depending on the density contrast between the stellar wind and ISM, and the velocities of the wind, dust, star, and ISM, instabilities can form that affect the overall geometry of the interaction regions, and the structure within the regions (e.g., width and clumpiness; see C2012 for a detailed discussion). The discrepancy emphasises the importance of modelling the interaction regions explicitly using hydrodynamical models~\citep[e.g., C2012;][]{villaveretal2012}. 
In this context, it is worth noting that we generally can not reproduce the observed shapes in the wind-ISM interaction regions using wilkinoids in our 3-dimensional radiative transfer models -- for any given stand-off distance, the wilkinoid becomes too wide at increasing angles $\theta$ (Fig.~\ref{f:ellipseparams}). While the ellipsoid manages to reproduce the overall structures, it naturally fails to produce the extended tails that develop behind the stars. However, up to an angle of $\approx$100\textdegree\,(relative to the direction of the stand-off distance) we find that ellipsoids follow the shape of the interaction regions better than wilkinoids. For the purpose of this paper, it is important to place the observable dust in our models at the correct distance from the star. As such, our ellipsoids are \emph{ad hoc} descriptions of the observed structures without underlying physical processes, but allowing us to model the dust radiative transfer of the observed structures. 

\section{Discussion}
\label{s:discussion}

\subsection{Structure and densities}
\label{s:structure}

The set of PACS observations of the wind-ISM interaction regions around AGB stars show a variety of morphologies (C2012). The different morphologies can generally be explained with the interaction of the stellar wind with the ISM, and/or the interaction between different periods of the stellar wind with varying mass-loss rates and expansion velocities. A full description of the structures requires hydrodynamical modelling of the winds and the surrounding material~\citep[Sect.~\ref{s:coxetal} and, e.g.,][]{steffenco2000,mattssonetal2007,wareingetal2007,coxetal2012,villaveretal2012}. Keeping that in mind, the wind-ISM interaction regions modelled here can be overall well reproduced using an ellipsoid, and the derived stand-off distance is still connected to relevant physical parameters, such as the stellar- and wind-velocities, the mass-loss rate, and the density in the surrounding ISM.

The derived full-width-at-half-maximum (FWHM) of the interaction regions, \dRia\, is 10\%--15\% of the stand-off distance. This is consistent with the detached shells around carbon stars created by wind-wind interaction~\citep[e.g.,][]{schoieretal2005,mattssonetal2007,maerckeretal2010,olofssonetal2010}. Note, however, that for some objects a clear distinction can be made between the wind-wind interaction due to a thermal pulse and wind-ISM interaction. The carbon stars U~Cam, R~Scl, S~Sct show clear wind-ISM interaction regions at further distances from the star than the spherical detached shells, indicating a predominantly radial wind-wind interaction due to a thermal pulse inside the wind-ISM interaction region. In the case of spherical shells, the origin of the observed structures may hence be due to wind-ISM interaction, and/or wind-wind interaction following a thermal pulse.

\cite{libertetal2007} present a simple model in order to describe the properties of the detached dust shell around the carbon star Y~CVn. In their model the interaction with the ISM consists of a spherical shell with external material swept up from the ISM by the bow-shock, separated from circumstellar material by a contact discontinuity~\citep[see Fig.~5 in][]{libertetal2007}. They estimate the total mass of the shell to be dominated by circumstellar material, the mass swept up from the ISM only constituting a few percent of the total mass.

In order to get a rough estimate of the timescales and densities for the sources discussed here, we assume a simplified approximation of the wind-ISM interaction regions along the lines of the work by~\cite{libertetal2007}. For each source we assume a spherical half-shell with a radius \Rso\,and width \dRia, and estimate the amount of mass that could have been swept up by the surrounding ISM in the corresponding volume

\begin{equation}
$\Via$=0.5\times\frac{4\pi}{3}($\Rso$+0.5$\dRia$)^3.
\label{e:volume}
\end{equation}

The total mass from the ISM swept up by this volume is estimated using the densities of the surrounding ISM, \rhoism,

\begin{equation}
$\Mism$=$\rhoism$ \,$\Via$=\mu_{\rm{H}}\,m_{\rm{H}}\,n_{\rm{H}}\,$\Via$,
\label{e:ismmass}
\end{equation}

\noindent
where $\mu_{\rm{H}}=1.4$ is the mean nucleus number per hydrogen atom, $m_{\rm{H}}$ is the mass of the hydrogen atom, and $n_{\rm{H}}$ is the hydrogen density in the ISM. The surrounding $n_{\rm{H}}$ can be calculated assuming a relation between the particle density and the distance to the galactic plane $z$~\citep[following the same procedure as C2012, based on][]{spitzer1978,mihalasco1981,loupetal1993},

\begin{equation}
n_{\rm{H}}(z)=2.0\times\rm{e}^{-{{|z|} \over {100\rm{pc}}}},
\label{e:ismdens}
\end{equation}
 where $z(\rm{pc}) = \emph{d}\, \rm{sin}(\emph{b}) + 15$, and $d$ is the distance and $b$ the Galactic latitude. Assuming a gas-to-dust ratio in the ISM of 1000, we find that the amount of swept-up mass in the wind-ISM interaction regions is less than 1\% of the observed mass. 

Finally, if we assume that the stars have been losing mass at constant rates \Mdotdust~and constant velocities \vterminal, we can calculate the time \tia~it takes the stellar wind to build up the mass in the interaction region, and the crossing time \tcross~(the time it takes for the material in the stellar wind to reach the interaction region):

\begin{equation}
$\tia$=$\Mia$ / $\Mdotdust$,
\end{equation}

\begin{equation}
$\tcross$=$\Rso$ / $\vterminal$.
\end{equation}

Table~\ref{t:models} shows the resulting build-up and crossing times. Note that both estimates are very uncertain because of uncertain present-day dust mass-loss rates \Mdotdustpd~(which are not well constrained in the SED models), and the assumption that the dust has (and retains) the same velocity as the gas (Sect.~\ref{s:radmc3d}). The estimated crossing times and build-up times are therefore only upper limits. However, for most sources \tcross\,is significantly less that \tia (changes in the dust velocity would affect both to the same degree), and the dust particles remain in the wind-ISM interaction regions on time-scales of a few 10000 --100000 years. The timescales are hence consistent with the build-up of the mass in the interaction region during AGB evolution. The average estimated gas density $n_{\rm{IA}}^{H}$ is on the order of a few 1000\,$\rm{cm^{-3}}$ (assuming purely atomic hydrogen and a gas-to-dust ratio of 1000), corresponding to the densities in regular circumstellar envelopes at distances of a few 1\,000\,AU from the star. The densities are consistent with hydrodynamical models of interaction regions by~\cite{coxetal2012}, and are consistent or slightly higher than the densities predicted by~\cite{wareingetal2007} and~\cite{villaveretal2012}. The derived densities are comparable to the number densities found in giant molecular clouds, dark clouds, and star forming clumps ($10^3 - 10^5$\,cm$^{-3}$), and three to four orders of magnitude larger than the cold and warm ISM~\citep[e.g.,][]{lequeux2005}.

It must be kept in mind that these estimates are based on over-simplifying assumptions. However, the estimates give an order-of-magnitude indication of the parameters determined, and indicate that the wind-ISM interaction regions contain significant amounts of dust and gas that is consistent with the build-up dominated by material from the AGB stars. The mass resides in the shell for a significant period of time (several 10\,000\,yr to a few 100\,000\,yr) at relatively high densities and cold temperatures.

\subsection{Constraints on grain sizes}
\label{s:grainsizes}

Although comparatively high, the derived densities when taken at face value are likely not high enough and the timescales are too short to allow for significant further grain growth~\citep[requiring densities $>10^5$\percm\, and timescales of $\sim$10$^5$ years;][]{ossenkopfco1994}. However, observations in the submm of thermal emission from the dust in detached shell sources indicate the presence of relatively large grains (0.1\,\micron\,--2.0\,\micron) compared to what generally is assumed in the inner AGB wind \citep[$\approx$0.1\,--\,0.5\,\micron, e.g.,][]{hofner2008,mattssonco2011,norrisetal2012}, indicating possible continued growth and processing of the dust grains in the wind-wind interaction. The densities in the shells are an order of magnitude larger than what is derived here, but the timescales are significantly shorter (a few 1000 years) compared to the time the dust resides in the wind-ISM interaction. It is possible that increased densities in clumps caused by instabilities, and/or increased densities in connection with the shock front allow grains to grow to larger sizes.  Additionally, the models using \agrain=0.1\,\micron~generally do not reproduce the observed \Fratio~because the temperatures of the grains are too high (Sect.~\ref{s:fluxmodelcomp}).

We therefore investigate the constraints observations in the FIR can provide on the grain sizes, and whether different grain sizes can reproduce the observed \Fratio. For a given grain type, the ratio between the fluxes at \Pblue\,and \Pred\, depends on the grain size, as the ratio between the absorption at optical and NIR wavelength and the emission in the FIR changes, effectively changing the grain temperature. We determine the FIR model flux ratios for the different grain sizes for each object, and compare the models to the observed ratios in the wind-ISM interaction regions, using opacities for \agrain=\,0.01\,\micron, 0.1\,\micron, 0.25\,\micron, 0.5\,\micron, 0.75\,\micron, 1.0\,\micron, 2.0\,\micron, and 5.0\,\micron, for both amorphous carbon and silicate grains~\citep{suh1999,suh2000}.

\begin{figure*}[t]
\centering
\includegraphics[width=16cm]{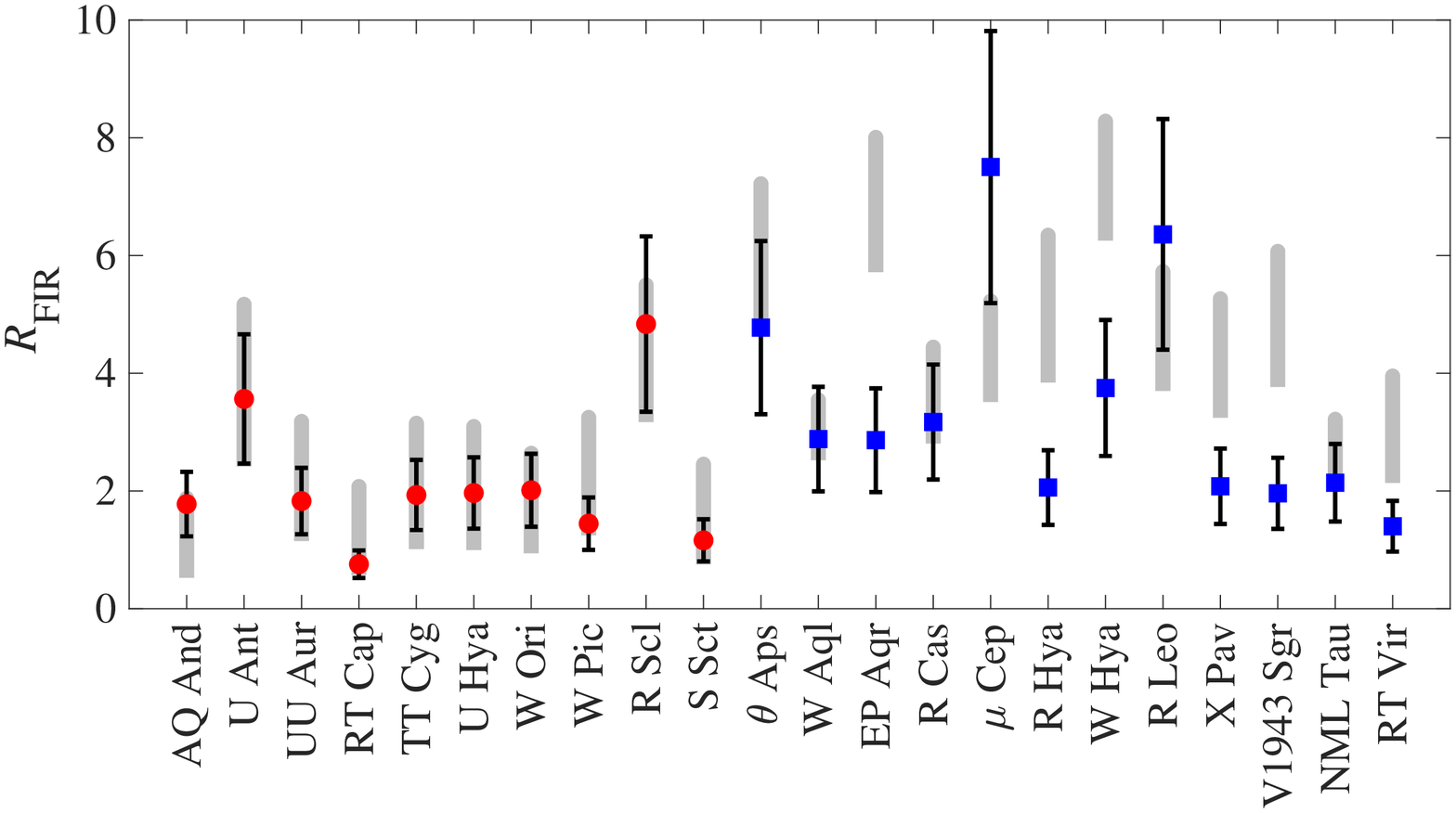}\\
\includegraphics[width=8cm]{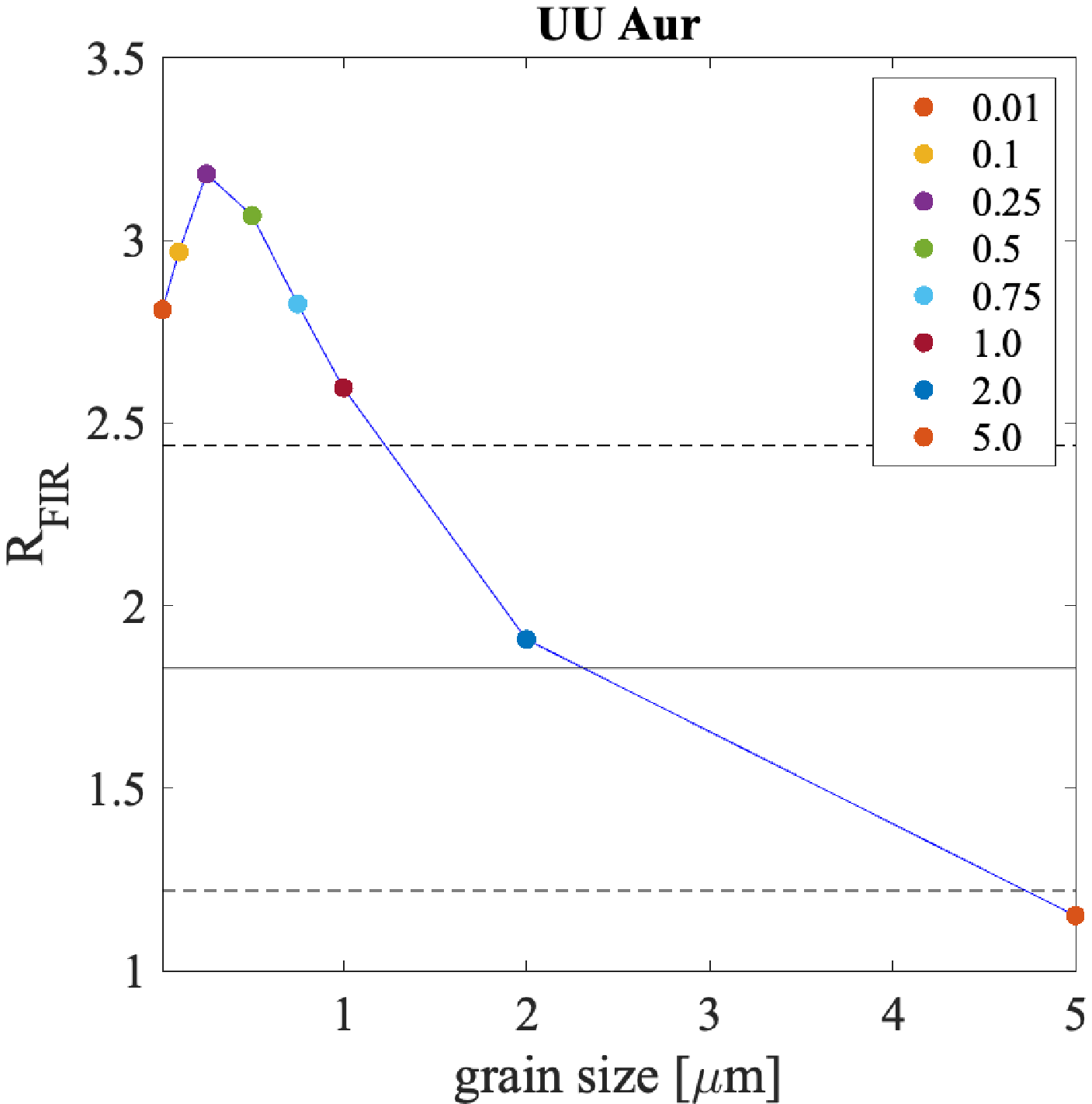}
\includegraphics[width=8cm]{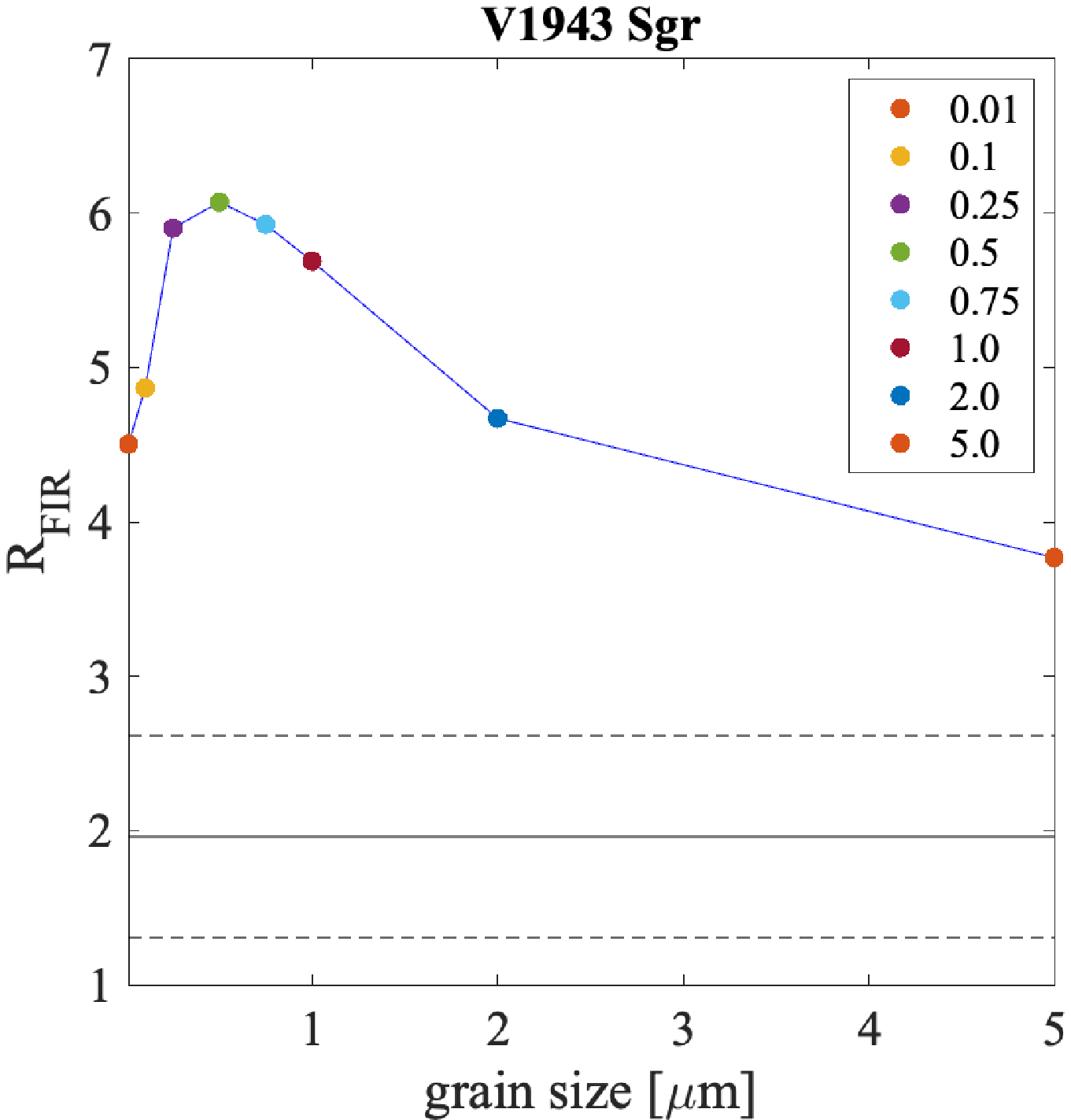}
\caption{\emph{Top panel:} The observed values for \Fratio~for C-rich sources (red dots) and O-rich sources (blue squares) and their uncertainty $\Delta$\Fratio. The grey bars show the range of \Fratio~ that the models produce for grain sizes between 0.01\micron~and 5.0\micron. \emph{Bottom panels:} Examples of the model \Fratio~vs. grain size for the carbon star UU~Aur (left) and the O-rich star V1943~Sgr (right). The horizontal solid lines show the observed \Fratio, the dashed horizontal lines indicate the uncertainty $\Delta$\Fratio.}
\label{f:fratios}
\end{figure*}
Figure~\ref{f:fratios} shows the range of modelled \Fratio\, for the different grain sizes for each source compared to the observed \Fratio values, separated into C-rich and O-rich objects, as well as examples of the modelled \Fratio~ vs. grain size for a C-rich and O-rich object, respectively. For the sources with amorphous carbon grains, the observed \Fratio\,constrains the grain sizes for grains larger than $\approx$0.75--1.0\,\micron, while for smaller grains there is an ambiguity between sizes $<0.25$\,\micron~on the one hand and 0.25\,\micron~to 0.75\,\micron~on the other (that is, grains smaller than 0.25\,\micron\, and grains between 0.25\,\micron\, and 0.75\,\micron\, produce the same \Fratio). For silicate grains the ambiguity between small and larger grains is even larger. 

For the carbon-rich AGB stars, the observed ratios indicate relatively large grains ($\gtrsim$1\,\micron). Note that we are only modelling grains of one constant size, while a distribution in grain sizes would be more realistic. However, given the observational information, such a distribution would be difficult to constrain. The effects of using different optical constants were investigated in the case of detached shells by~\cite{brunneretal2018}. They do not find a significant difference between different optical constants. Taken at face value, the results here indicate that a significant amount of dust must be present in relatively large grains in the wind-ISM interaction regions around the carbon AGB stars.

Although the uncertainties in the measured fluxes do not constrain the grain sizes for the C-rich objects very well, generally the models produce \Fratio~values within the observed range. For the models with silicate grains for O-rich stars it is significantly more difficult to reproduce the observed \Fratio. In half of the cases the observed \Fratio~is a factor 1.5--2 lower than the smallest predicted ratio. In the remaining cases the uncertainty in the observed \Fratio~is so large that the grain size cannot be constrained. This may partially be caused by bad measurements of the flux in the \Pred\, images. However, comparing the observed values of \Fratio~to the modelled values, we systematically over-predict \Fratio\, for models with silicate dust, while the models with amorphous carbon are comparable to the observed ratios. Unless this is a coincidence, it is more likely that the reason for the discrepancy in the O-rich sources lies in the details of the dust modelling.

\subsection{Assumptions in the dust modelling and their effects on the modelled \Fratio}
\label{s:effects}

The factors that determine \Fratio\, are the temperature of the grain, and the slope (and value) of the opacities in the FIR. In addition to the ratio between absorption in the optical and NIR and emission in the FIR, the temperature of the grain is also affected by the shape of the SED (that is, the incident light that the grain receives), in particular in the optical and NIR. This opens for (at least) two problems. The optical properties of the grains are not measured but are extrapolated at $\lambda>100$\,\micron. Failing to reproduce the observed ratios in the FIR may hence be due to unknown dust properties. A second issue is that we assume a stellar black body with a given luminosity and effective temperature in all our models. However, molecular and atomic absorption in the optical and NIR may significantly reduce the SED at short wavelengths compared to a plain black body.

\begin{figure}
\centering
\includegraphics[width=8cm]{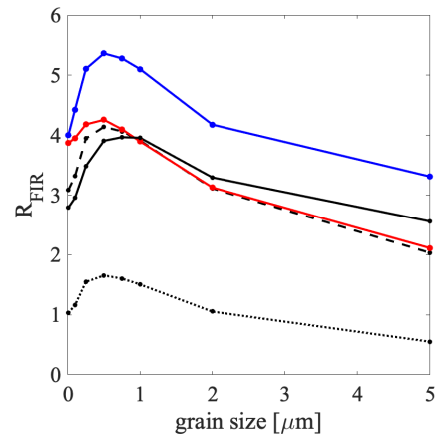}
\caption{Comparison of the \Fratio\,for different grain sizes and different opacities: silicate grains (blue), amorphous carbon grains (red), silicate grains with $\lambda^{-1.5}$ slope at $\lambda$\,>\,100\,\micron\,(black-dashed), opacities that represent the opacities presented by~\cite[][black dotted]{fanciulloetal2020}, and models with silicate grains but with a modified stellar black body (black-solid). The models are calculated for the test model. See text for details)}
\label{f:Rcompare}
\end{figure}

In order to test to what extent these different mechanisms affect the observed \Fratio, we construct a test model with representative parameters for the observed sources and wind-ISM interaction regions (Table~\ref{t:testmodel}). Figure~\ref{f:Rcompare} shows model FIR ratios derived for our test model. The models are identical in their geometrical setup and density distribution, and are calculated using the standard silicate grains and amorphous carbon grains as used for all other sources. We additionally artificially changed the slope of the silicate grains to follow a $\lambda^{-1.5}$ law instead of $\lambda^{-2}$ at $\lambda$\,>\,100\,\micron, more similar to the amorphous carbon grains (corresponding to a direct change in dust properties). We also calculated models where we reduced the stellar black body to 25\% at $\lambda$\,<\,2\,\micron~(corresponding to the molecular and atomic absorption in the star's atmosphere). Both decreasing the stellar SED, and changing the grain properties at long wavelengths, would change the model \Fratio~to be consistent with observations. 

\begin{table}[t]
  \centering
  \caption{Parameters for the test model. For all test models the provided values were left fixed, and only grain properties and inclination angles were changes (see text for details). The values for \Rso, \Rp, and \dRia~are given for 0 inclination.}
\begin{tabular}{lll}
\hline\hline 
Parameter & &Value \\
\hline 
\textbf{Stellar} &&\\
Luminosity& \Lstar     & 6000\,\Lsun \\
Effective temperature& \Tstar     & 2300\,K \\
Distance& D     & 200\,pc \\
&\\
\textbf{Present-day mass-loss}&& \\
Dust mass-loss rate& \Mdotdustpd & $5\times10^{-9}$\,\Msunyr \\
Wind expansion velocity&\vterminal & 10\,\kms \\
&\\
\textbf{Wind-ISM interaction}&&\\
Dust mass&\Mmod & $10^{-5}$\,\Msun \\
Stand-off distance& \Rso  & 65\arcsec \\
Perpendicular size&\Rp   & 80\arcsec \\
Width & \dRia & 10\arcsec \\
Opening angle & $\theta_{\rm{lim}}$ & 100$^{o}$\\
Density taper &  A & 1\\
Peak density & \rhoiapeak & 2.0$\times10^{-24}$\gcm \\
\hline
\label{t:testmodel}
\end{tabular}%

\end{table}

\cite{fanciulloetal2020} recently summarised lab measurements of opacities in the FIR of silicate dust grains. They found that compared to the extrapolation from shorter wavelengths the measured opacities have values that are larger by a factor of $\approx$\,10. Significantly larger values of the opacity at FIR wavelengths would strongly affect the dust temperature, and hence \Fratio. In order to test this, we increased the values of the opacities of our silicate dust at $\lambda$\,>\,50\,\micron~by a factor of 10, and again changed the slope of the opacities to follow a $\lambda^{-1.5}$ law, making them similar to the grains in~\cite{fanciulloetal2020} (and references therein). The result on \Fratio\, is shown in Fig.~\ref{f:Rcompare} and has by far the strongest effect on the observed ratios. However, the grains summarised in~\cite{fanciulloetal2020} are representative of dust observed in the cold ISM. They are formed under much colder conditions (20 -- 100 \,K) than is typical for AGB stars (where the dust forms at $\sim$1000\,K), and they contain a mix of silicate and carbon dust. Whether these opacities are representative of the dust that is observed in the wind-ISM interaction regions is hence unclear.

The silicate grains here include Fe, which affects the opacities at short wavelengths and hence the temperature of the grains. The composition of the dust around O-rich AGB stars is debated. Generally, grains that include Fe form too far from the star in order to accelerate the stellar wind, since they would be too warm to form in the acceleration region (however, they may still form further out in the wind). An alternative explanation for the driving of the stellar wind from O-rich AGB stars is the radiation pressure through scattering of photons off relatively large ($\approx$0.5\,\micron) iron-free silicate grains~\citep{hofner2008}. Such grains would decrease the temperature, making them more consistent with the observed \Fratio.

Finally, \cite{ysardetal2018} investigated the effect of different grain properties on the opacities for silicate, amorphous carbon, and hydrogenated carbon grains. Effects that they studied were porosity, shape (spherical vs. oblate), size, and aggregates, and combinations of these parameters. Depending on the parameters changed, and the different combinations, opacities can change significantly from the optical to the FIR, hence also likely affecting the temperature of the grains and the observed emission at different wavelengths. 

The detailed properties of the grains are hence very uncertain, and the derived values are strongly affected by a combination of our assumptions on the composition 
(e.g. Fe-free silicates), the stellar SEDs (to account for the absorption in the atmosphere at short wavelengths), the structure (spherical vs. oblate, porous, aggregates), and the sizes. The FIR observations here alone are not able to constrain the dust properties in the wind-ISM interaction regions around AGB stars. Additional spatially resolved spectroscopic and imaging observations in the optical, near- and far-infrared of dust towards the wind-ISM interaction regions, combined with direct observations of the stellar SED, are necessary to constrain the radiation field, grain composition, and grain structures.

While it is not possible to derive tight constraints on the grain properties based on the observations here, taking our results \emph{at face value} it appears that the grains in the wind-ISM regions may be \emph{different} from the grains in the circumstellar envelopes around AGB stars. Generally, when explicitly modelling observations of the dust in winds around AGB stars, the same assumptions are often made as in this study: stellar black body, spherical solid grains, one grain size. This may be more easily possible since current observations do not provide high enough spatial resolution to constrain the dust distribution from the inner to the outer CSE, hence allowing more freedom in the dust temperature. Any effect described above that might affect the dust models in the wind-ISM interaction regions, would also affect the models of dust in circumstellar winds. However, these effects might influence the two regions differently, and whether the discrepancy between the CSE and wind-ISM interaction region remains needs to be determined through further radiative transfer modelling.

\section{Conclusions}
\label{s:conclusions}

We investigated the far-infrared emission from wind-ISM interaction regions around AGB stars observed with the PACS instrument aboard Herschel at \Pblue~and \Pred. We derive the temperatures and dust masses in the observed regions directly from the observed ratios between the two wavelengths (\Fratio), as well as through explicit radiative transfer modelling of the observed structures, assuming amorphous carbon grains for C-rich AGB stars, and silicate grains for O-rich (M- and S-type) AGB stars.

When assuming grain-sizes of \agrain\,=\,0.1\,\micron, the models produce grains with relatively high temperatures and fail to reproduce the observed \Fratio. For amorphous carbon grains models with \agrain\,$\gtrsim$\,1\,\micron~can explain the observed \Fratio, indicating the presence of relatively large grains in the wind-ISM interaction regions. For O-rich and S-type AGB stars, the observed \Fratio~can not be reproduced for most sources, irrespective of the grain size. In this case it is likely that a combination of grain properties (composition, structure, and size) and the detailed stellar SED affect the temperatures of the grains. We discuss several possible mechanisms that could explain the discrepancies between modelled and observed \Fratio~for the O-rich sources:

\begin{itemize}

\item The slope in the opacities for silicate grains at $\lambda$\,>\,100\,\micron.
\item A reduction in the incident SED at $\lambda$\,<\,2\,\micron~as an effect of atomic and molecular absorption in the stellar atmosphere.
\item The possibility of grains with the properties of cold interstellar grains at $\lambda$\,>\,50\,\micron.
\item Possible contamination of the grains with Fe.
\item Changes in the opacities owing to porosity, shape, size, and aggregates. 
\end{itemize}
It is not clear which of these mechanisms dominate, and in particular the implications for the grain properties affect our understanding of the dust that is released into the ISM.

The derived dust masses in the wind-ISM interaction regions are consistent with the build-up of dust through the stellar mass-loss, with swept-up ISM dust only constituting $\approx$1\% of the total observed dust mass. The indication of large grains is consistent with previous studies of the dust in detached shells, and indicates that dust particles from AGB stars continue to grow to significantly larger grains compared to what is derived for the inner AGB wind. However, the derived typical densities in the wind-ISM interaction regions do not seem to be high enough to allow for (obvious) grain growth, unless the interaction regions contain density enhancements with up to two orders of magnitude. 

This has implications for the properties of  particles that are released into the ISM, where the AGB dust acts as seed particles for continued grain growth in the ISM. Generally it is either assumed that the dust grains released in to the ISM are unchanged from the original dust formed, or that they are sublimated by the interstellar radiation field and supernova shocks and do not contribute to the dust (growth) in the ISM at all. However, while we can not conclusively constrain the dust properties, our results indicate that assumptions that are often made to model the dust in the inner CSE do not successfully reproduce the observations of wind-ISM interaction regions. There is additionally a tendency towards larger grains compared to what is generally assumed. This increases the chance of grain survival, and affects the role dust grains from AGB stars play in forming the seeds for grain growth in the ISM. The detection of pre-solar grains from AGB stars is consistent with this picture.

While our models give rough estimates of the grain sizes and dust masses in the wind-interaction regions, and explicitly model the 3D structure of the regions, they are still very basic. In order to fully derive the conditions in the wind-ISM interaction regions, and the properties of the dust released into the ISM, it is necessary to obtain both spectroscopic and imaging observations of the dust, spatially resolved, in order to determine the grain compositions and temperatures, as well as direct observations of the stellar SEDs. Full hydrodynamical models of the interaction regions aimed at explicitly reproducing the observed cases are necessary to determine the physical parameters of the regions. Such investigations will be necessary to constrain models of grain growth and processing in the wind-ISM interaction regions to derive the detailed properties (size, composition, shape, porosity) of the grains released into the ISM.

\begin{acknowledgements}
M.Maercker acknowledges support from the Swedish Research Council under grant number 2016-03402. T.K. acknowledges support from the Swedish Research Council under grant number 2019-03777. E.D.B. acknowledges funding by the Swedish National Spaceboard.
\end{acknowledgements}

\newpage
\bibliographystyle{aa} 
\bibliography{../../../bib/maercker}

\begin{thebibliography}{50}
\expandafter\ifx\csname natexlab\endcsname\relax\def\natexlab#1{#1}\fi

\bibitem[{{Bocchio} {et~al.}(2016){Bocchio}, {Marassi}, {Schneider}, {Bianchi},
  {Limongi}, \& {Chieffi}}]{bocchioetal2016}
{Bocchio}, M., {Marassi}, S., {Schneider}, R., {et~al.} 2016, A\&A, 587, A157

\bibitem[{{Brunner} {et~al.}(2018){Brunner}, {Maercker}, {Mecina}, {Khouri}, \&
  {Kerschbaum}}]{brunneretal2018}
{Brunner}, M., {Maercker}, M., {Mecina}, M., {Khouri}, T., \& {Kerschbaum}, F.
  2018, A\&A, 614, A17

\bibitem[{{Cox} {et~al.}(2012){Cox}, {Kerschbaum}, {van Marle}, {Decin},
  {Ladjal}, {Mayer}, {Groenewegen}, {van Eck}, {Royer}, {Ottensamer}, {Ueta},
  {Jorissen}, {Mecina}, {Meliani}, {Luntzer}, {Blommaert}, {Posch},
  {Vandenbussche}, \& {Waelkens}}]{coxetal2012}
{Cox}, N.~L.~J., {Kerschbaum}, F., {van Marle}, A.-J., {et~al.} 2012, A\&A,
  537, A35

\bibitem[{{de Vries} {et~al.}(2010){de Vries}, {Min}, {Waters}, {Blommaert}, \&
  {Kemper}}]{devriesetal2010}
{de Vries}, B.~L., {Min}, M., {Waters}, L.~B.~F.~M., {Blommaert}, J.~A.~D.~L.,
  \& {Kemper}, F. 2010, A\&A, 516, A86

\bibitem[{{Draine}(2009)}]{draine2009}
{Draine}, B.~T. 2009, in Astronomical Society of the Pacific Conference Series,
  Vol. 414, Cosmic Dust - Near and Far, ed. T.~{Henning}, E.~{Gr{\"u}n}, \&
  J.~{Steinacker}, 453

\bibitem[{{Dullemond}(2012)}]{dullemond2012}
{Dullemond}, C.~P. 2012, {RADMC-3D: A multi-purpose radiative transfer tool},
  astrophysics Source Code Library

\bibitem[{{Fanciullo} {et~al.}(2020){Fanciullo}, {Kemper}, {Scicluna},
  {Dharmawardena}, \& {Srinivasan}}]{fanciulloetal2020}
{Fanciullo}, L., {Kemper}, F., {Scicluna}, P., {Dharmawardena}, T.~E., \&
  {Srinivasan}, S. 2020, MNRAS, 499, 4666

\bibitem[{{Ferrara} {et~al.}(2016){Ferrara}, {Viti}, \&
  {Ceccarelli}}]{ferraraetal2016}
{Ferrara}, A., {Viti}, S., \& {Ceccarelli}, C. 2016, MNRAS, 463, L112

\bibitem[{{Ferrarotti} \& {Gail}(2006)}]{ferrarottico2006}
{Ferrarotti}, A.~S. \& {Gail}, H.-P. 2006, A\&A, 447, 553

\bibitem[{{Ginolfi} {et~al.}(2018){Ginolfi}, {Graziani}, {Schneider},
  {Marassi}, {Valiante}, {Dell'Agli}, {Ventura}, \& {Hunt}}]{ginolfietal2018}
{Ginolfi}, M., {Graziani}, L., {Schneider}, R., {et~al.} 2018, MNRAS, 473, 4538

\bibitem[{{Groenewegen} {et~al.}(2009){Groenewegen}, {Sloan}, {Soszy{\'n}ski},
  \& {Petersen}}]{groenewegenetal2009}
{Groenewegen}, M.~A.~T., {Sloan}, G.~C., {Soszy{\'n}ski}, I., \& {Petersen},
  E.~A. 2009, A\&A, 506, 1277

\bibitem[{{Groenewegen} {et~al.}(2011){Groenewegen}, {Waelkens}, {Barlow},
  {Kerschbaum}, {Garcia-Lario}, {Cernicharo}, {Blommaert}, {Bouwman}, {Cohen},
  {Cox}, {Decin}, {Exter}, {Gear}, {Gomez}, {Hargrave}, {Henning},
  {Hutsem{\'e}kers}, {Ivison}, {Jorissen}, {Krause}, {Ladjal}, {Leeks}, {Lim},
  {Matsuura}, {Naz{\'e}}, {Olofsson}, {Ottensamer}, {Polehampton}, {Posch},
  {Rauw}, {Royer}, {Sibthorpe}, {Swinyard}, {Ueta}, {Vamvatira-Nakou},
  {Vandenbussche}, {van de Steene}, {van Eck}, {van Hoof}, {van Winckel},
  {Verdugo}, \& {Wesson}}]{groenewegenetal2011}
{Groenewegen}, M.~A.~T., {Waelkens}, C., {Barlow}, M.~J., {et~al.} 2011, A\&A,
  526, A162

\bibitem[{Heck {et~al.}(2020)Heck, Greer, K{\"o}{\"o}p, Trappitsch, Gyngard,
  Busemann, Maden, {\'A}vila, Davis, \& Wieler}]{hecketal2020}
Heck, P.~R., Greer, J., K{\"o}{\"o}p, L., {et~al.} 2020, PNAS, 117, 1884

\bibitem[{{H{\"o}fner}(2008)}]{hofner2008}
{H{\"o}fner}, S. 2008, A\&A, 491, L1

\bibitem[{{H{\"o}fner} \& {Olofsson}(2018)}]{hofnerco2018}
{H{\"o}fner}, S. \& {Olofsson}, H. 2018, A\&ARv, 26, 1

\bibitem[{{Inoue}(2011)}]{inoue2011}
{Inoue}, A.~K. 2011, Earth, Planets, and Space, 63, 1027

\bibitem[{{Khouri} {et~al.}(2016){Khouri}, {Maercker}, {Waters}, {Vlemmings},
  {Kervella}, {de Koter}, {Ginski}, {De Beck}, {Decin}, {Min}, {Dominik},
  {O'Gorman}, {Schmid}, {Lombaert}, \& {Lagadec}}]{khourietal2016a}
{Khouri}, T., {Maercker}, M., {Waters}, L.~B.~F.~M., {et~al.} 2016, A\&A, 591,
  A70

\bibitem[{{Khouri} {et~al.}(2020){Khouri}, {Vlemmings}, {Paladini}, {Ginski},
  {Lagadec}, {Maercker}, {Kervella}, {De Beck}, {Decin}, {de Koter}, \&
  {Waters}}]{khourietal2020}
{Khouri}, T., {Vlemmings}, W.~H.~T., {Paladini}, C., {et~al.} 2020, A\&A, 635,
  A200

\bibitem[{{Ladjal} {et~al.}(2010){Ladjal}, {Justtanont}, {Groenewegen},
  {Blommaert}, {Waelkens}, \& {Barlow}}]{ladjaletal2010}
{Ladjal}, D., {Justtanont}, K., {Groenewegen}, M.~A.~T., {et~al.} 2010, A\&A,
  513, A53

\bibitem[{{Lequeux}(2005)}]{lequeux2005}
{Lequeux}, J. 2005, {The Interstellar Medium} (The interstellar medium,
  Translation from the French language edition of: Le Milieu Interstellaire by
  James Lequeux, EDP Sciences, 2003 Edited by J. Lequeux. Astronomy and
  astrophysics library, Berlin: Springer, 2005)

\bibitem[{{Li} \& {Draine}(2001)}]{lico2001}
{Li}, A. \& {Draine}, B.~T. 2001, ApJ, 554, 778

\bibitem[{{Libert} {et~al.}(2007){Libert}, {G{\'e}rard}, \& {Le
  Bertre}}]{libertetal2007}
{Libert}, Y., {G{\'e}rard}, E., \& {Le Bertre}, T. 2007, MNRAS, 380, 1161

\bibitem[{{Loup} {et~al.}(1993){Loup}, {Forveille}, {Omont}, \&
  {Paul}}]{loupetal1993}
{Loup}, C., {Forveille}, T., {Omont}, A., \& {Paul}, J.~F. 1993, A\&AS, 99, 291

\bibitem[{{Maercker} {et~al.}(2018){Maercker}, {Khouri}, {De Beck}, {Brunner},
  {Mecina}, \& {Jaldehag}}]{maerckeretal2018b}
{Maercker}, M., {Khouri}, T., {De Beck}, E., {et~al.} 2018, A\&A, 620, A106

\bibitem[{{Maercker} {et~al.}(2010){Maercker}, {Olofsson}, {Eriksson},
  {Gustafsson}, \& {Sch{\"o}ier}}]{maerckeretal2010}
{Maercker}, M., {Olofsson}, H., {Eriksson}, K., {Gustafsson}, B., \&
  {Sch{\"o}ier}, F.~L. 2010, A\&A, 511, A37+

\bibitem[{{Mattsson} \& {H{\"o}fner}(2011)}]{mattssonco2011}
{Mattsson}, L. \& {H{\"o}fner}, S. 2011, A\&A, 533, A42

\bibitem[{{Mattsson} {et~al.}(2007){Mattsson}, {H{\"o}fner}, \&
  {Herwig}}]{mattssonetal2007}
{Mattsson}, L., {H{\"o}fner}, S., \& {Herwig}, F. 2007, A\&A, 470, 339

\bibitem[{{Micha{\l}owski} {et~al.}(2010){Micha{\l}owski}, {Watson}, \&
  {Hjorth}}]{michalowskietal2010}
{Micha{\l}owski}, M.~J., {Watson}, D., \& {Hjorth}, J. 2010, ApJ, 712, 942

\bibitem[{{Mihalas} \& {Binney}(1981)}]{mihalasco1981}
{Mihalas}, D. \& {Binney}, J. 1981, {Galactic astronomy. Structure and
  kinematics}

\bibitem[{{Norris} {et~al.}(2012){Norris}, {Tuthill}, {Ireland}, {Lacour},
  {Zijlstra}, {Lykou}, {Evans}, {Stewart}, \& {Bedding}}]{norrisetal2012}
{Norris}, B. R.~M., {Tuthill}, P.~G., {Ireland}, M.~J., {et~al.} 2012, Nature,
  484, 220

\bibitem[{{Olofsson} {et~al.}(2010){Olofsson}, {Maercker}, {Eriksson},
  {Gustafsson}, \& {Sch{\"o}ier}}]{olofssonetal2010}
{Olofsson}, H., {Maercker}, M., {Eriksson}, K., {Gustafsson}, B., \&
  {Sch{\"o}ier}, F. 2010, A\&A, 515, A27

\bibitem[{{Ossenkopf} \& {Henning}(1994)}]{ossenkopfco1994}
{Ossenkopf}, V. \& {Henning}, T. 1994, A\&A, 291, 943

\bibitem[{{Pilbratt} {et~al.}(2010){Pilbratt}, {Riedinger}, {Passvogel},
  {Crone}, {Doyle}, {Gageur}, {Heras}, {Jewell}, {Metcalfe}, {Ott}, \&
  {Schmidt}}]{pilbrattetal2010}
{Pilbratt}, G.~L., {Riedinger}, J.~R., {Passvogel}, T., {et~al.} 2010, A\&A,
  518, L1

\bibitem[{{Poglitsch} {et~al.}(2010){Poglitsch}, {Waelkens}, {Geis},
  {Feuchtgruber}, {Vandenbussche}, {Rodriguez}, {Krause}, {Renotte}, {van
  Hoof}, {Saraceno}, {Cepa}, {Kerschbaum}, {Agn{\`e}se}, {Ali}, {Altieri},
  {Andreani}, {Augueres}, {Balog}, {Barl}, {Bauer}, {Belbachir}, {Benedettini},
  {Billot}, {Boulade}, {Bischof}, {Blommaert}, {Callut}, {Cara}, {Cerulli},
  {Cesarsky}, {Contursi}, {Creten}, {De Meester}, {Doublier}, {Doumayrou},
  {Duband}, {Exter}, {Genzel}, {Gillis}, {Gr{\"o}zinger}, {Henning},
  {Herreros}, {Huygen}, {Inguscio}, {Jakob}, {Jamar}, {Jean}, {de Jong},
  {Katterloher}, {Kiss}, {Klaas}, {Lemke}, {Lutz}, {Madden}, {Marquet},
  {Martignac}, {Mazy}, {Merken}, {Montfort}, {Morbidelli}, {M{\"u}ller},
  {Nielbock}, {Okumura}, {Orfei}, {Ottensamer}, {Pezzuto}, {Popesso},
  {Putzeys}, {Regibo}, {Reveret}, {Royer}, {Sauvage}, {Schreiber}, {Stegmaier},
  {Schmitt}, {Schubert}, {Sturm}, {Thiel}, {Tofani}, {Vavrek}, {Wetzstein},
  {Wieprecht}, \& {Wiezorrek}}]{poglitschetal2010}
{Poglitsch}, A., {Waelkens}, C., {Geis}, N., {et~al.} 2010, A\&A, 518, L2

\bibitem[{{Ramstedt} {et~al.}(2009){Ramstedt}, {Sch{\"o}ier}, \&
  {Olofsson}}]{ramstedtetal2009}
{Ramstedt}, S., {Sch{\"o}ier}, F.~L., \& {Olofsson}, H. 2009, A\&A, 499, 515

\bibitem[{{Ramstedt} {et~al.}(2008){Ramstedt}, {Sch{\"o}ier}, {Olofsson}, \&
  {Lundgren}}]{ramstedtetal2008}
{Ramstedt}, S., {Sch{\"o}ier}, F.~L., {Olofsson}, H., \& {Lundgren}, A.~A.
  2008, A\&A, 487, 645

\bibitem[{{Sandin} \& {H{\"o}fner}(2003)}]{sandinco2003}
{Sandin}, C. \& {H{\"o}fner}, S. 2003, A\&A, 404, 789

\bibitem[{{Sandin} \& {Mattsson}(2020)}]{sandinco2020}
{Sandin}, C. \& {Mattsson}, L. 2020, MNRAS, 499, 1531

\bibitem[{{Schneider} {et~al.}(2014){Schneider}, {Valiante}, {Ventura},
  {dell'Agli}, {Di Criscienzo}, {Hirashita}, \& {Kemper}}]{schneideretal2014}
{Schneider}, R., {Valiante}, R., {Ventura}, P., {et~al.} 2014, MNRAS, 442, 1440

\bibitem[{{Sch{\"o}ier} {et~al.}(2005){Sch{\"o}ier}, {Lindqvist}, \&
  {Olofsson}}]{schoieretal2005}
{Sch{\"o}ier}, F.~L., {Lindqvist}, M., \& {Olofsson}, H. 2005, A\&A, 436, 633

\bibitem[{{Spitzer}(1978)}]{spitzer1978}
{Spitzer}, Lyman, J. 1978, JRASC, 72, 349

\bibitem[{{Steffen} \& {Sch{\"o}nberner}(2000)}]{steffenco2000}
{Steffen}, M. \& {Sch{\"o}nberner}, D. 2000, A\&A, 357, 180

\bibitem[{{Suh}(1999)}]{suh1999}
{Suh}, K.-W. 1999, MNRAS, 304, 389

\bibitem[{{Suh}(2000)}]{suh2000}
{Suh}, K.-W. 2000, MNRAS, 315, 740

\bibitem[{{Villaver} {et~al.}(2012){Villaver}, {Manchado}, \&
  {Garc{\'\i}a-Segura}}]{villaveretal2012}
{Villaver}, E., {Manchado}, A., \& {Garc{\'\i}a-Segura}, G. 2012, ApJ, 748, 94

\bibitem[{{Wareing} {et~al.}(2007){Wareing}, {Zijlstra}, \&
  {O'Brien}}]{wareingetal2007}
{Wareing}, C.~J., {Zijlstra}, A.~A., \& {O'Brien}, T.~J. 2007, MNRAS, 382, 1233

\bibitem[{{Wilkin}(1996)}]{wilkin1996}
{Wilkin}, F.~P. 1996, ApJL, 459, L31

\bibitem[{{Woitke}(2006)}]{woitke2006}
{Woitke}, P. 2006, A\&A, 460, L9

\bibitem[{{Ysard} {et~al.}(2018){Ysard}, {Jones}, {Demyk}, {Bout{\'e}raon}, \&
  {Koehler}}]{ysardetal2018}
{Ysard}, N., {Jones}, A.~P., {Demyk}, K., {Bout{\'e}raon}, T., \& {Koehler}, M.
  2018, A\&A, 617, A124

\bibitem[{{Zhukovska} {et~al.}(2016){Zhukovska}, {Dobbs}, {Jenkins}, \&
  {Klessen}}]{zhukovskaetal2016}
{Zhukovska}, S., {Dobbs}, C., {Jenkins}, E.~B., \& {Klessen}, R.~S. 2016, ApJ,
  831, 147

\end{thebibliography}

\newpage
%


\appendix

\section{Models and images for sources that were modelled}
\label{a:models}

\begin{figure*}
\centering
\includegraphics[width=8cm]{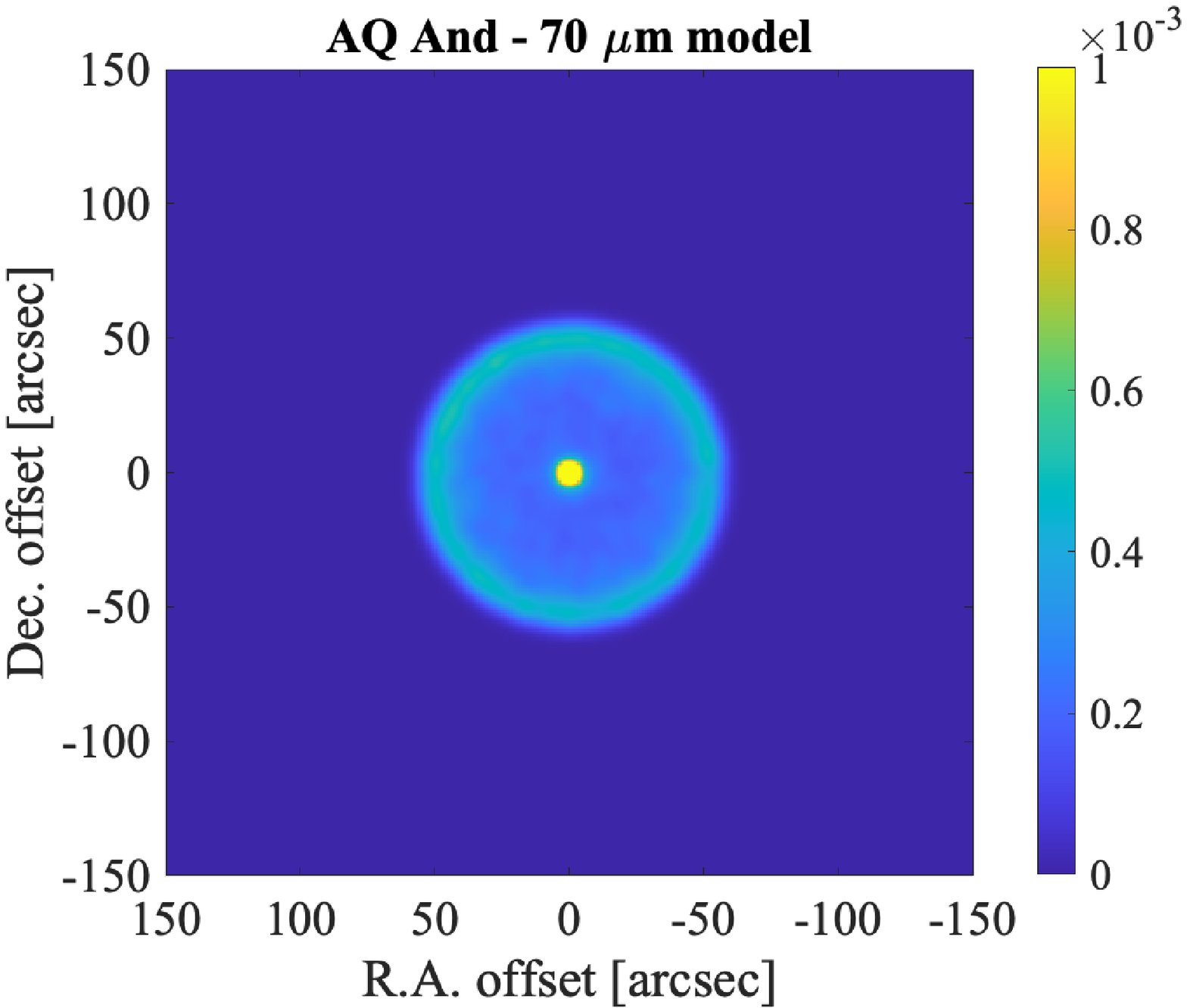}
\includegraphics[width=8cm]{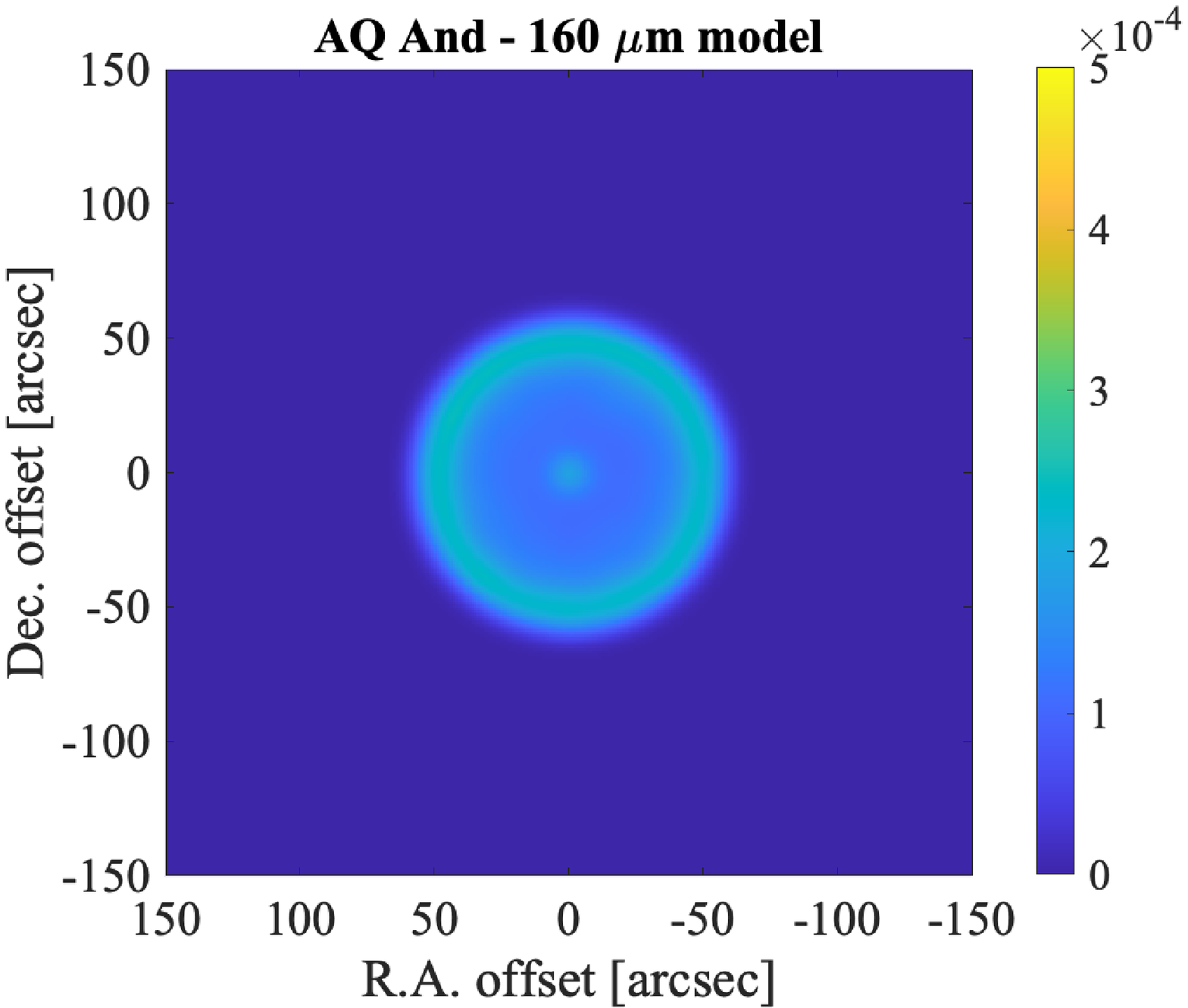}
\includegraphics[width=8cm]{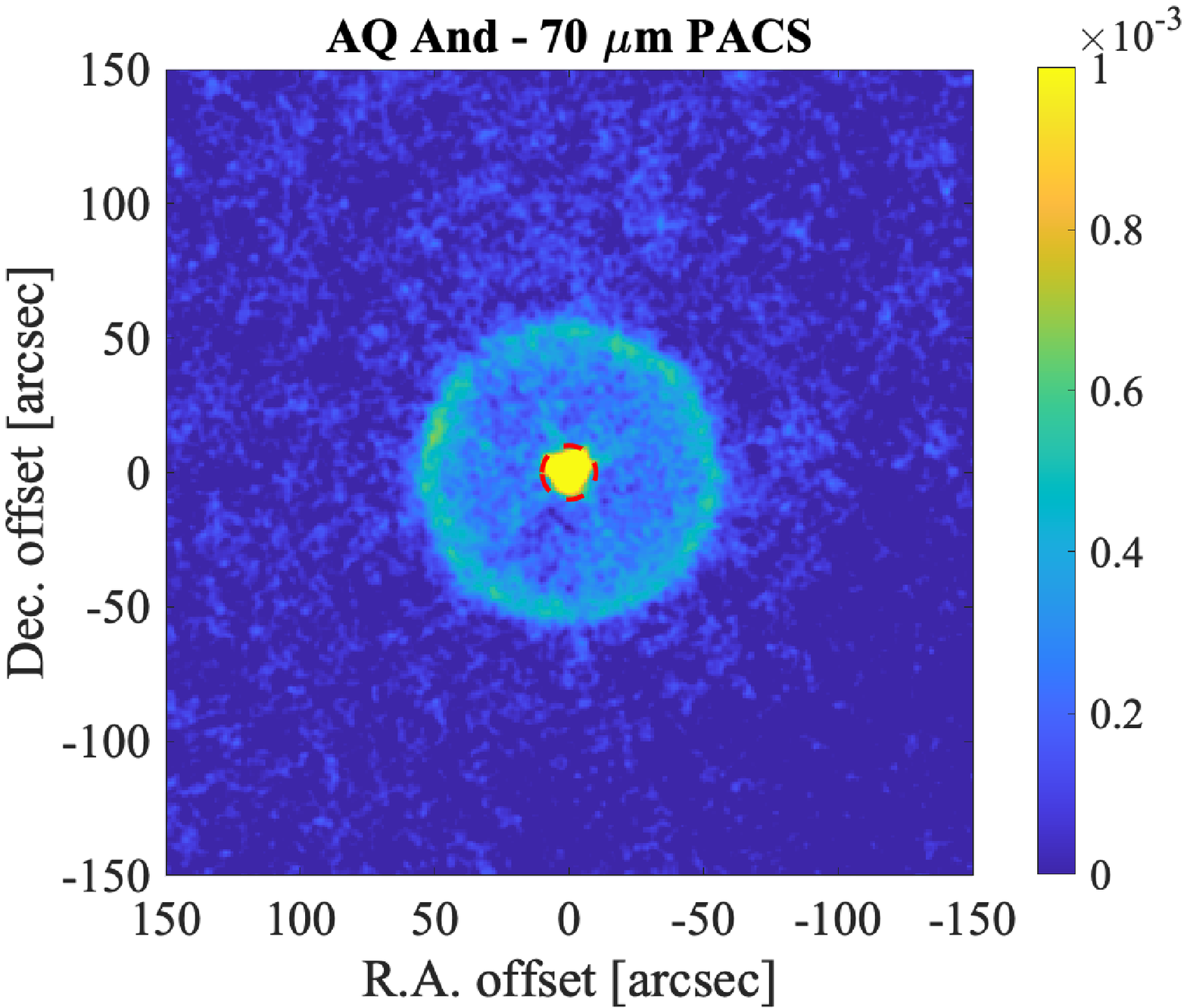}
\includegraphics[width=8cm]{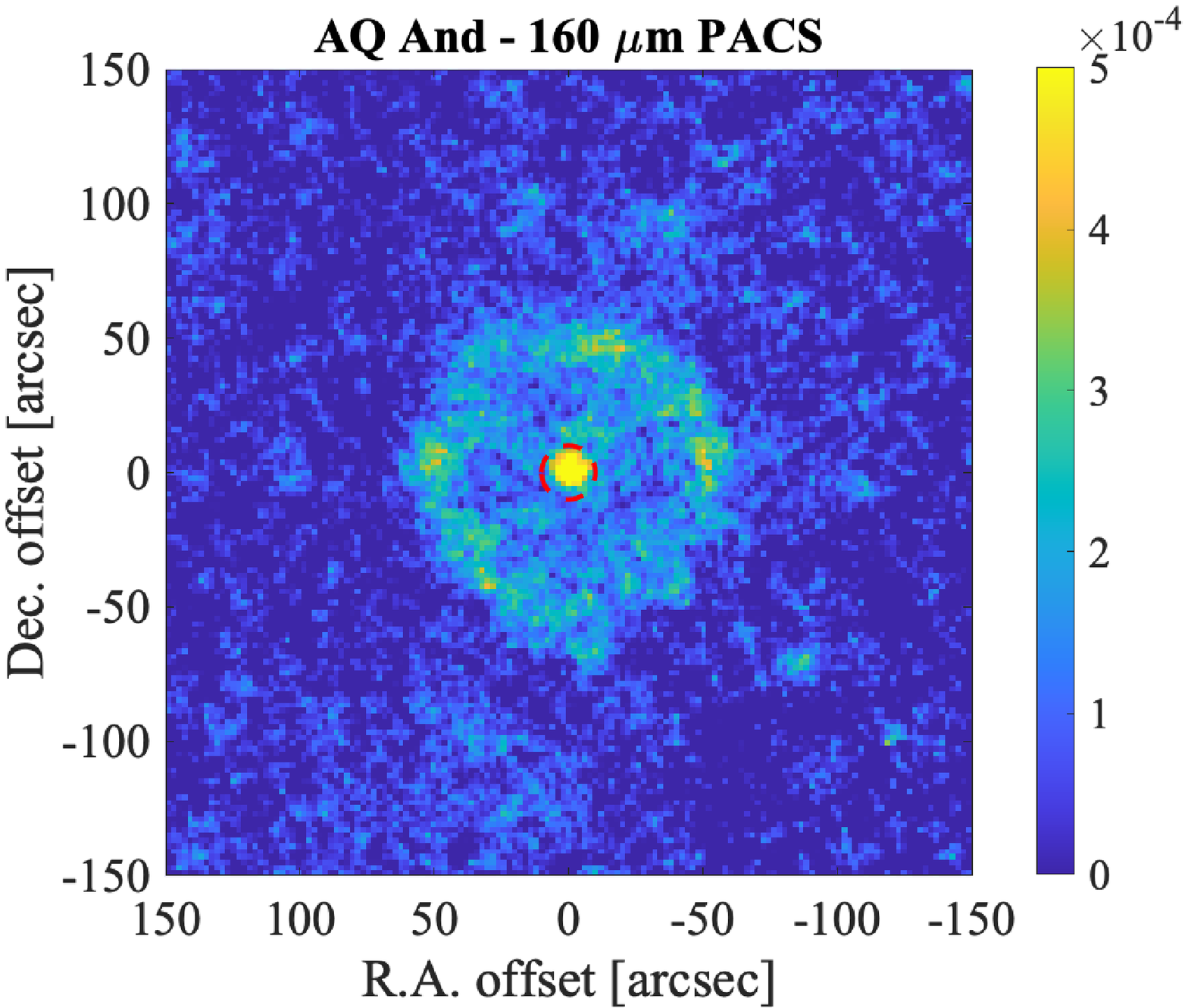}
\includegraphics[width=8cm]{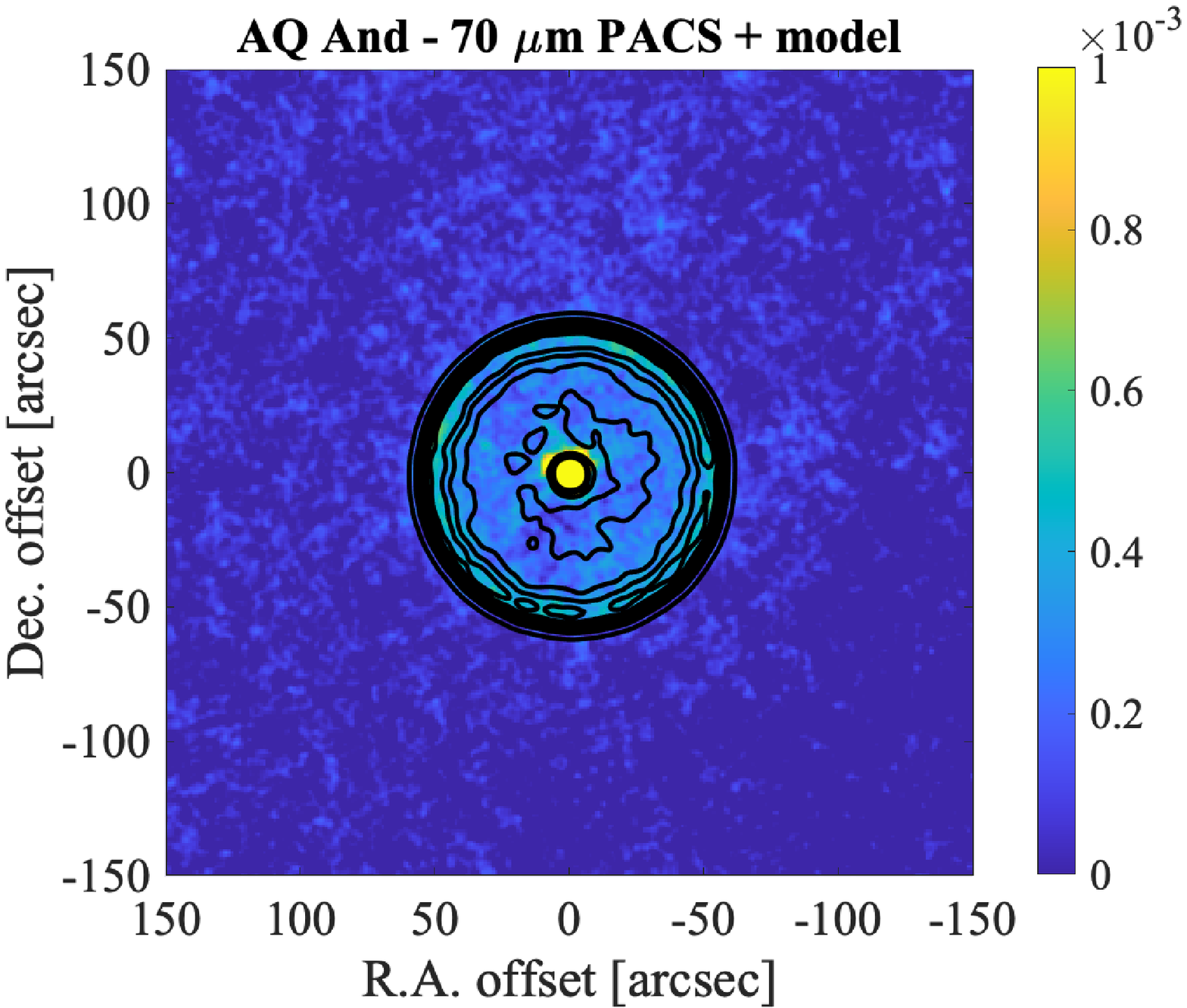}
\includegraphics[width=8cm]{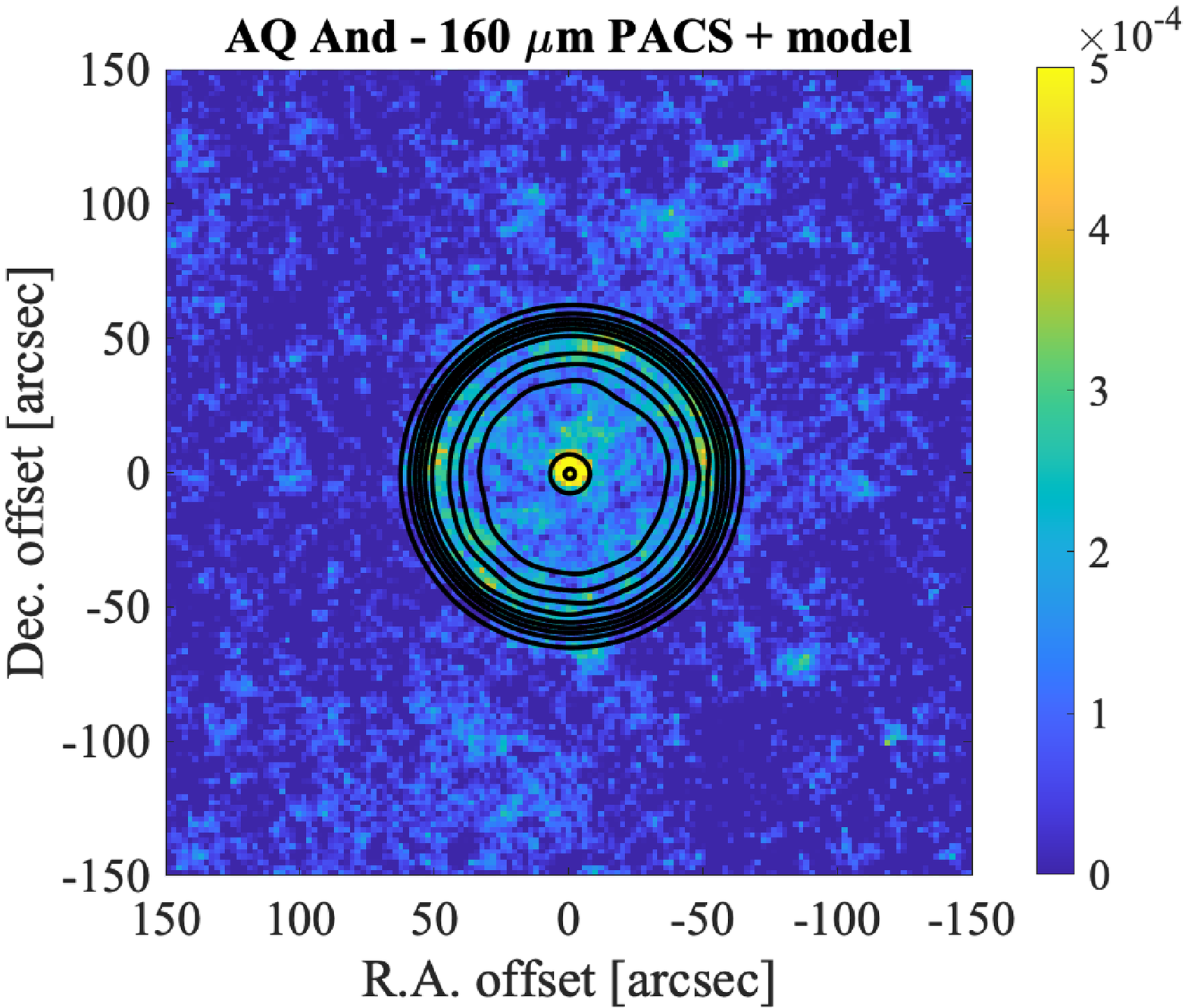}
\caption{AQ And: \emph{Top to bottom:} The Radmc3D model, the PACS image, and the PACS image with contours from the model. Images are for 70\,\micron~(left) and 160\,\micron~(right). Maximum contour levels are 0.4$\times10^{-3}$\,\Jyarcsec (70\,\micron) and 0.2$\times10^{-3}$\,\Jyarcsec (160\,\micron), respectively. Minimum contour levels are 10\% of maximum. The colour scale is in \Jyarcsec. The red dashed circle shows the mask used to measure the flux from the star and present-day mass-loss.}
\label{f:aqand}
\end{figure*}

\begin{figure*}
\centering
\includegraphics[width=8cm]{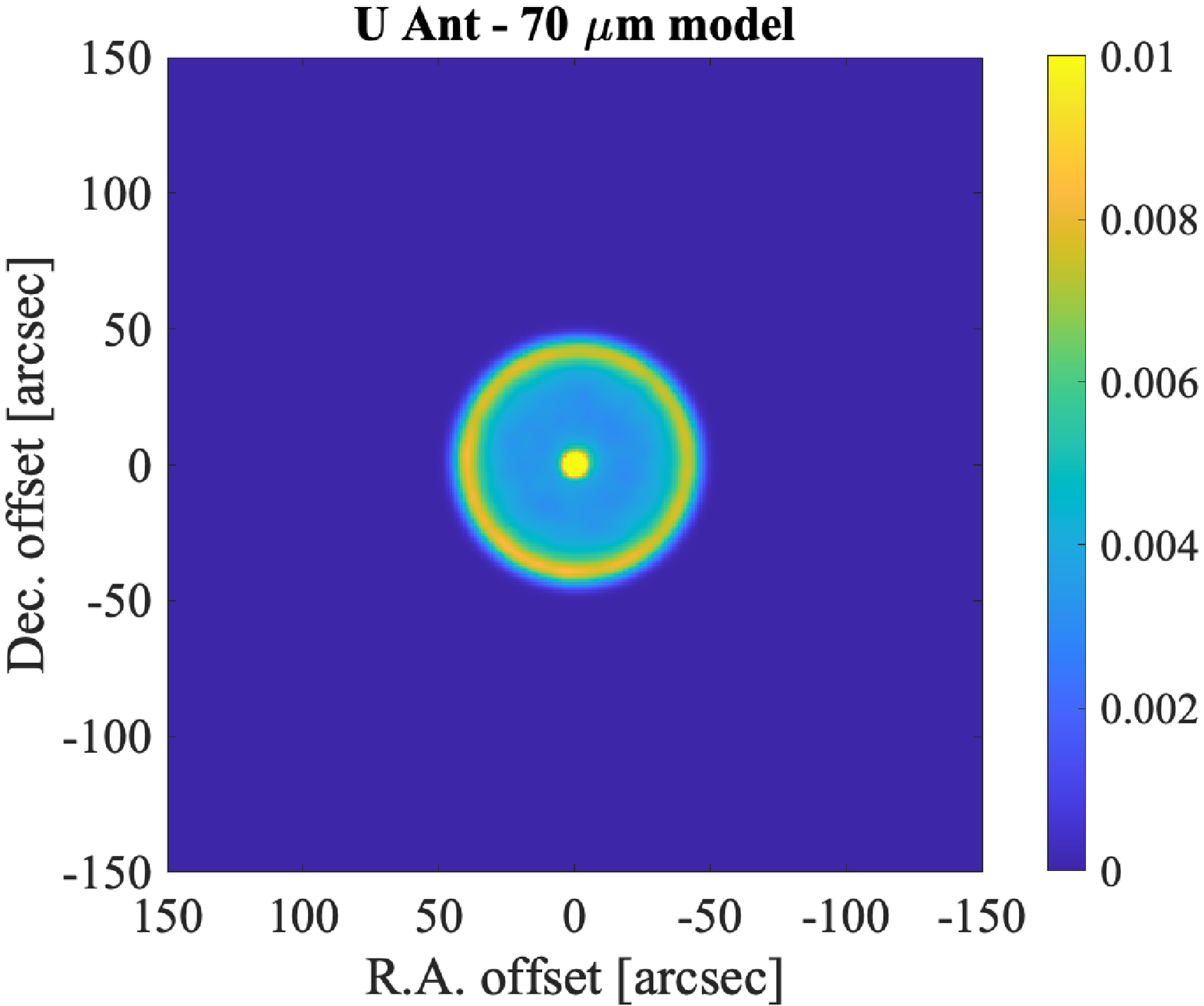}
\includegraphics[width=8cm]{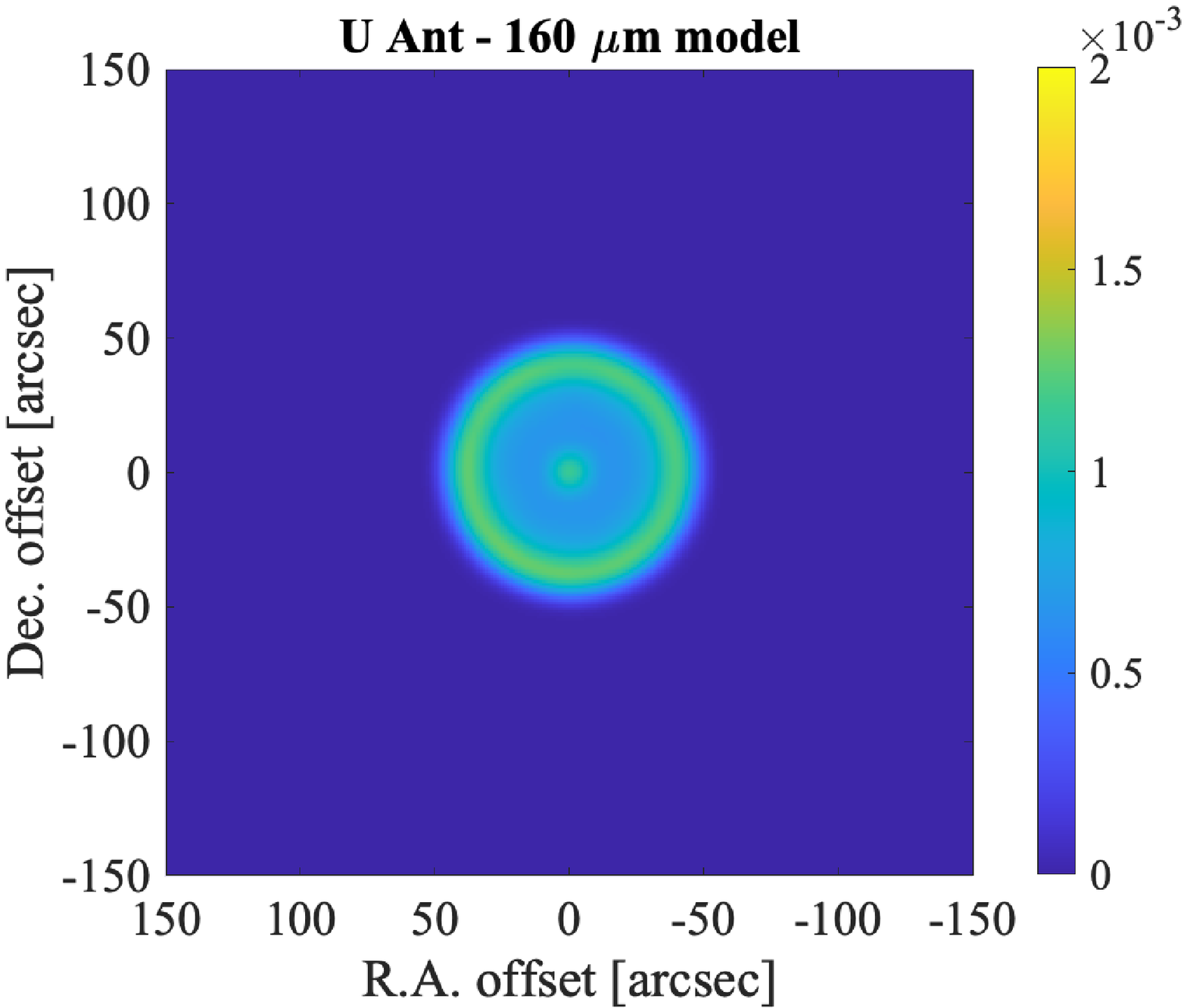}
\includegraphics[width=8cm]{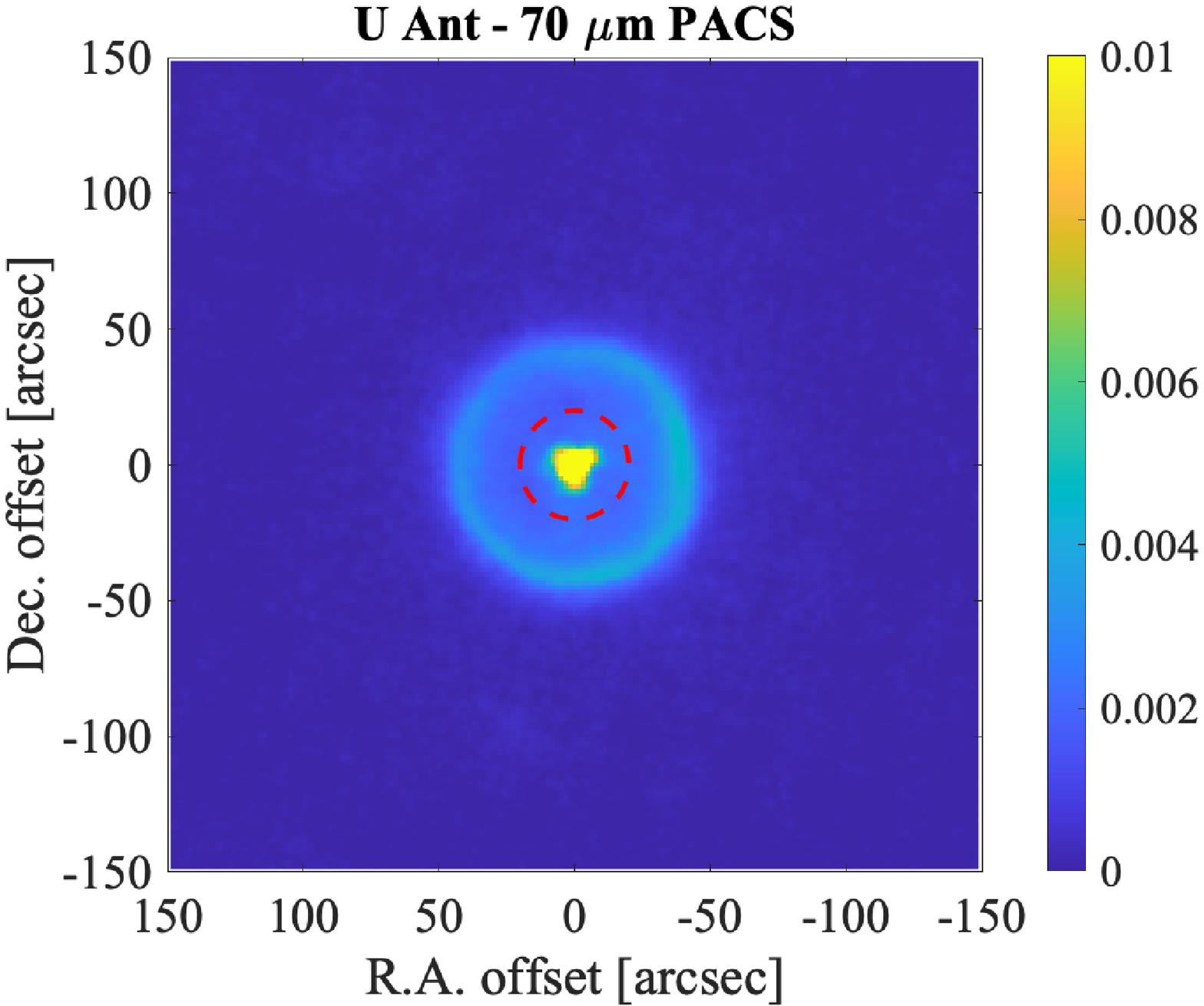}
\includegraphics[width=8cm]{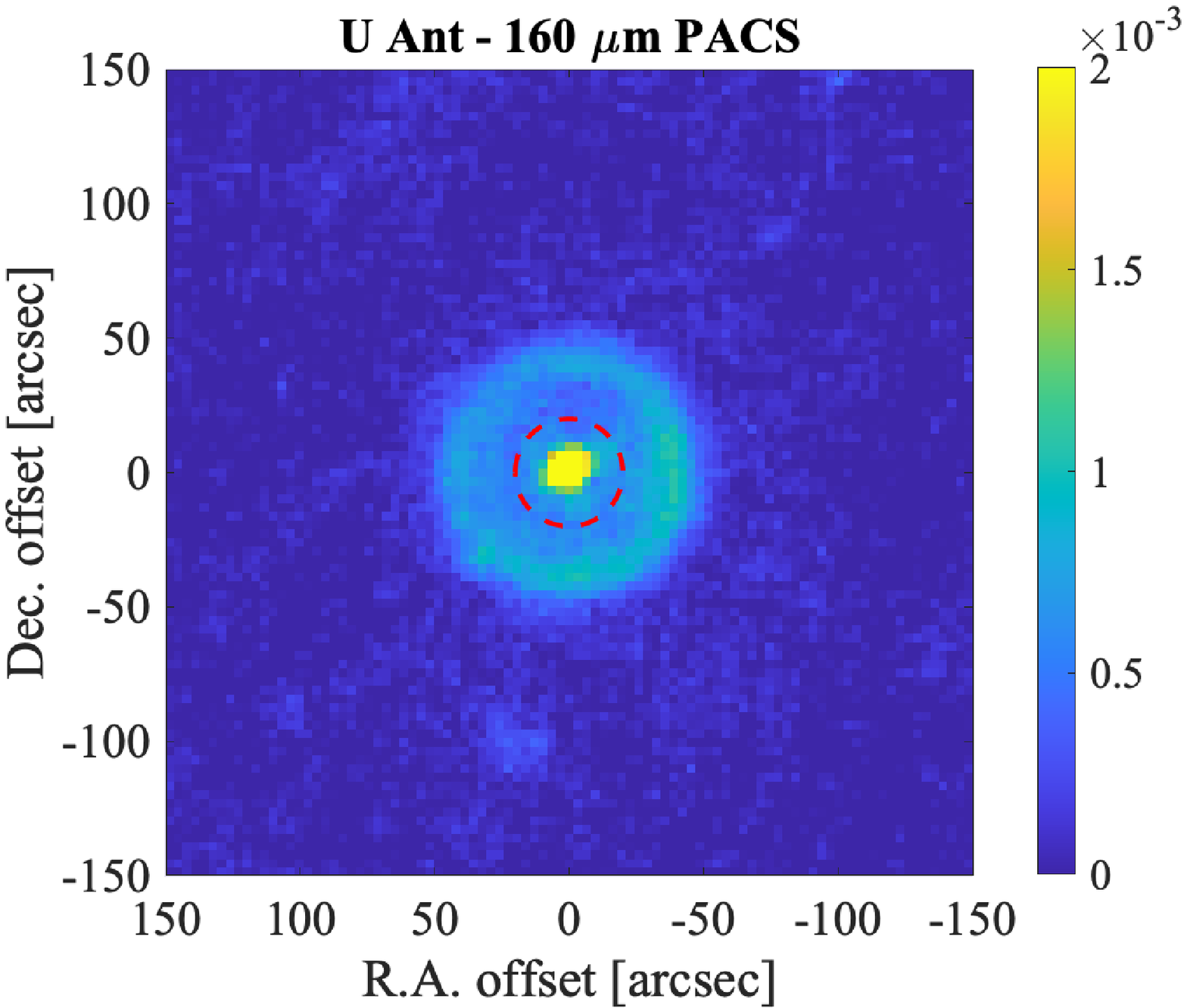}
\includegraphics[width=8cm]{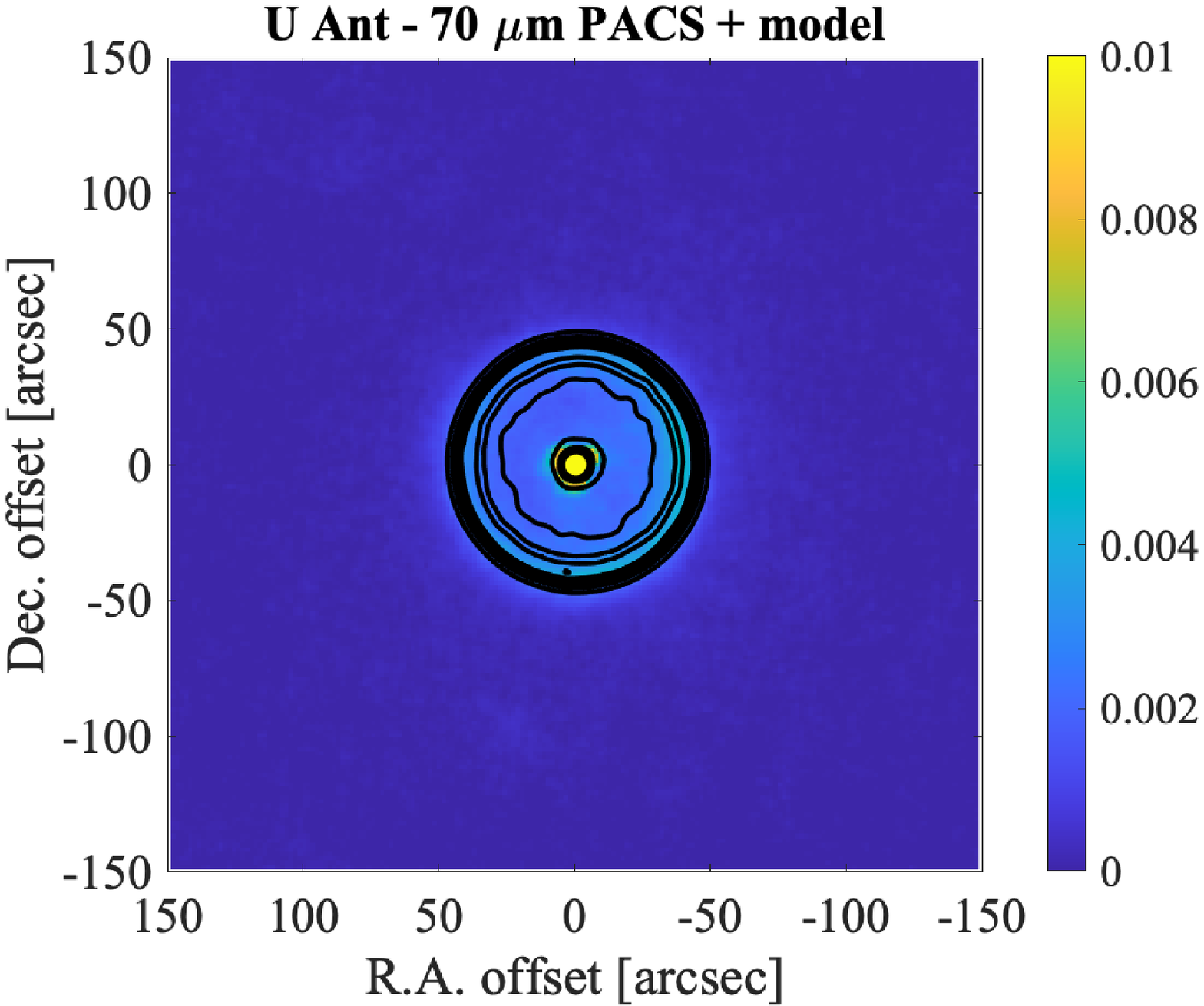}
\includegraphics[width=8cm]{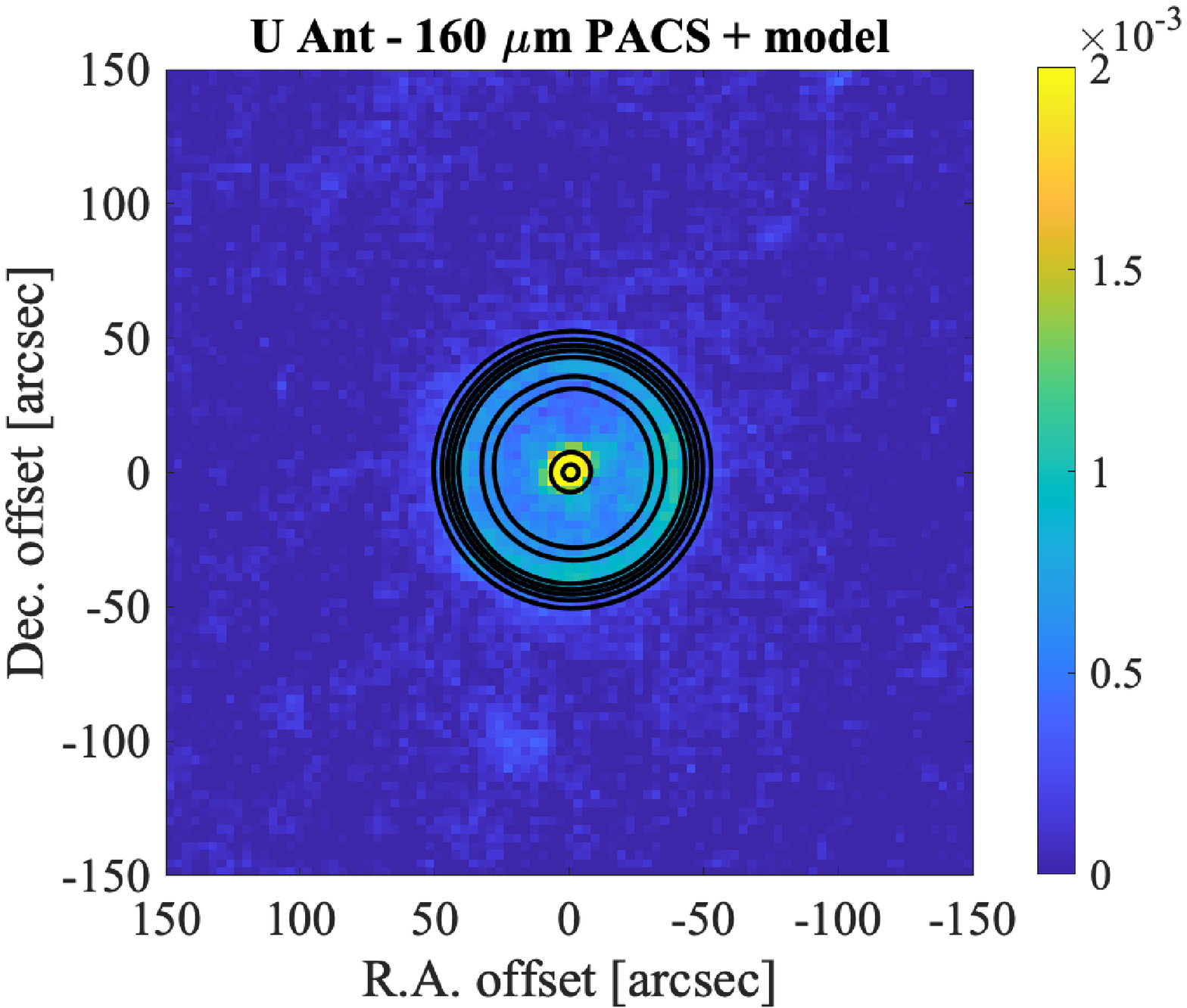}
\caption{U Ant: \emph{Top to bottom:} The Radmc3D model, the PACS image, and the PACS image with contours from the model. Images are for 70\,\micron~(left) and 160\,\micron~(right). Maximum contour levels are 7.6$\times10^{-3}$\,\Jyarcsec (70\,\micron) and 1.2$\times10^{-3}$\,\Jyarcsec (160\,\micron), respectively. Minimum contour levels are 10\% of maximum. The colour scale is in \Jyarcsec. The red dashed circle shows the mask used to measure the flux from the star and present-day mass-loss.}
\label{f:uant}
\end{figure*}

\begin{figure*}
\centering
\includegraphics[width=8cm]{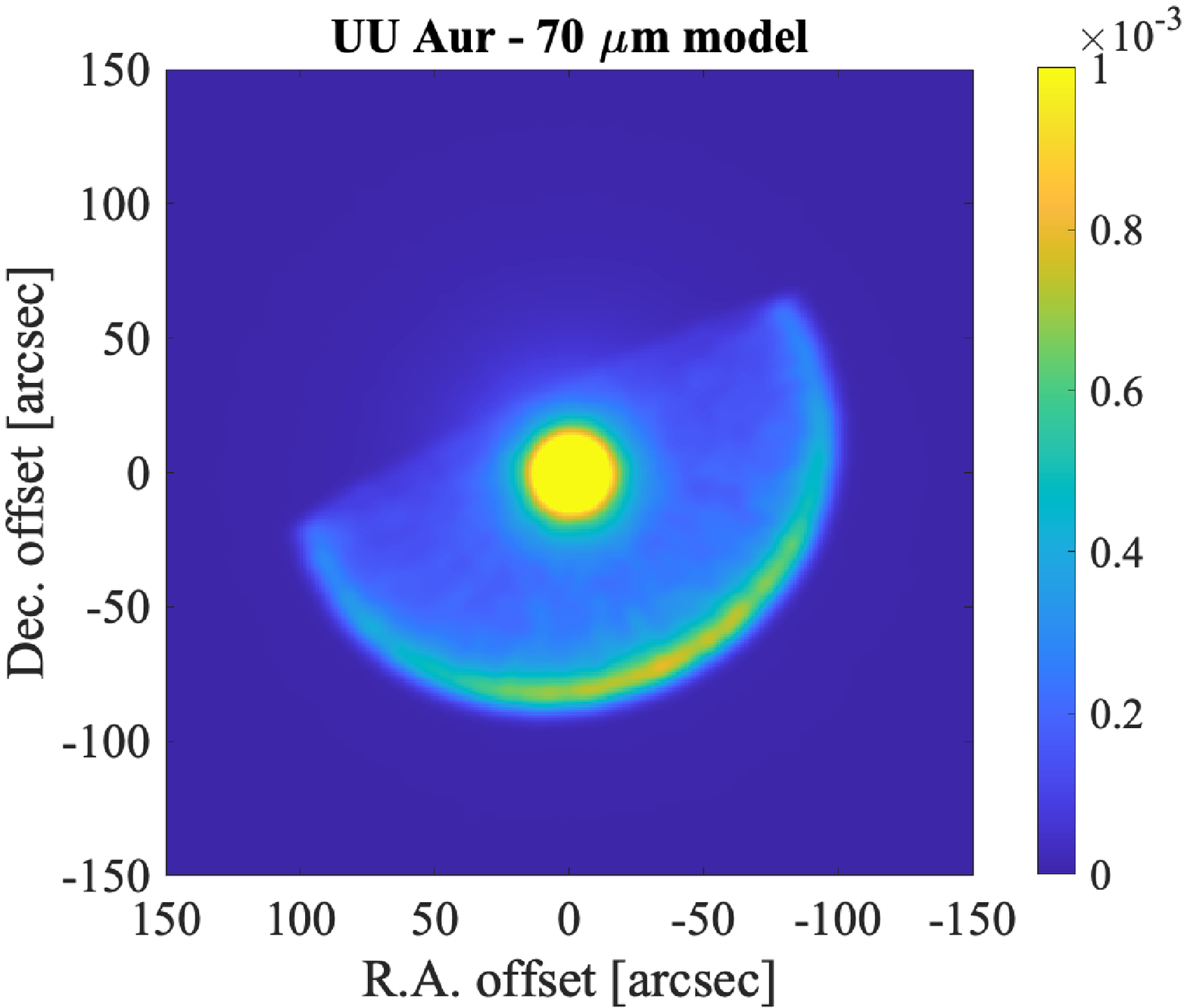}
\includegraphics[width=8cm]{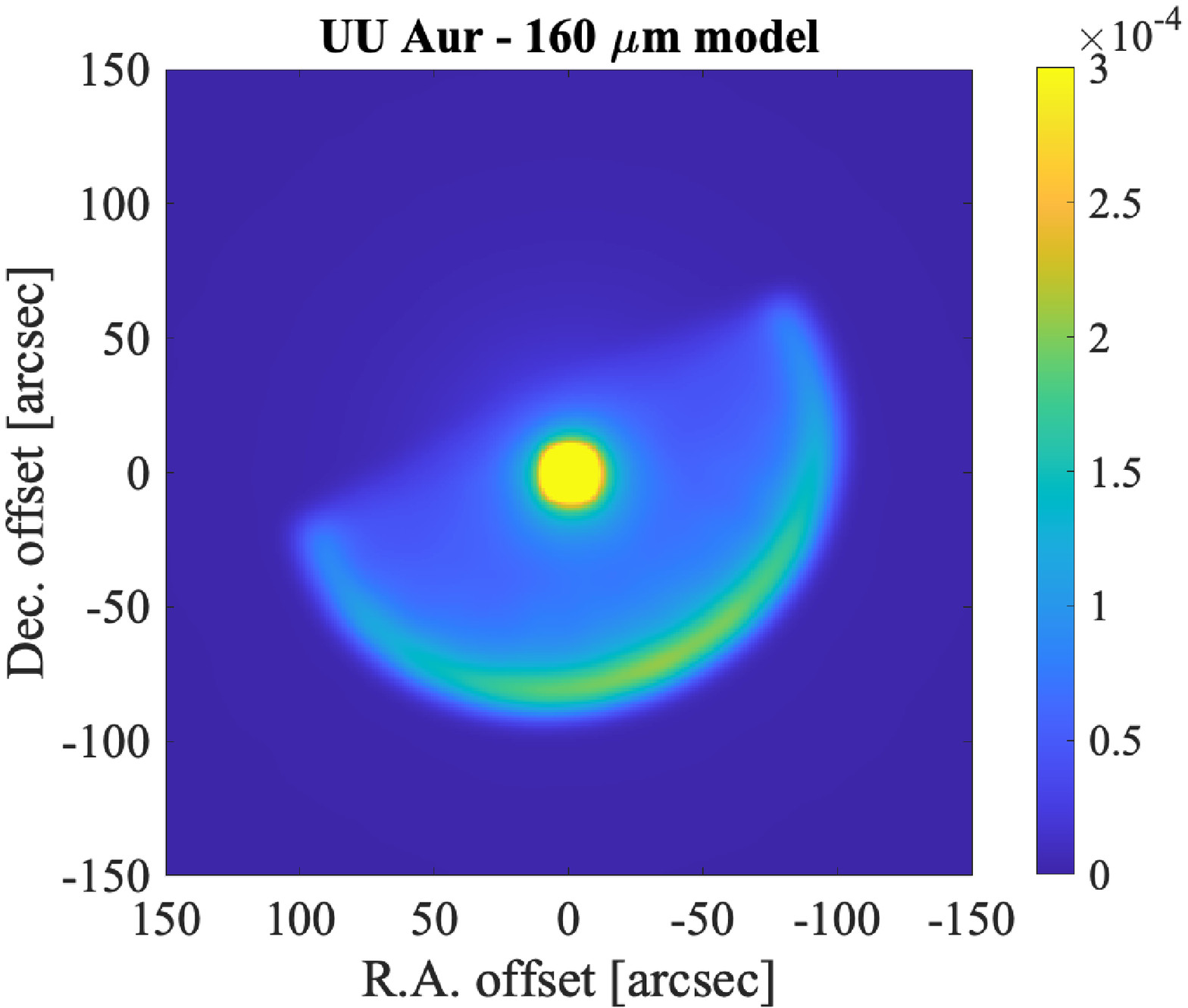}
\includegraphics[width=8cm]{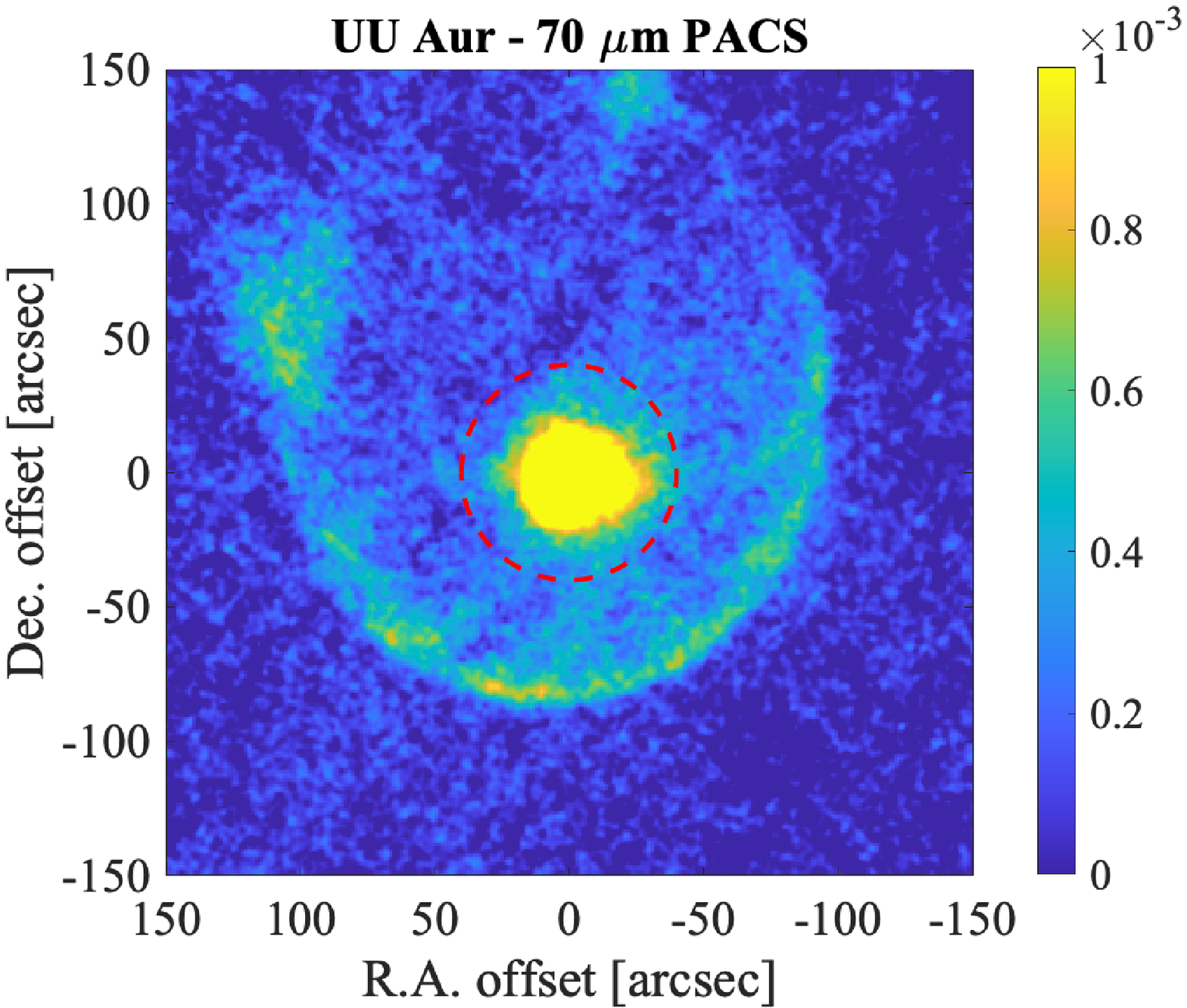}
\includegraphics[width=8cm]{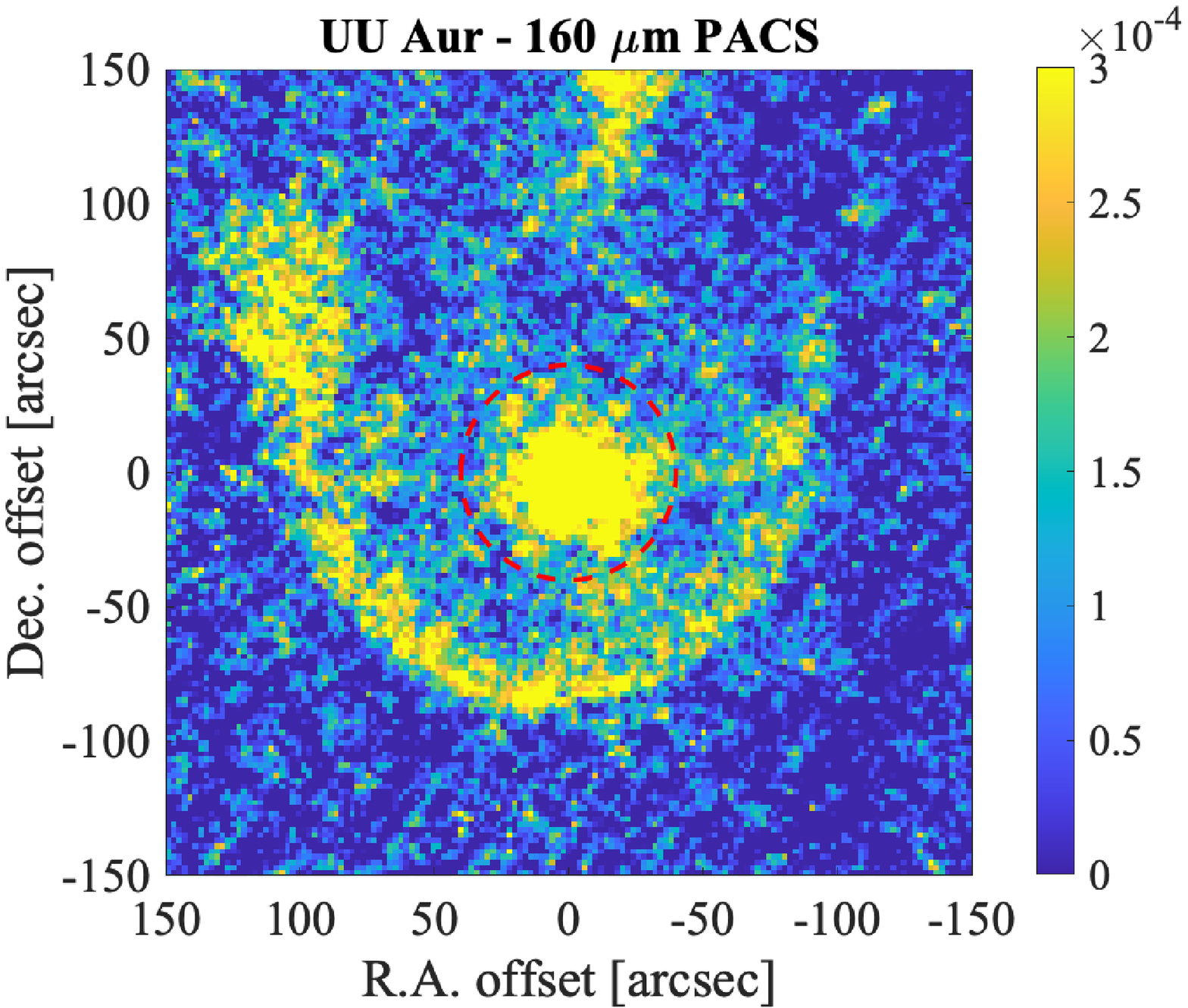}
\includegraphics[width=8cm]{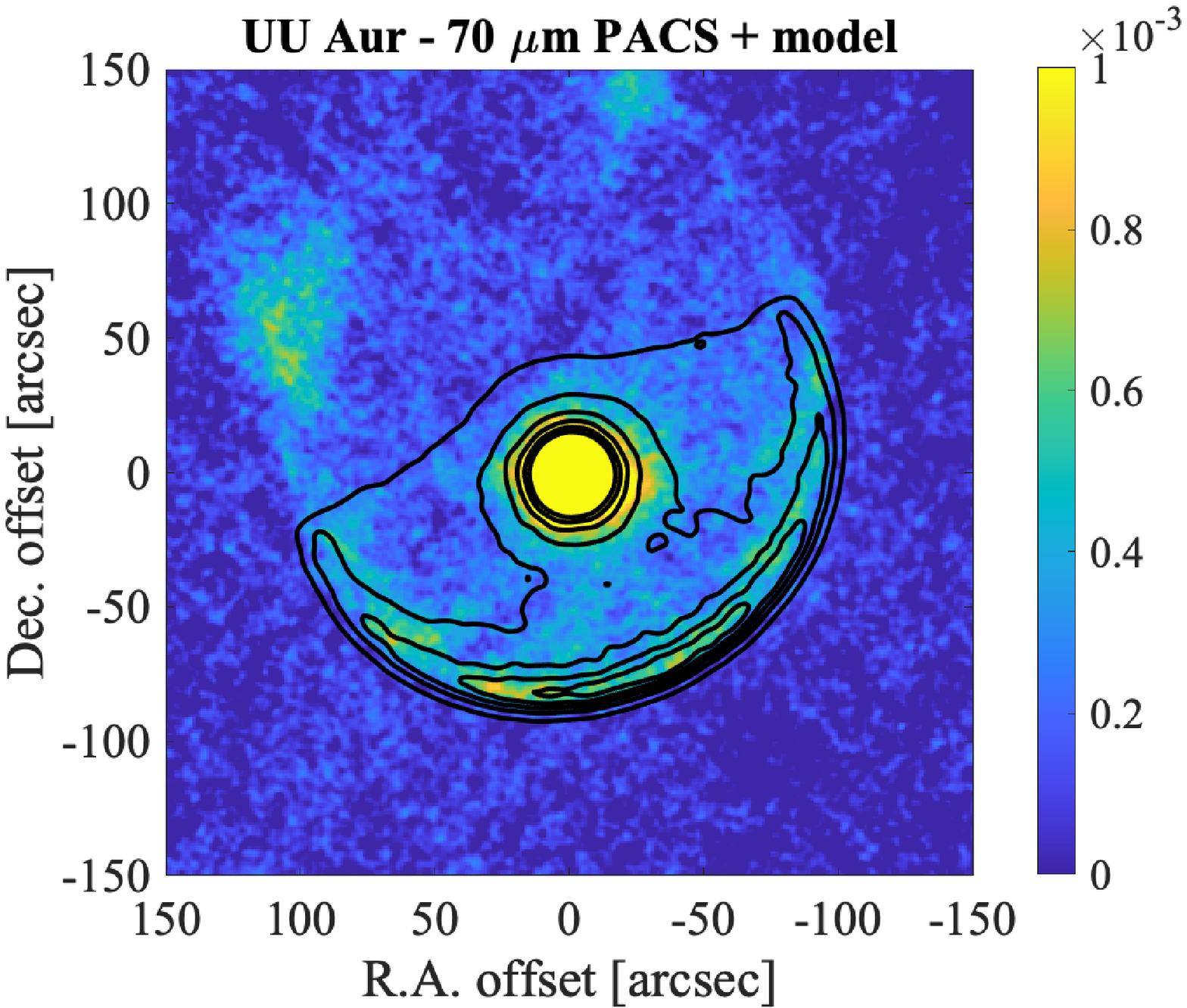}
\includegraphics[width=8cm]{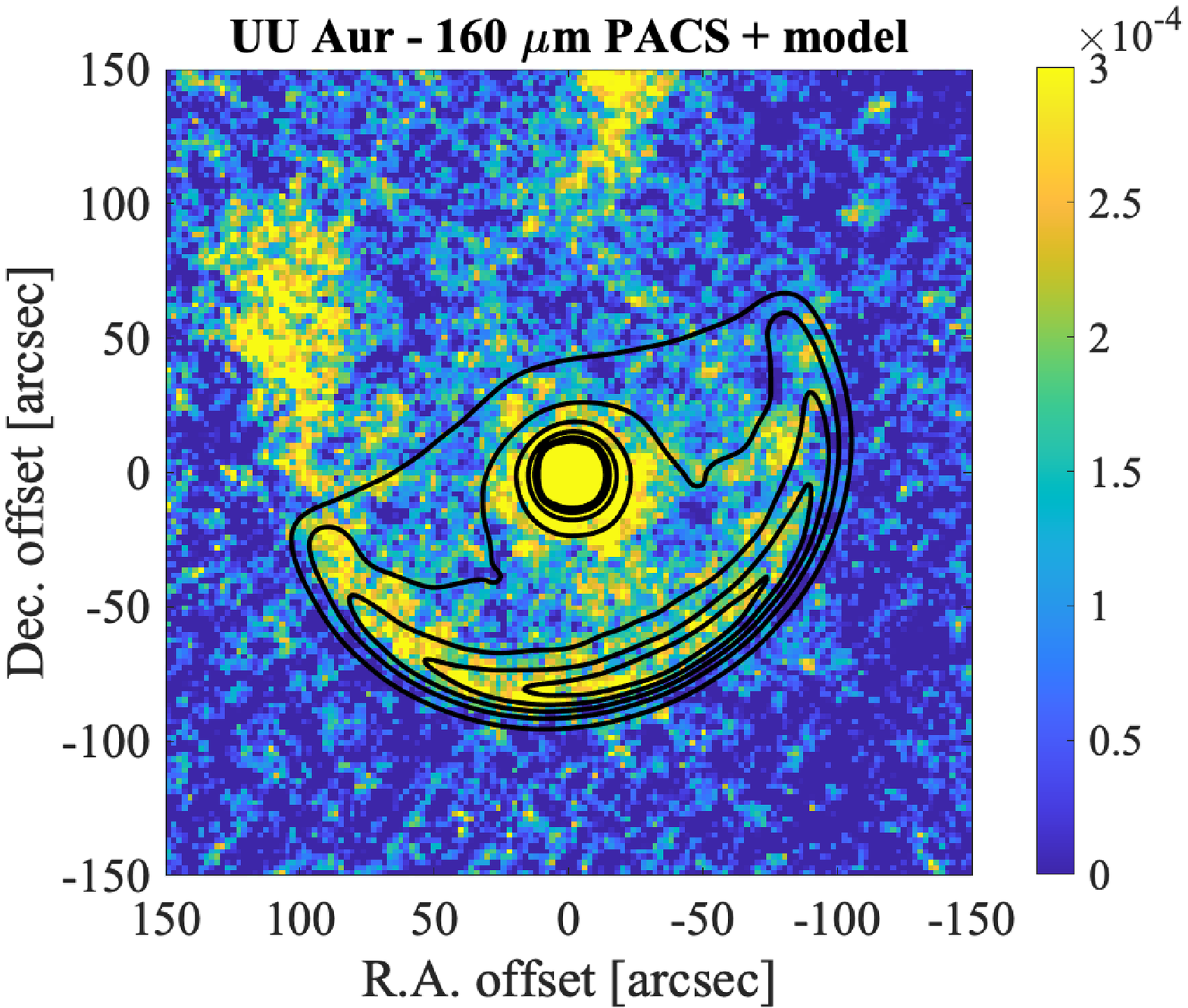}
\caption{UU Aur: \emph{Top to bottom:} The Radmc3D model, the PACS image, and the PACS image with contours from the model. Images are for 70\,\micron~(left) and 160\,\micron~(right). Maximum contour levels are 0.75$\times10^{-3}$\,\Jyarcsec (70\,\micron) and 0.2$\times10^{-3}$\,\Jyarcsec (160\,\micron), respectively. Minimum contour levels are 10\% of maximum. The colour scale is in \Jyarcsec. The red dashed circle shows the mask used to measure the flux from the star and present-day mass-loss.}
\label{f:uuaur}
\end{figure*}

\begin{figure*}
\centering
\includegraphics[width=8cm]{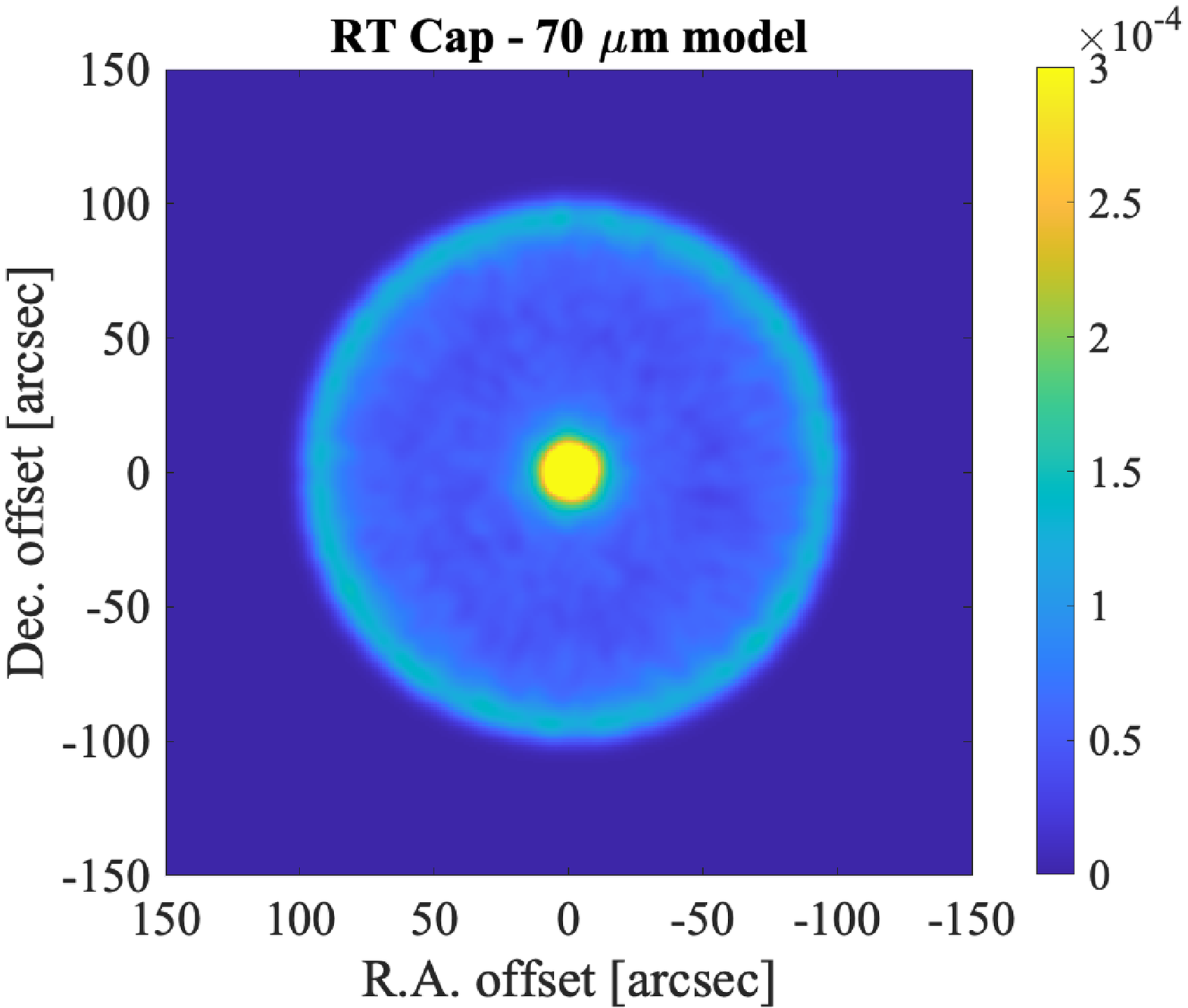}
\includegraphics[width=8cm]{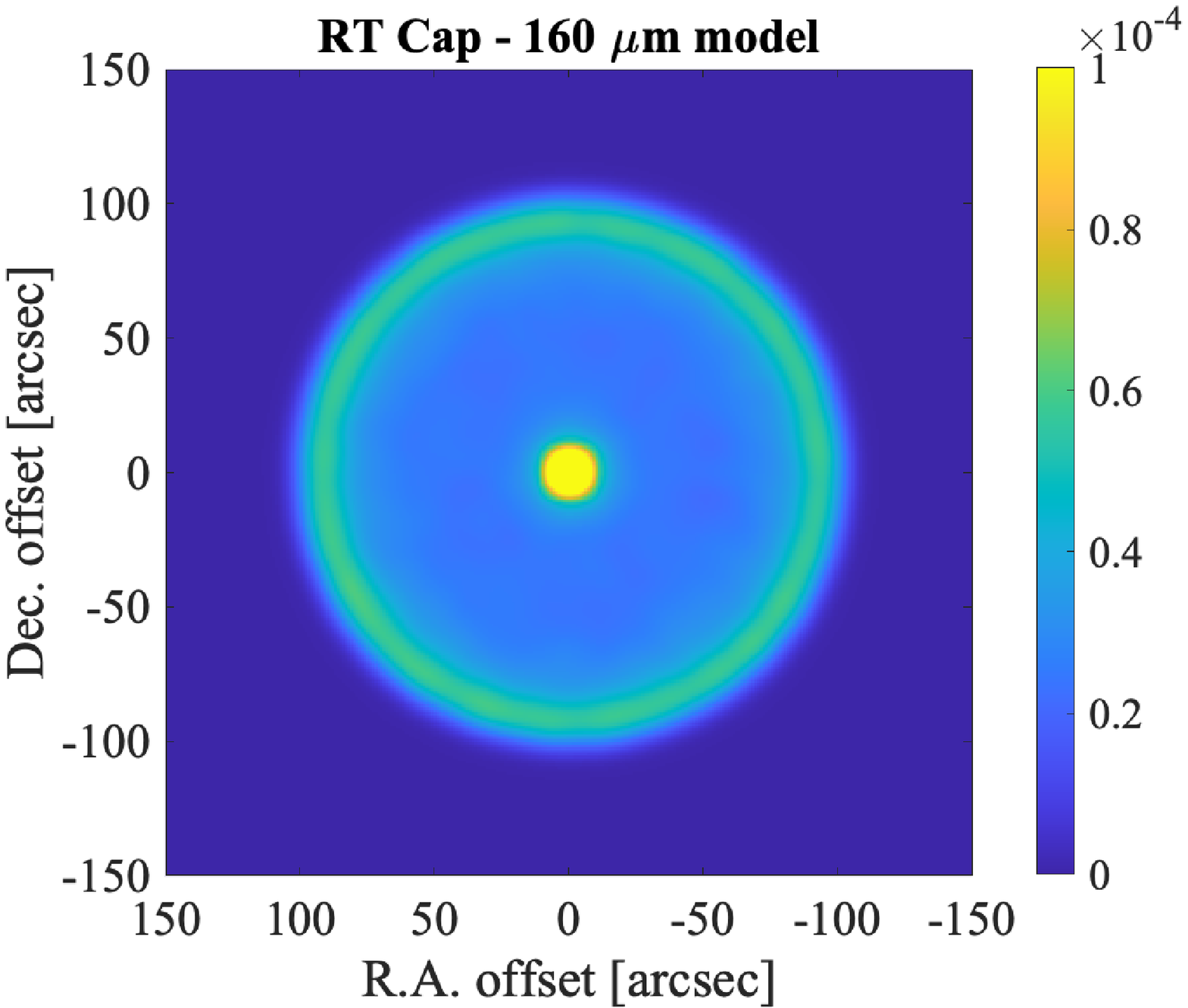}
\includegraphics[width=8cm]{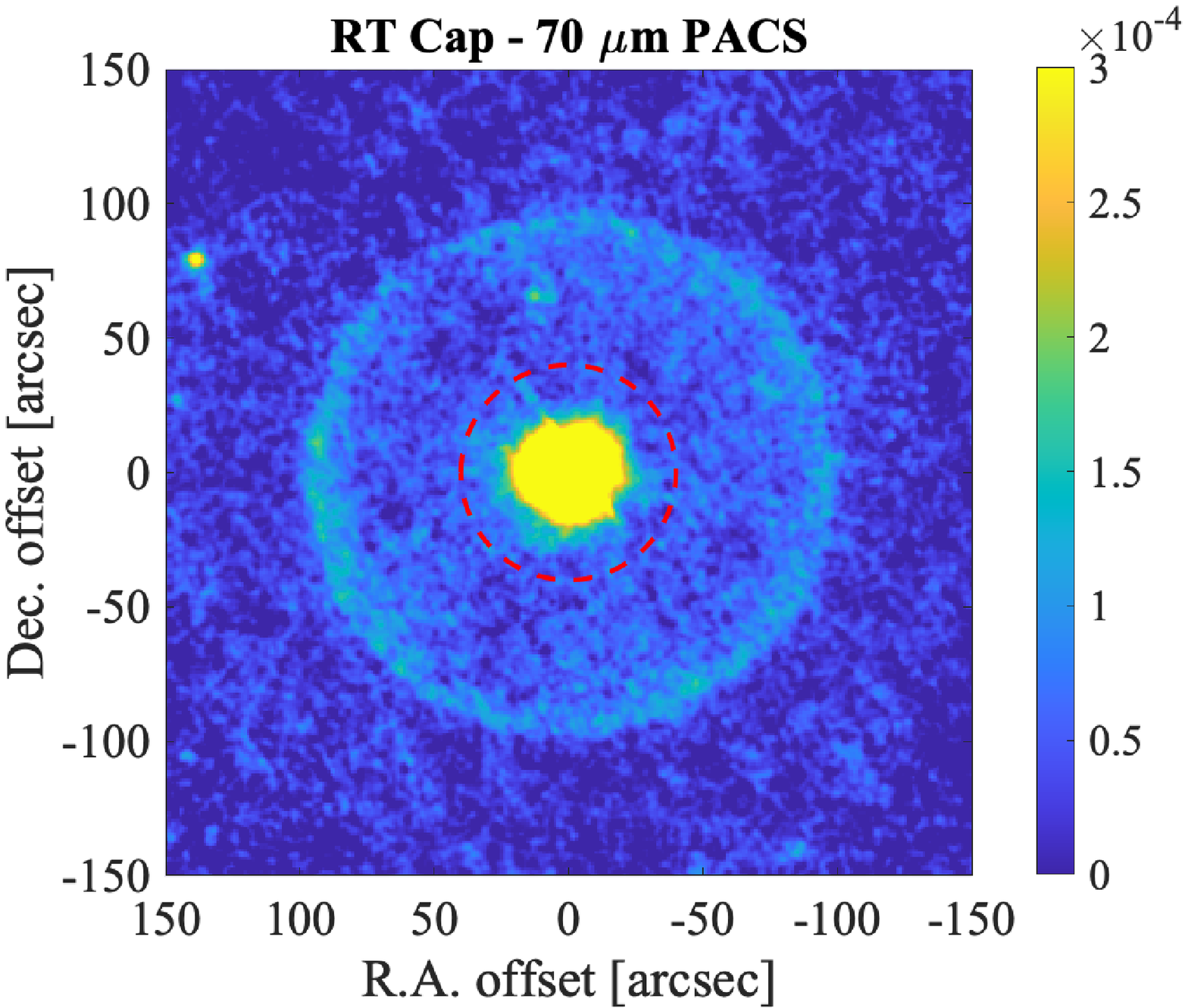}
\includegraphics[width=8cm]{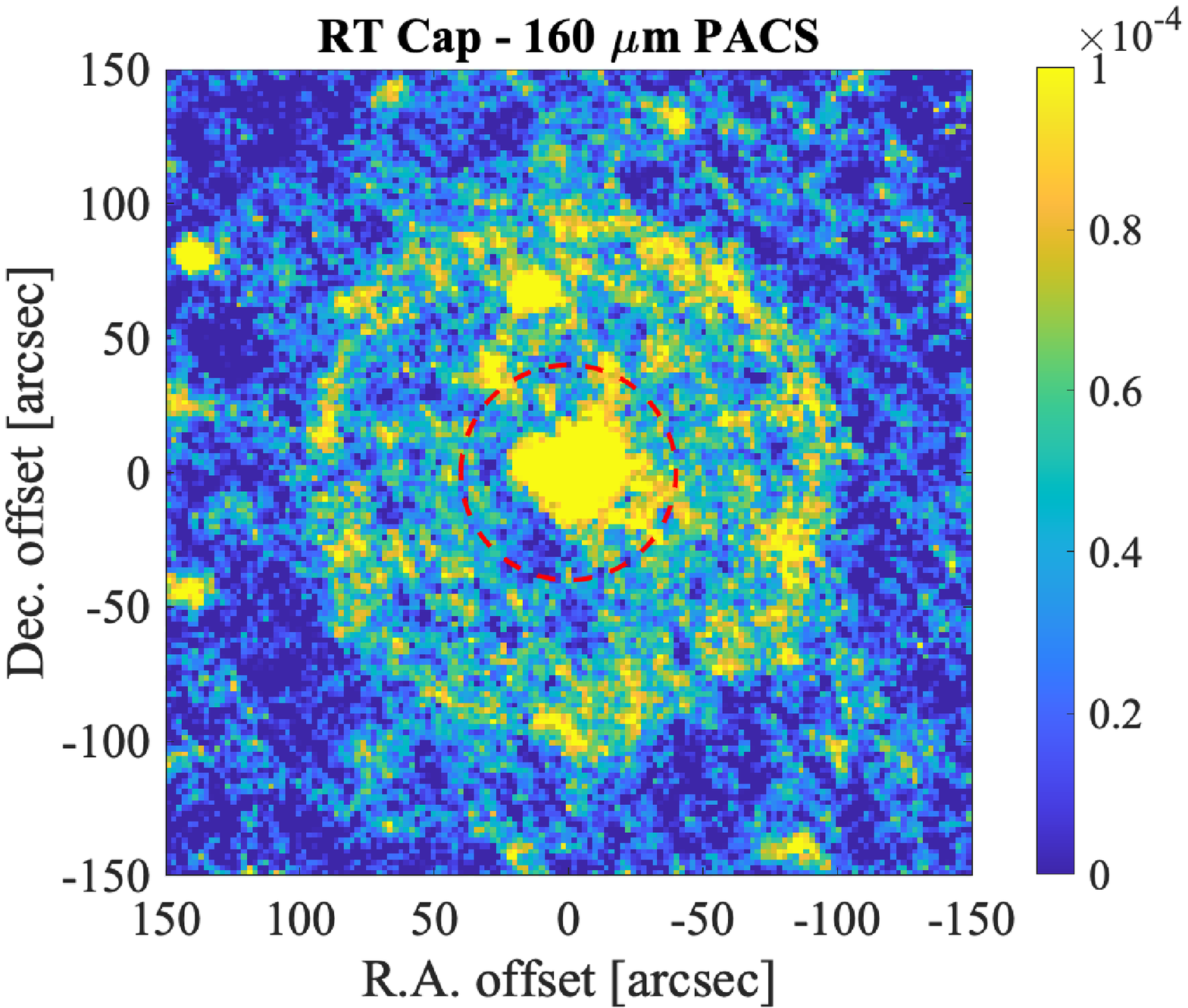}
\includegraphics[width=8cm]{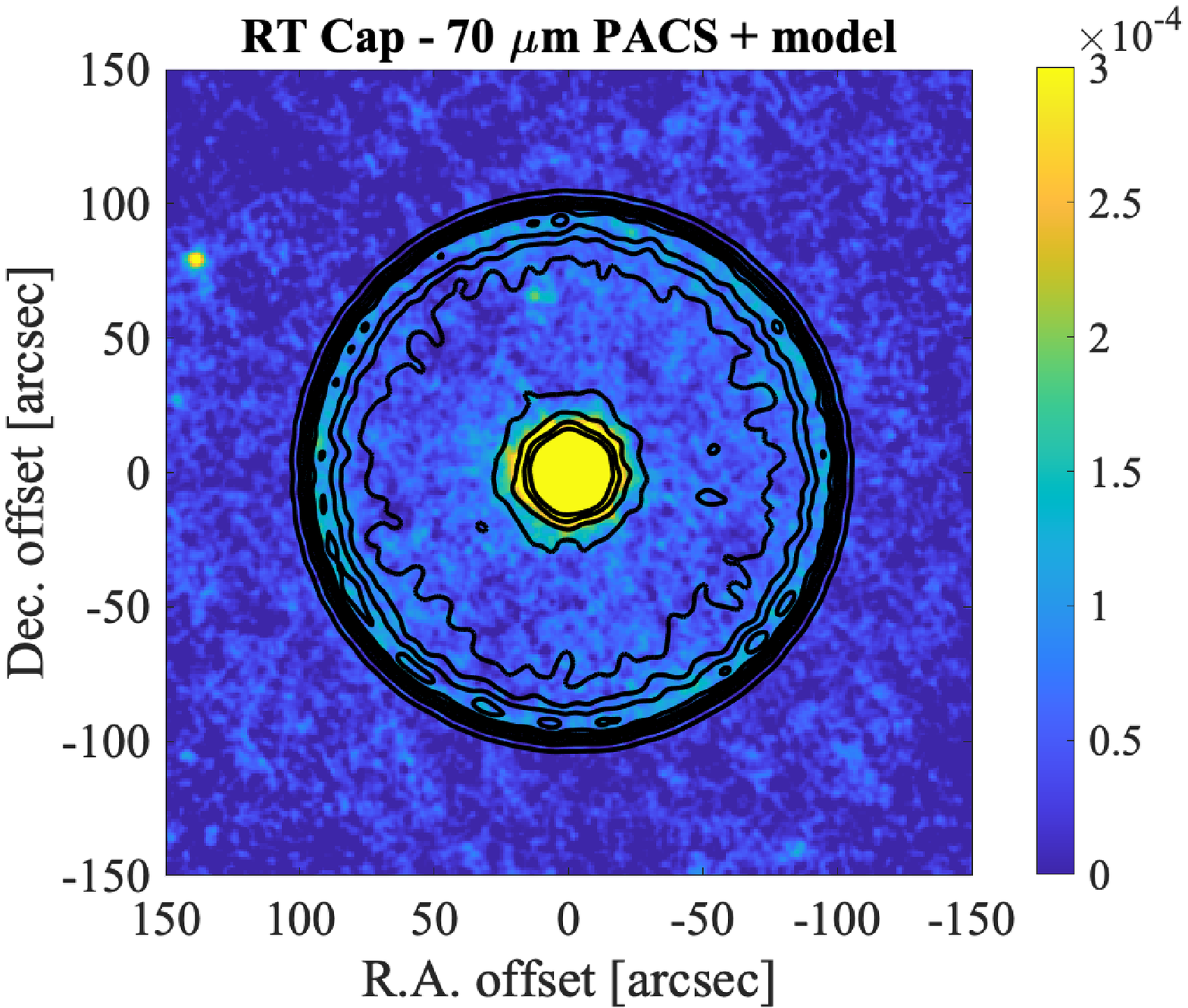}
\includegraphics[width=8cm]{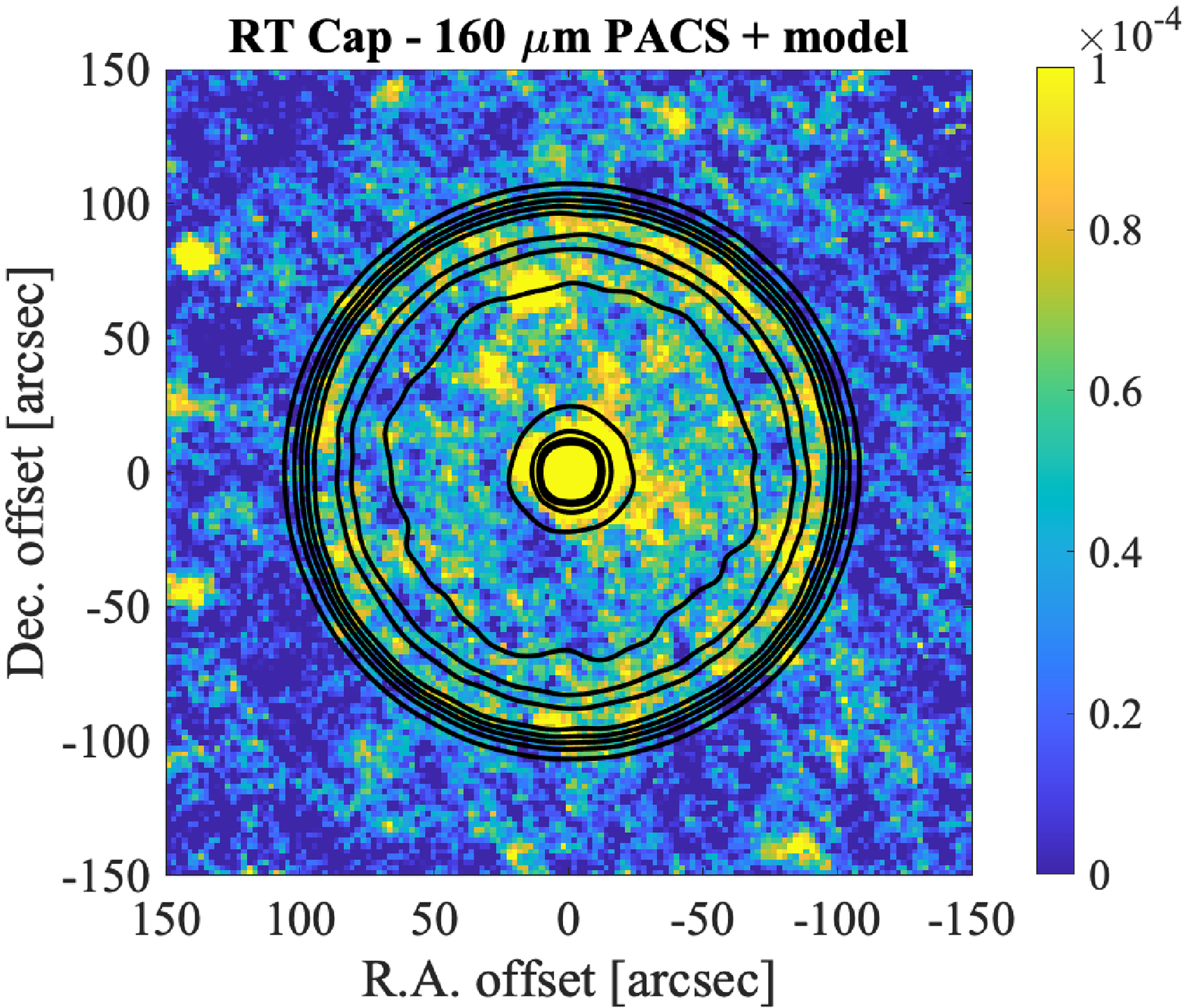}
\caption{RT Cap: \emph{Top to bottom:} The Radmc3D model, the PACS image, and the PACS image with contours from the model. Images are for 70\,\micron~(left) and 160\,\micron~(right). Maximum contour levels are 0.12$\times10^{-3}$\,\Jyarcsec (70\,\micron) and 0.056$\times10^{-3}$\,\Jyarcsec (160\,\micron), respectively. Minimum contour levels are 10\% of maximum. The colour scale is in \Jyarcsec. The red dashed circle shows the mask used to measure the flux from the star and present-day mass-loss.}
\label{f:rtcap}
\end{figure*}

\begin{figure*}
\centering
\includegraphics[width=8cm]{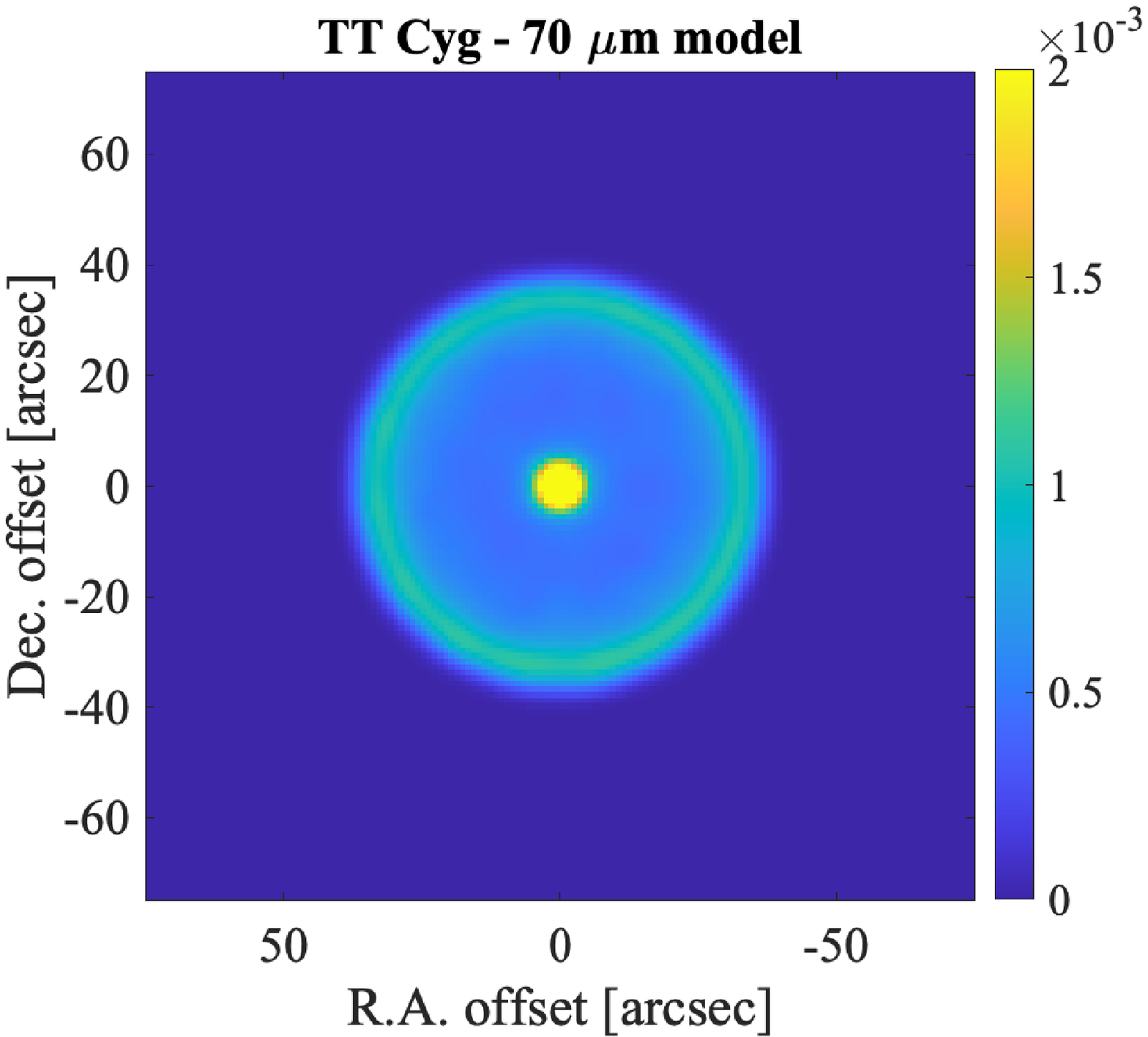}
\includegraphics[width=8cm]{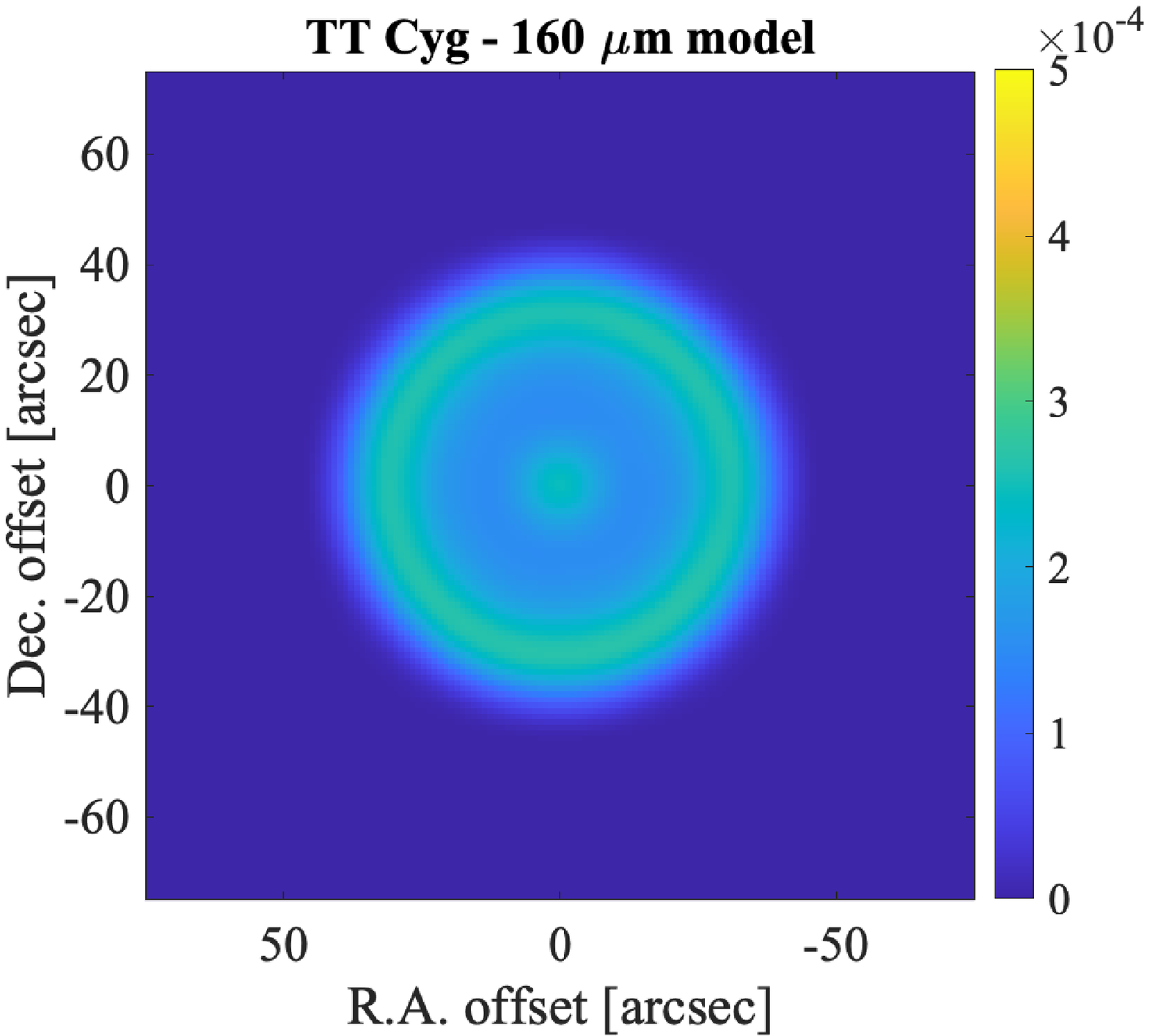}
\includegraphics[width=8cm]{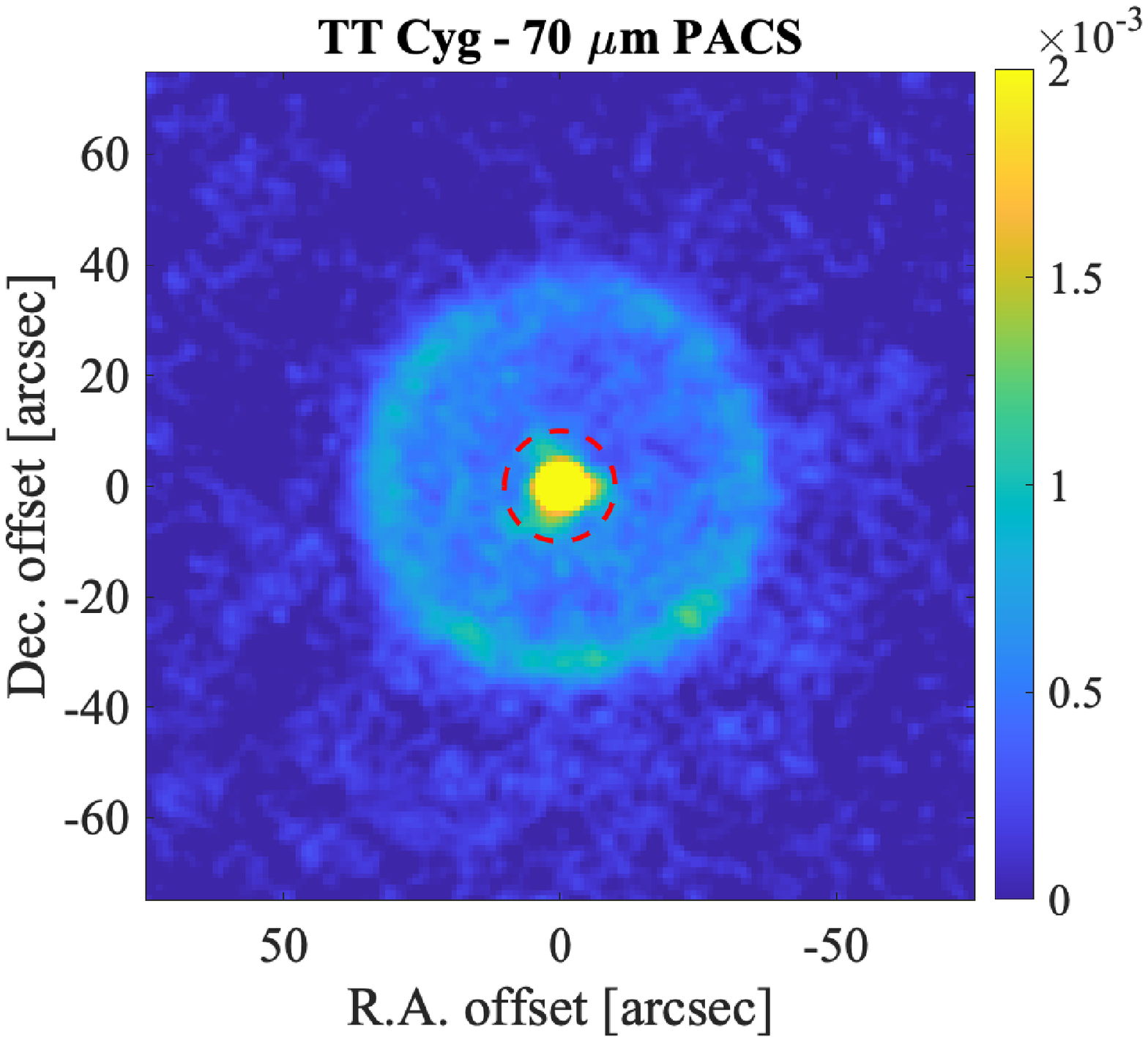}
\includegraphics[width=8cm]{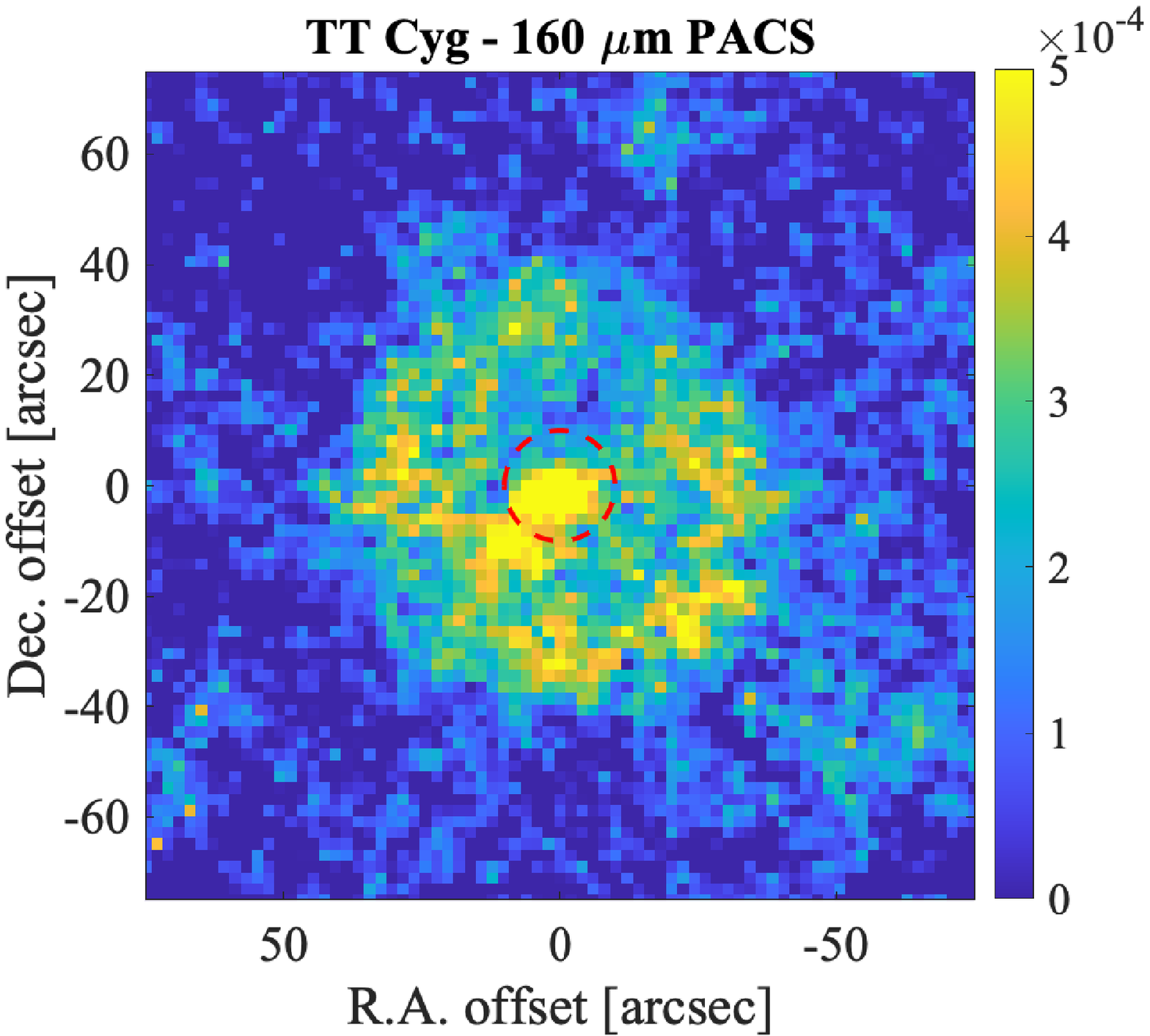}
\includegraphics[width=8cm]{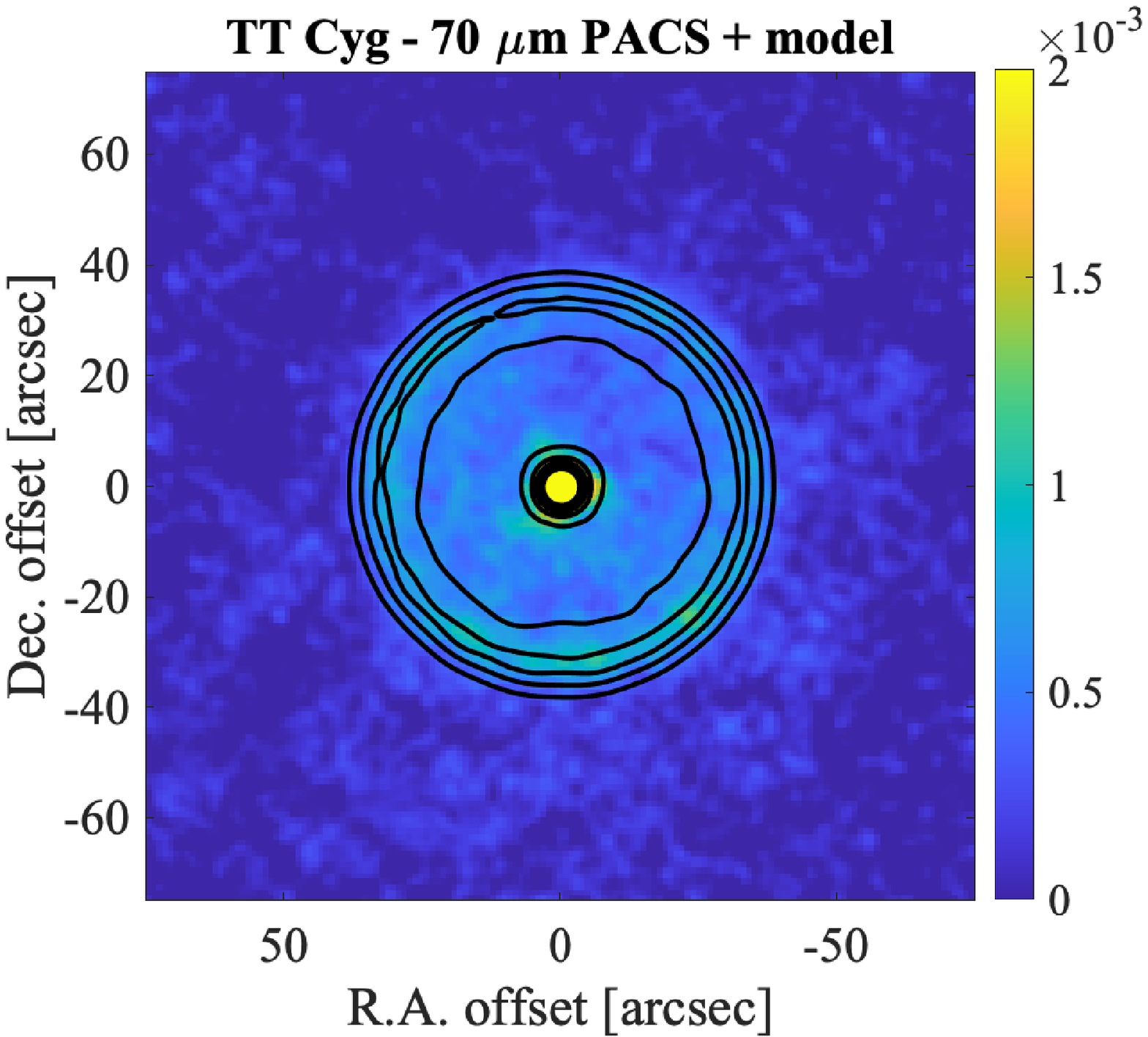}
\includegraphics[width=8cm]{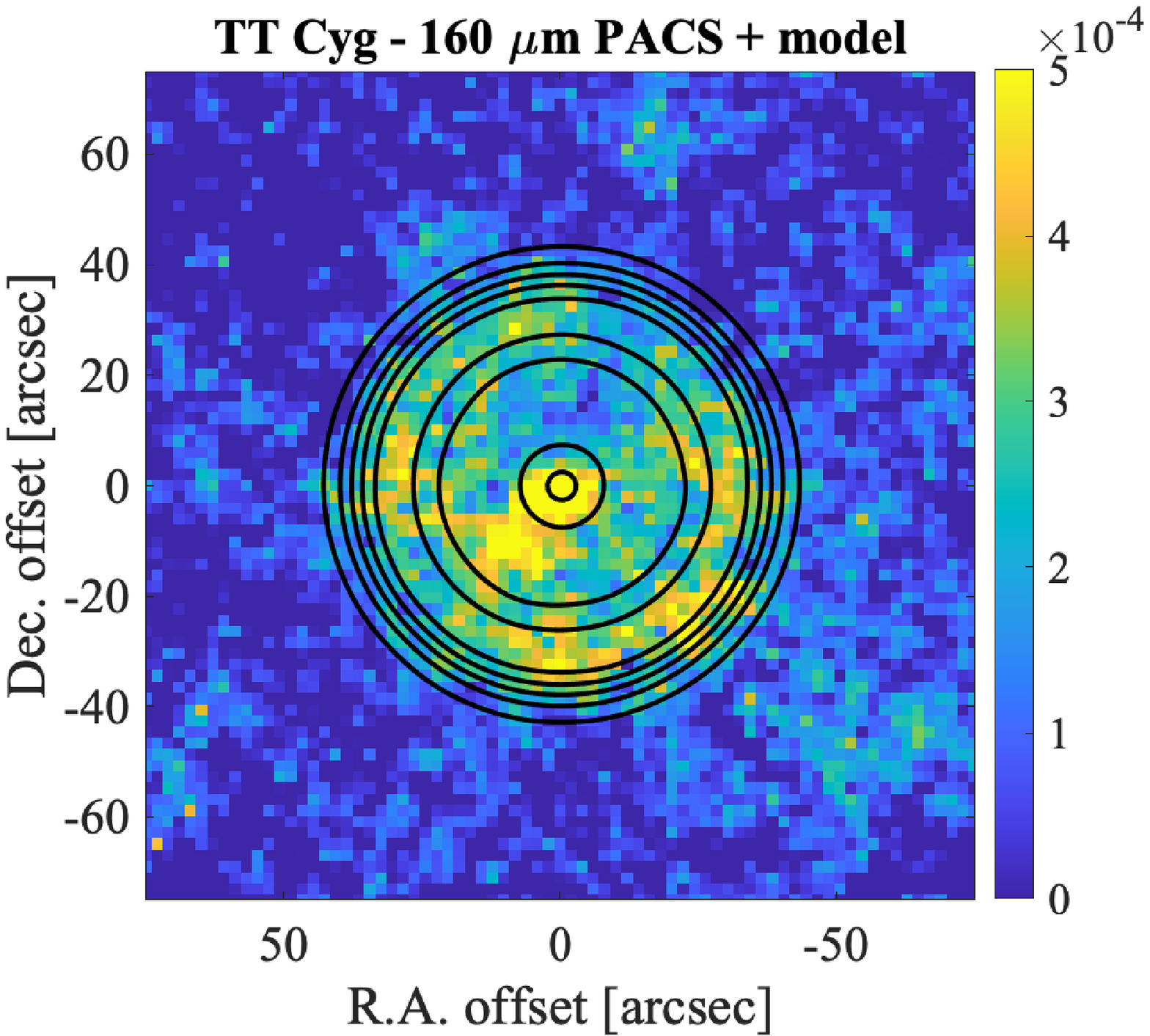}
\caption{TT Cyg: \emph{Top to bottom:} The Radmc3D model, the PACS image, and the PACS image with contours from the model. Images are for 70\,\micron~(left) and 160\,\micron~(right). Maximum contour levels are 2$\times10^{-3}$\,\Jyarcsec (70\,\micron) and 0.26$\times10^{-3}$\,\Jyarcsec (160\,\micron), respectively. Minimum contour levels are 10\% of maximum. The colour scale is in \Jyarcsec. The red dashed circle shows the mask used to measure the flux from the star and present-day mass-loss.}
\label{f:ttcyg}
\end{figure*}

\begin{figure*}
\centering
\includegraphics[width=8cm]{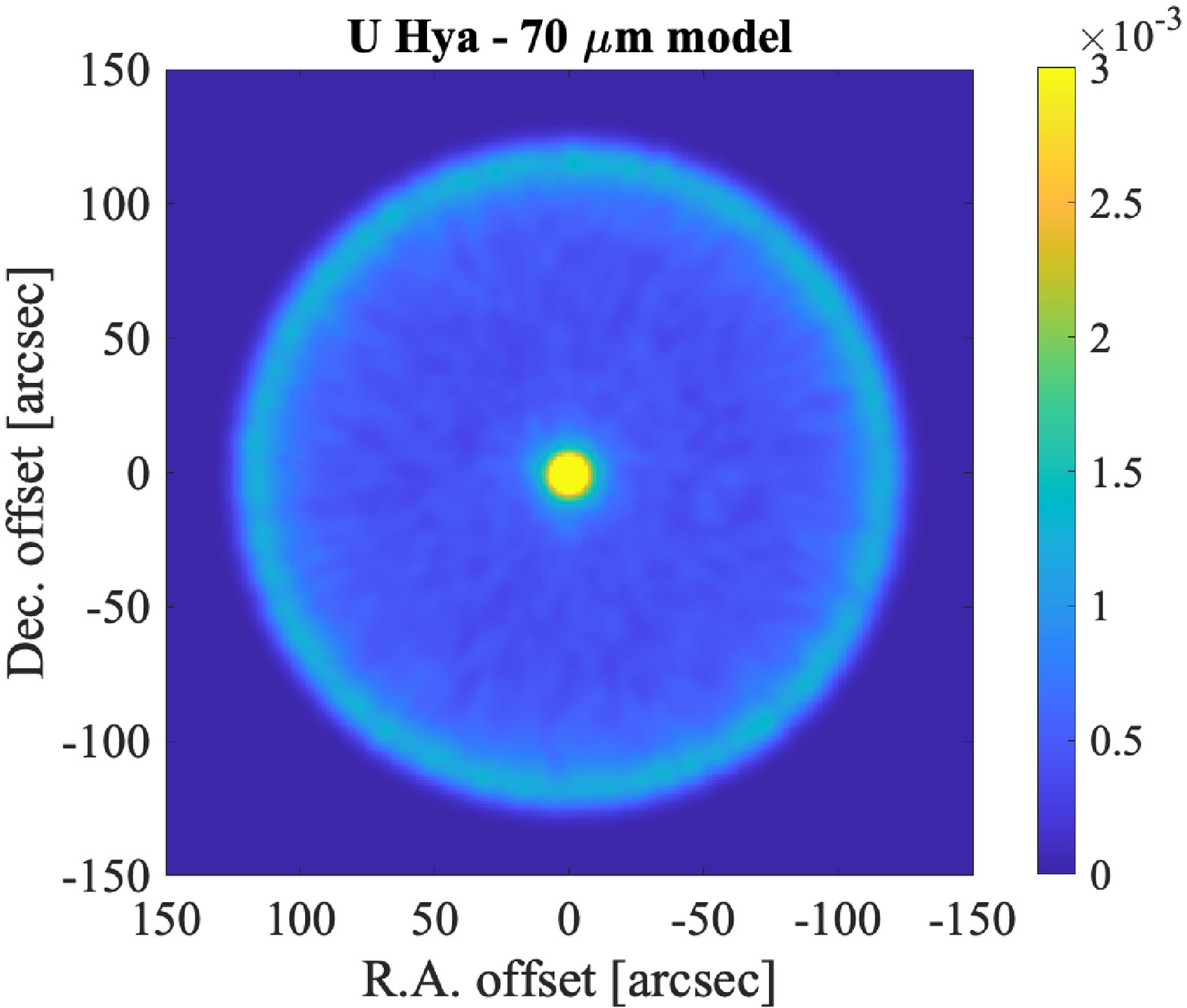}
\includegraphics[width=8cm]{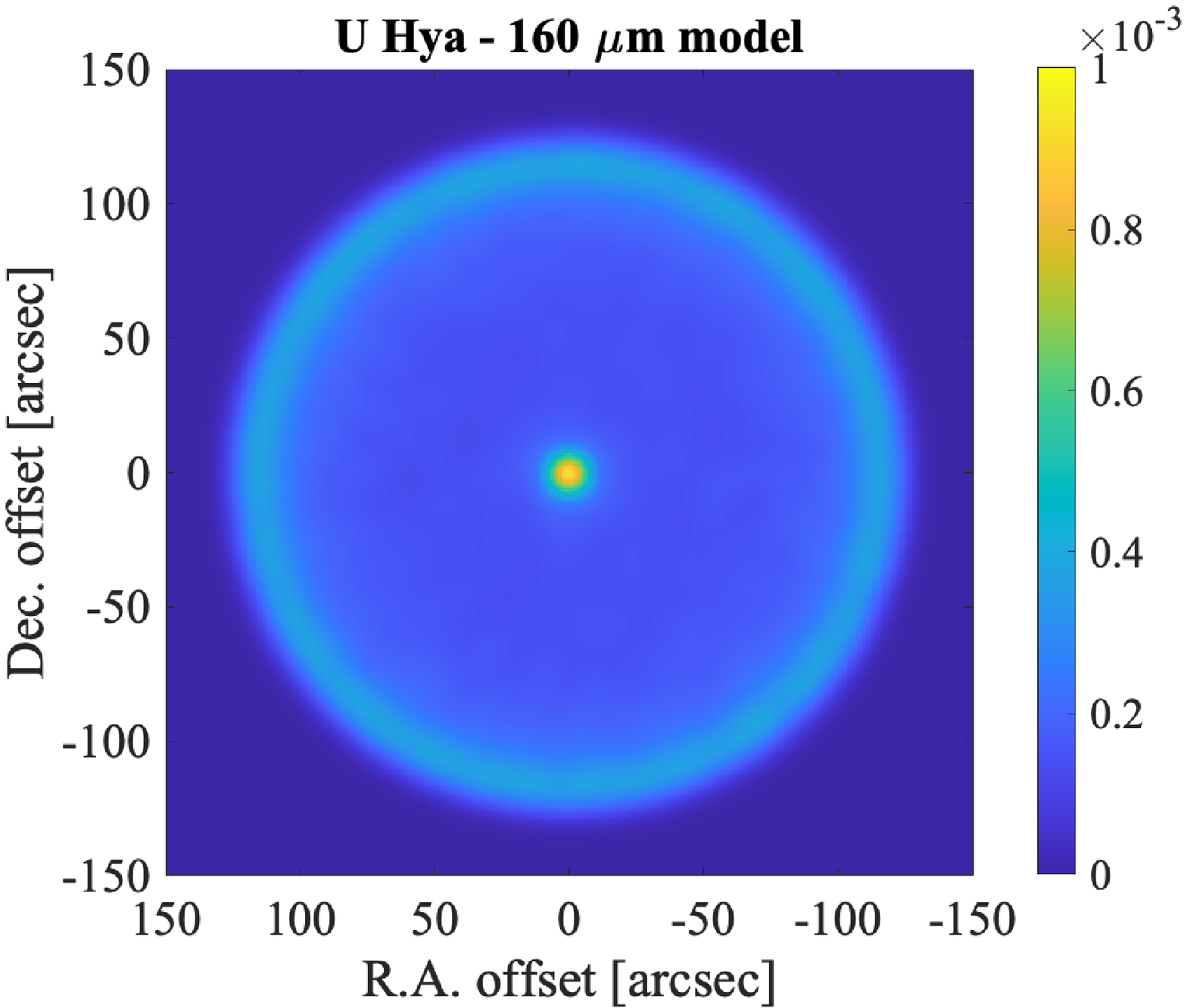}
\includegraphics[width=8cm]{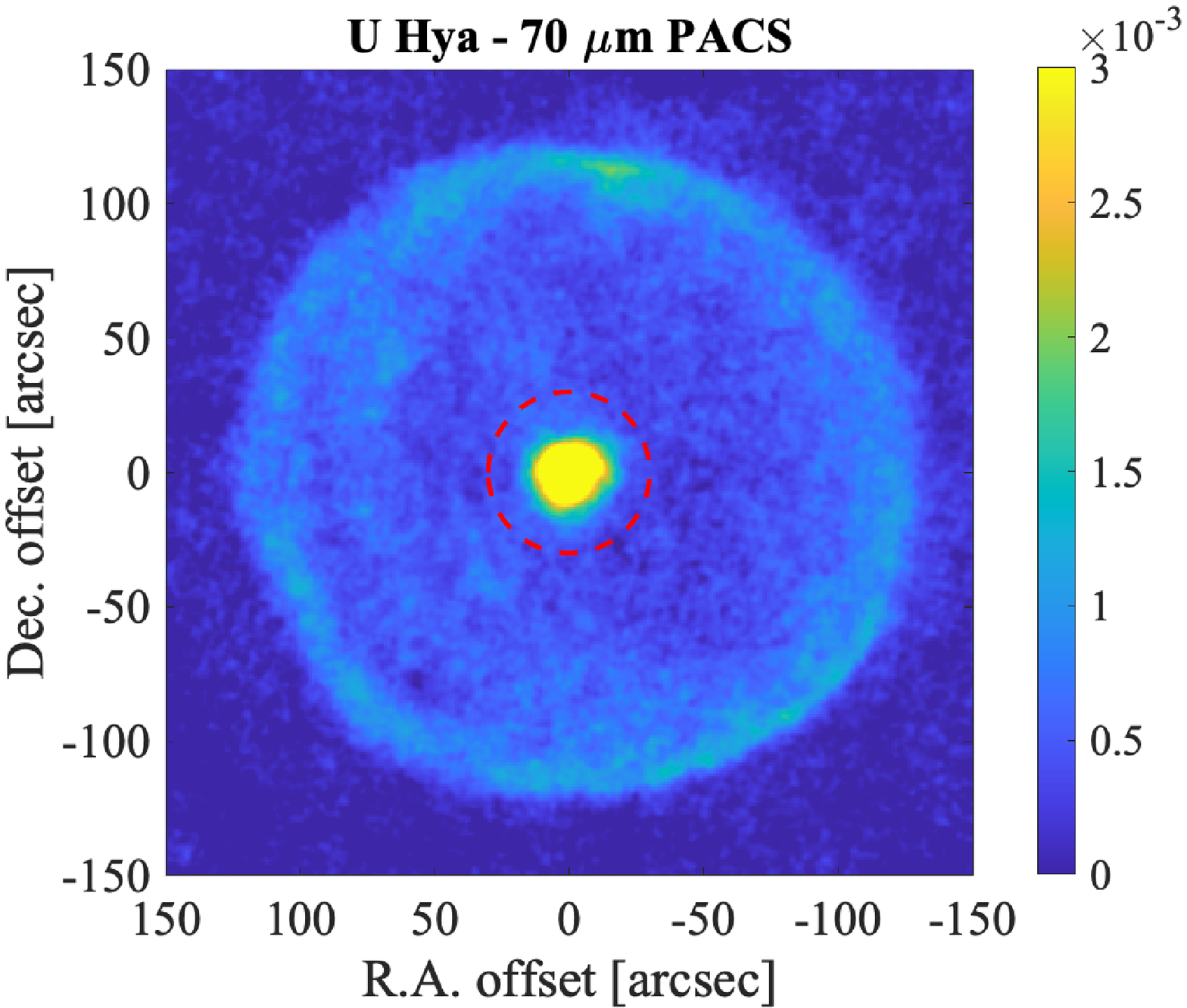}
\includegraphics[width=8cm]{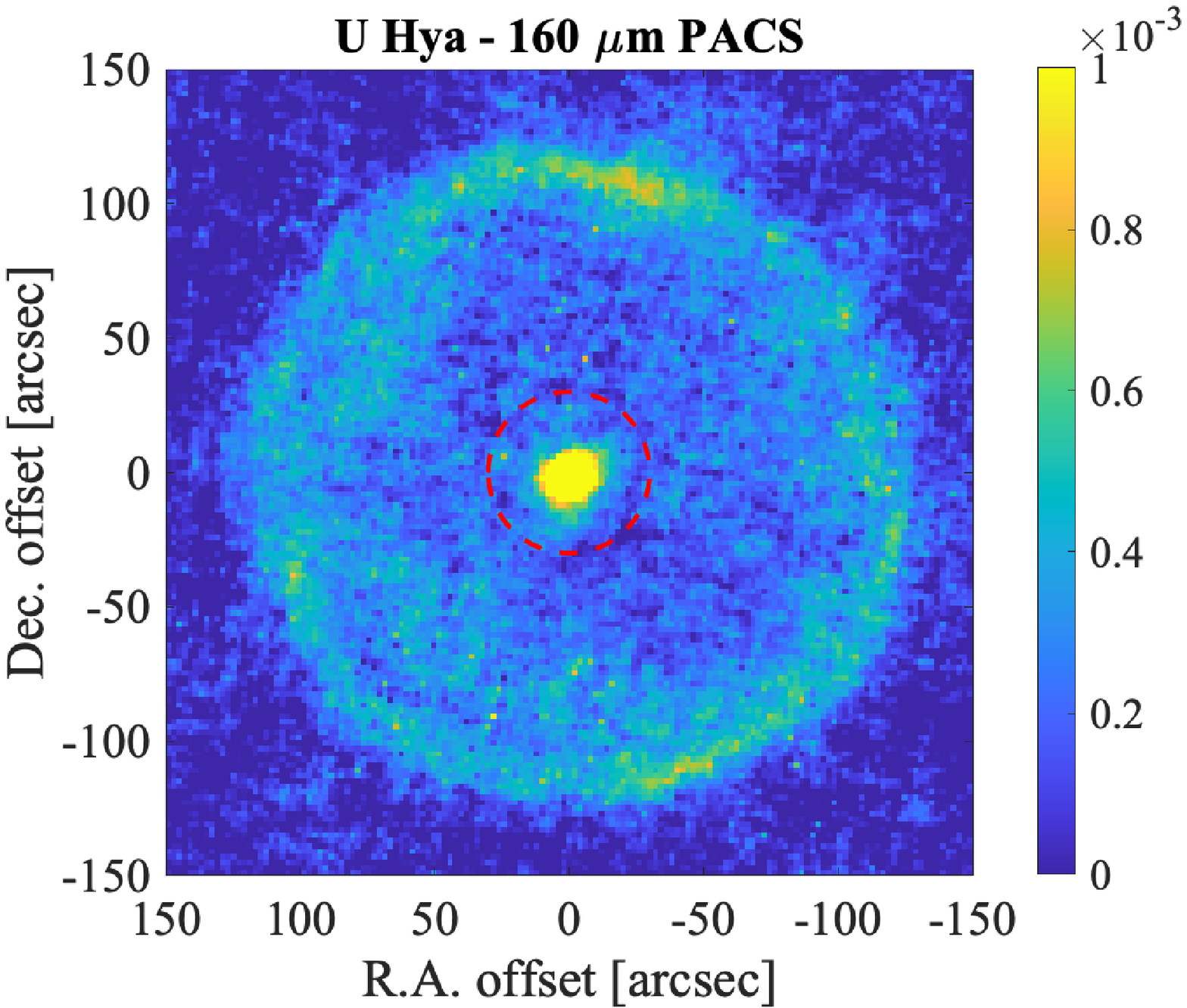}
\includegraphics[width=8cm]{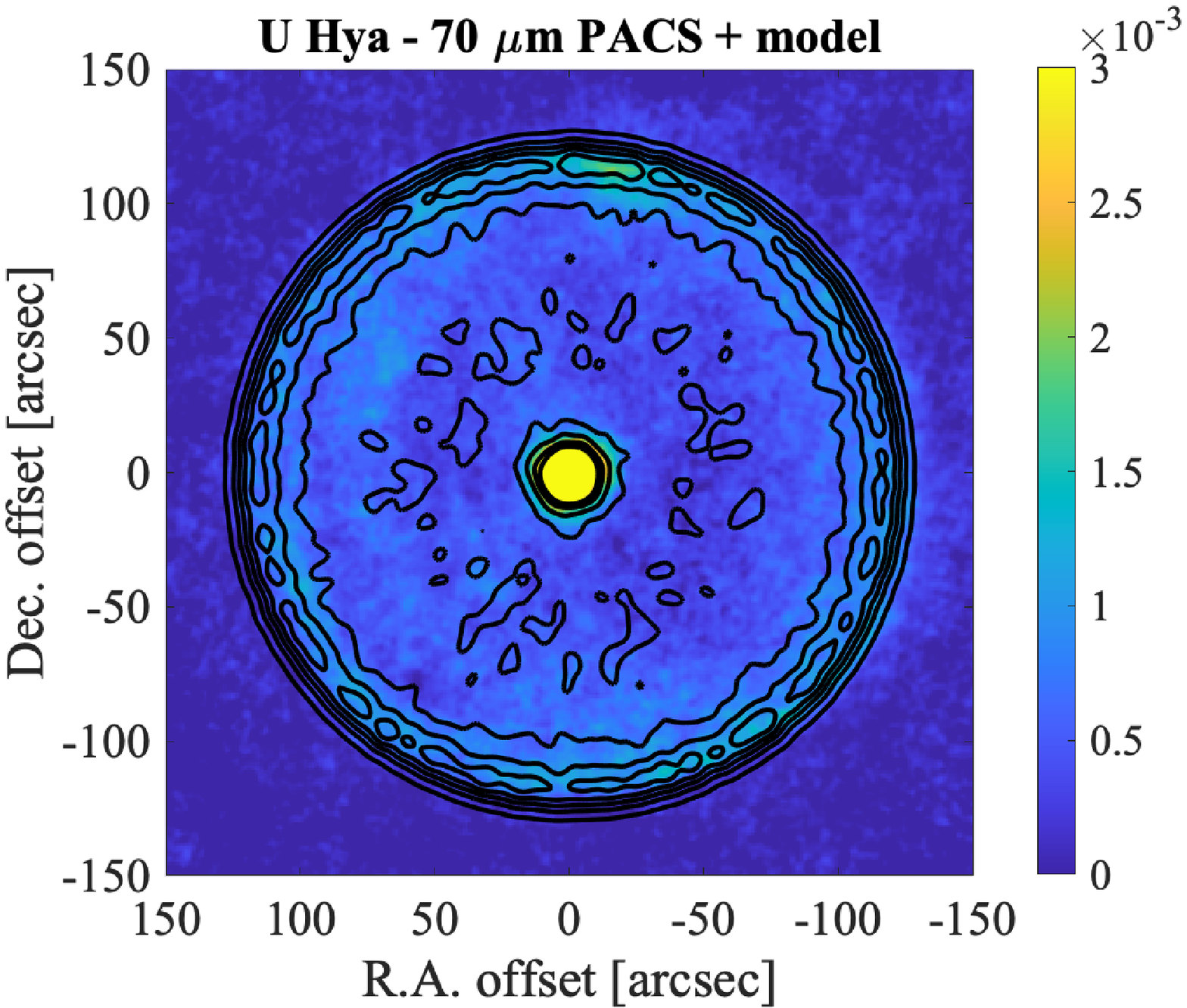}
\includegraphics[width=8cm]{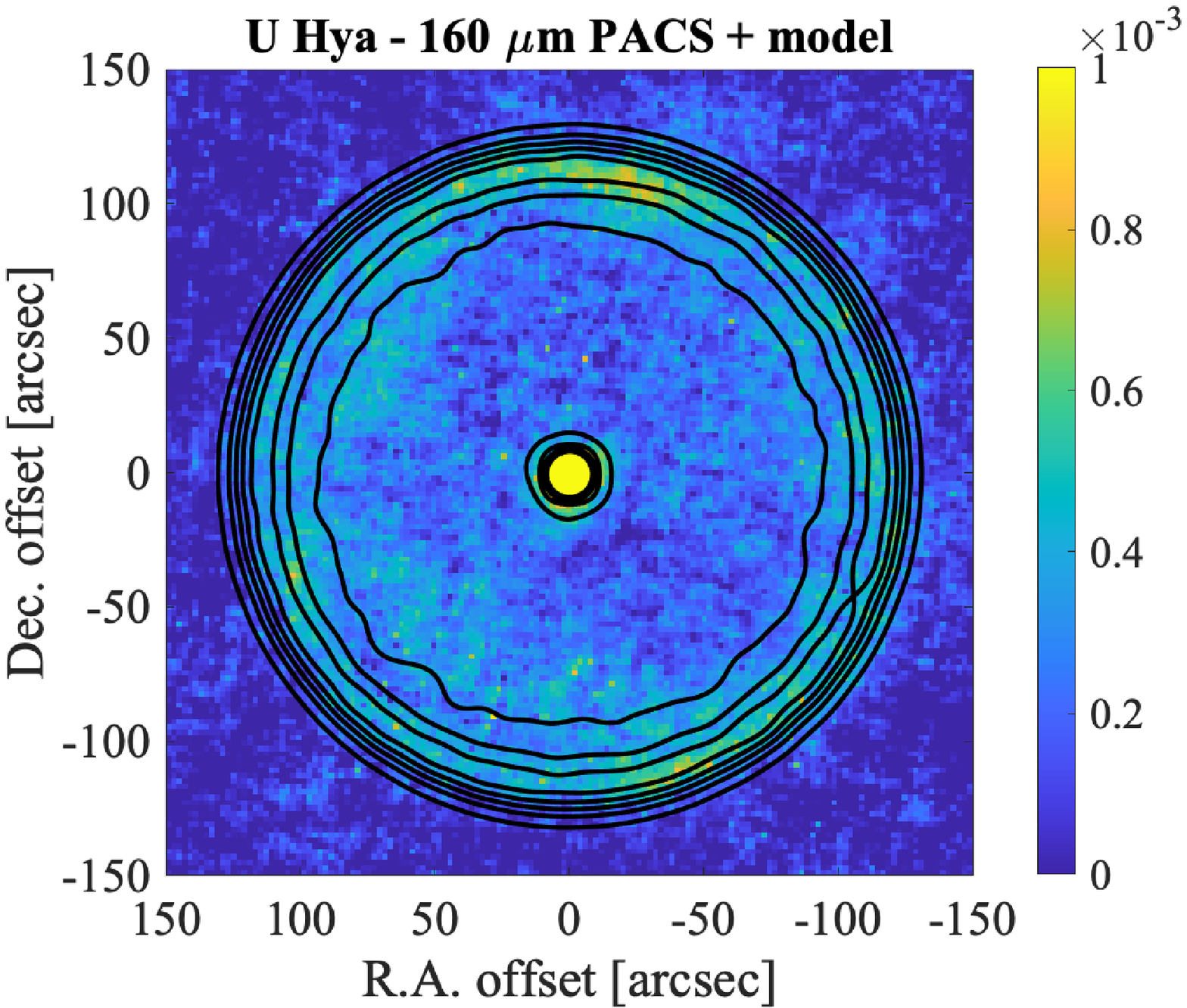}
\caption{U Hya: \emph{Top to bottom:} The Radmc3D model, the PACS image, and the PACS image with contours from the model. Images are for 70\,\micron~(left) and 160\,\micron~(right). Maximum contour levels are 1.3$\times10^{-3}$\,\Jyarcsec (70\,\micron) and 0.38$\times10^{-3}$\,\Jyarcsec (160\,\micron), respectively. Minimum contour levels are 10\% of maximum. The colour scale is in \Jyarcsec. The red dashed circle shows the mask used to measure the flux from the star and present-day mass-loss.}
\label{f:uhya}
\end{figure*}

\begin{figure*}
\centering
\includegraphics[width=8cm]{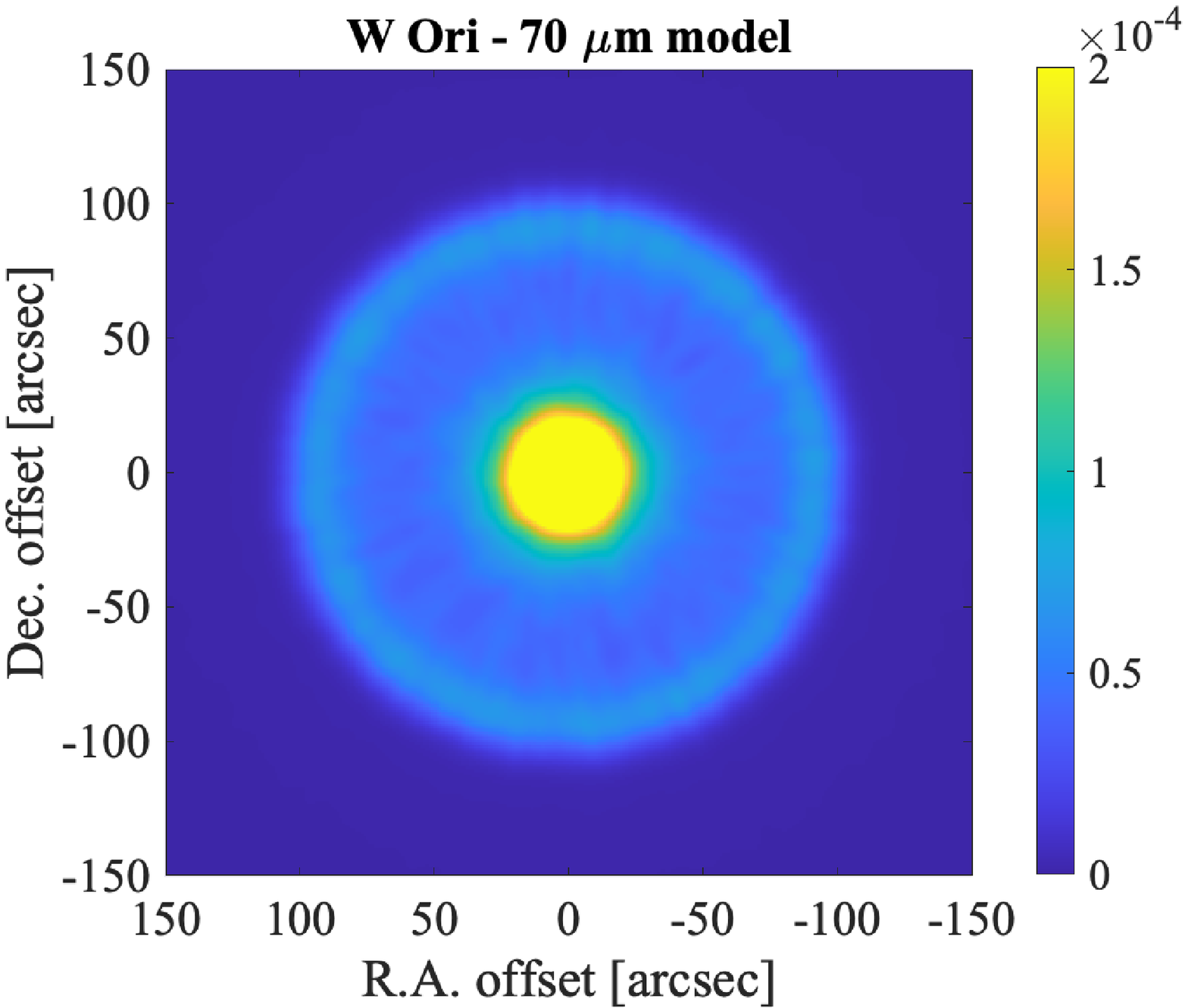}
\includegraphics[width=8cm]{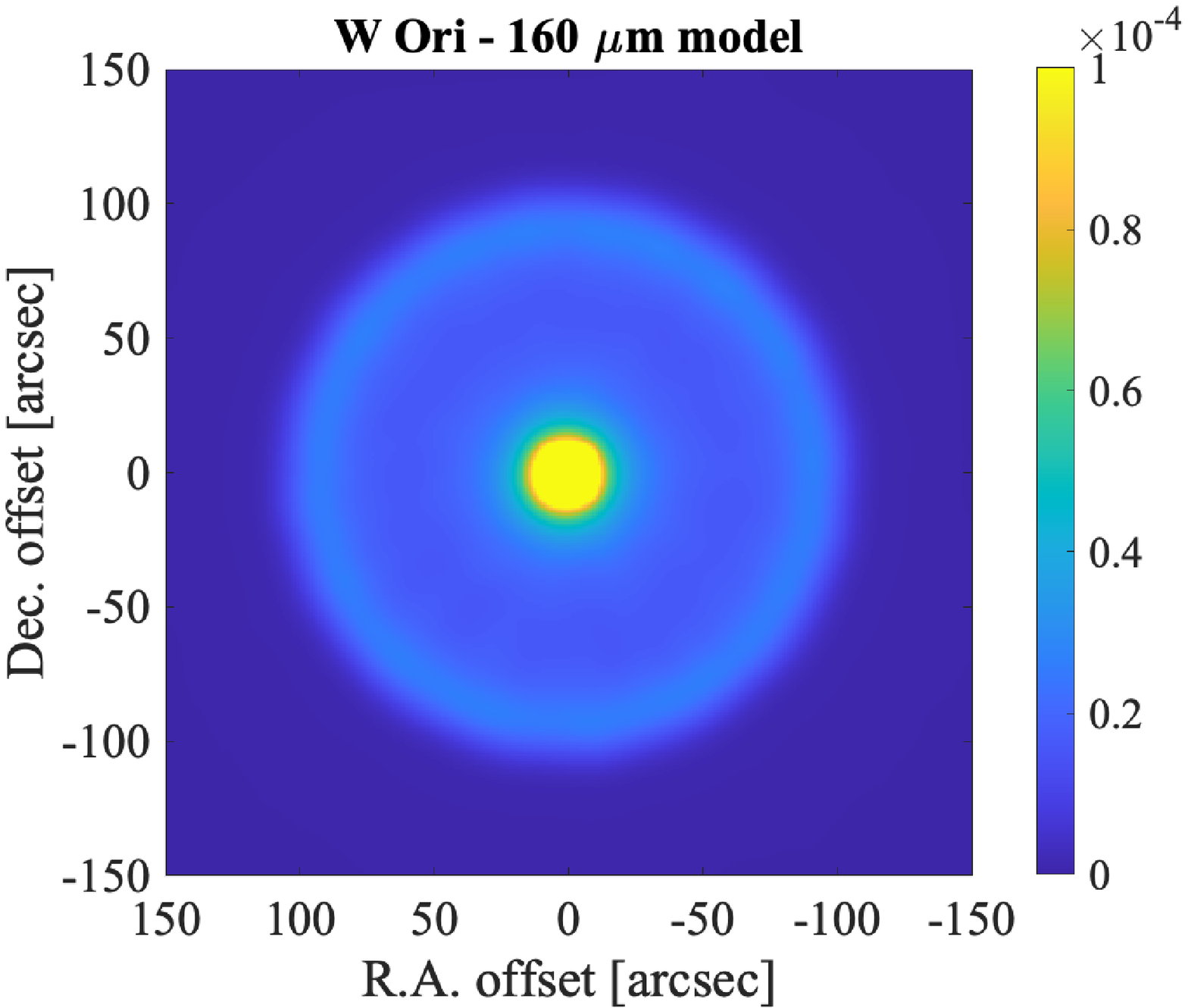}
\includegraphics[width=8cm]{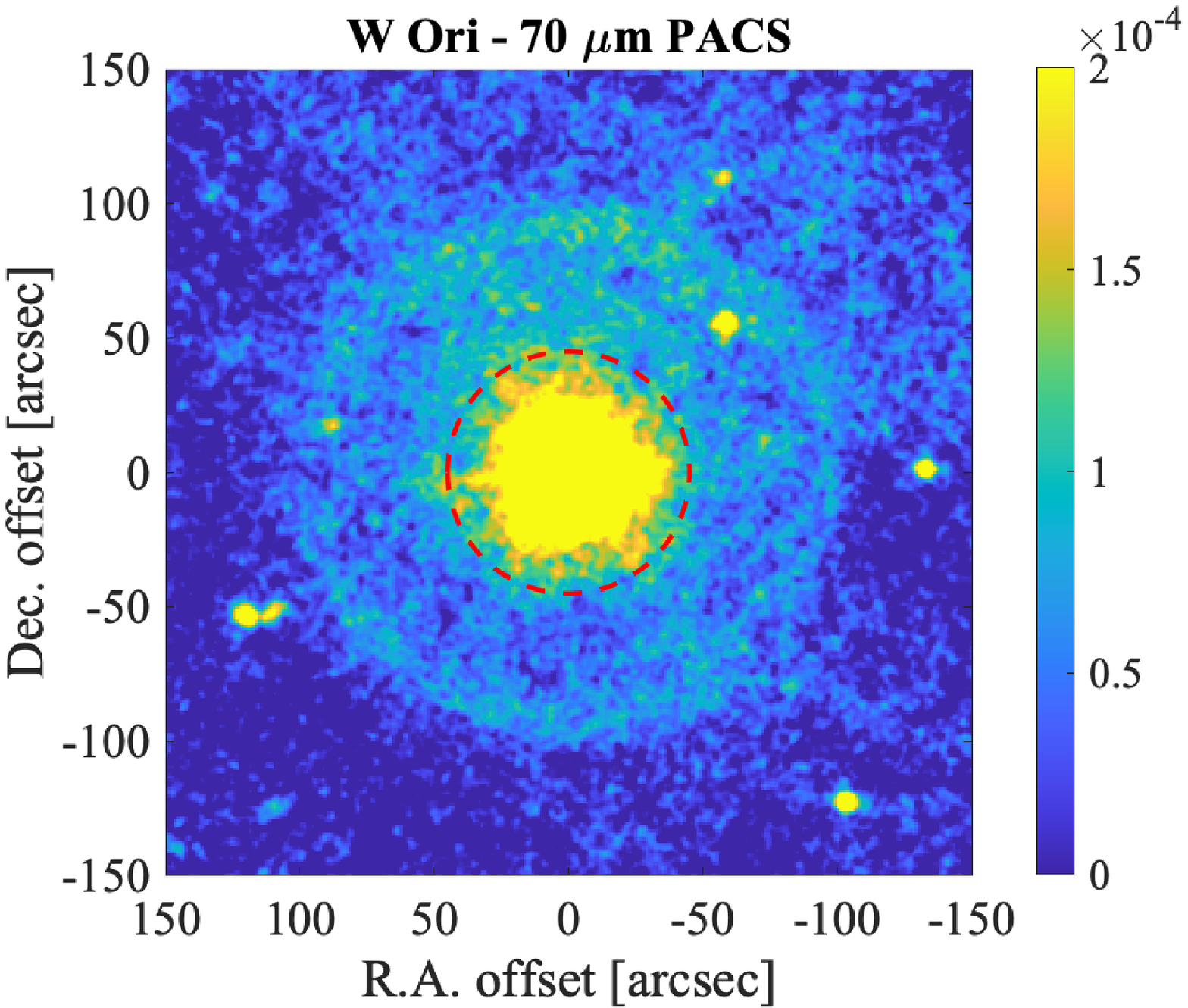}
\includegraphics[width=8cm]{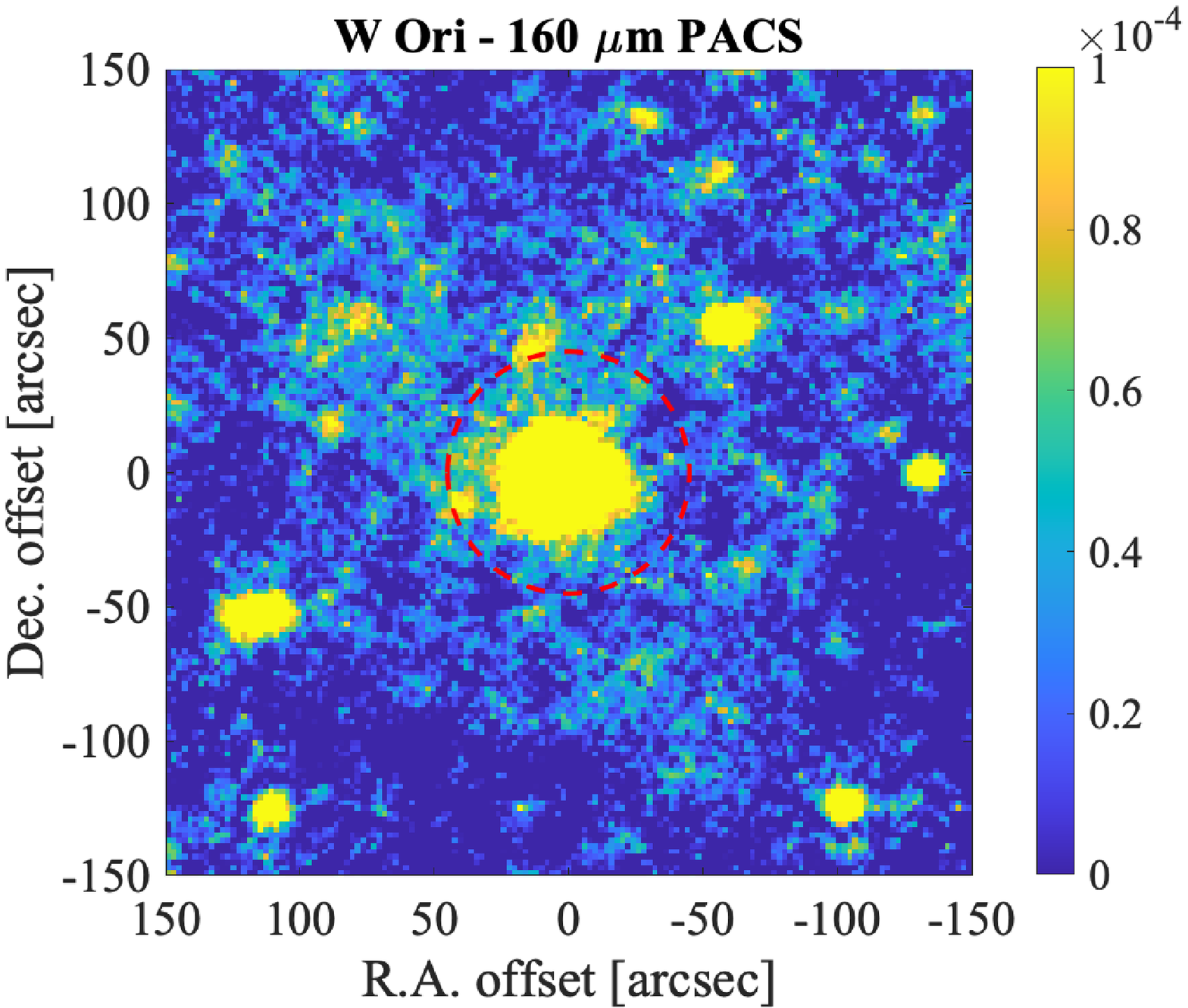}
\includegraphics[width=8cm]{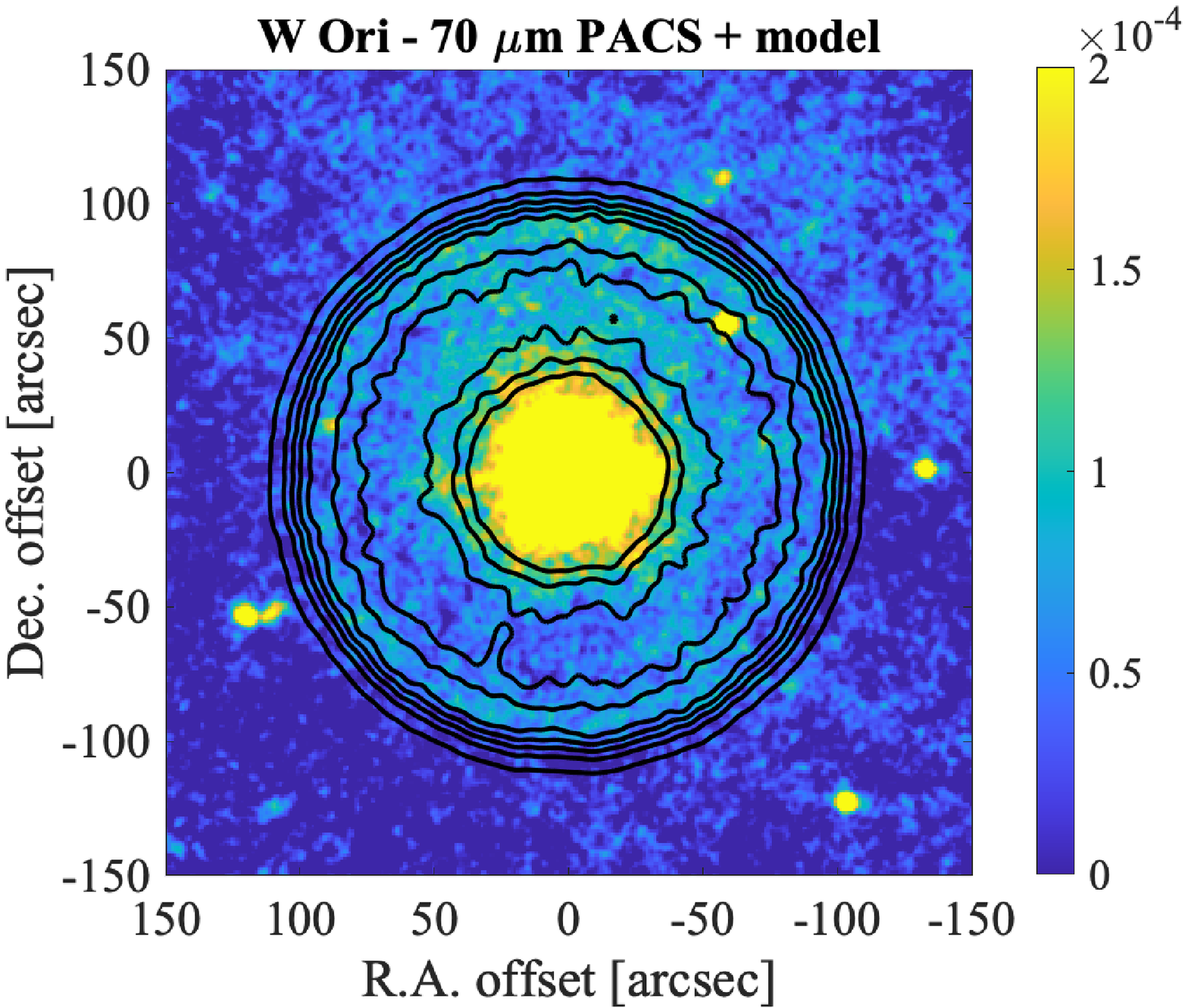}
\includegraphics[width=8cm]{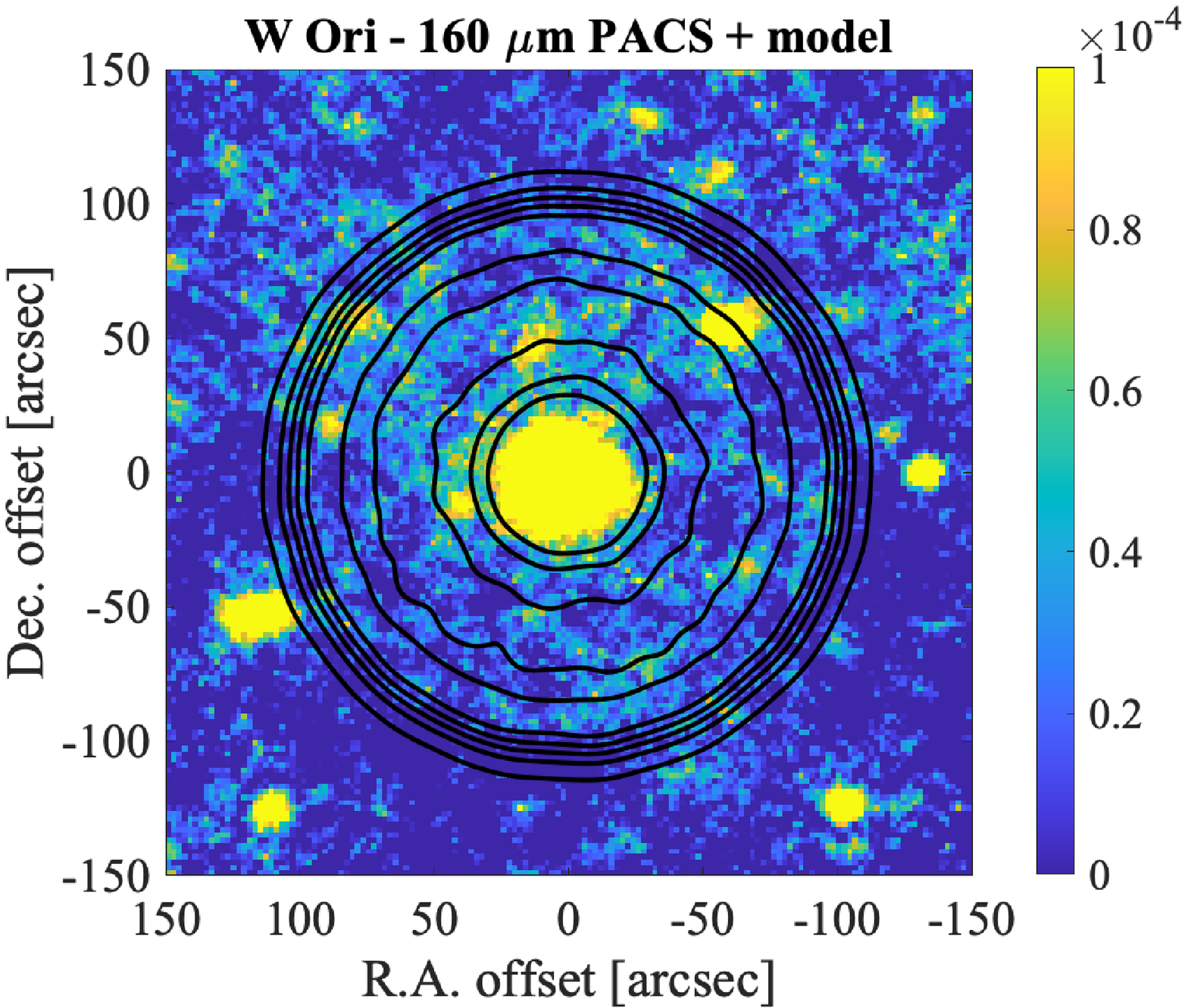}
\caption{W Ori: \emph{Top to bottom:} The Radmc3D model, the PACS image, and the PACS image with contours from the model. Images are for 70\,\micron~(left) and 160\,\micron~(right). Maximum contour levels are 0.066$\times10^{-3}$\,\Jyarcsec (70\,\micron) and 0.025$\times10^{-3}$\,\Jyarcsec (160\,\micron), respectively. Minimum contour levels are 10\% of maximum. The colour scale is in \Jyarcsec. The red dashed circle shows the mask used to measure the flux from the star and present-day mass-loss.}
\label{f:wori}
\end{figure*}

\begin{figure*}
\centering
\includegraphics[width=8cm]{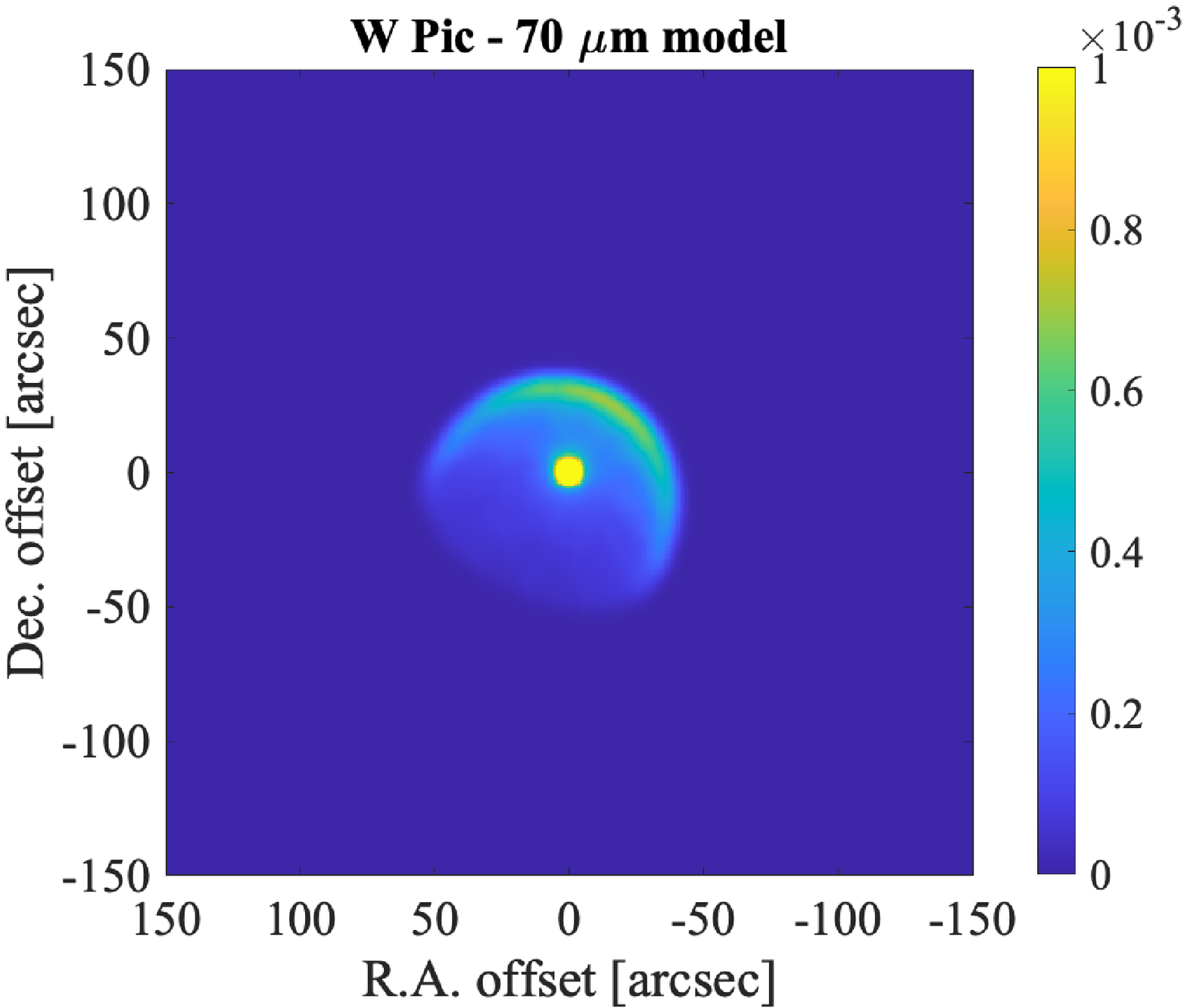}
\includegraphics[width=8cm]{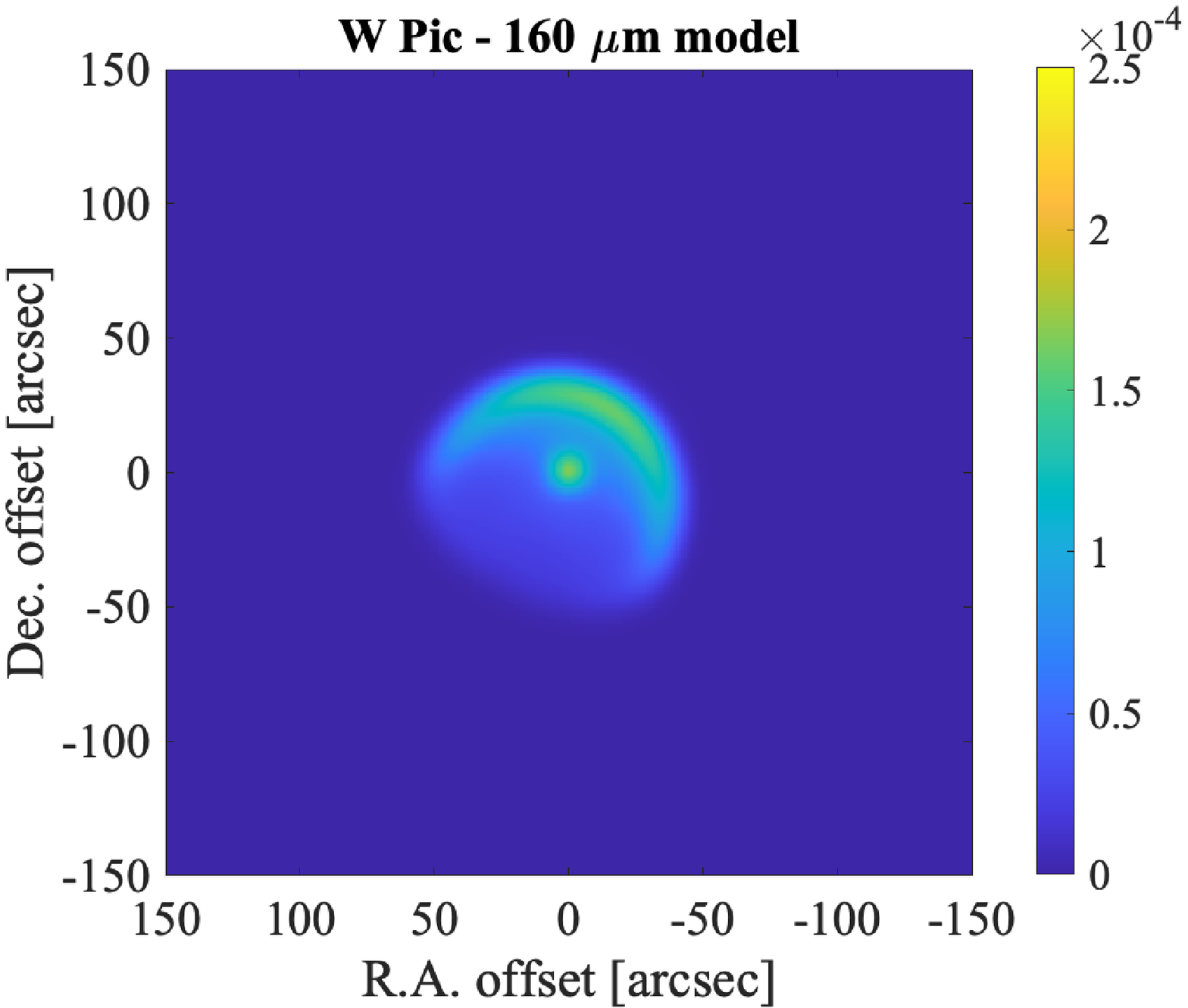}
\includegraphics[width=8cm]{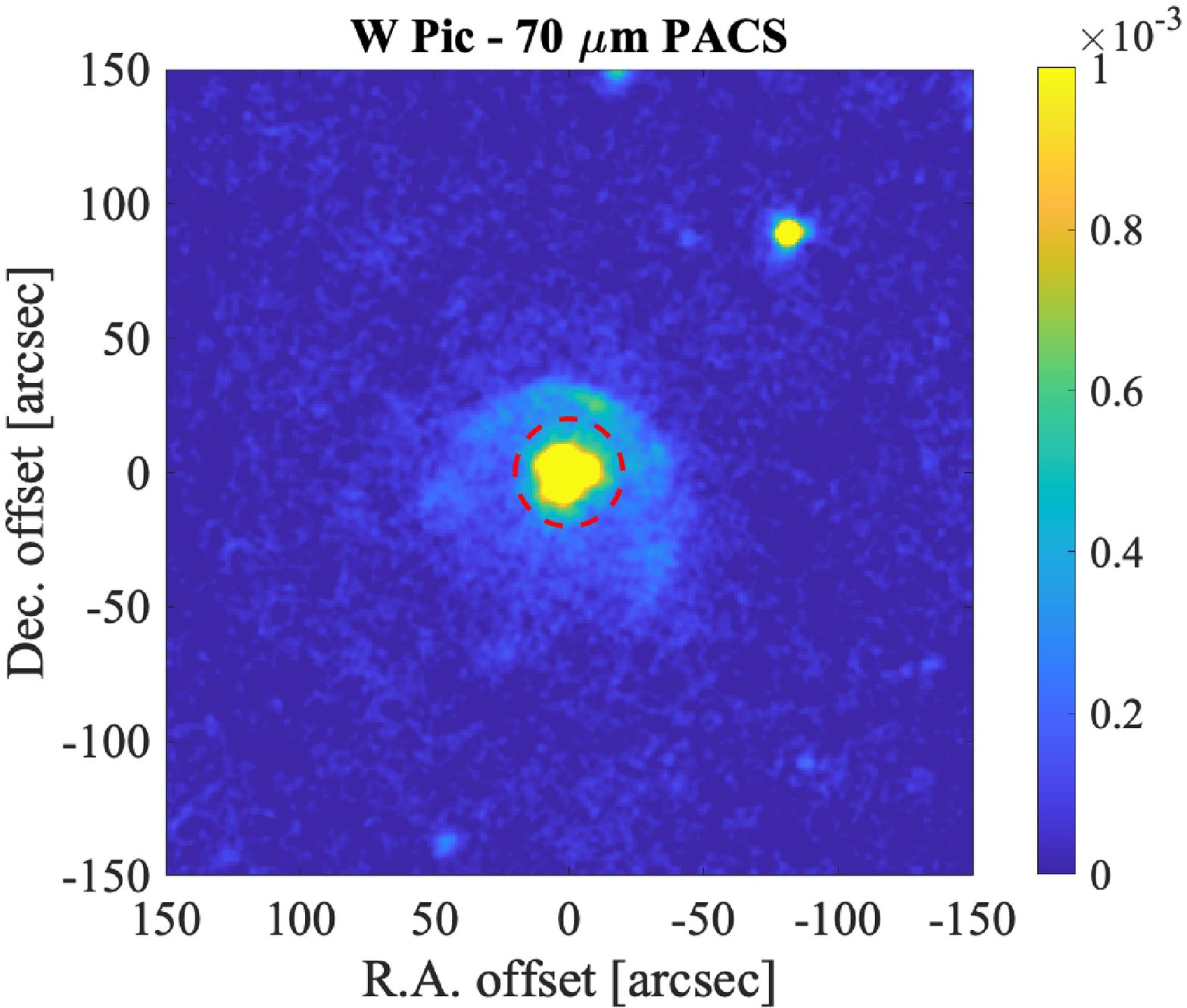}
\includegraphics[width=8cm]{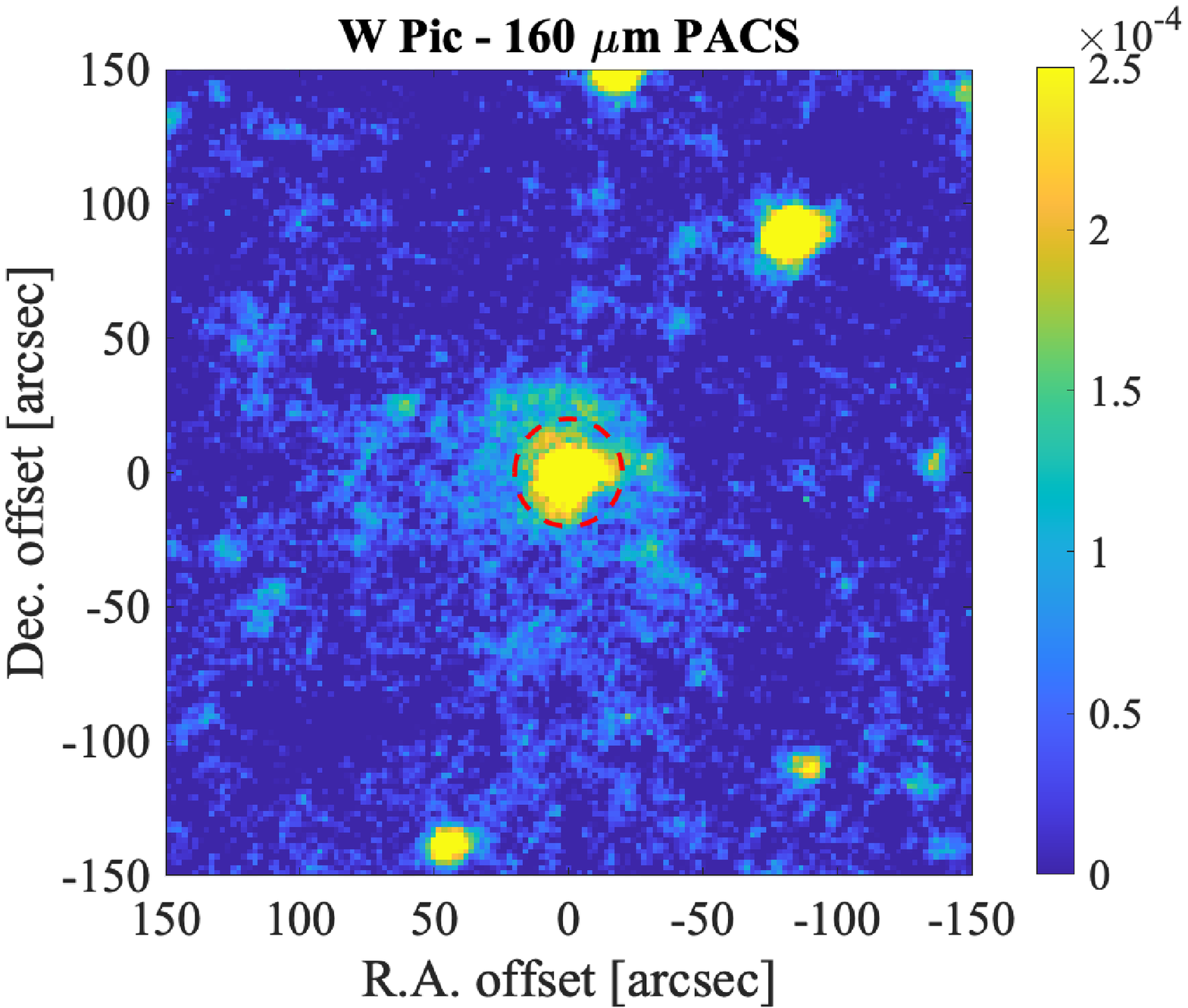}
\includegraphics[width=8cm]{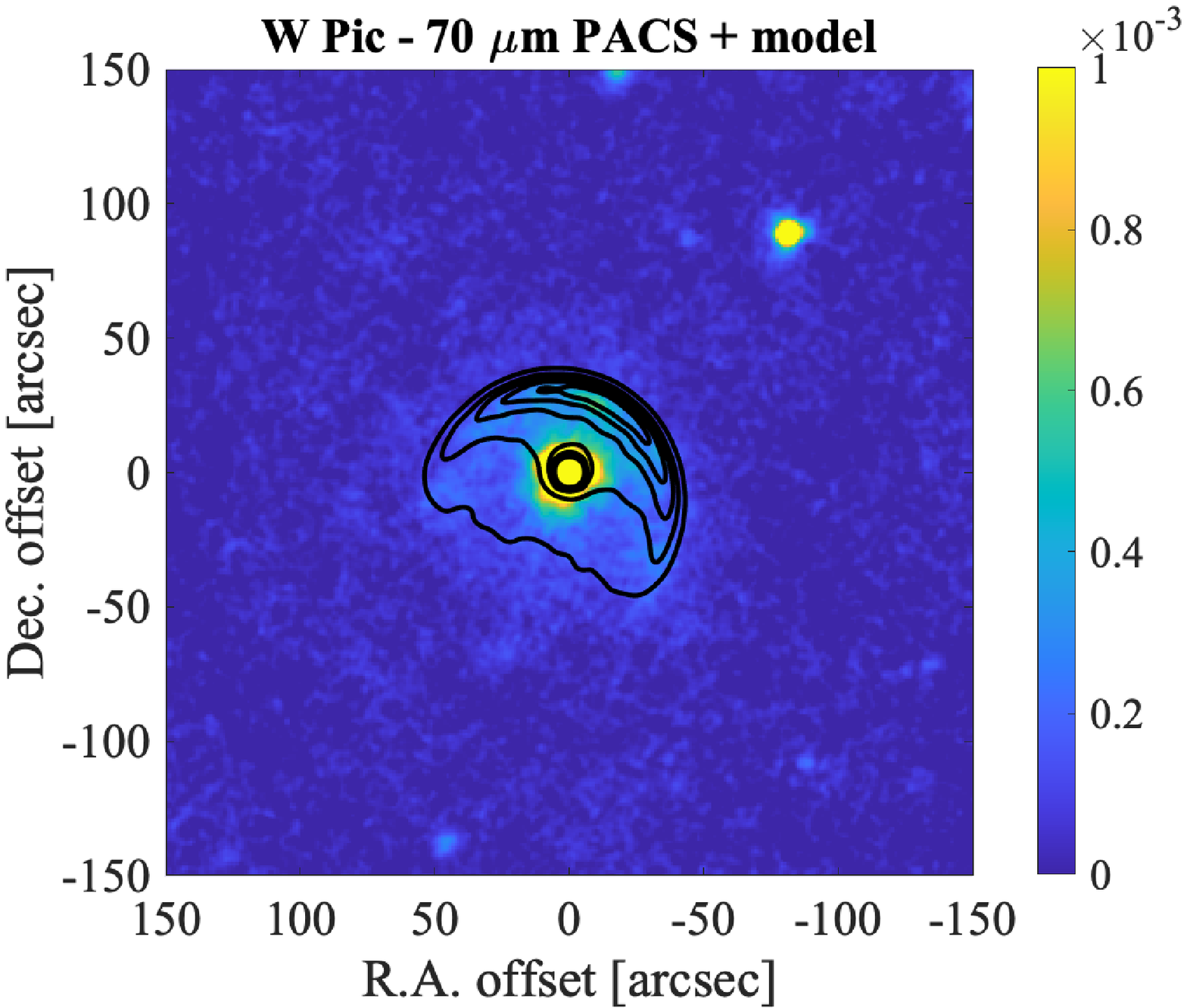}
\includegraphics[width=8cm]{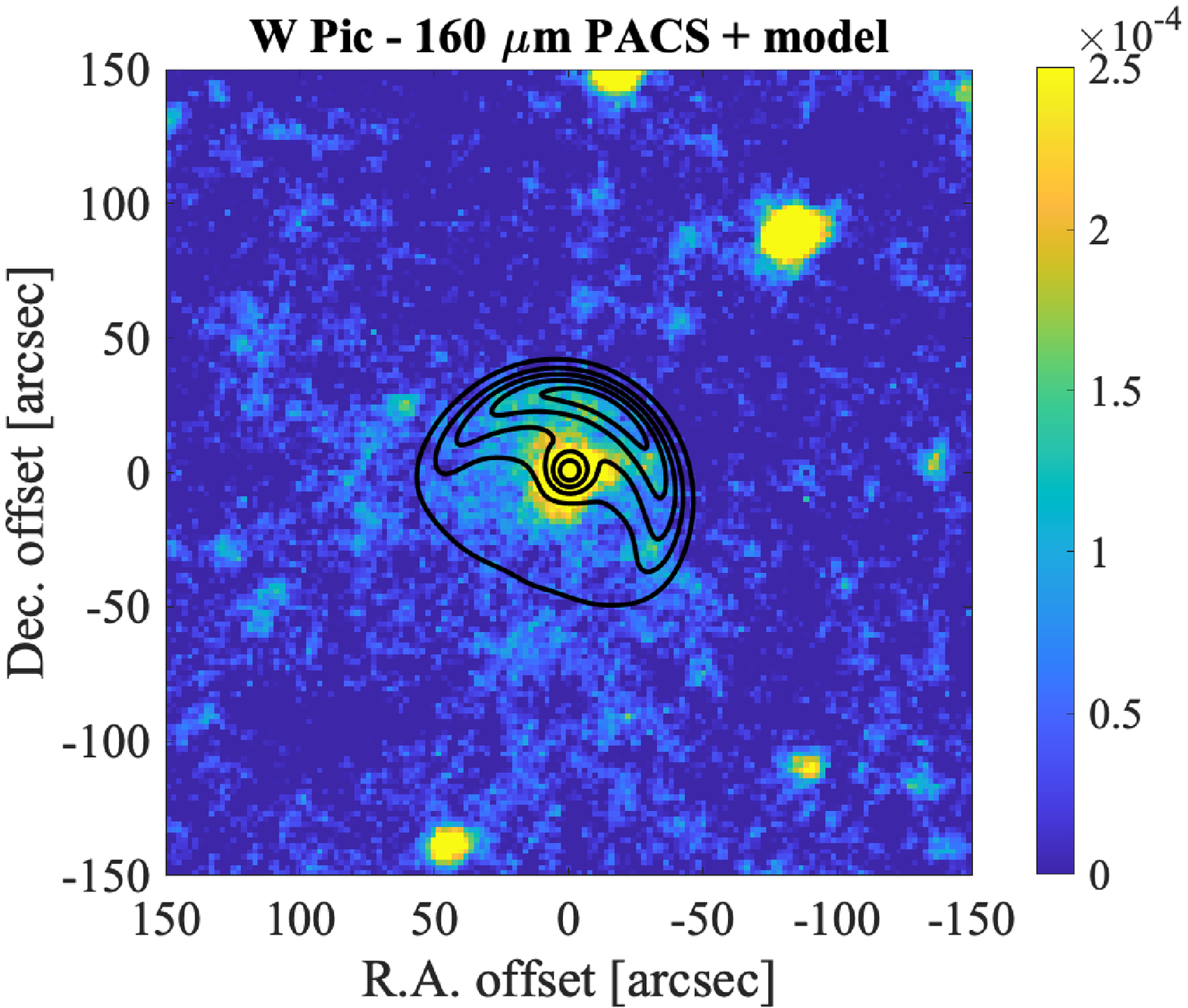}
\caption{W Pic: \emph{Top to bottom:} The Radmc3D model, the PACS image, and the PACS image with contours from the model. Images are for 70\,\micron~(left) and 160\,\micron~(right). Maximum contour levels are 0.68$\times10^{-3}$\,\Jyarcsec (70\,\micron) and 0.16$\times10^{-3}$\,\Jyarcsec (160\,\micron), respectively. Minimum contour levels are 10\% of maximum. The colour scale is in \Jyarcsec. The red dashed circle shows the mask used to measure the flux from the star and present-day mass-loss.}
\label{f:wpic}
\end{figure*}

\begin{figure*}
\centering
\includegraphics[width=8cm]{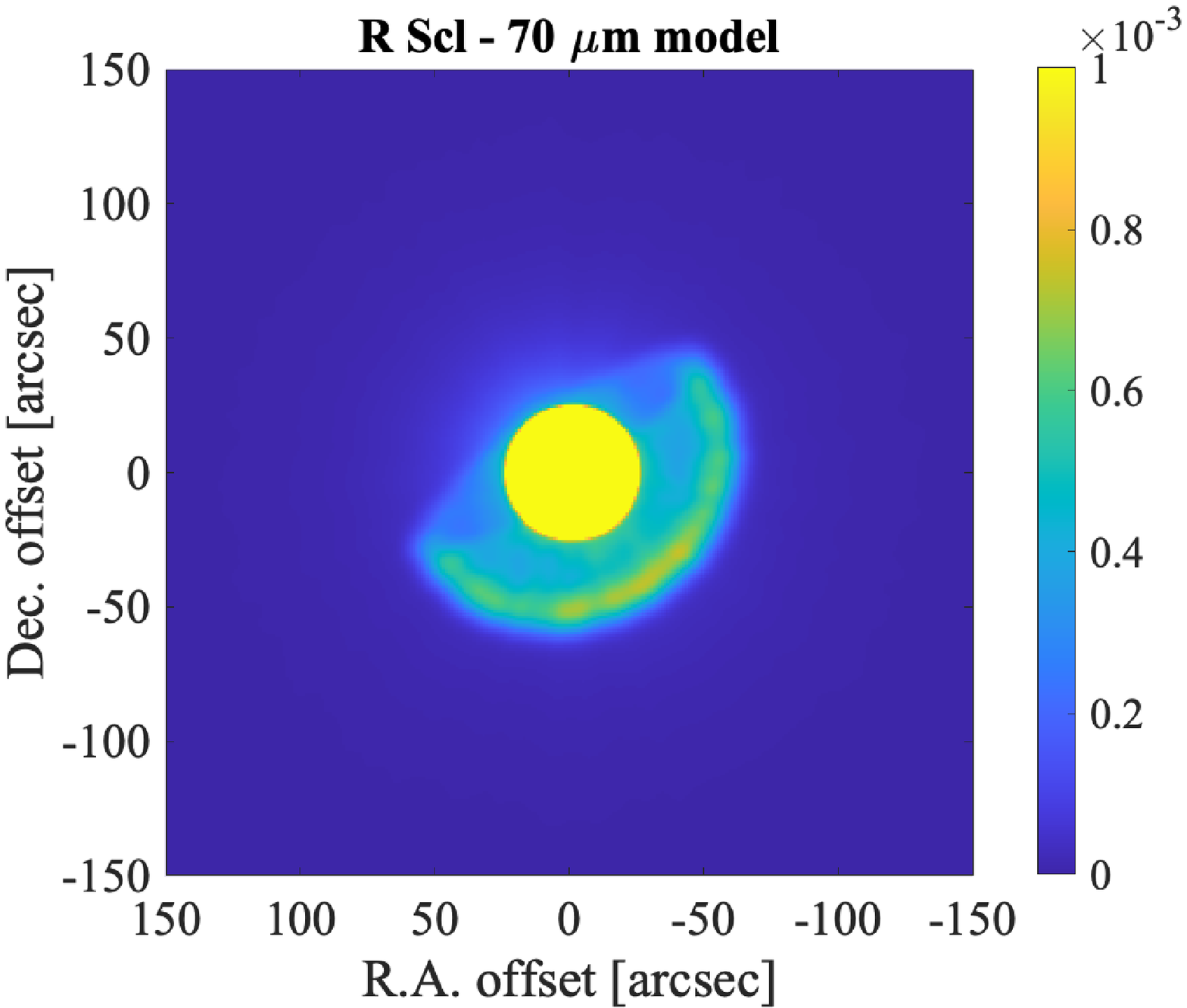}
\includegraphics[width=8cm]{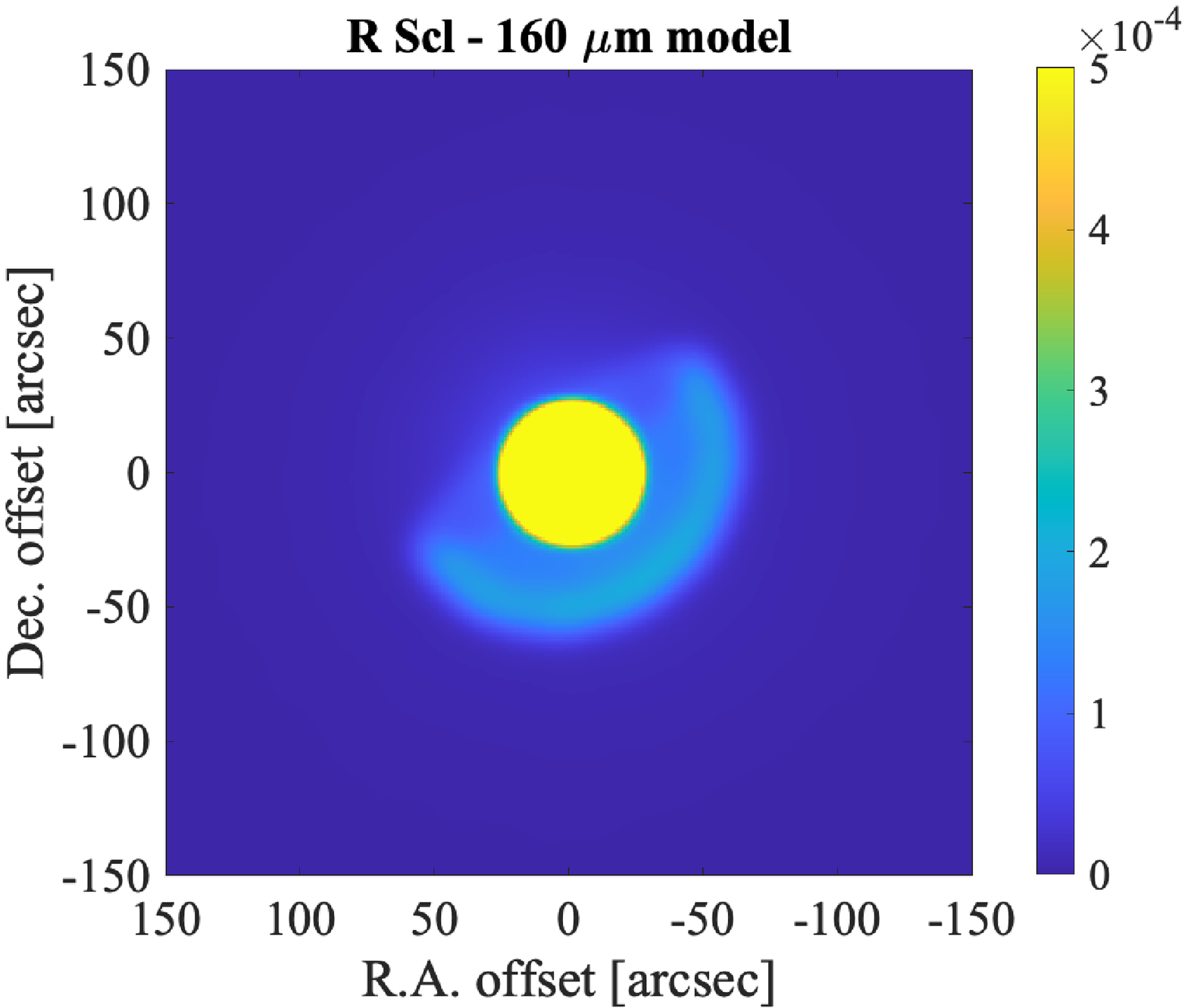}
\includegraphics[width=8cm]{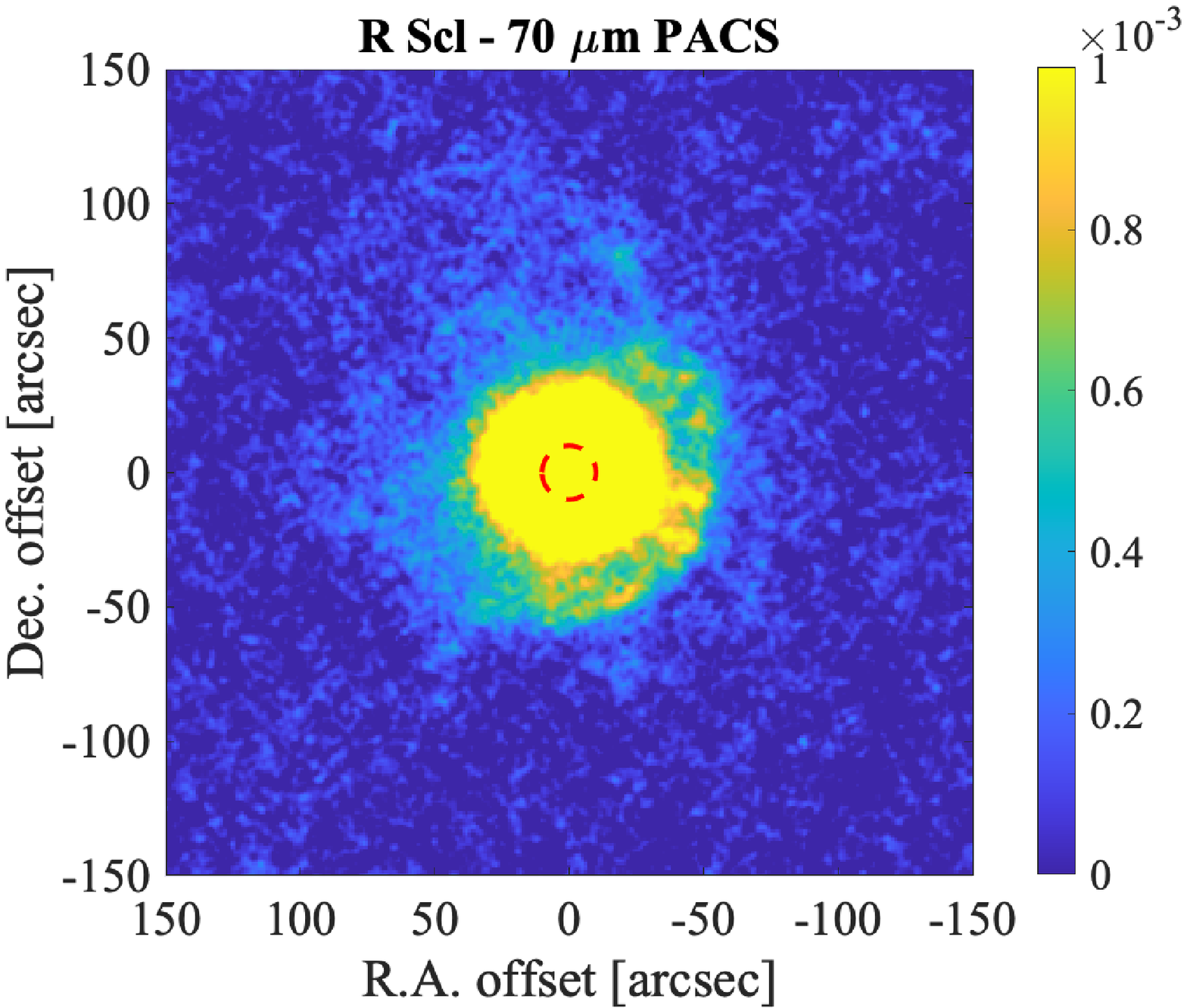}
\includegraphics[width=8cm]{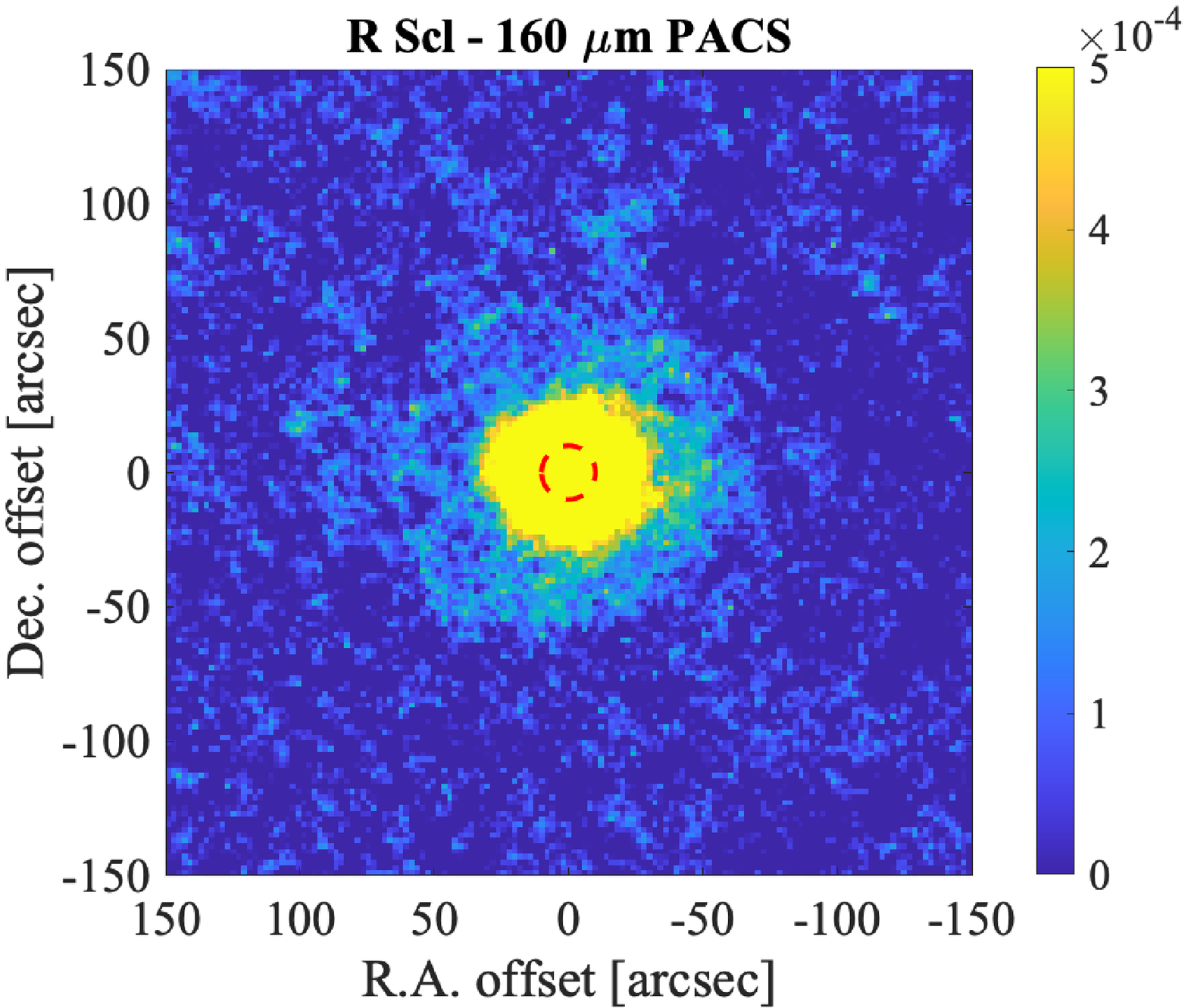}
\includegraphics[width=8cm]{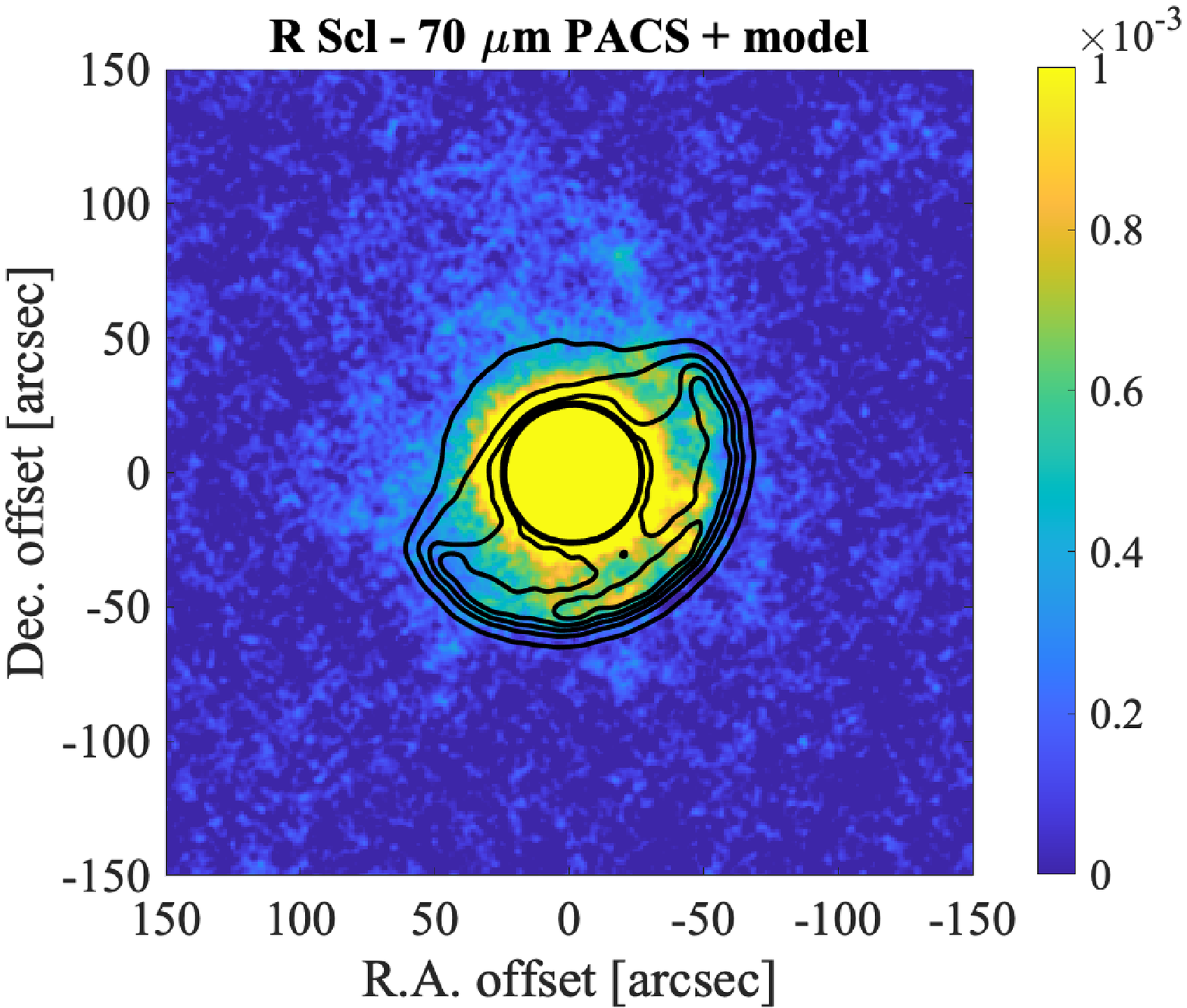}
\includegraphics[width=8cm]{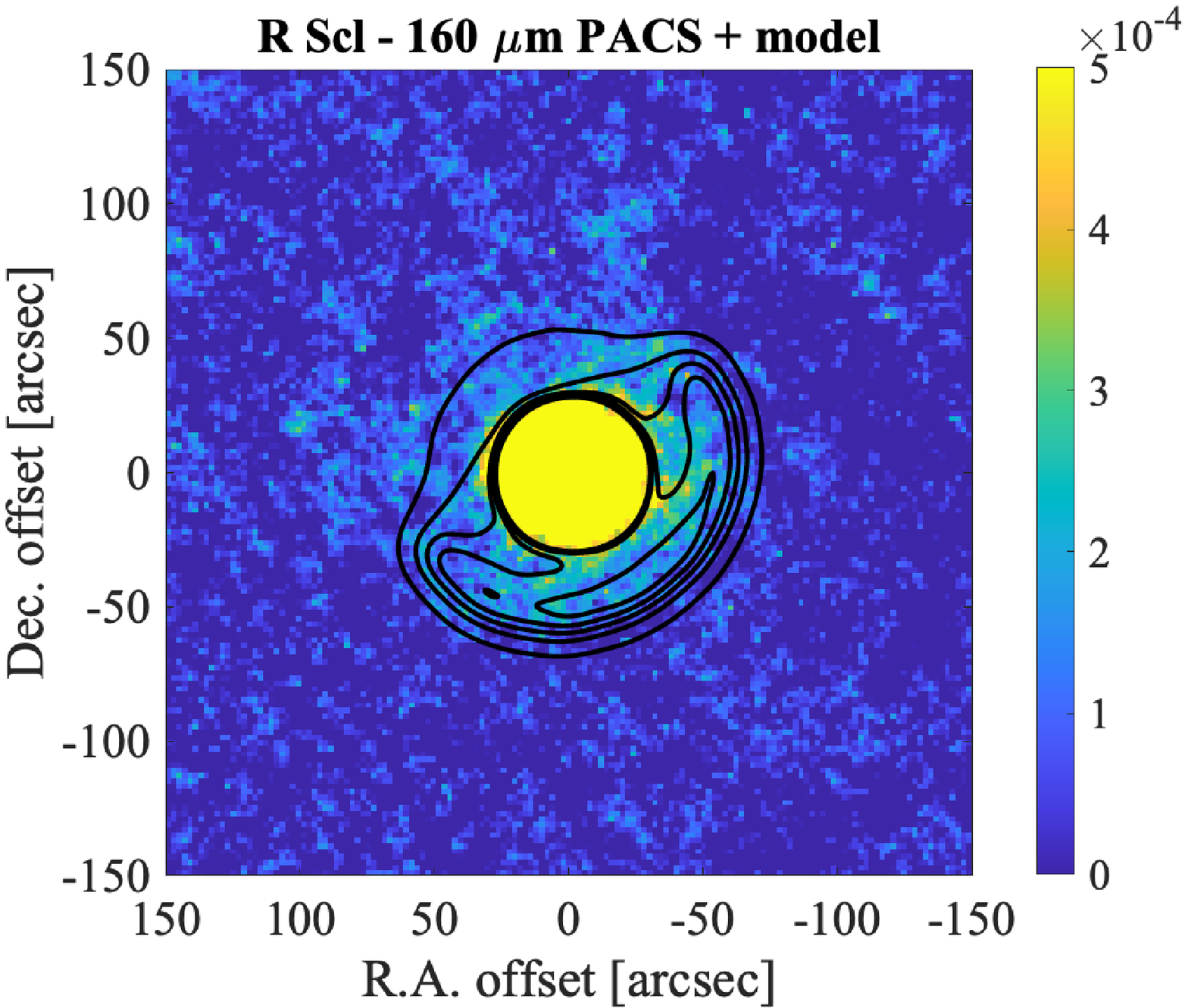}
\caption{R Scl: \emph{Top to bottom:} The Radmc3D model, the PACS image, and the PACS image with contours from the model. Images are for 70\,\micron~(left) and 160\,\micron~(right). Maximum contour levels are 0.7$\times10^{-3}$\,\Jyarcsec (70\,\micron) and 0.2$\times10^{-3}$\,\Jyarcsec (160\,\micron), respectively. Minimum contour levels are 10\% of maximum. The colour scale is in \Jyarcsec. The red dashed circle shows the mask used to measure the flux from the star and present-day mass-loss.}
\label{f:rscl}
\end{figure*}

\begin{figure*}
\centering
\includegraphics[width=8cm]{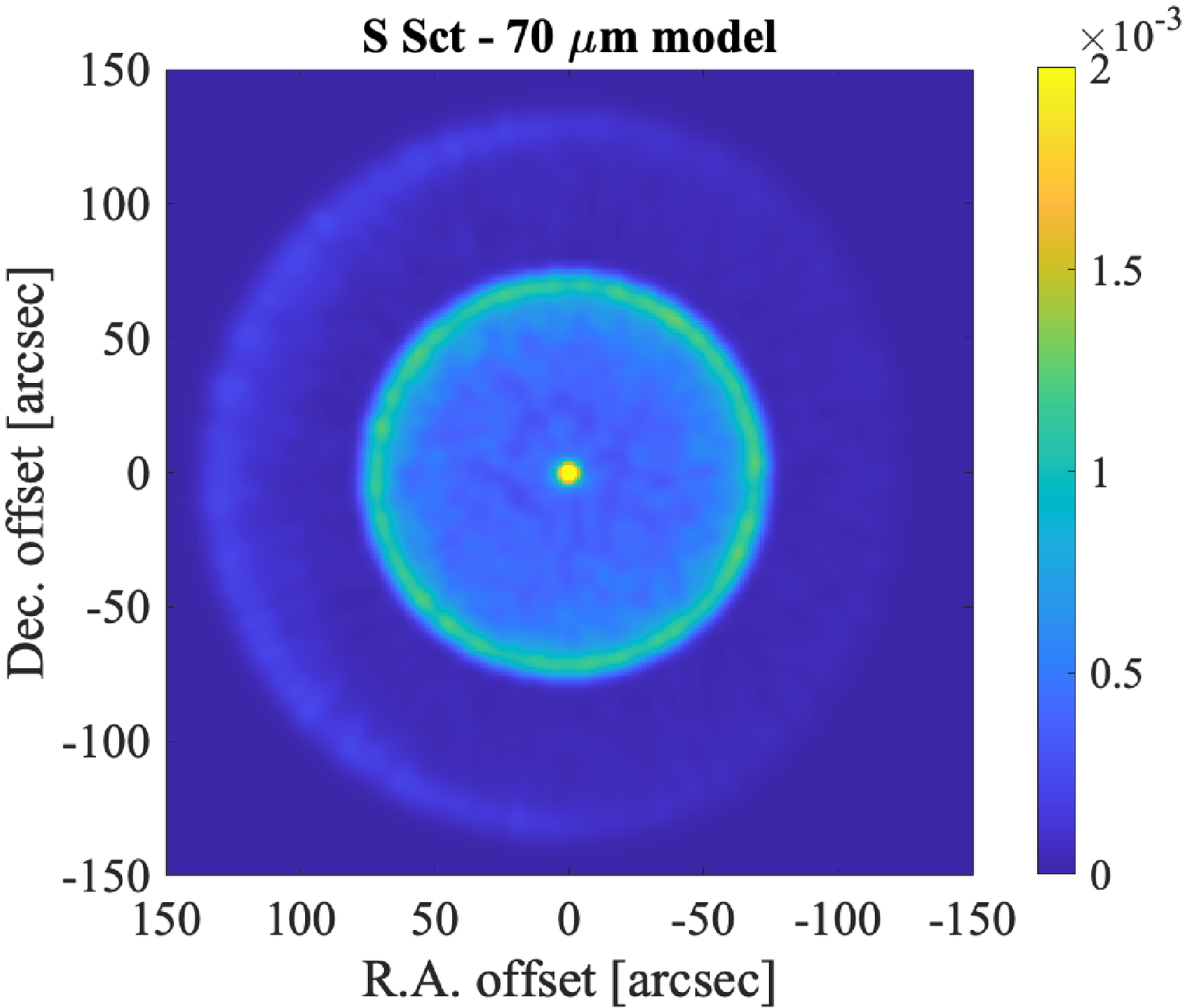}
\includegraphics[width=8cm]{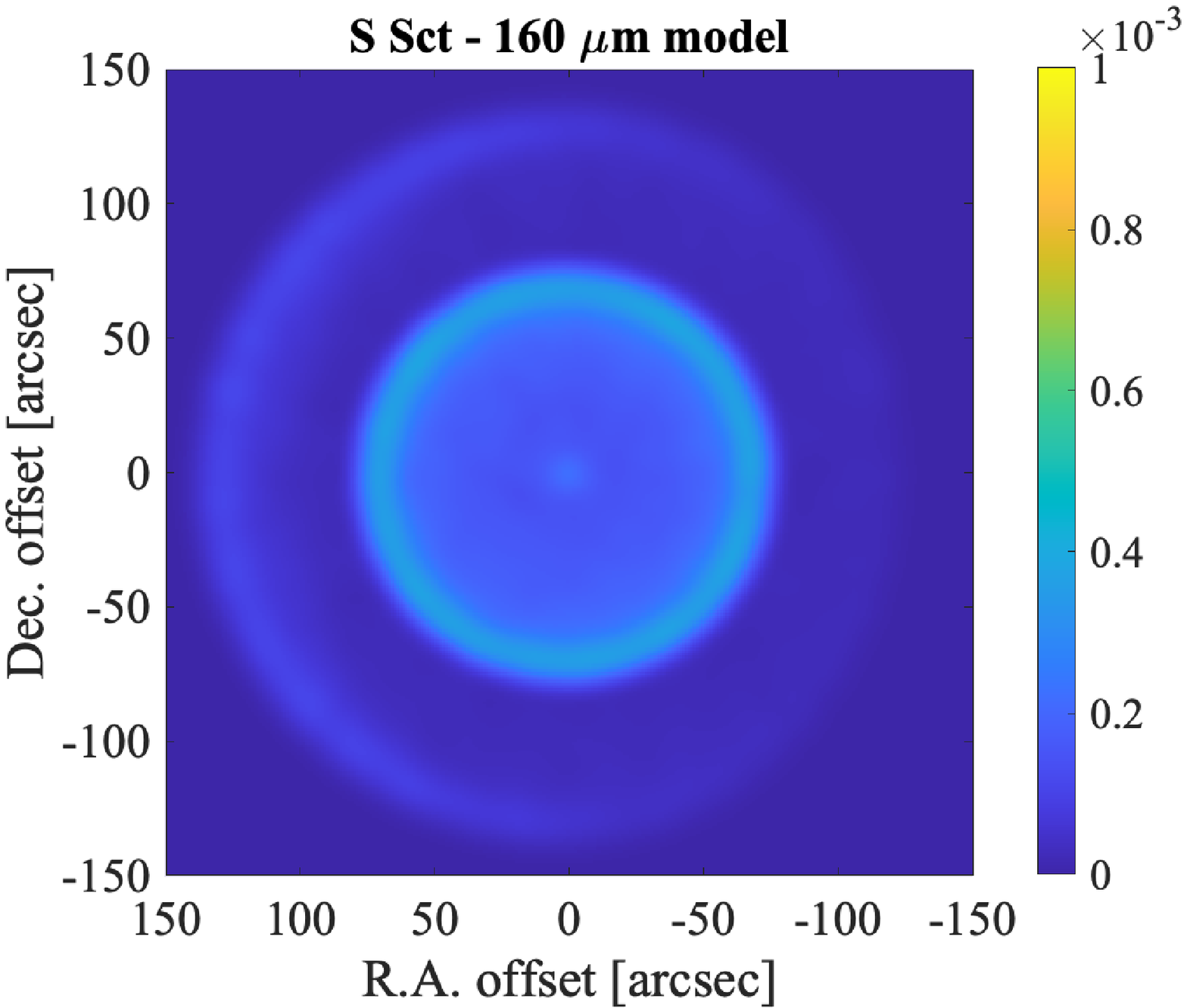}
\includegraphics[width=8cm]{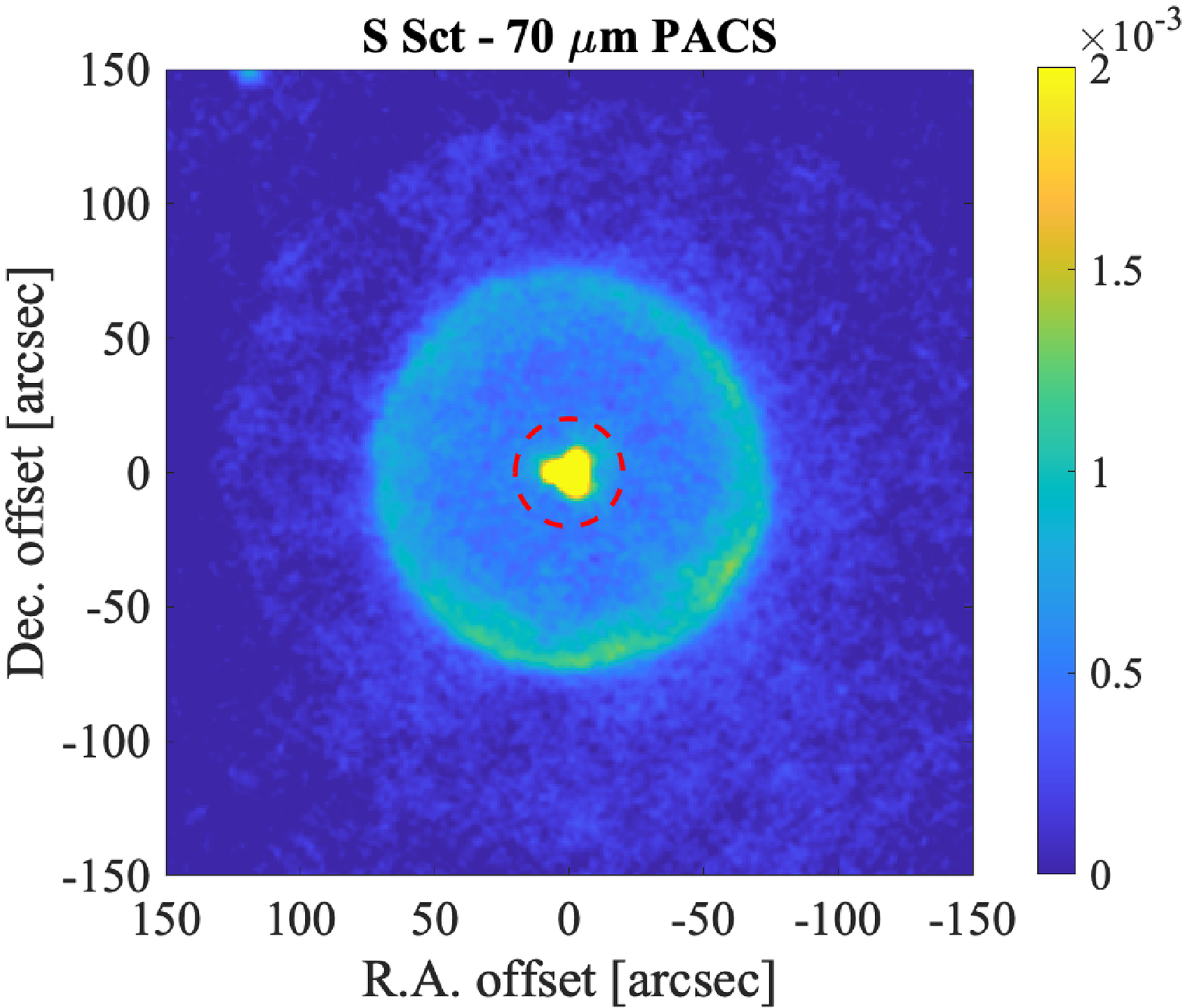}
\includegraphics[width=8cm]{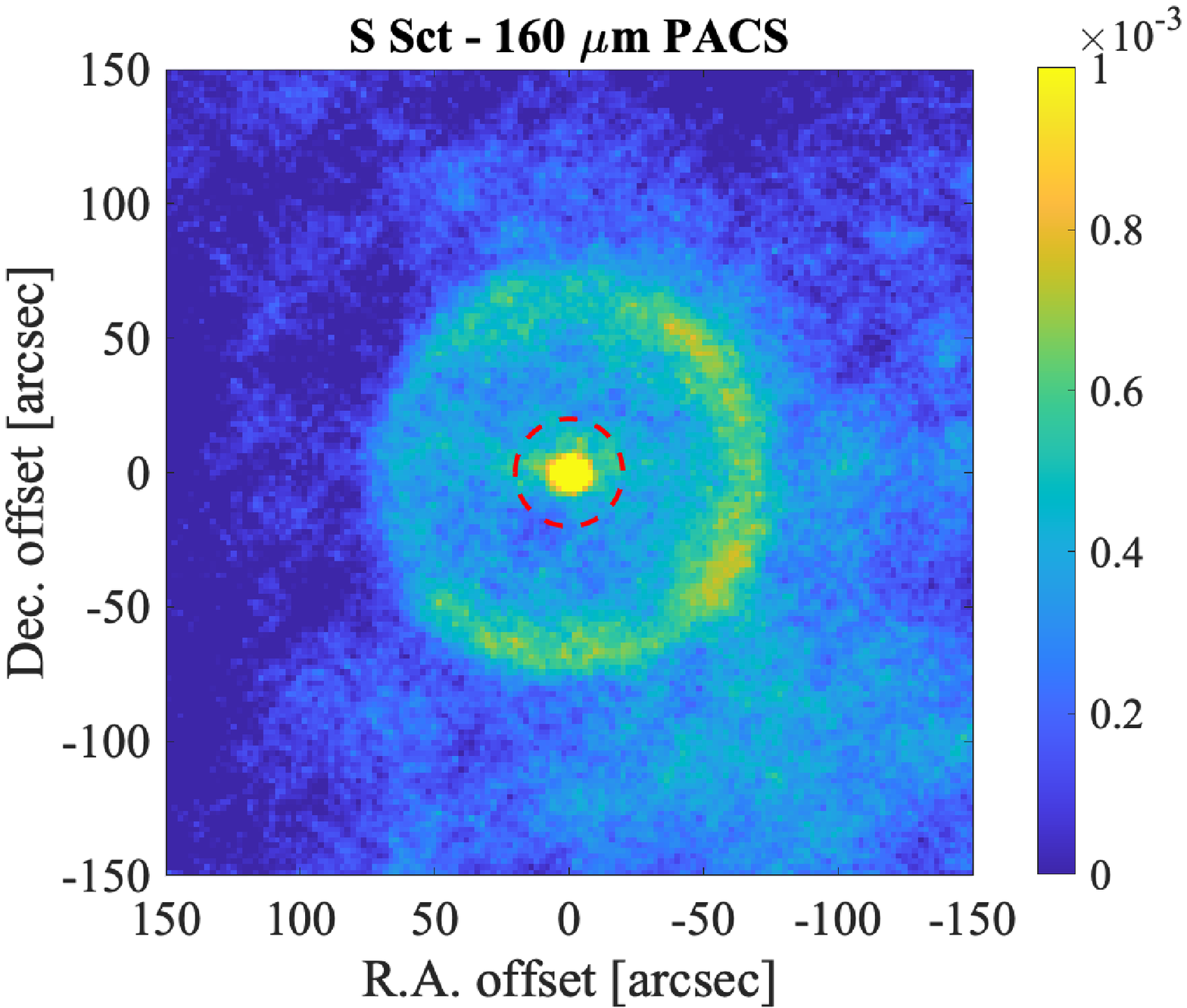}
\includegraphics[width=8cm]{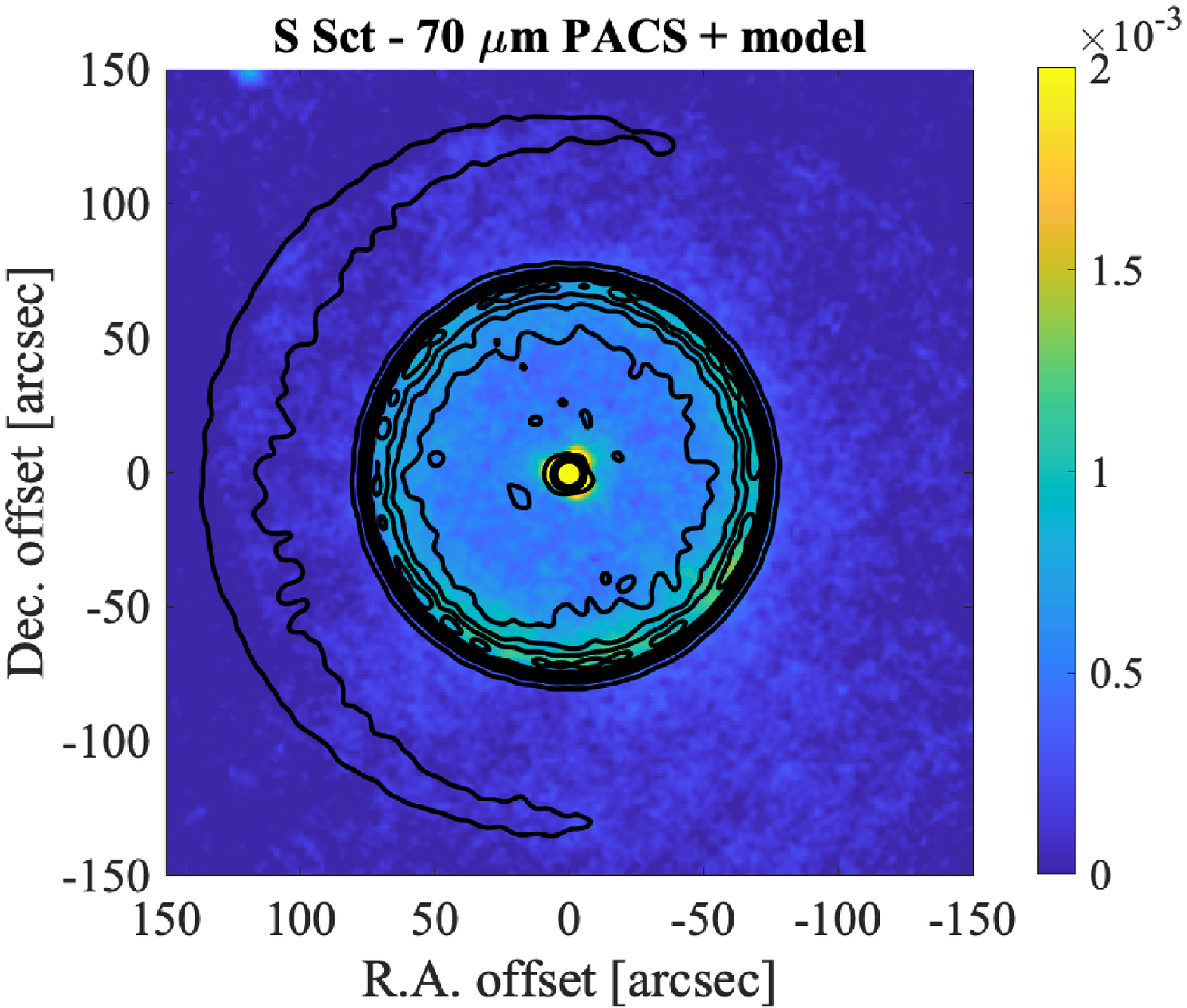}
\includegraphics[width=8cm]{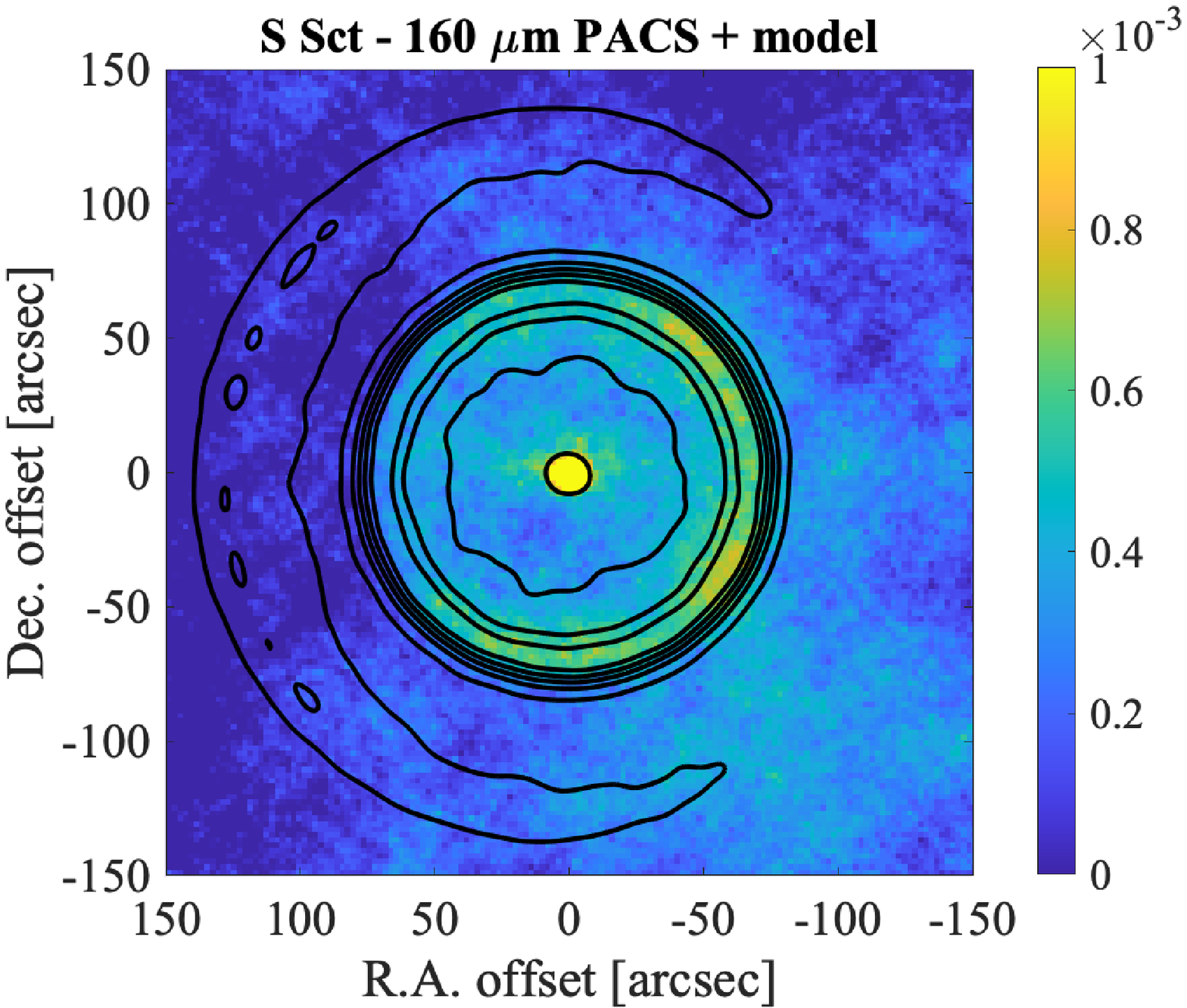}
\caption{S Sct: \emph{Top to bottom:} The Radmc3D model, the PACS image, and the PACS image with contours from the model. Images are for 70\,\micron~(left) and 160\,\micron~(right). Maximum contour levels are 1$\times10^{-3}$\,\Jyarcsec (70\,\micron) and 0.35$\times10^{-3}$\,\Jyarcsec (160\,\micron), respectively. Minimum contour levels are 10\% of maximum. The colour scale is in \Jyarcsec. The red dashed circle shows the mask used to measure the flux from the star and present-day mass-loss.}
\label{f:ssct}
\end{figure*}

\begin{figure*}
\centering
\includegraphics[width=8cm]{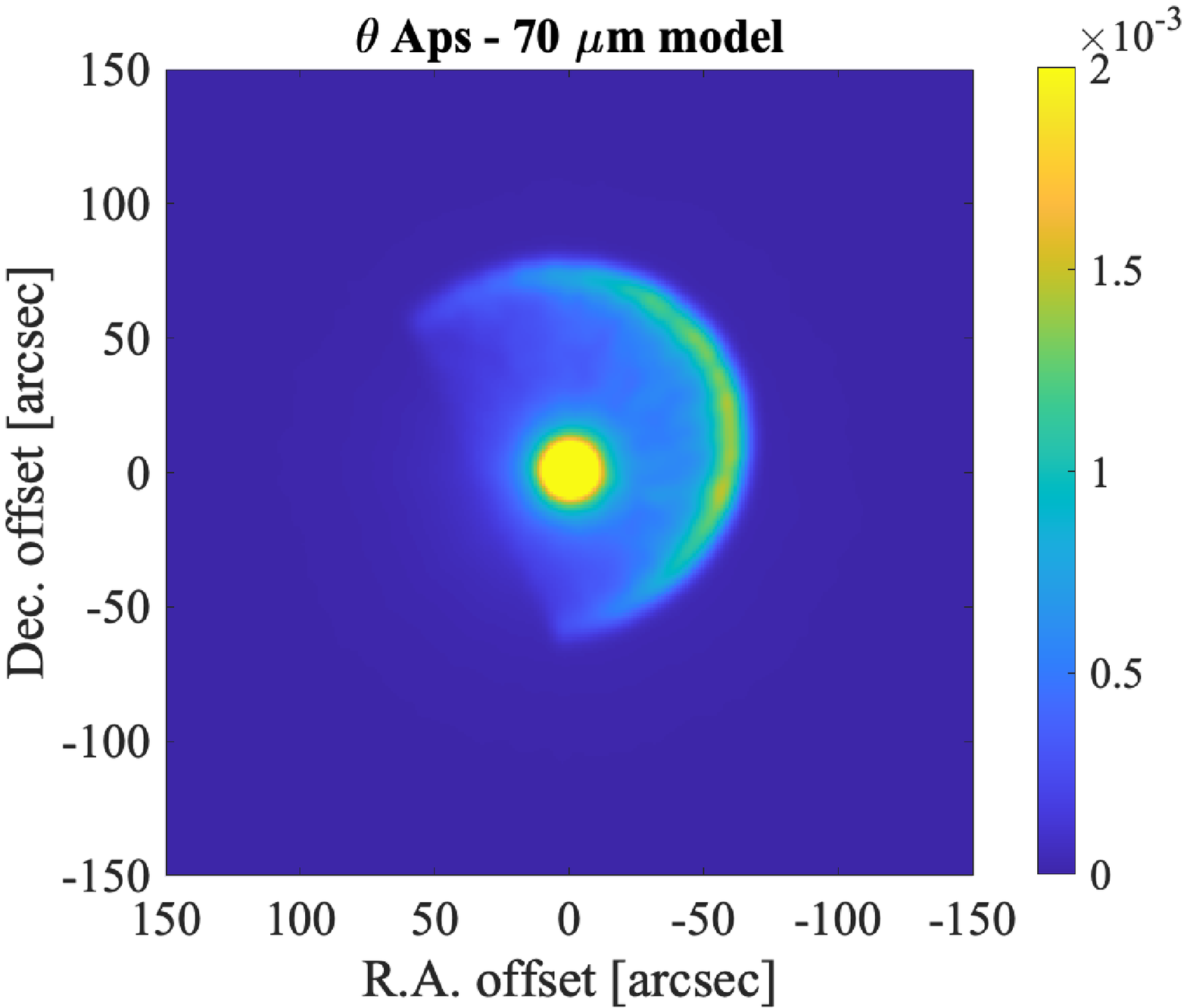}
\includegraphics[width=8cm]{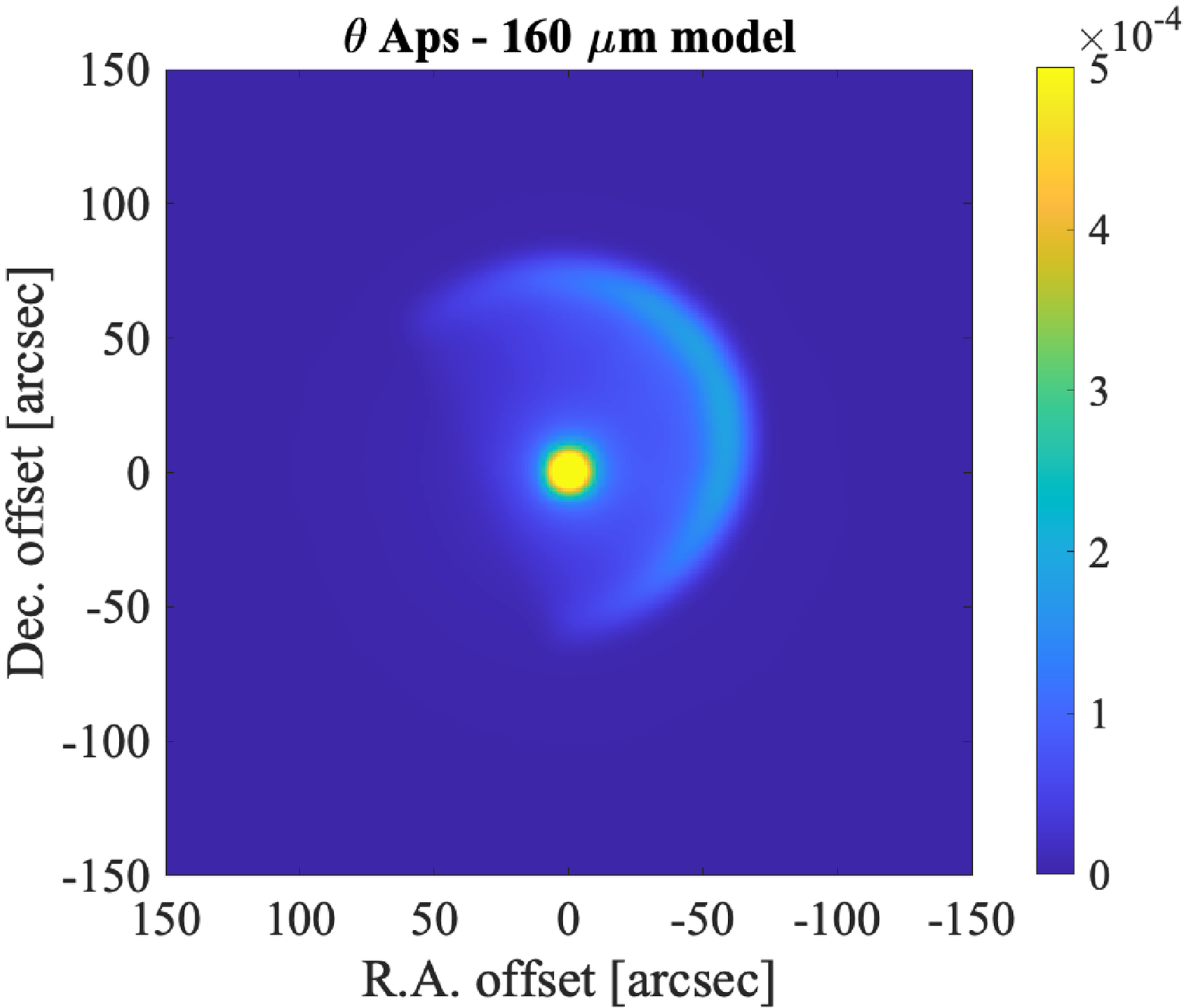}
\includegraphics[width=8cm]{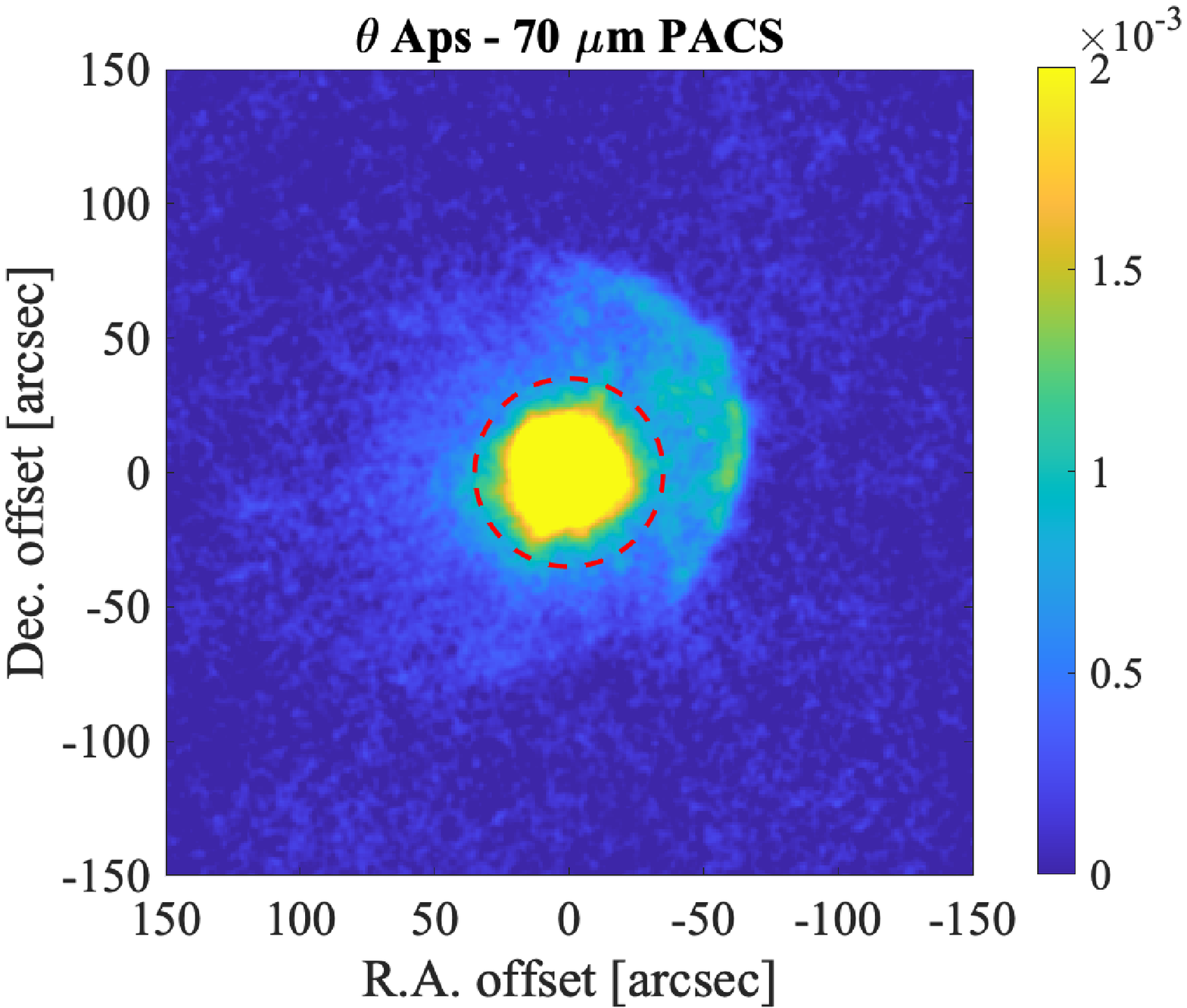}
\includegraphics[width=8cm]{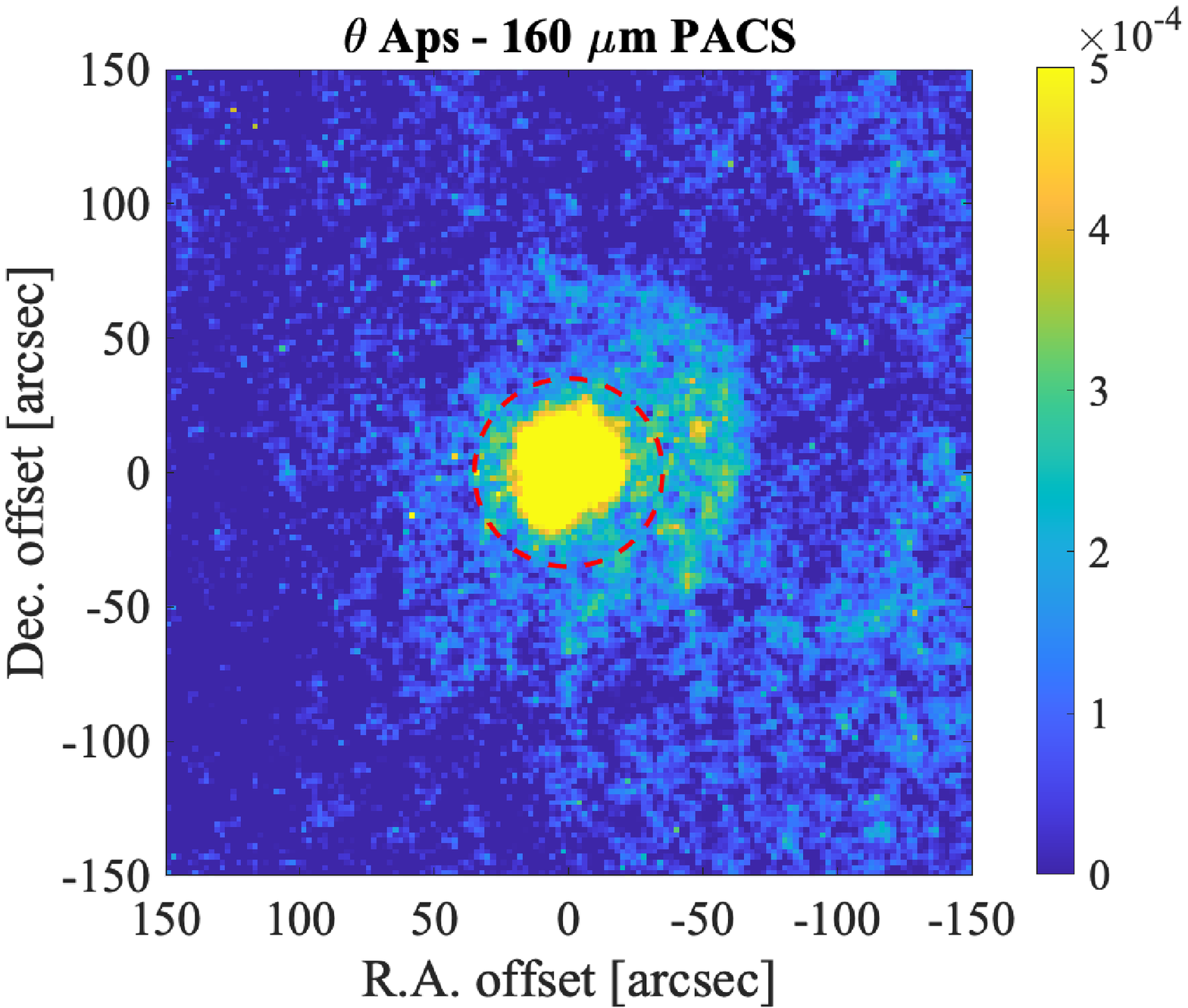}
\includegraphics[width=8cm]{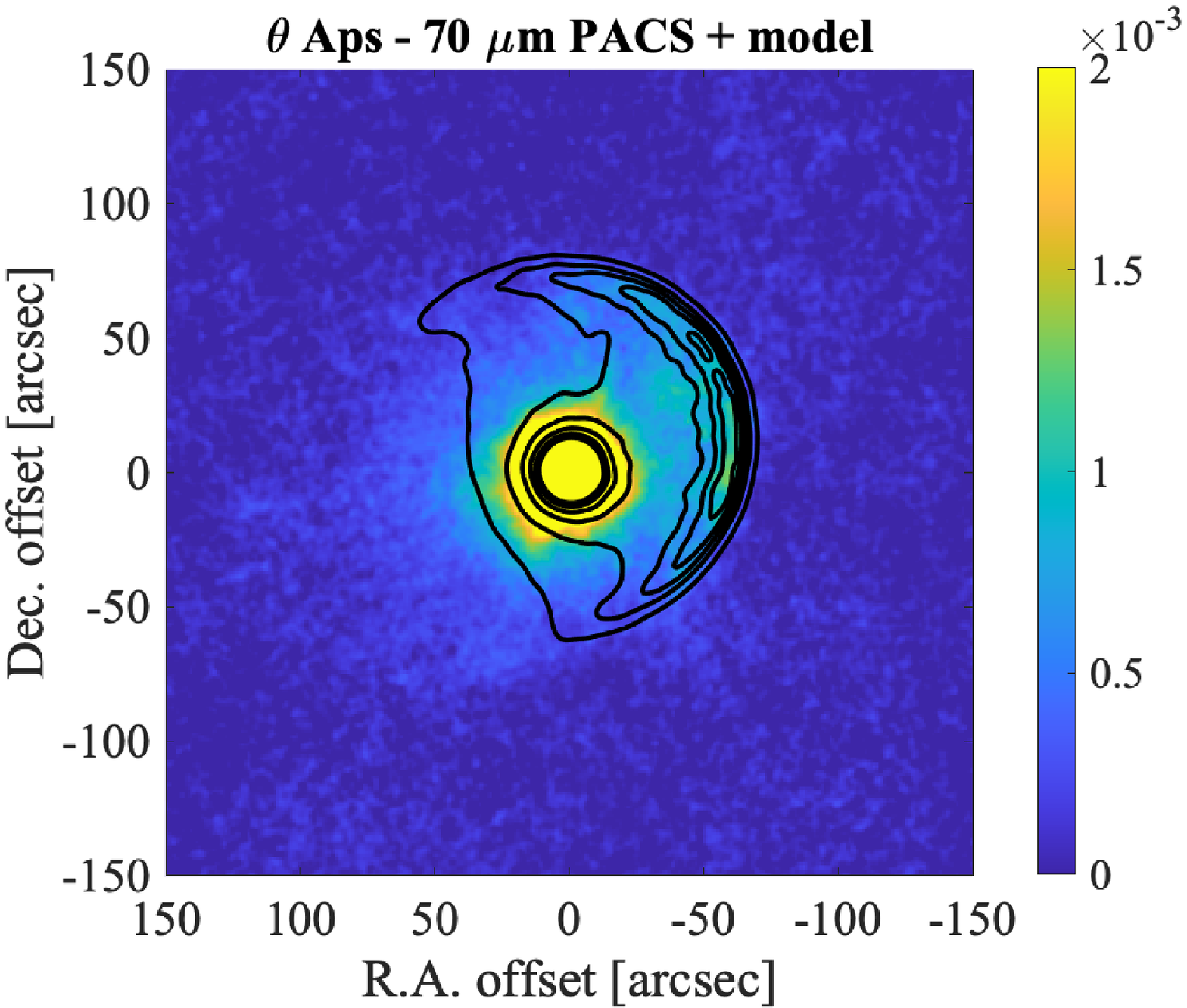}
\includegraphics[width=8cm]{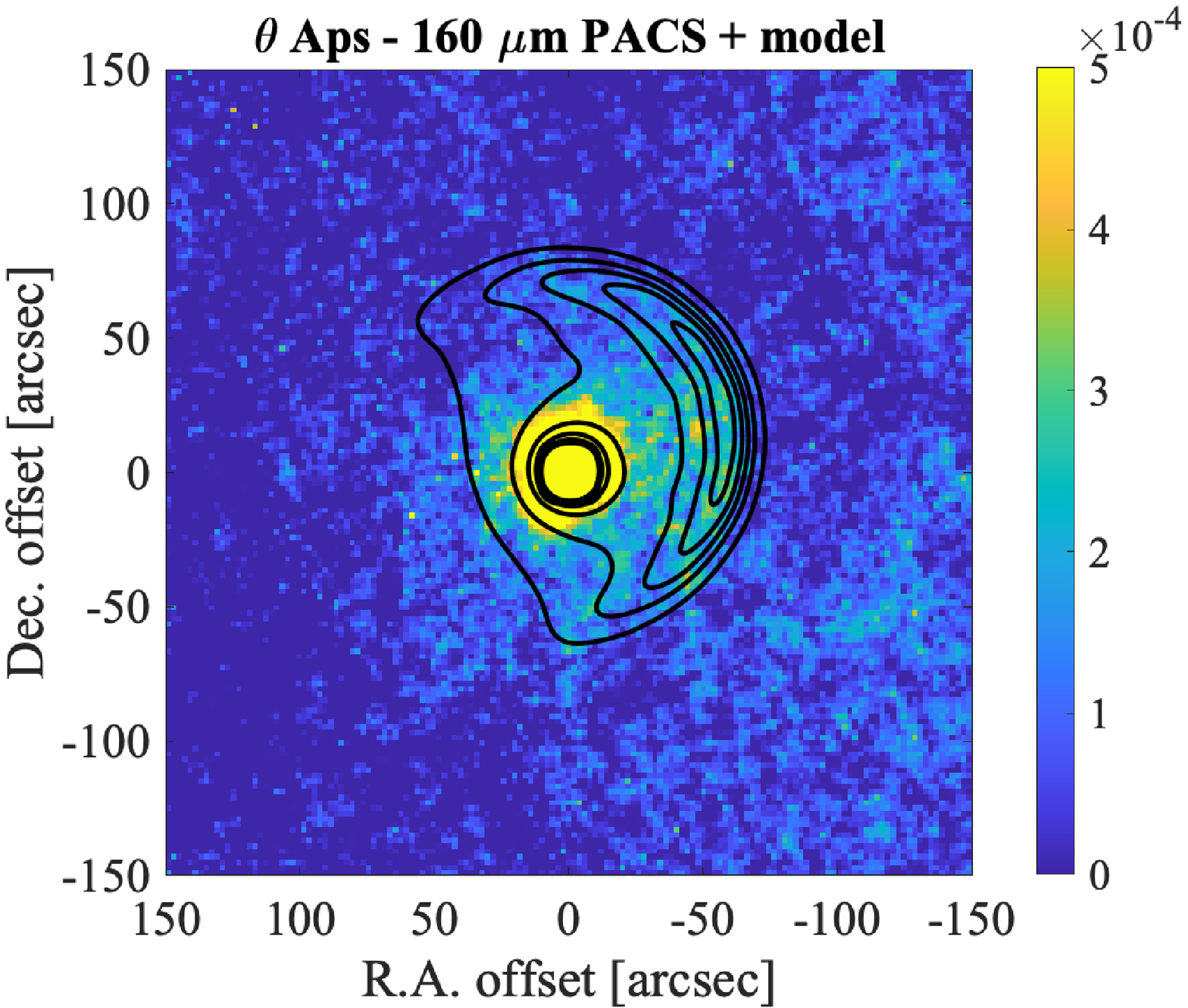}
\caption{$\theta$ Aps: \emph{Top to bottom:} The Radmc3D model, the PACS image, and the PACS image with contours from the model. Images are for 70\,\micron~(left) and 160\,\micron~(right). Maximum contour levels are 1.4$\times10^{-3}$\,\Jyarcsec (70\,\micron) and 0.19$\times10^{-3}$\,\Jyarcsec (160\,\micron), respectively. Minimum contour levels are 10\% of maximum. The colour scale is in \Jyarcsec. The red dashed circle shows the mask used to measure the flux from the star and present-day mass-loss.}
\label{f:thetaaps}
\end{figure*}

\begin{figure*}
\centering
\includegraphics[width=8cm]{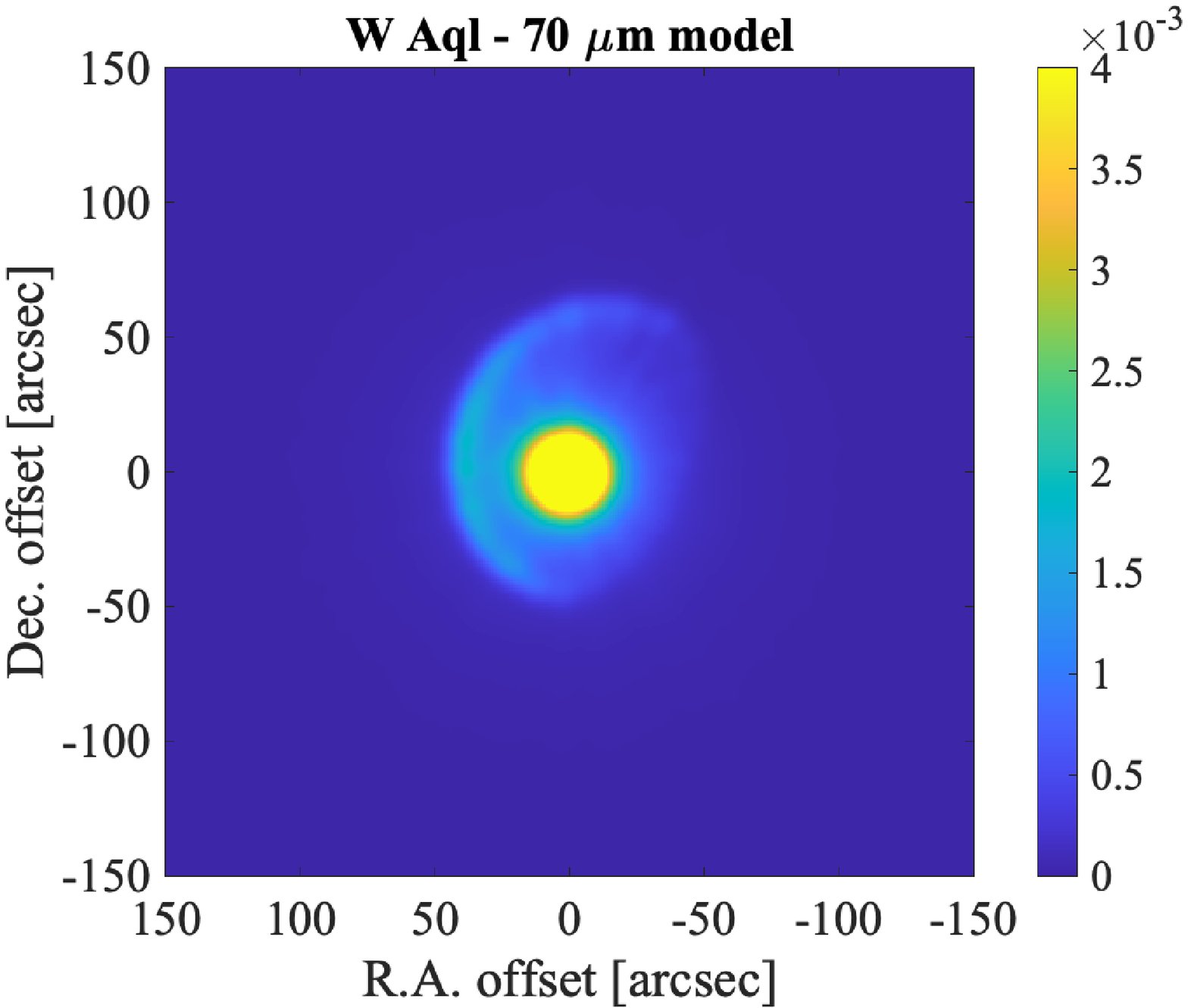}
\includegraphics[width=8cm]{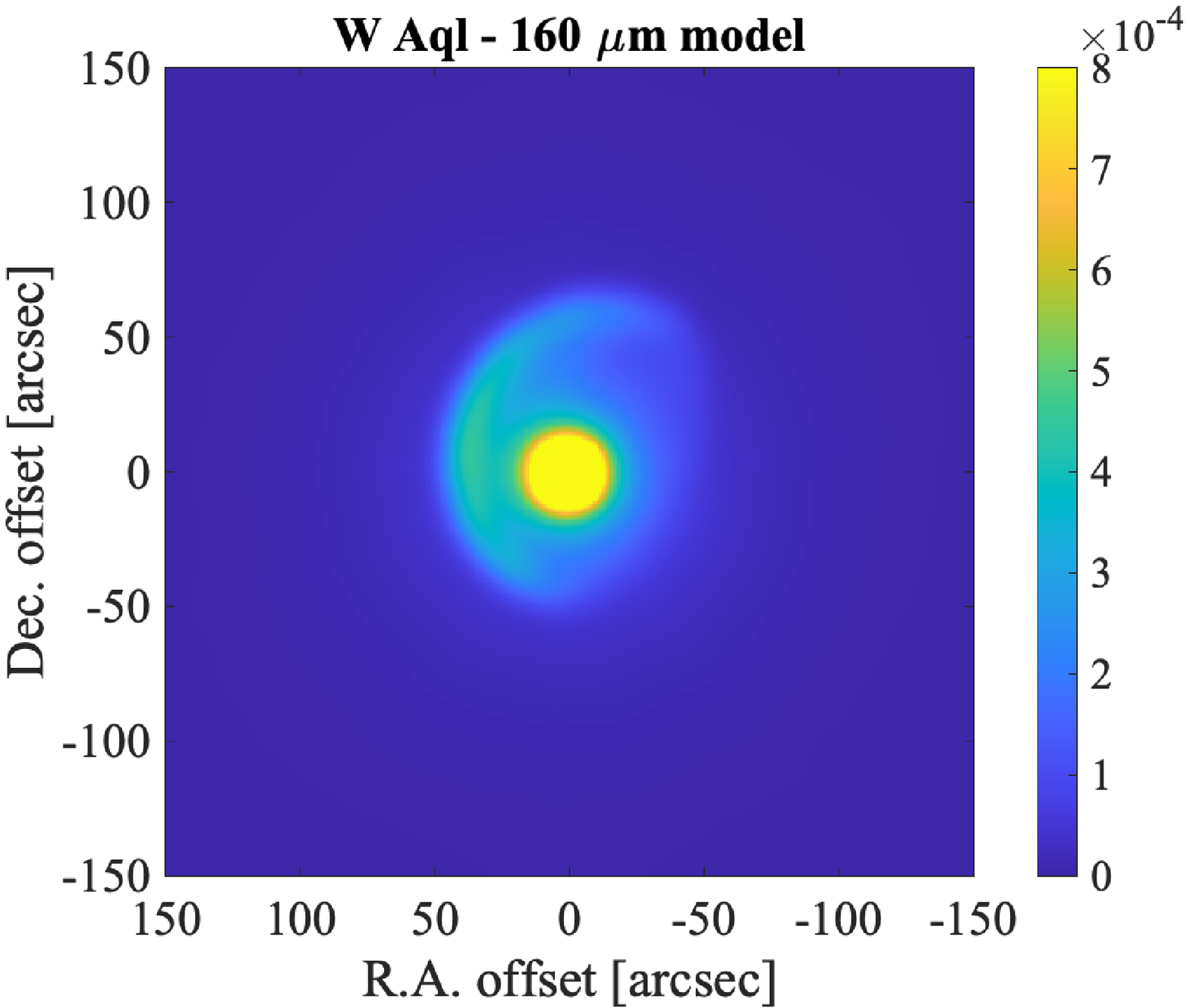}
\includegraphics[width=8cm]{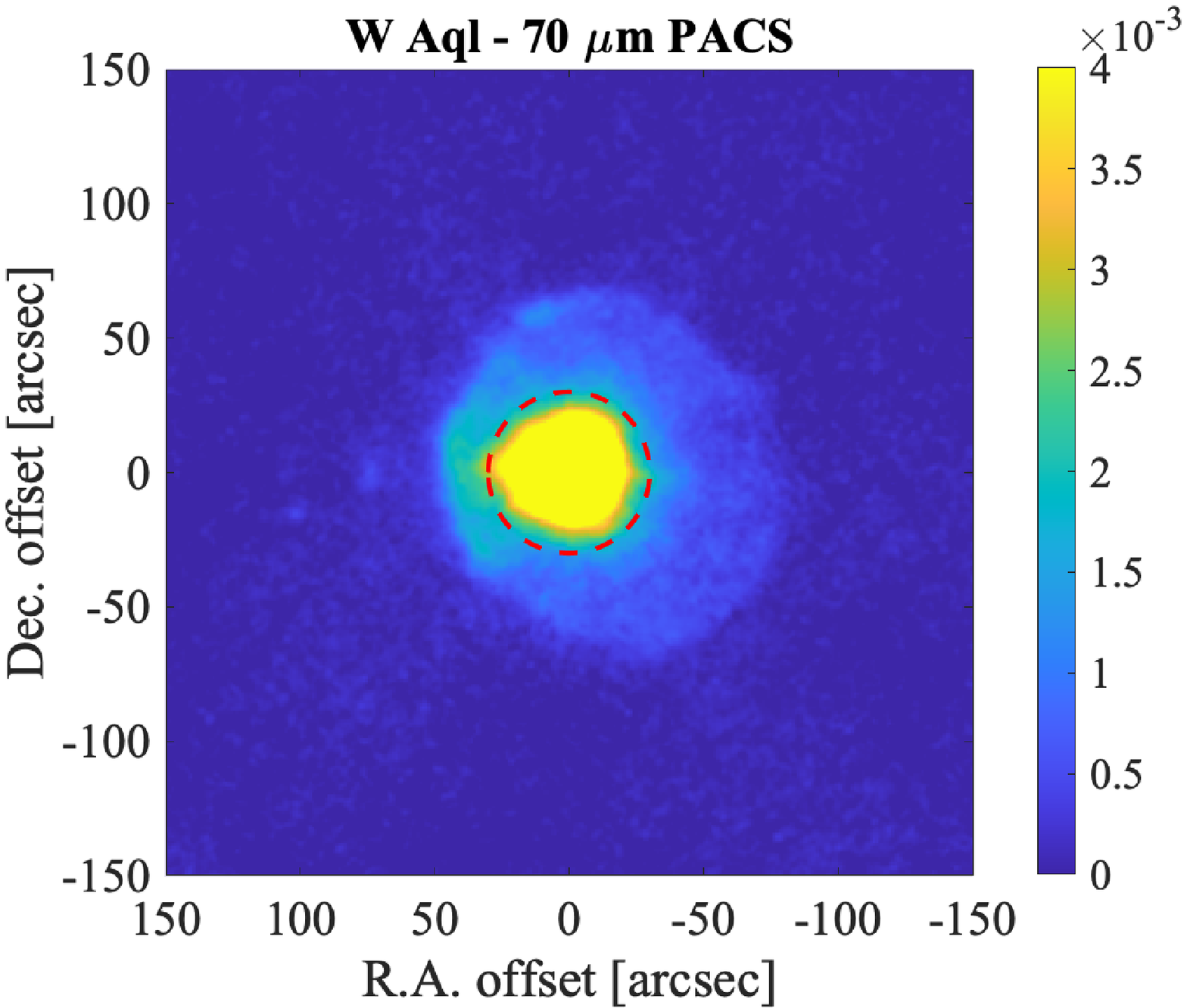}
\includegraphics[width=8cm]{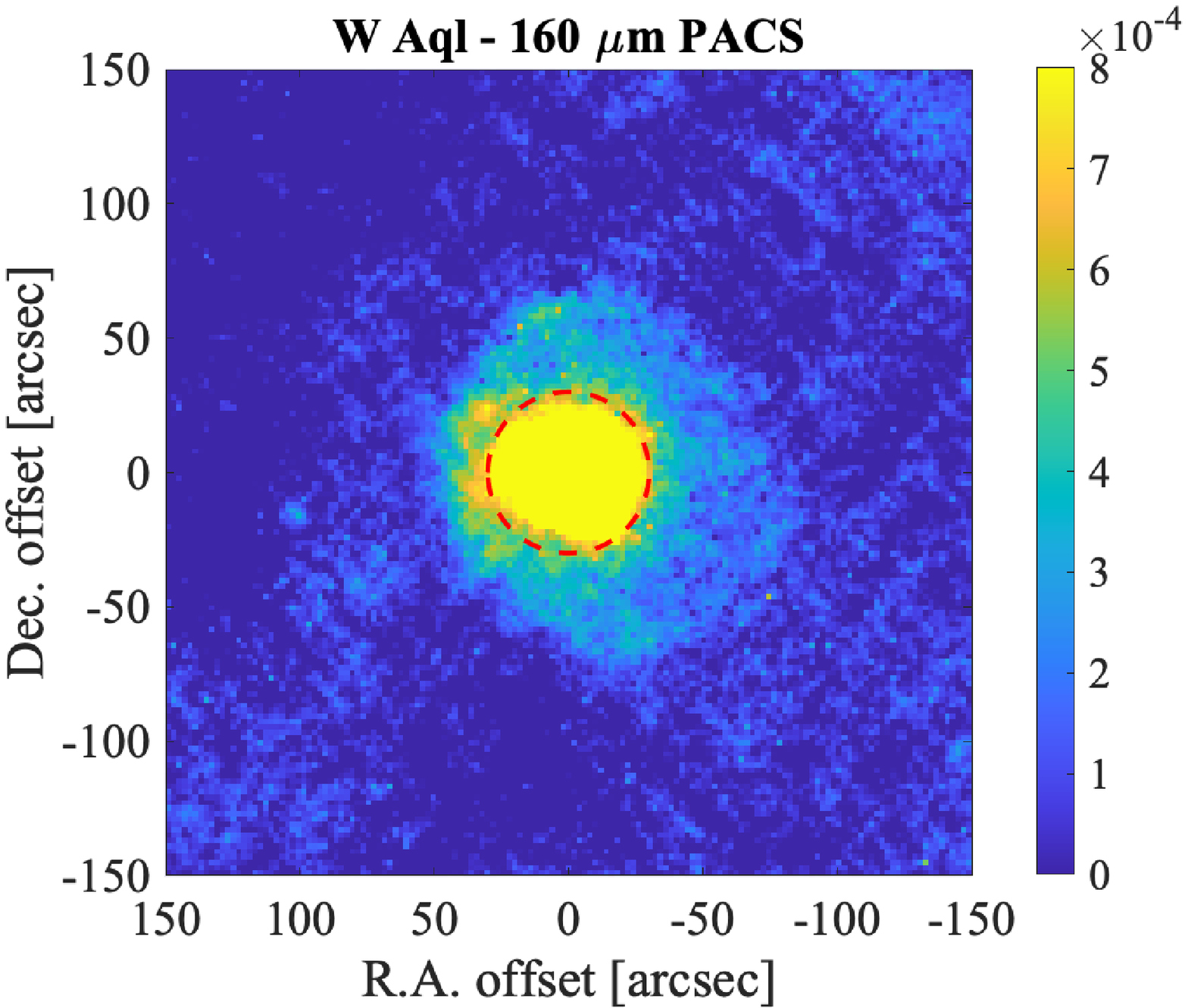}
\includegraphics[width=8cm]{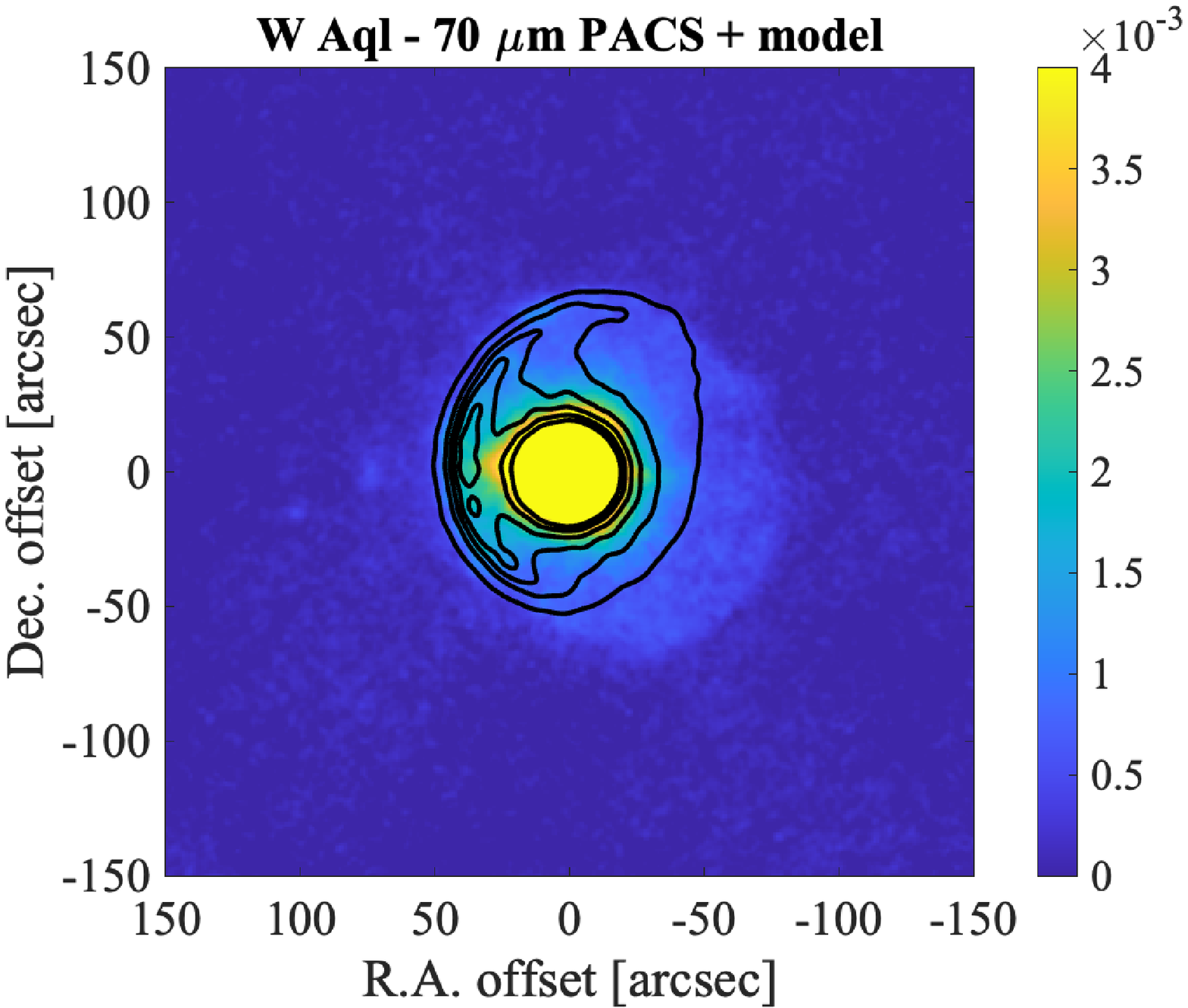}
\includegraphics[width=8cm]{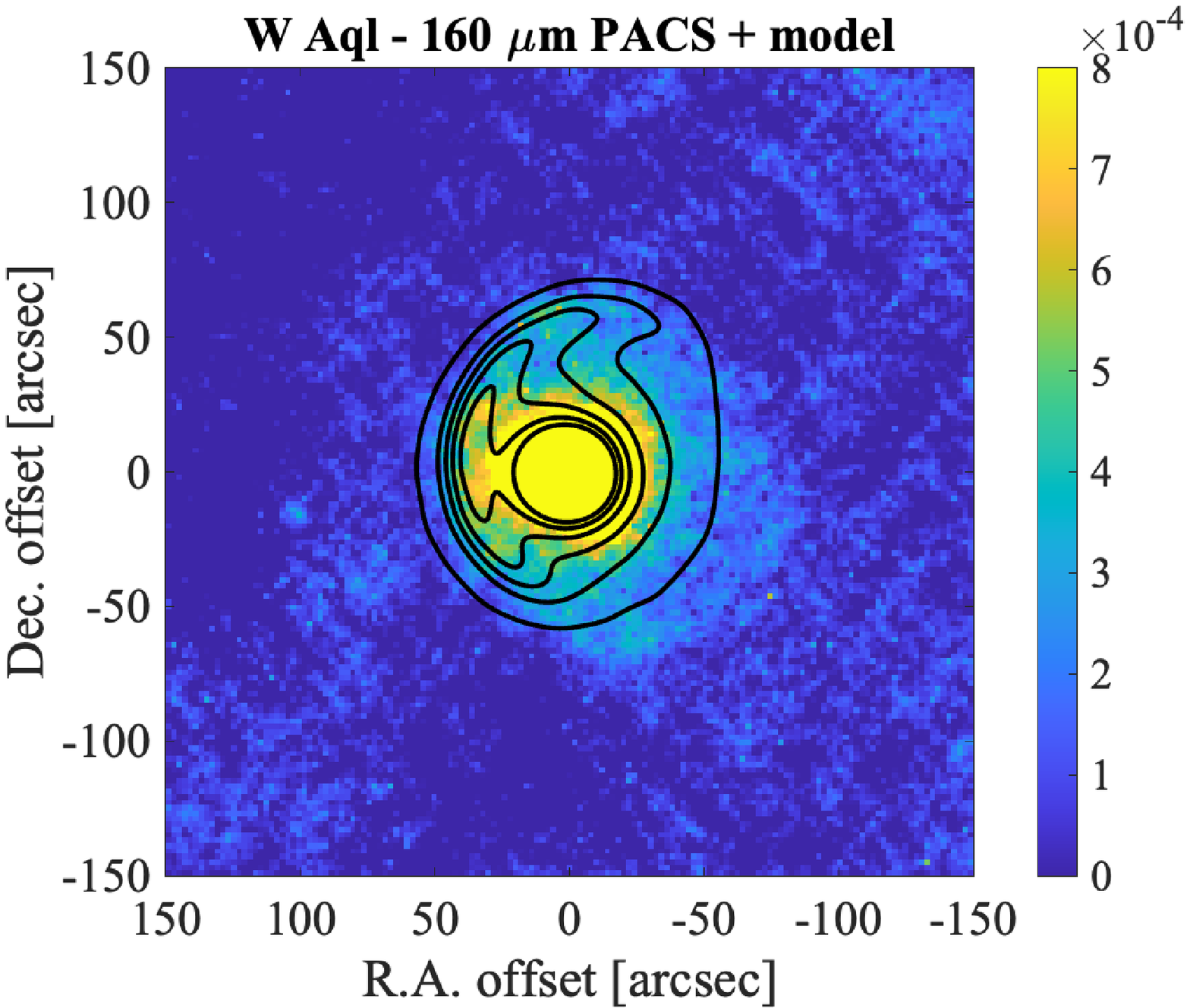}
\caption{W Aql: \emph{Top to bottom:} The Radmc3D model, the PACS image, and the PACS image with contours from the model. Images are for 70\,\micron~(left) and 160\,\micron~(right). Maximum contour levels are 1.8$\times10^{-3}$\,\Jyarcsec (70\,\micron) and 0.44$\times10^{-3}$\,\Jyarcsec (160\,\micron), respectively. Minimum contour levels are 10\% of maximum. The colour scale is in \Jyarcsec. The red dashed circle shows the mask used to measure the flux from the star and present-day mass-loss.}
\label{f:waql}
\end{figure*}

\begin{figure*}
\centering
\includegraphics[width=8cm]{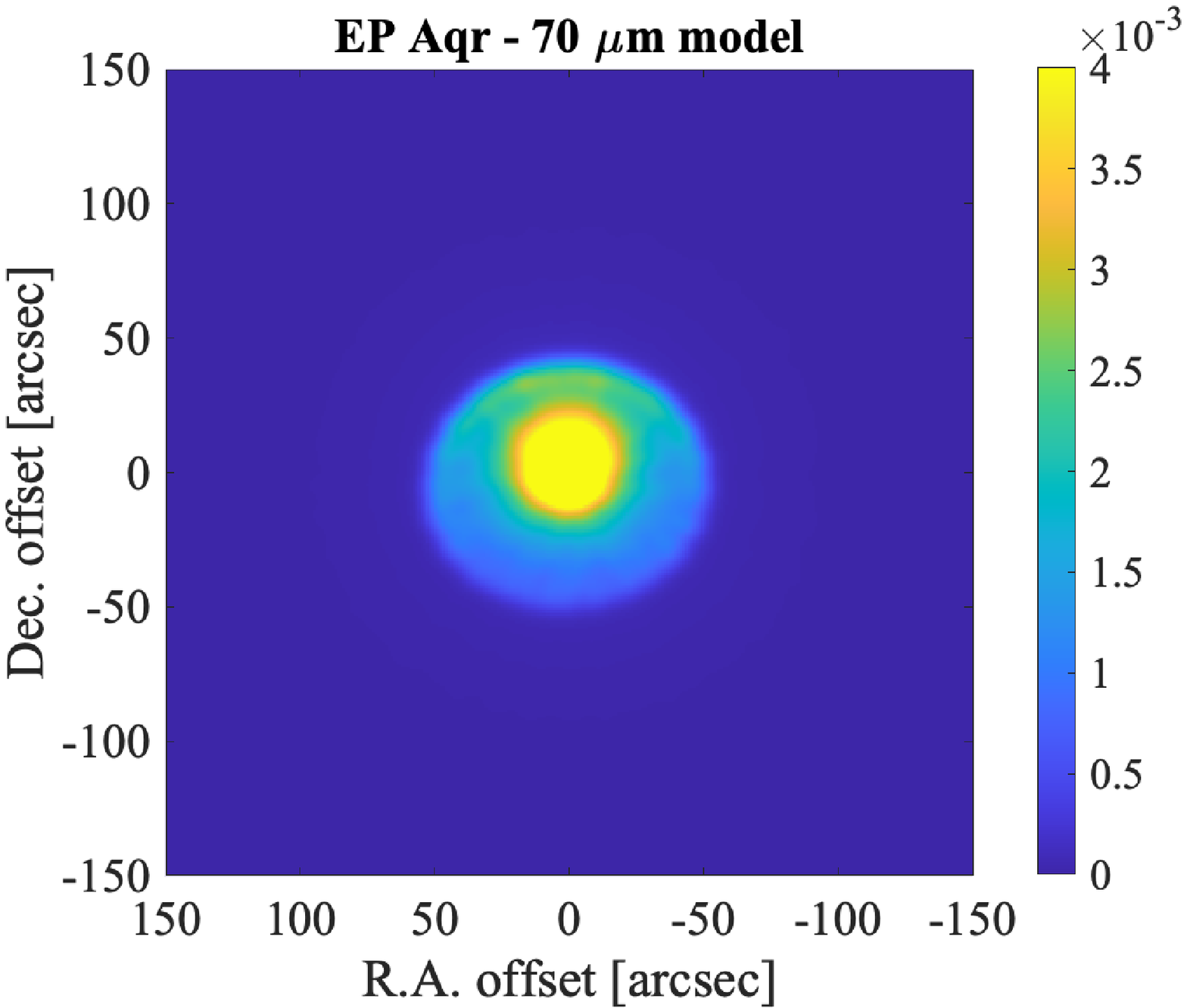}
\includegraphics[width=8cm]{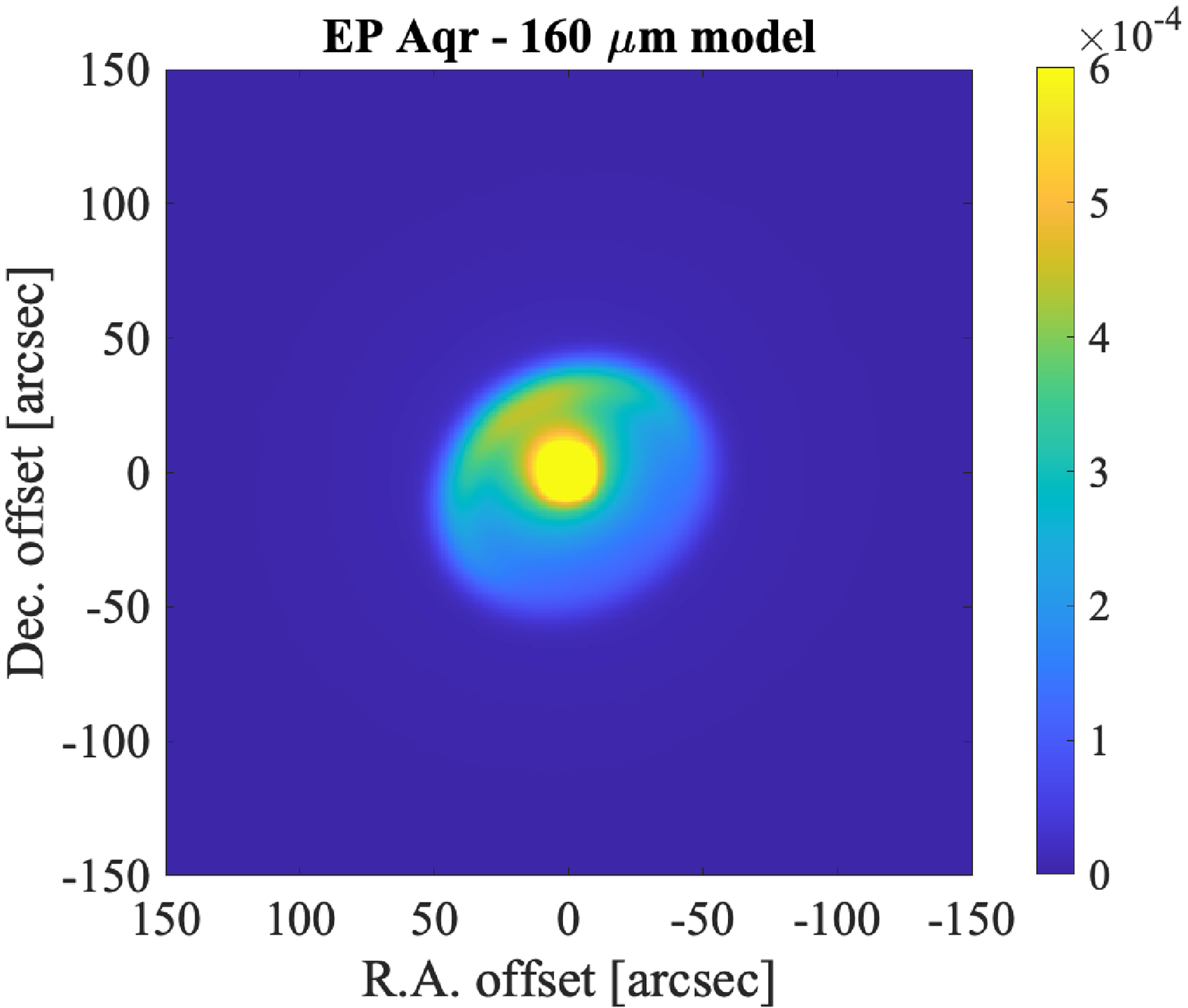}
\includegraphics[width=8cm]{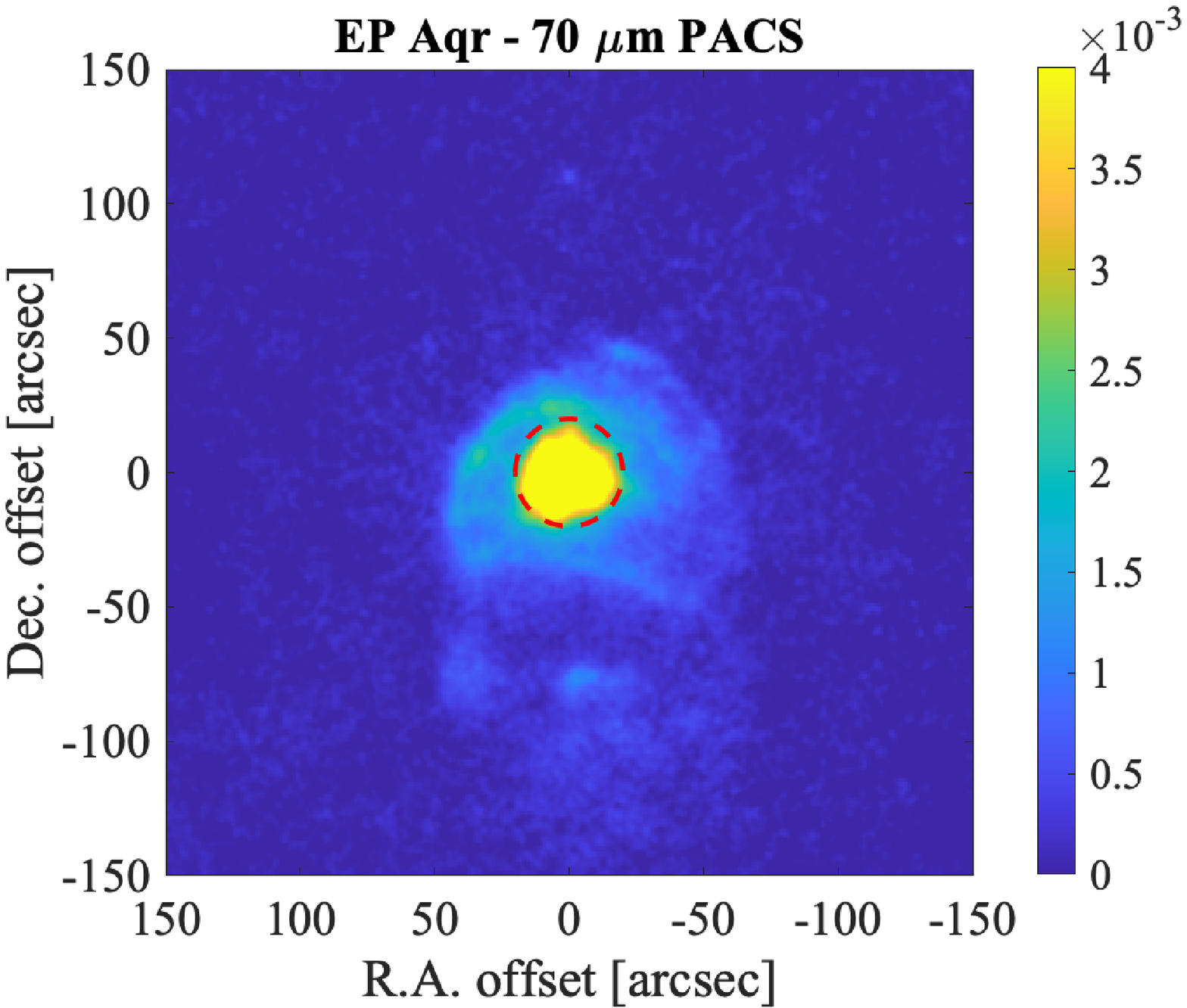}
\includegraphics[width=8cm]{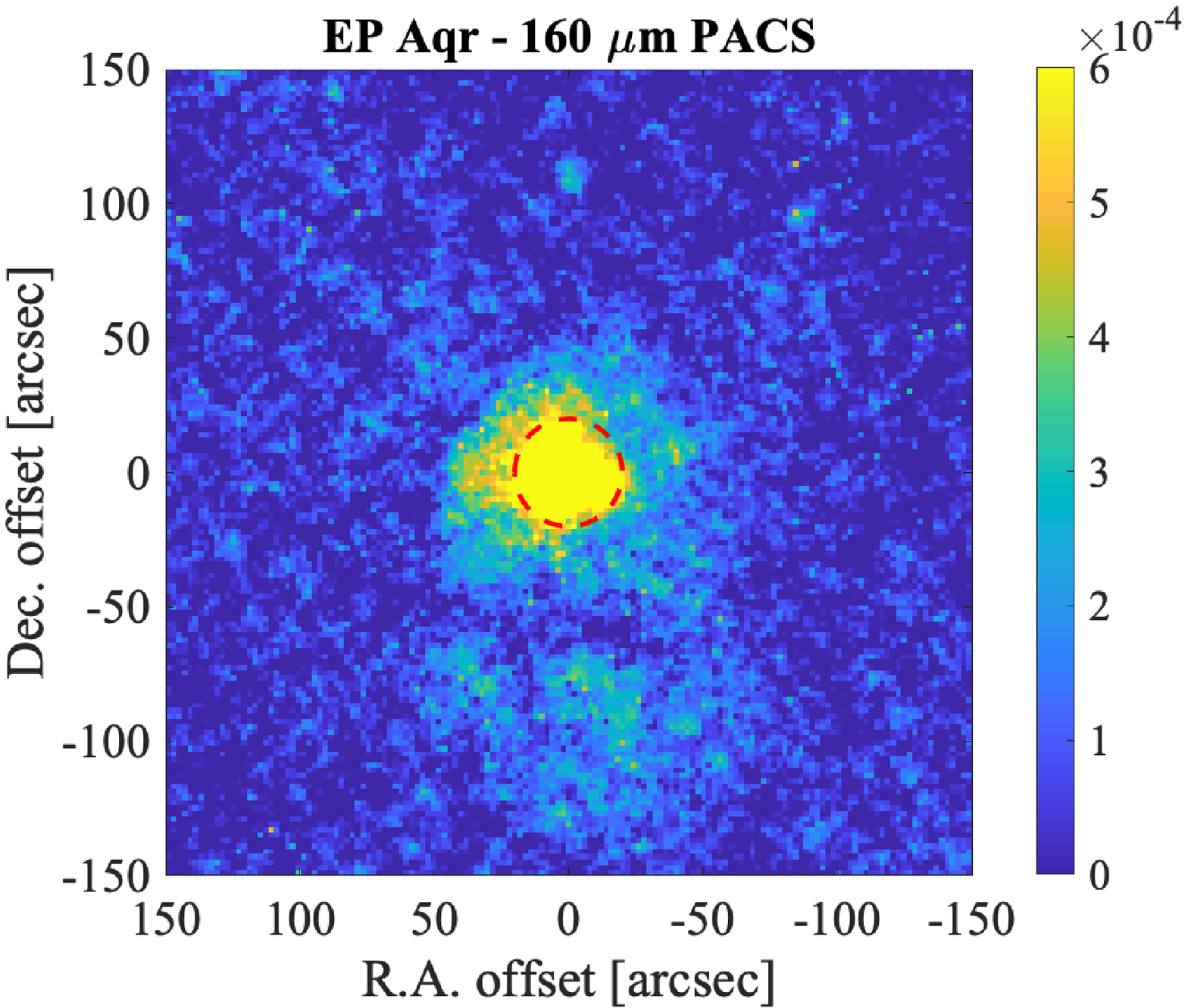}
\includegraphics[width=8cm]{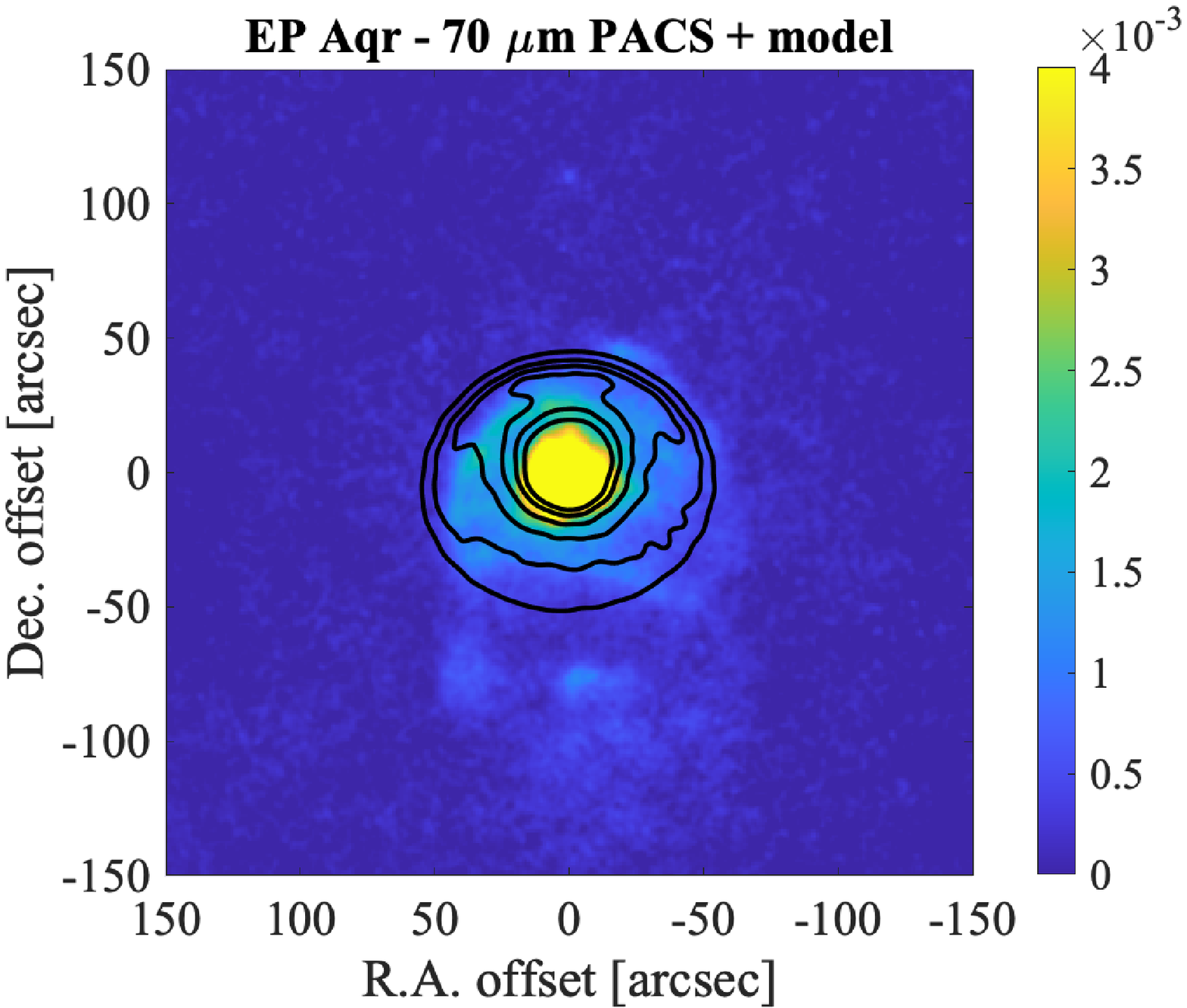}
\includegraphics[width=8cm]{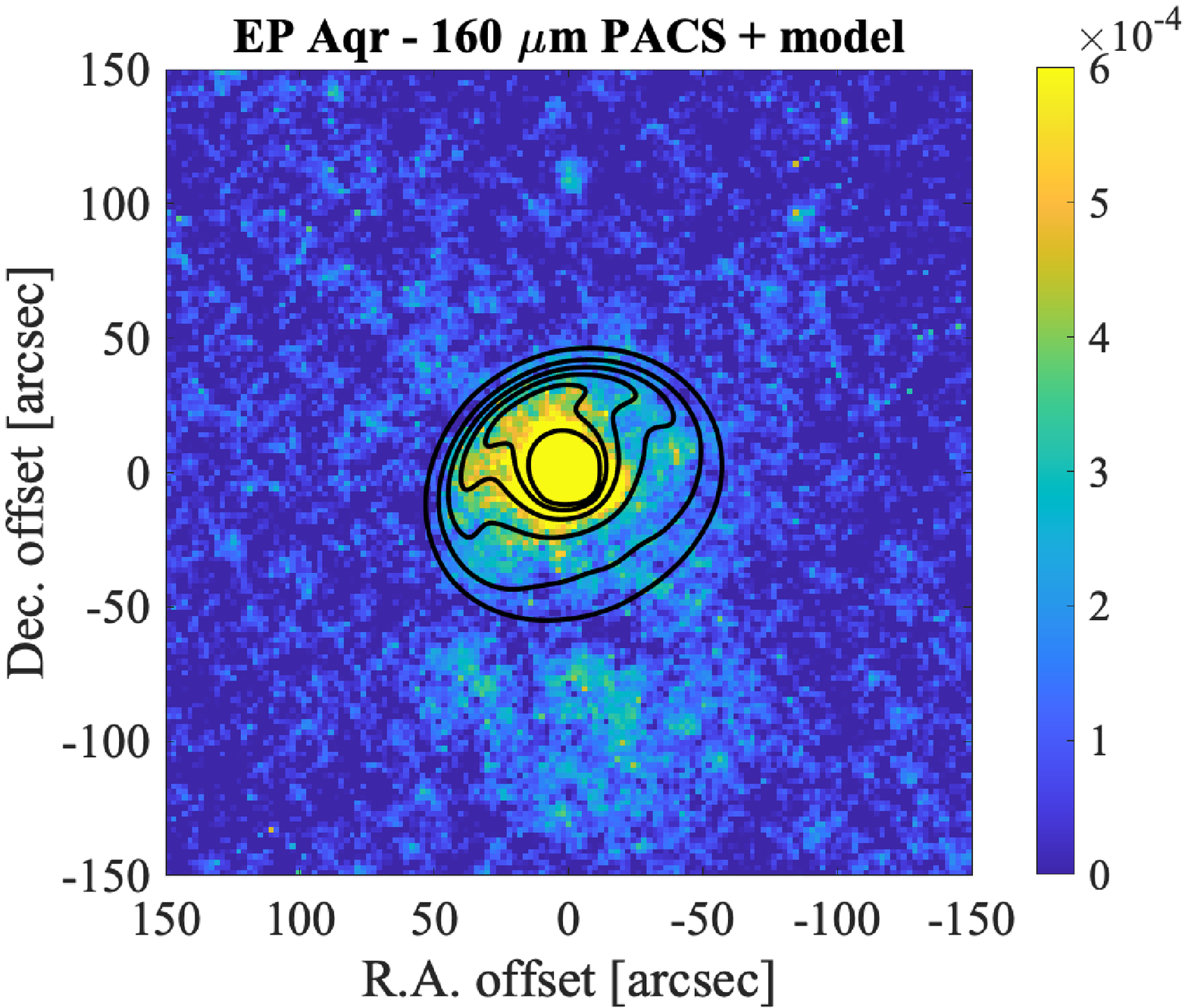}
\caption{EP Aqr: \emph{Top to bottom:} The Radmc3D model, the PACS image, and the PACS image with contours from the model. Images are for 70\,\micron~(left) and 160\,\micron~(right). Maximum contour levels are 3.5$\times10^{-3}$\,\Jyarcsec (70\,\micron) and 0.44$\times10^{-3}$\,\Jyarcsec (160\,\micron), respectively. Minimum contour levels are 10\% of maximum. The colour scale is in \Jyarcsec. The red dashed circle shows the mask used to measure the flux from the star and present-day mass-loss.}
\label{f:epaqr}
\end{figure*}

\begin{figure*}
\centering
\includegraphics[width=8cm]{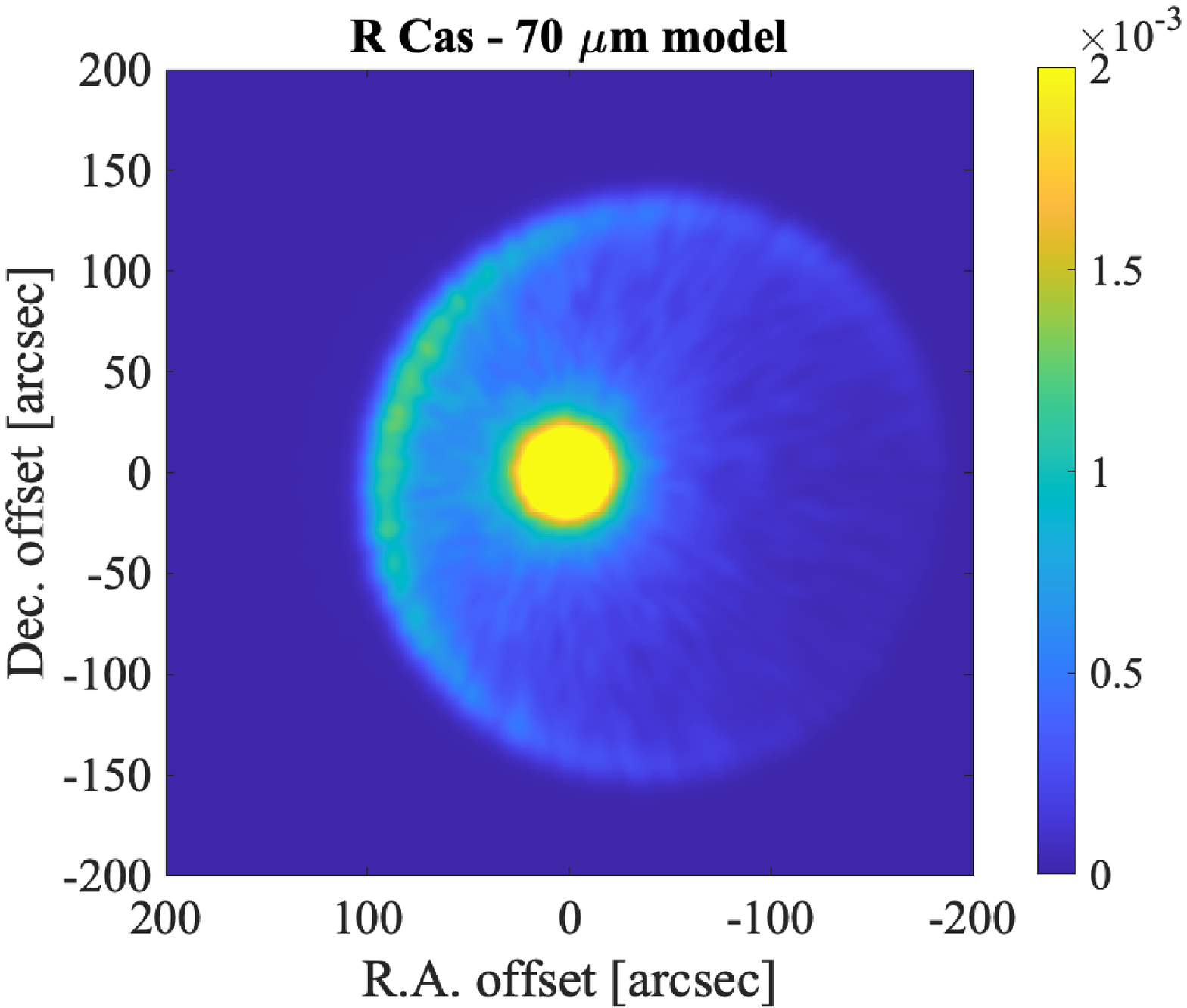}
\includegraphics[width=8cm]{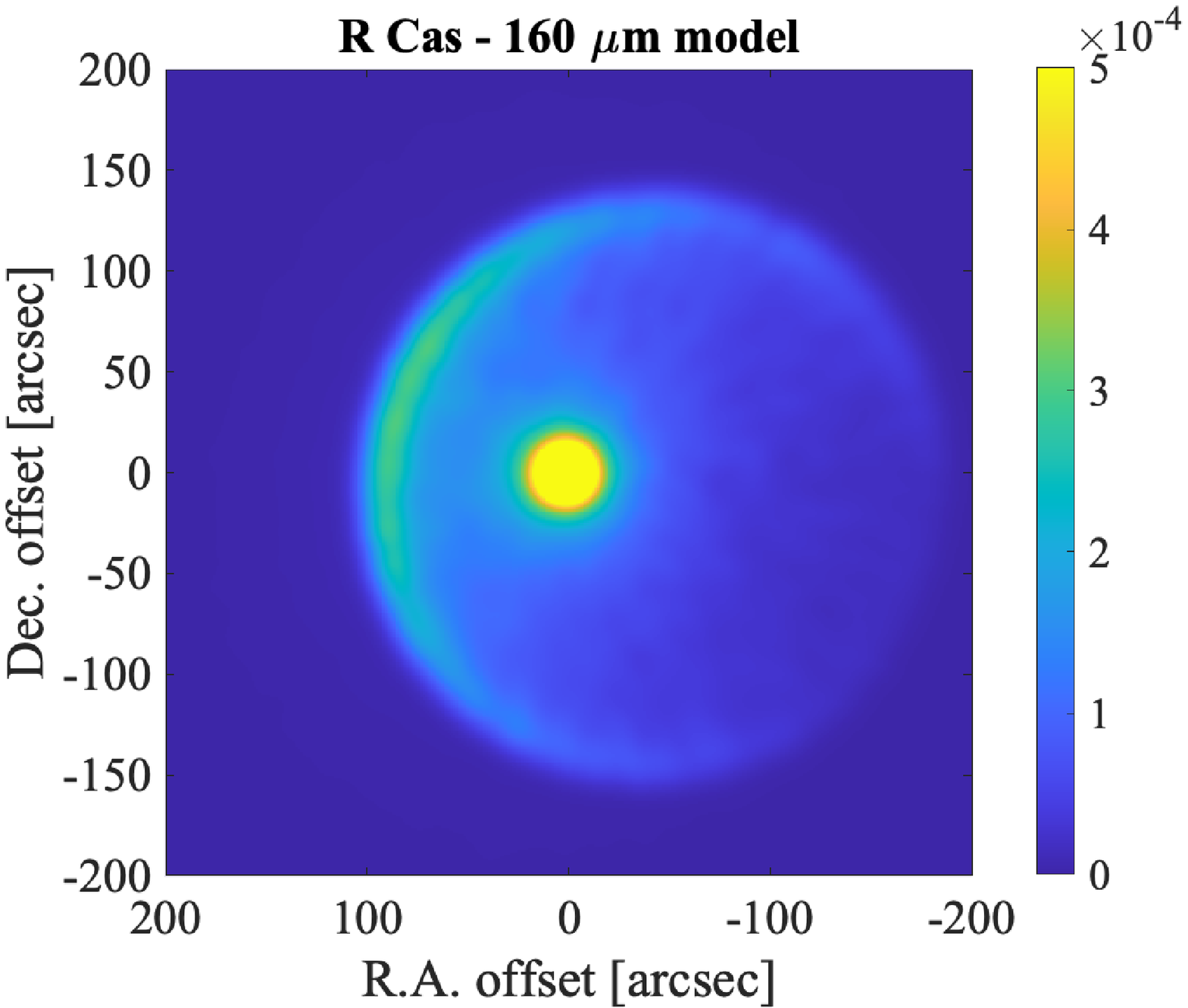}
\includegraphics[width=8cm]{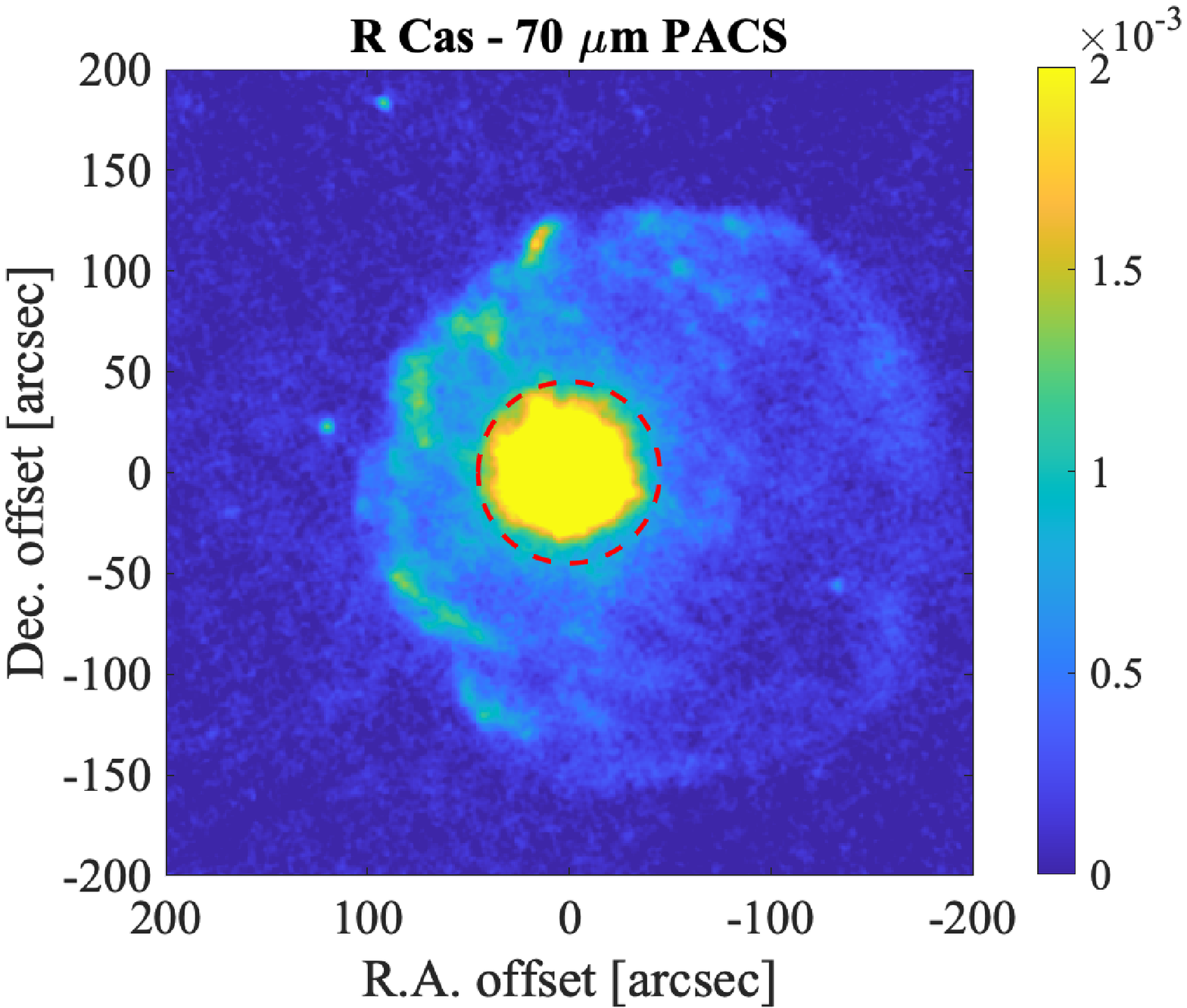}
\includegraphics[width=8cm]{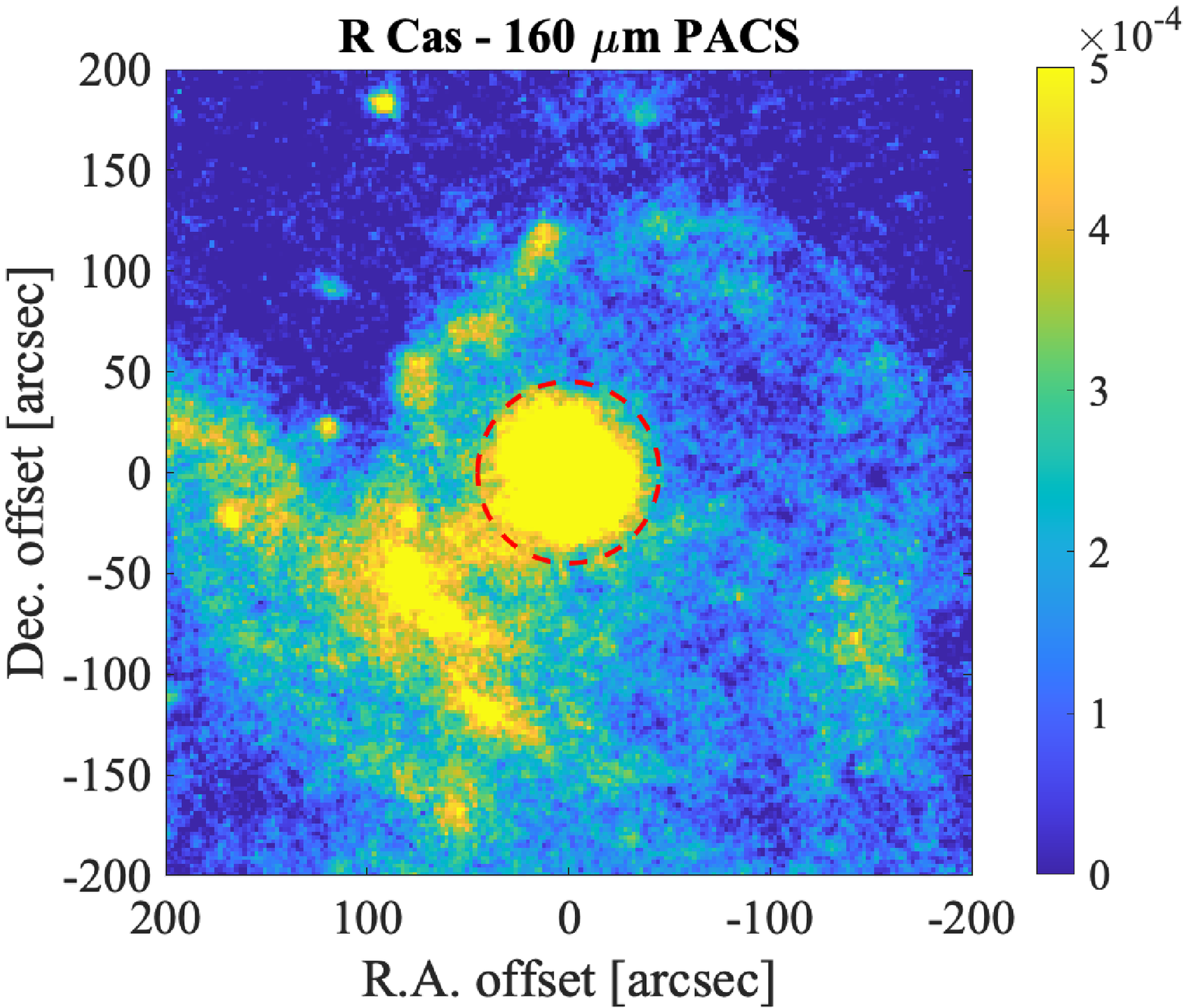}
\includegraphics[width=8cm]{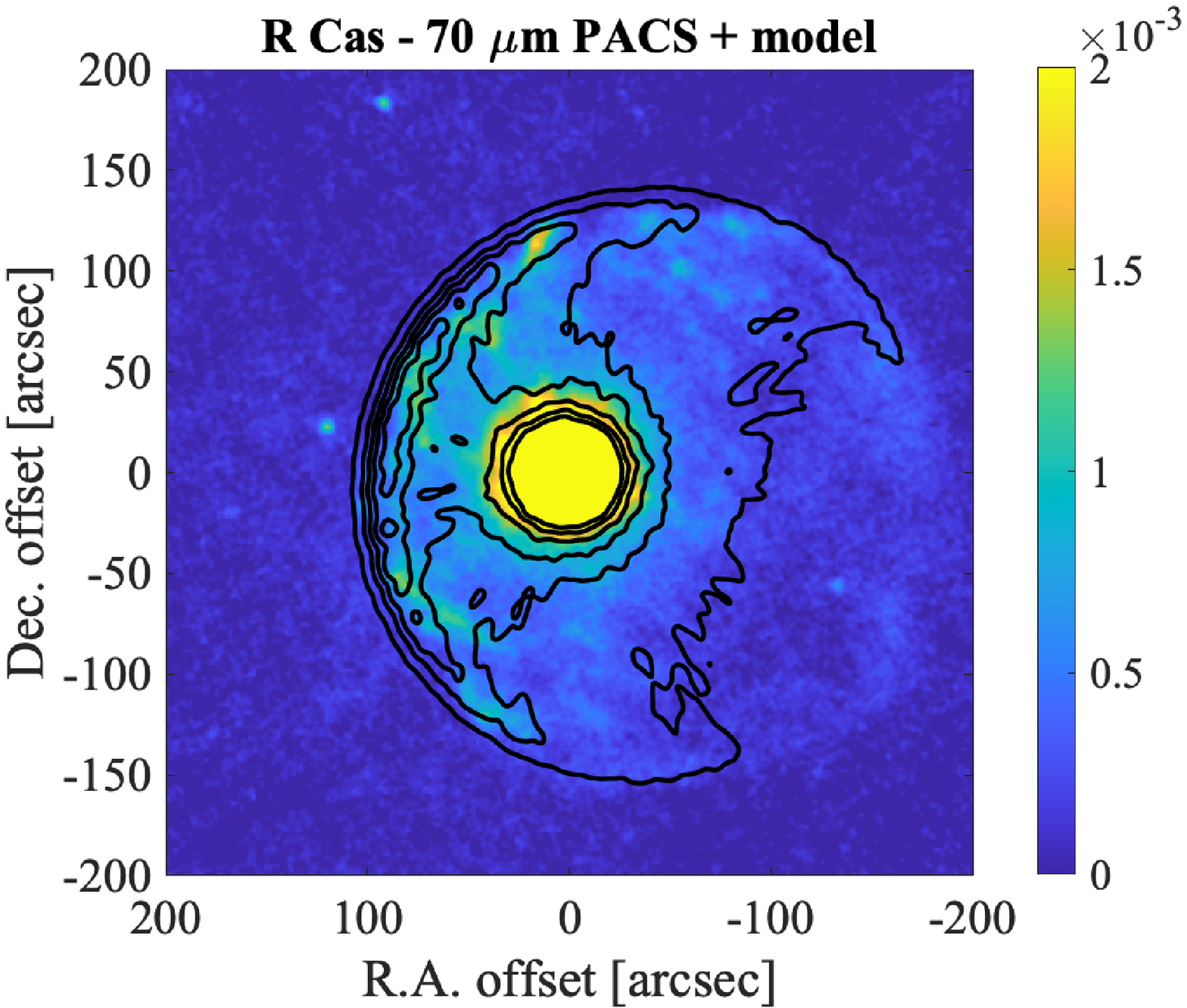}
\includegraphics[width=8cm]{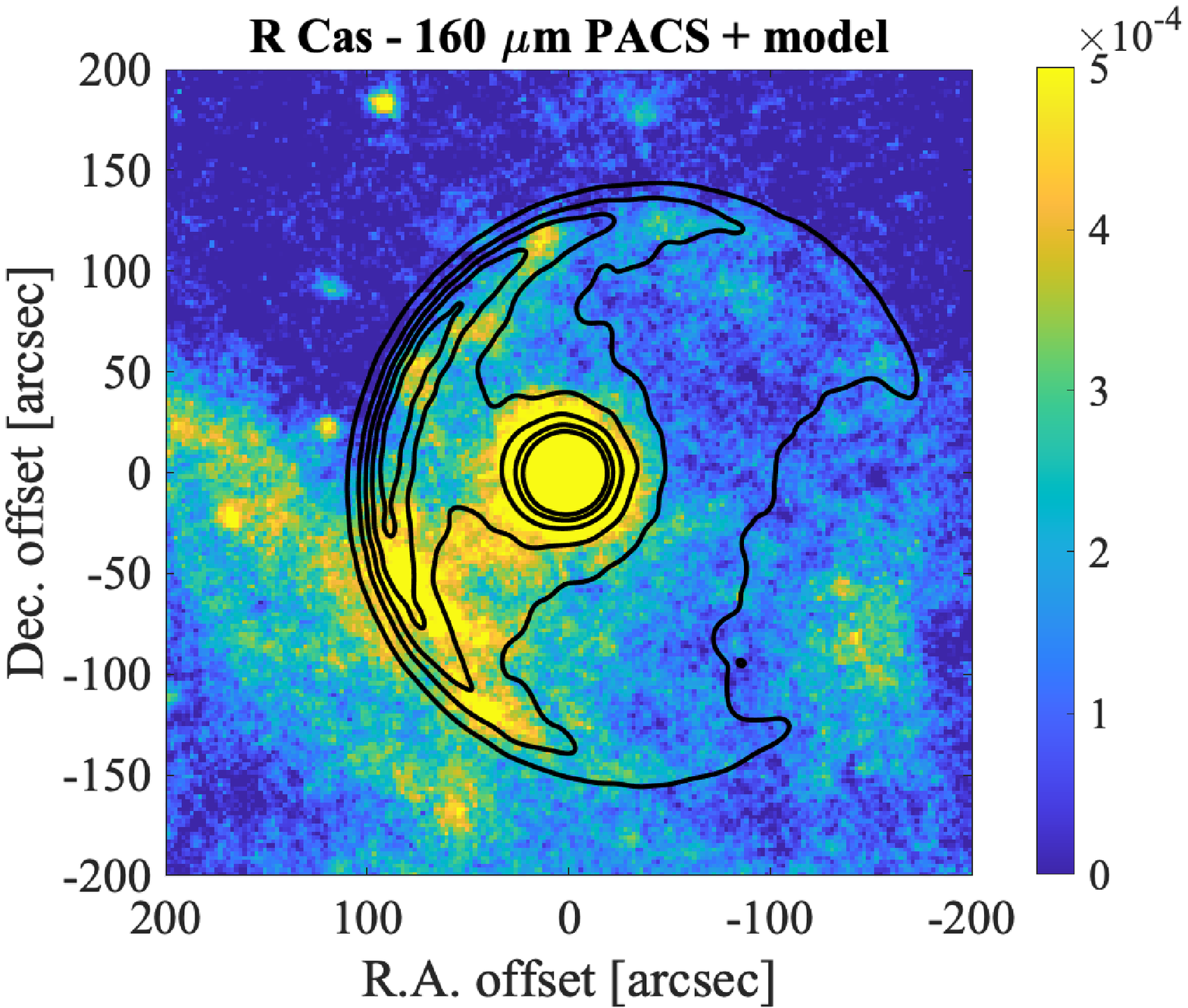}
\caption{R Cas: \emph{Top to bottom:} The Radmc3D model, the PACS image, and the PACS image with contours from the model. Images are for 70\,\micron~(left) and 160\,\micron~(right). Maximum contour levels are 1.2$\times10^{-3}$\,\Jyarcsec (70\,\micron) and 0.3$\times10^{-3}$\,\Jyarcsec (160\,\micron), respectively. Minimum contour levels are 10\% of maximum. The colour scale is in \Jyarcsec. The red dashed circle shows the mask used to measure the flux from the star and present-day mass-loss.}
\label{f:rcas}
\end{figure*}

\begin{figure*}
\centering
\includegraphics[width=8cm]{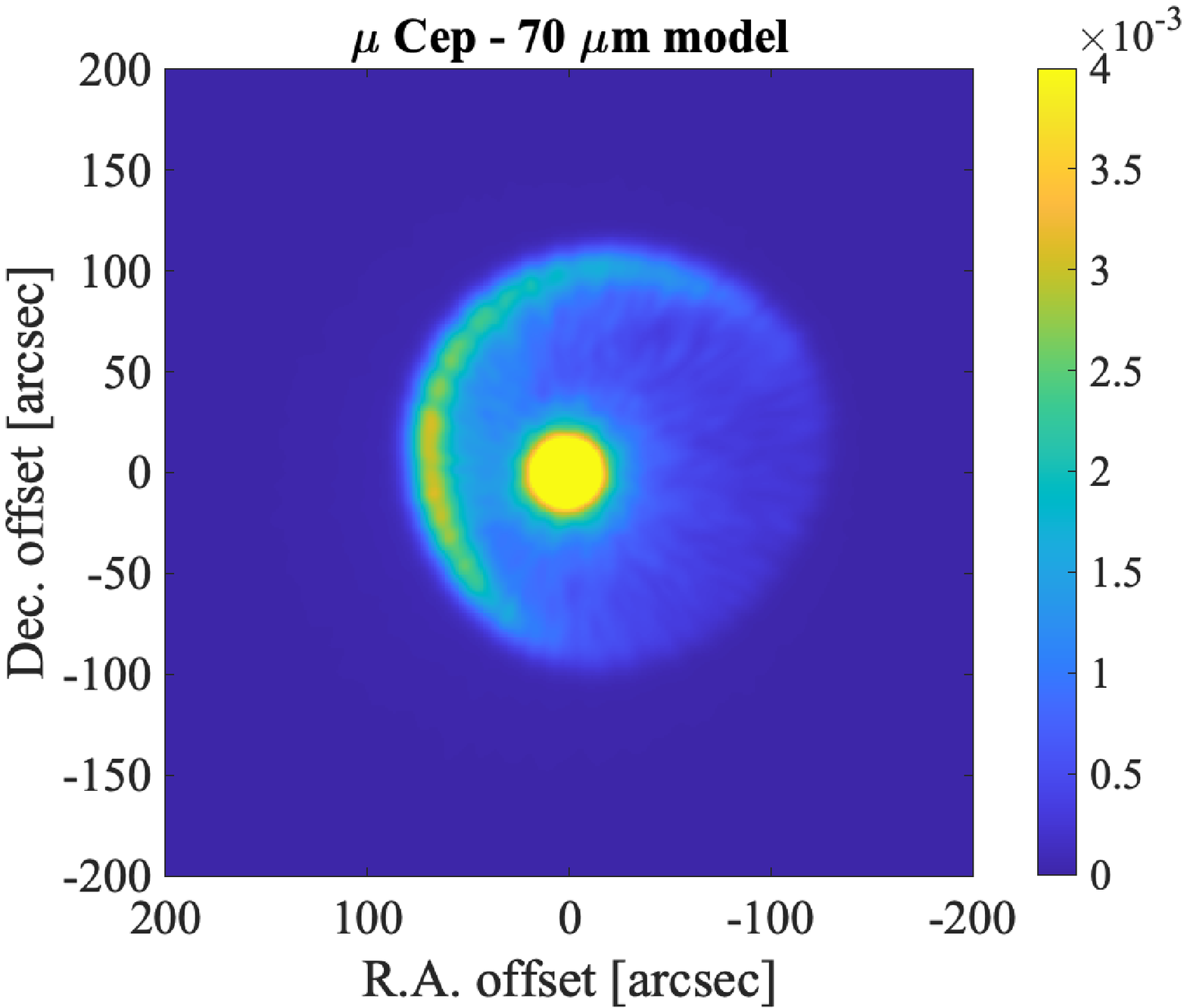}
\includegraphics[width=8cm]{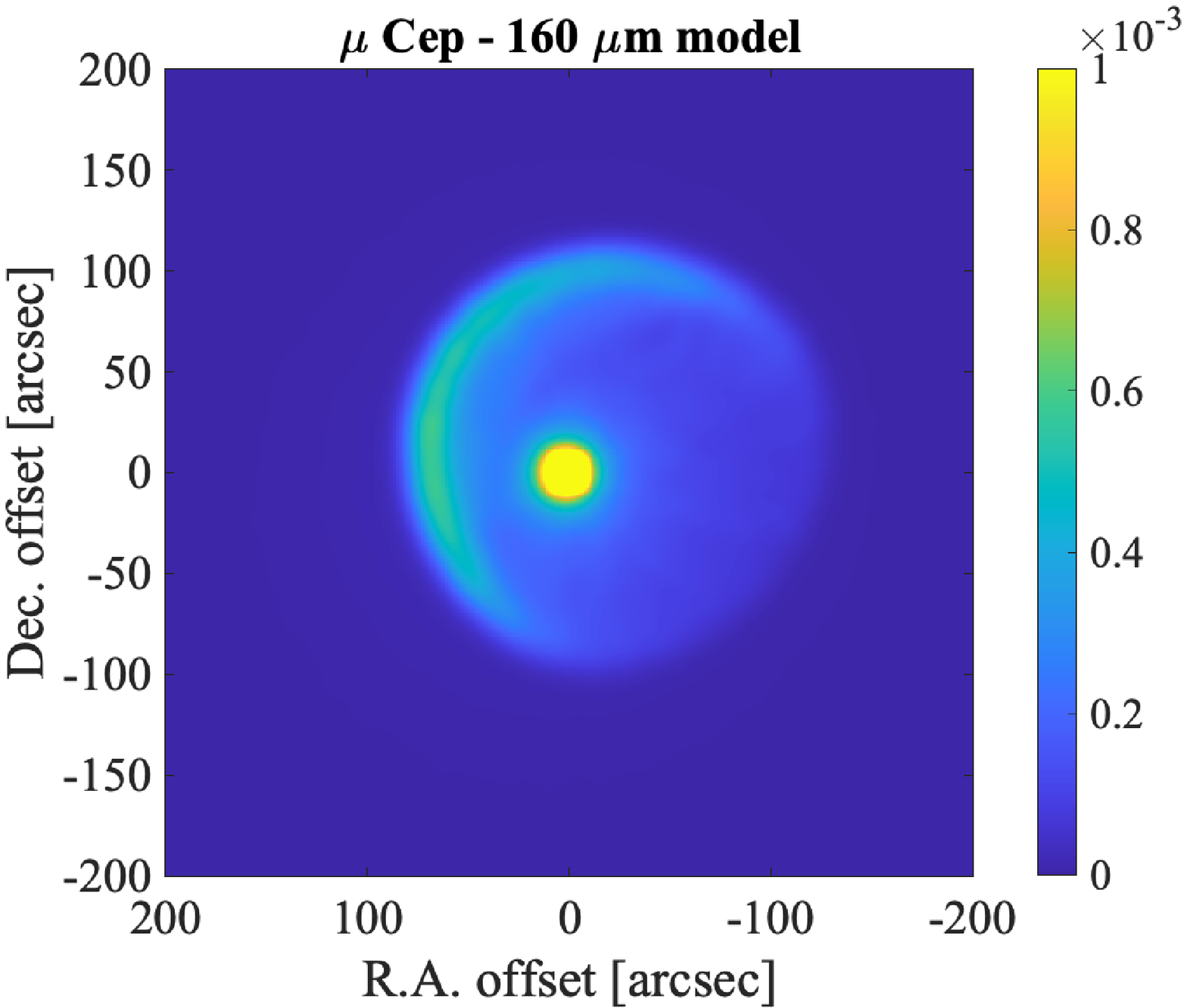}
\includegraphics[width=8cm]{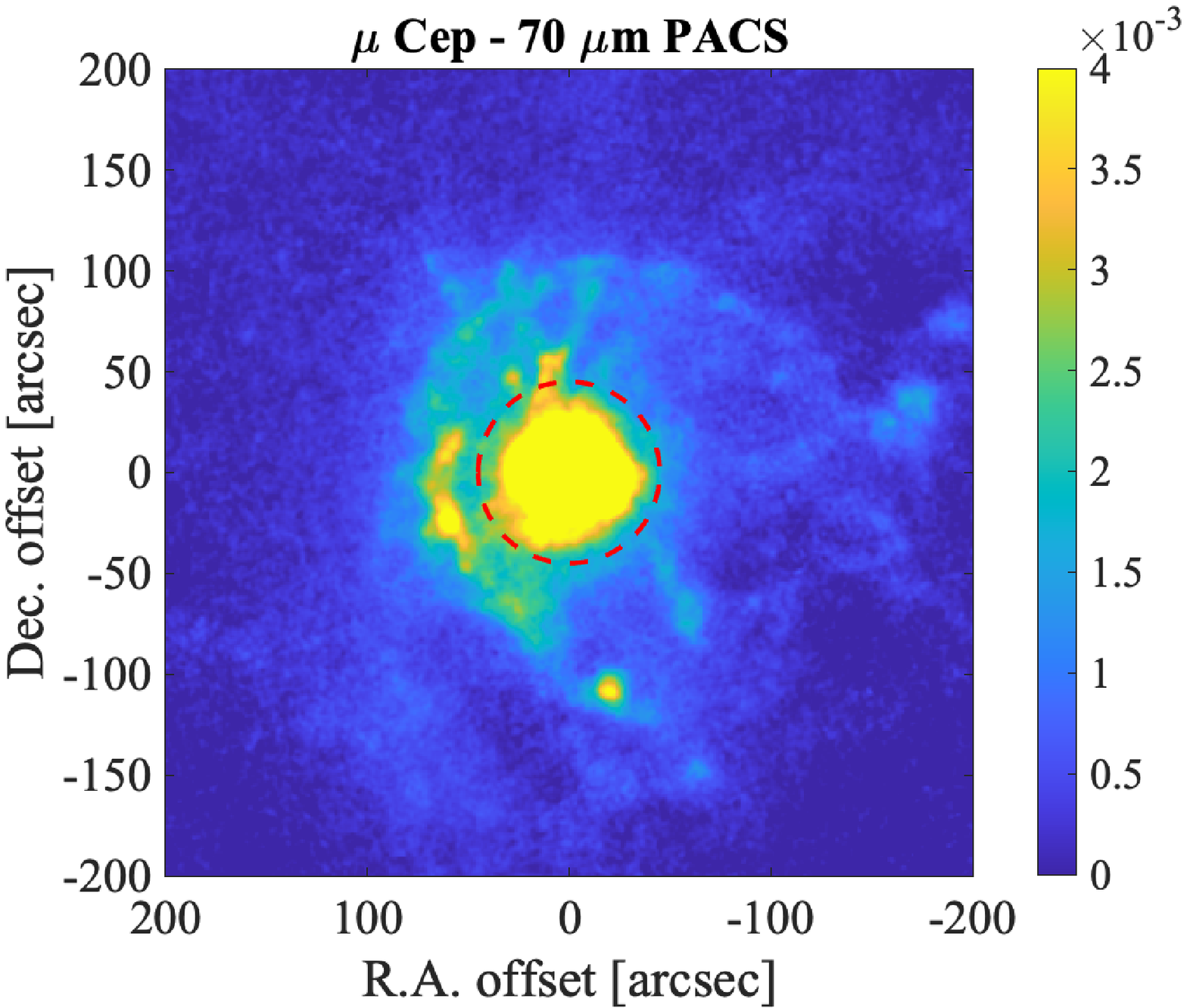}
\includegraphics[width=8cm]{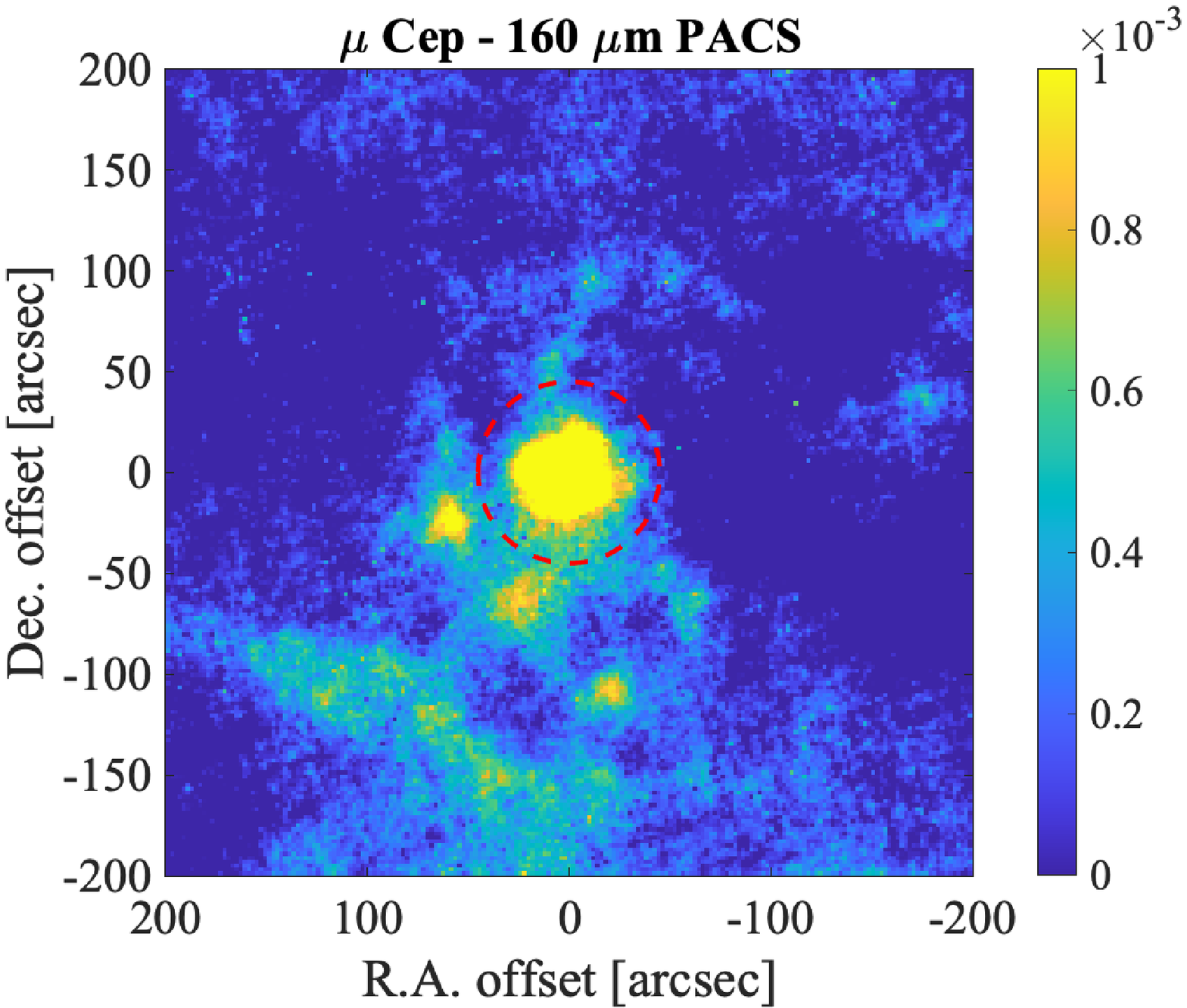}
\includegraphics[width=8cm]{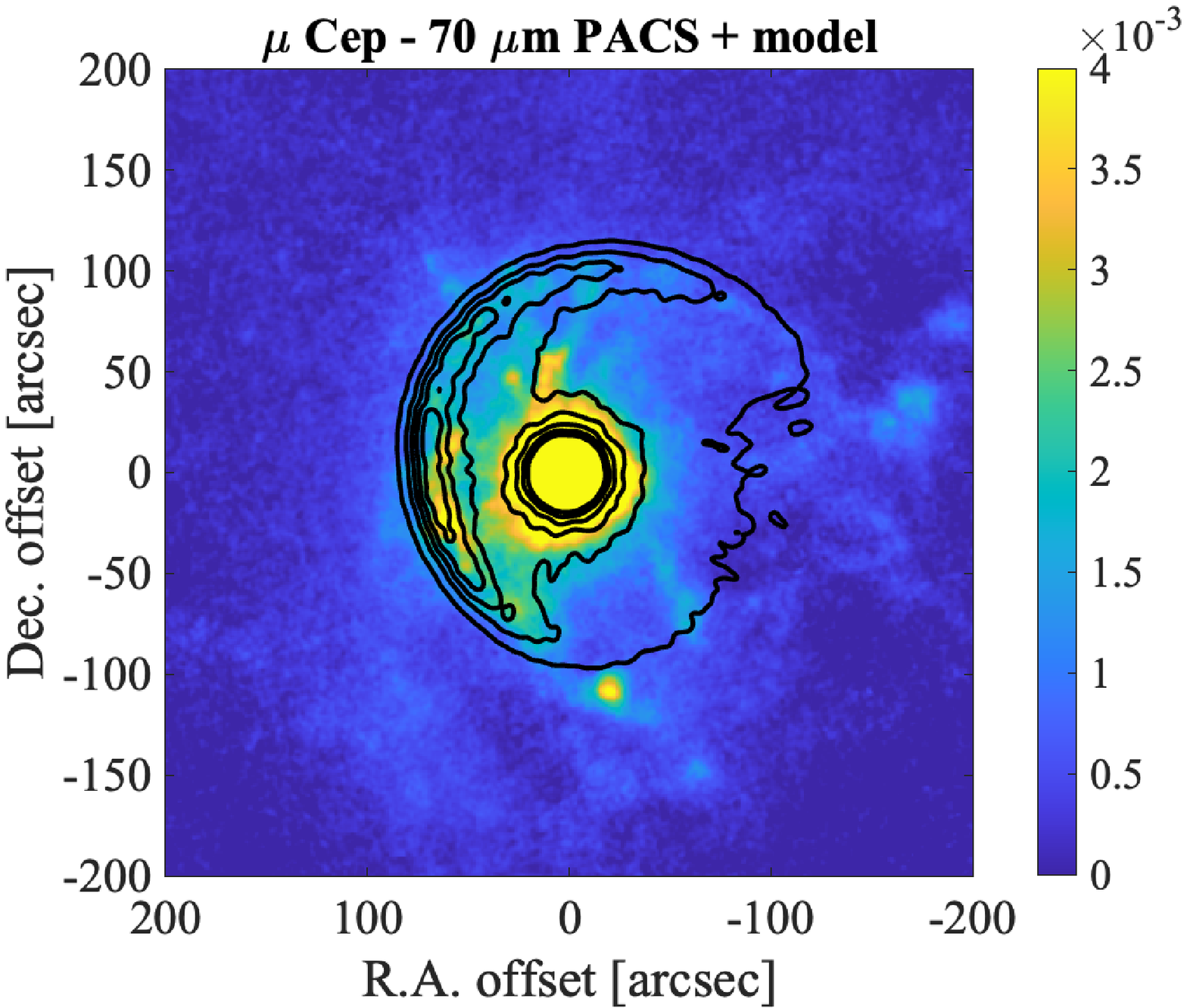}
\includegraphics[width=8cm]{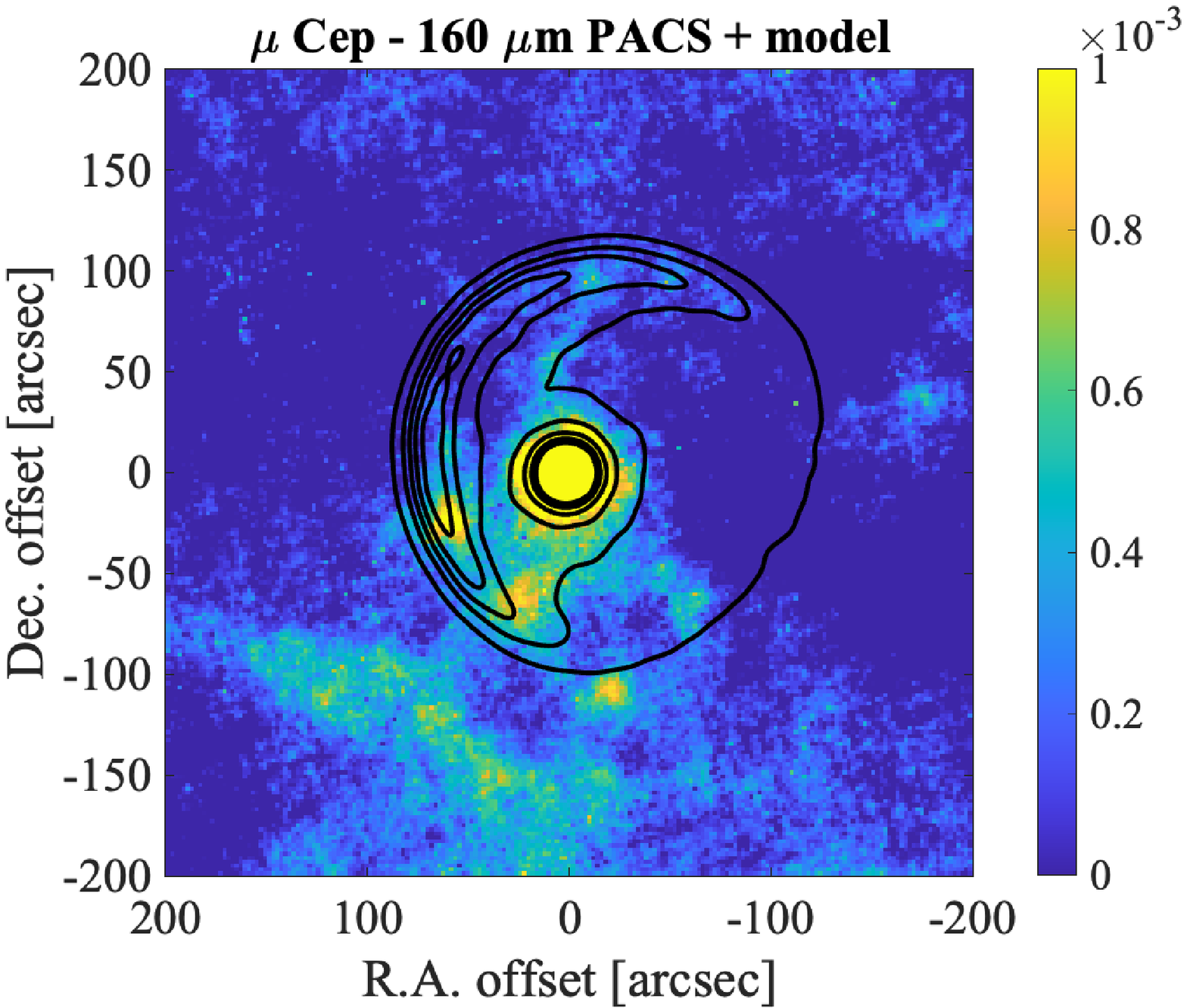}
\caption{$\mu$ Cep: \emph{Top to bottom:} The Radmc3D model, the PACS image, and the PACS image with contours from the model. Images are for 70\,\micron~(left) and 160\,\micron~(right). Maximum contour levels are 3$\times10^{-3}$\,\Jyarcsec (70\,\micron) and 0.58$\times10^{-3}$\,\Jyarcsec (160\,\micron), respectively. Minimum contour levels are 10\% of maximum. The colour scale is in \Jyarcsec. The red dashed circle shows the mask used to measure the flux from the star and present-day mass-loss.}
\label{f:mucep}
\end{figure*}

\begin{figure*}
\centering
\includegraphics[width=8cm]{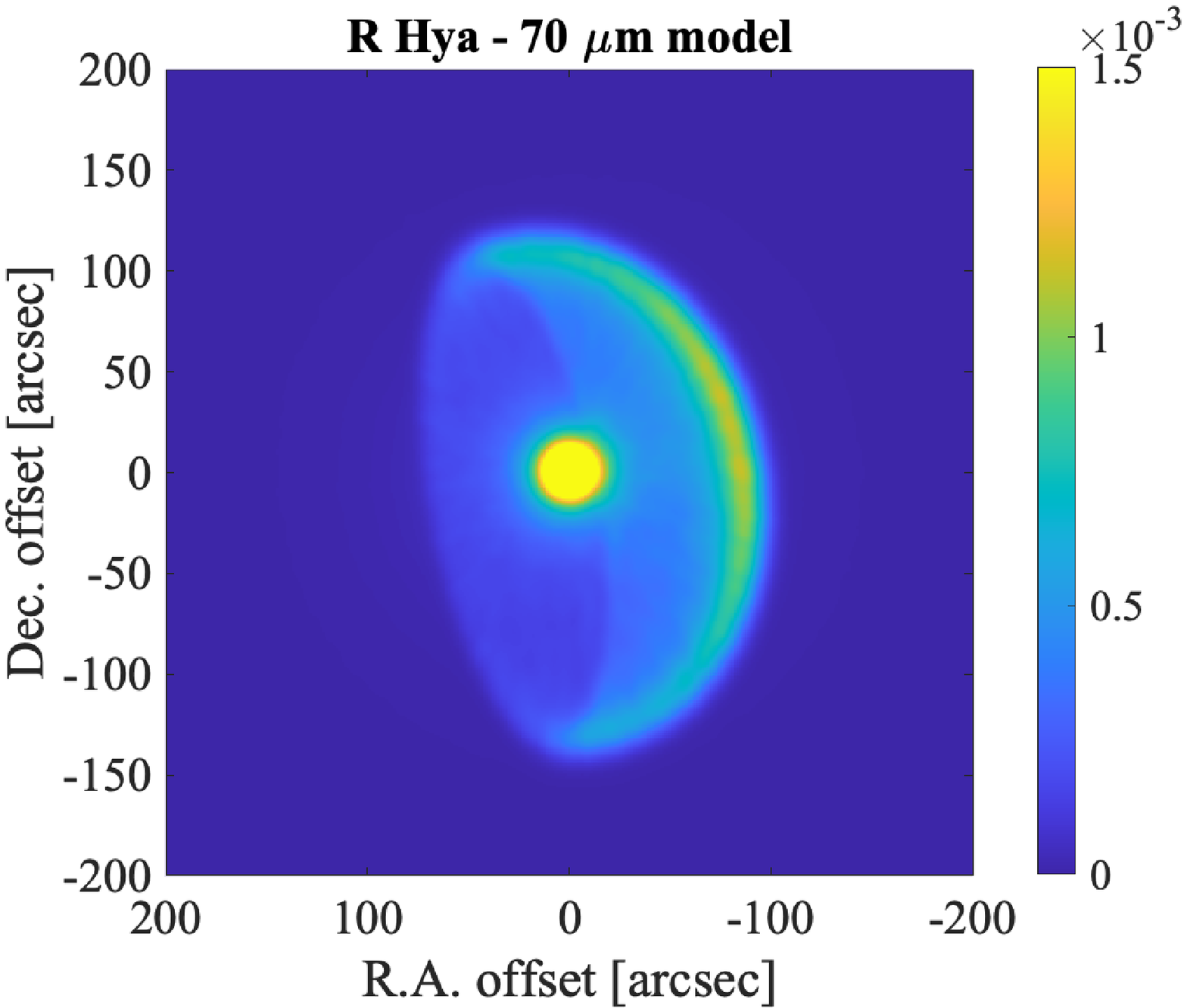}
\includegraphics[width=8cm]{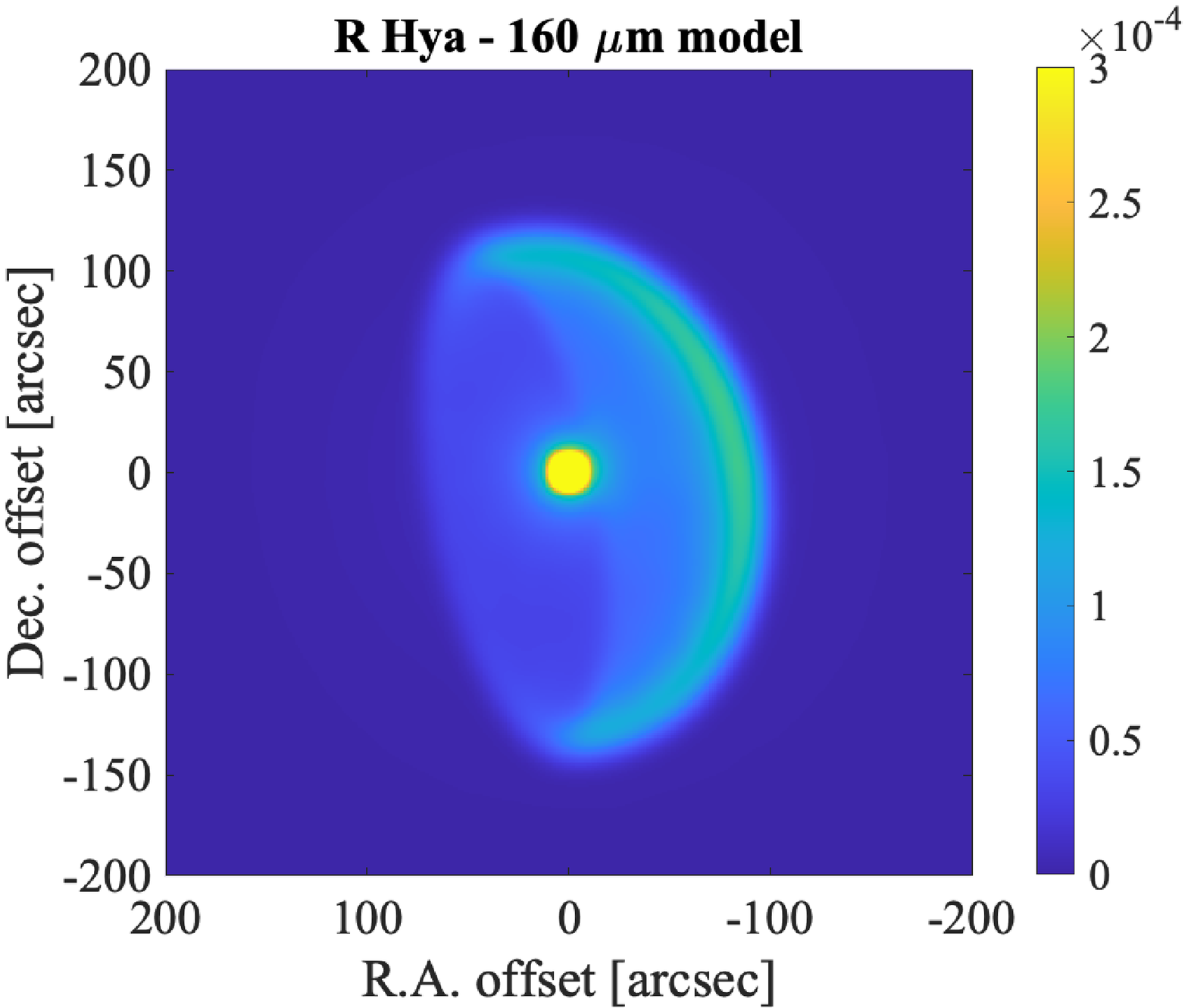}
\includegraphics[width=8cm]{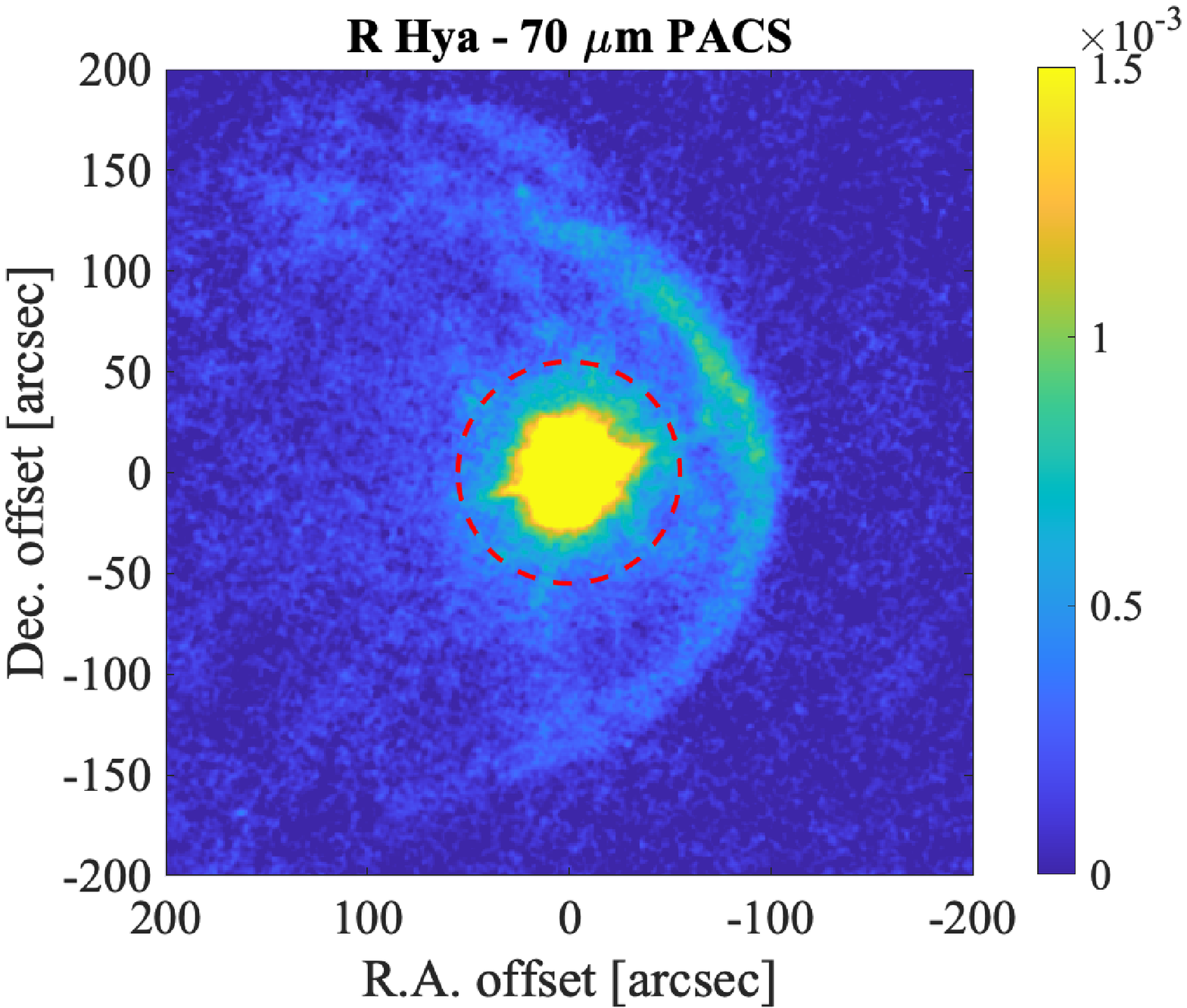}
\includegraphics[width=8cm]{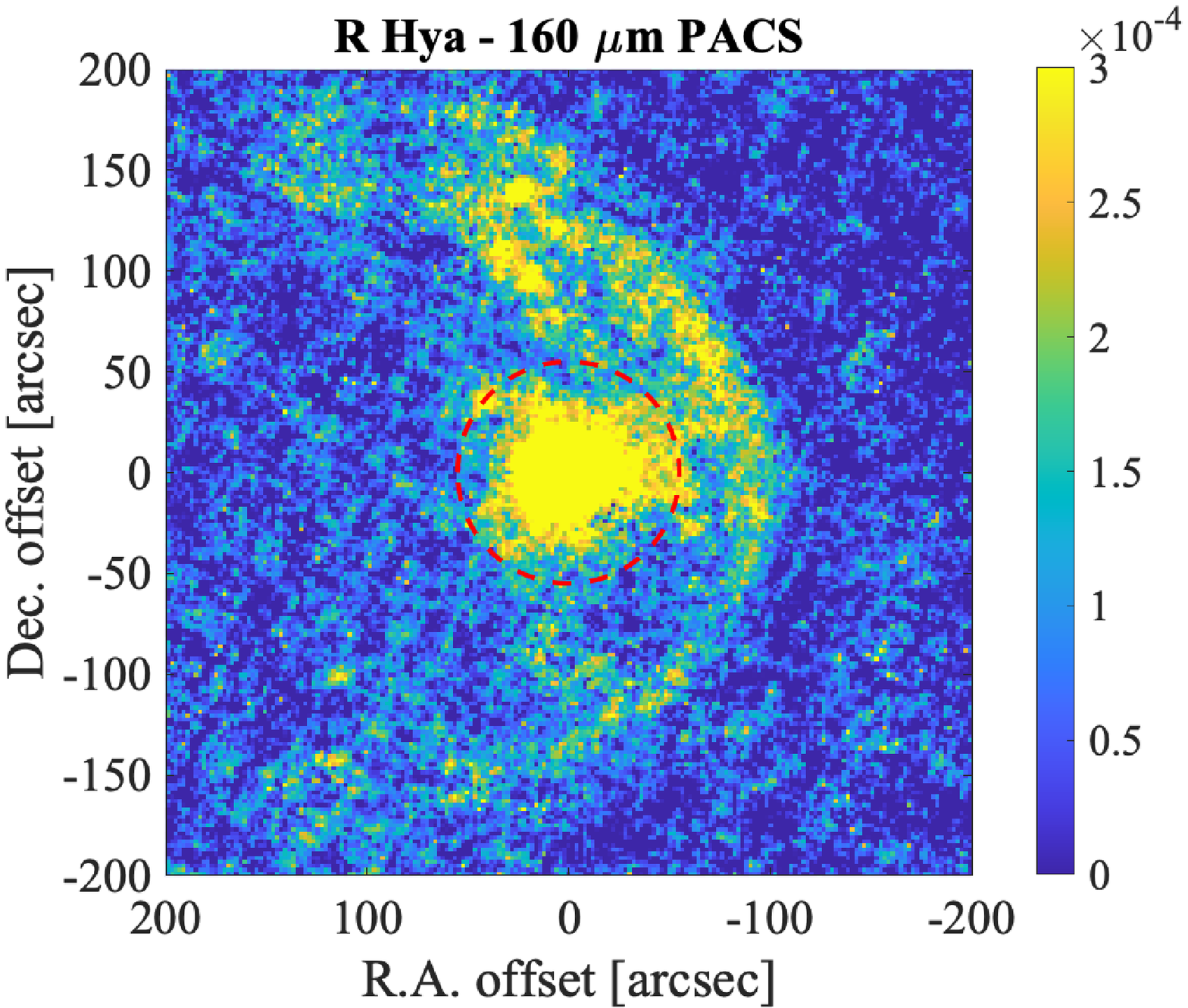}
\includegraphics[width=8cm]{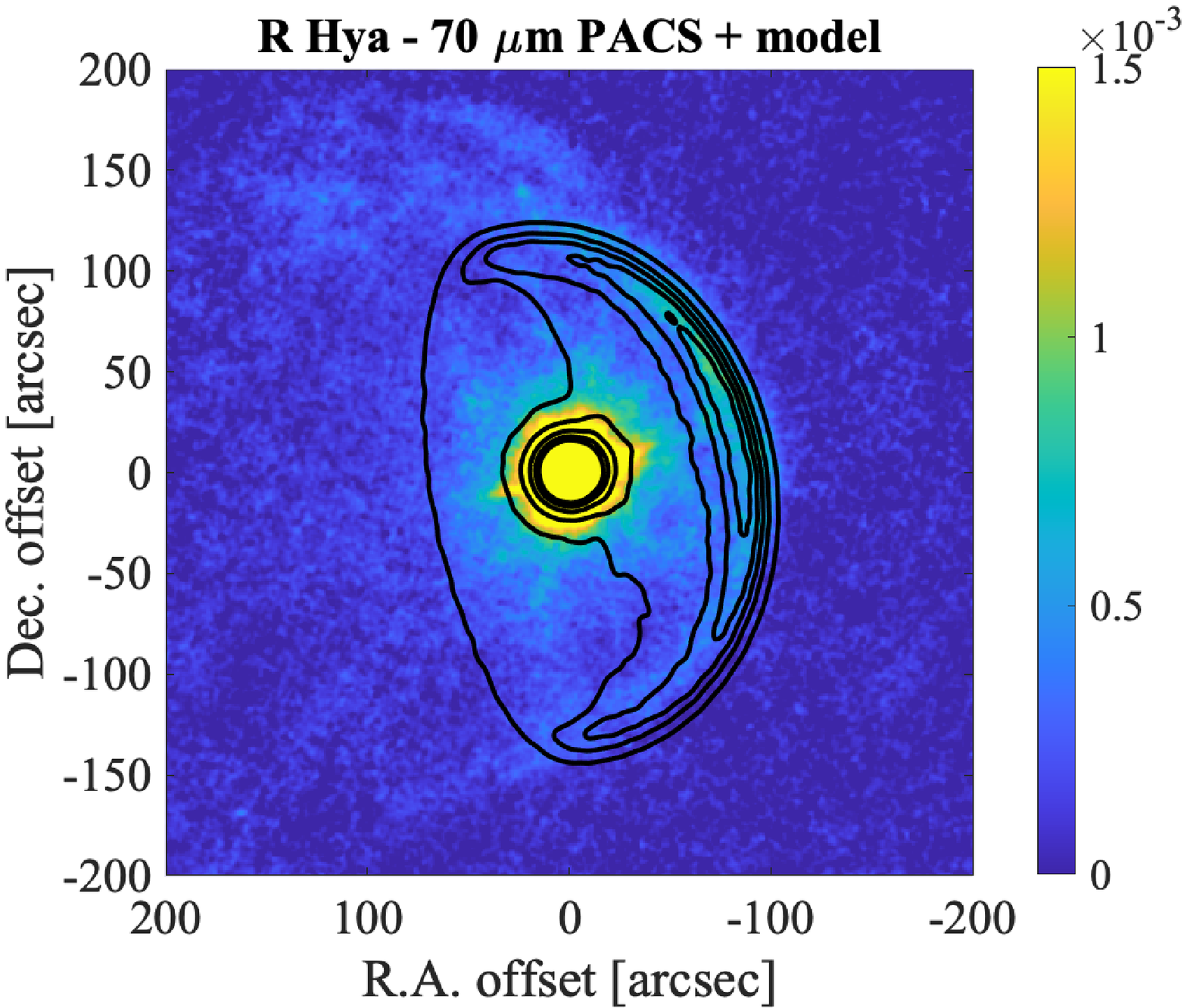}
\includegraphics[width=8cm]{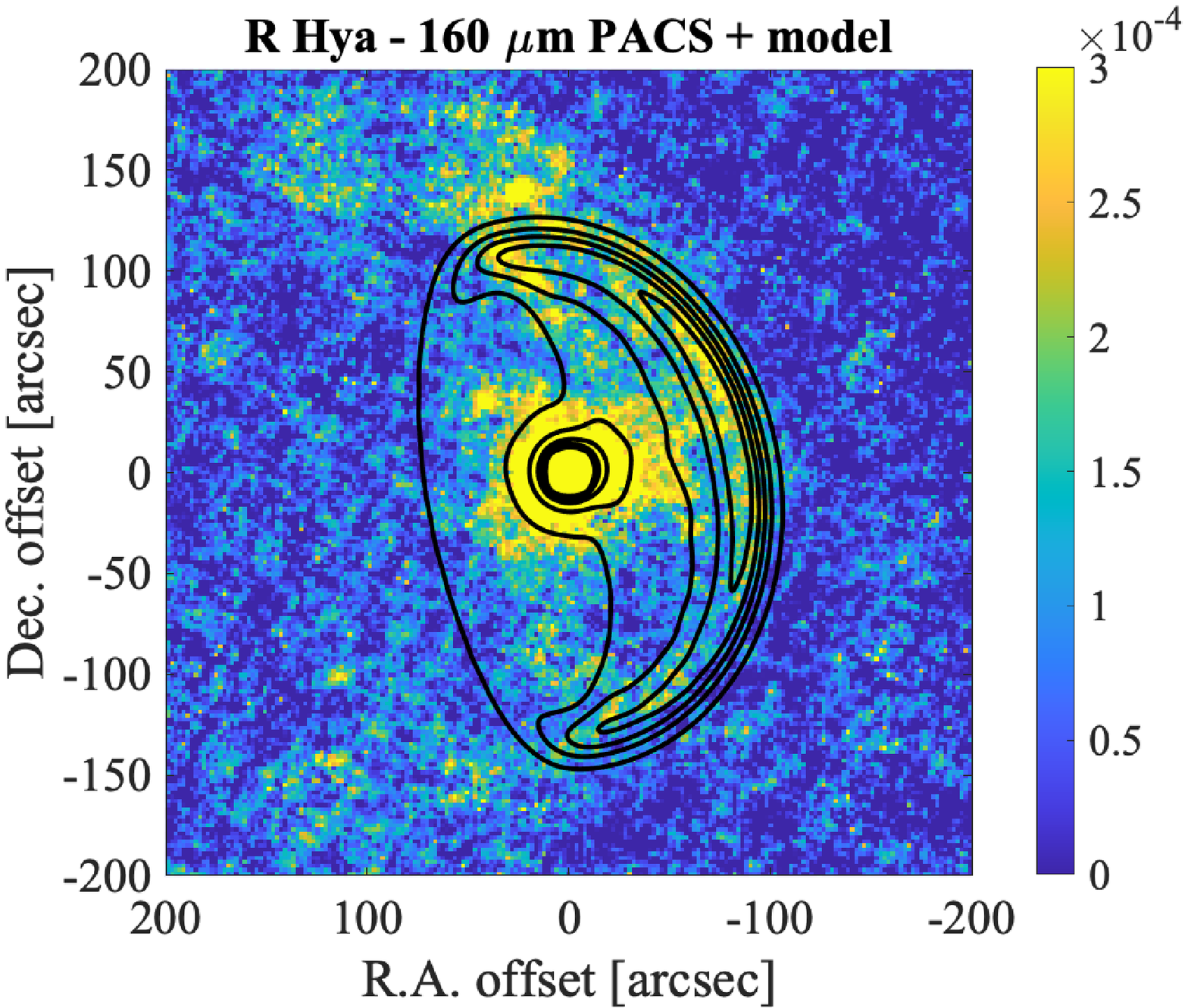}
\caption{R Hya: \emph{Top to bottom:} The Radmc3D model, the PACS image, and the PACS image with contours from the model. Images are for 70\,\micron~(left) and 160\,\micron~(right). Maximum contour levels are 1.1$\times10^{-3}$\,\Jyarcsec (70\,\micron) and 0.17$\times10^{-3}$\,\Jyarcsec (160\,\micron), respectively. Minimum contour levels are 10\% of maximum. The colour scale is in \Jyarcsec. The red dashed circle shows the mask used to measure the flux from the star and present-day mass-loss.}
\label{f:rhya}
\end{figure*}

\begin{figure*}
\centering
\includegraphics[width=8cm]{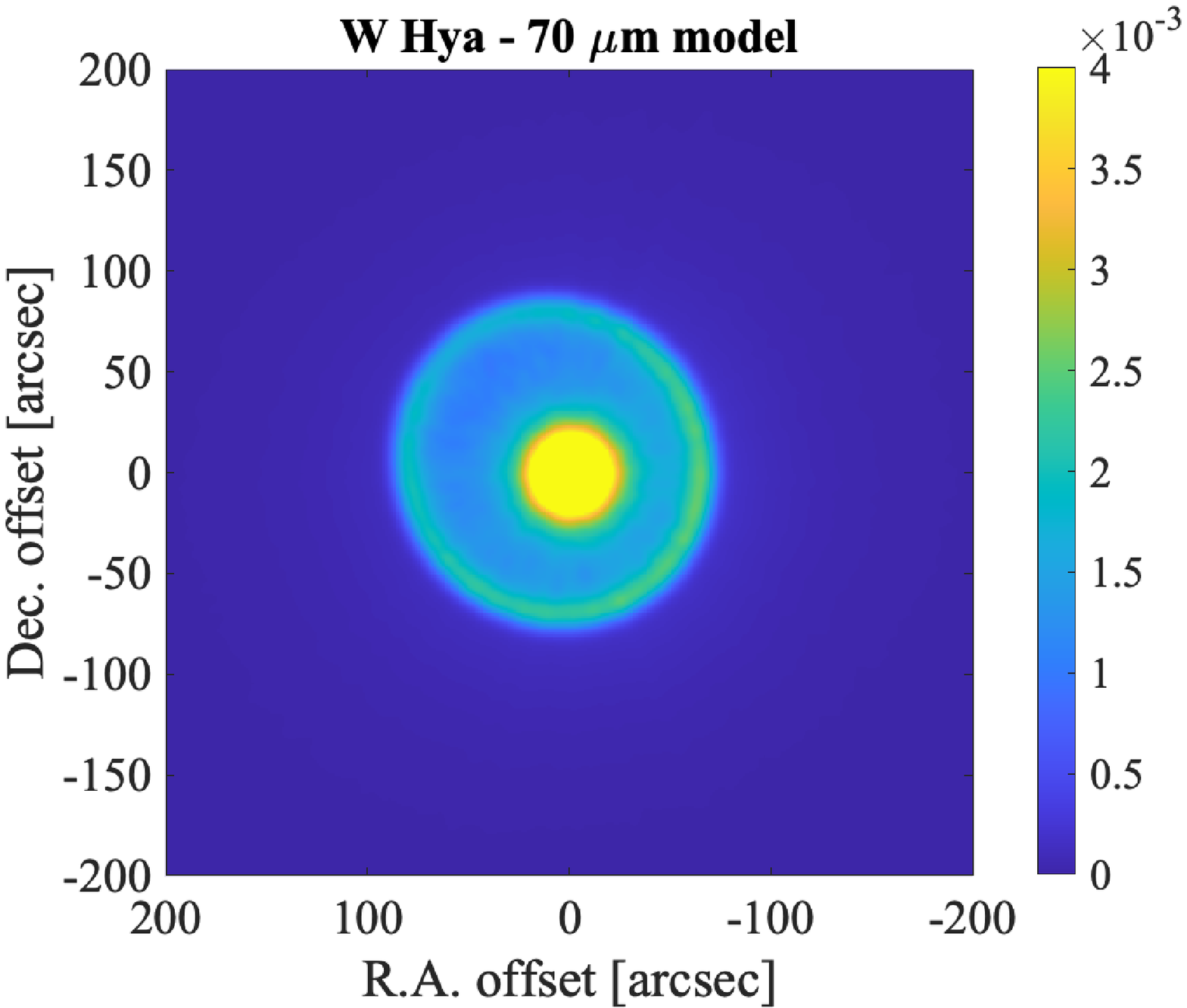}
\includegraphics[width=8cm]{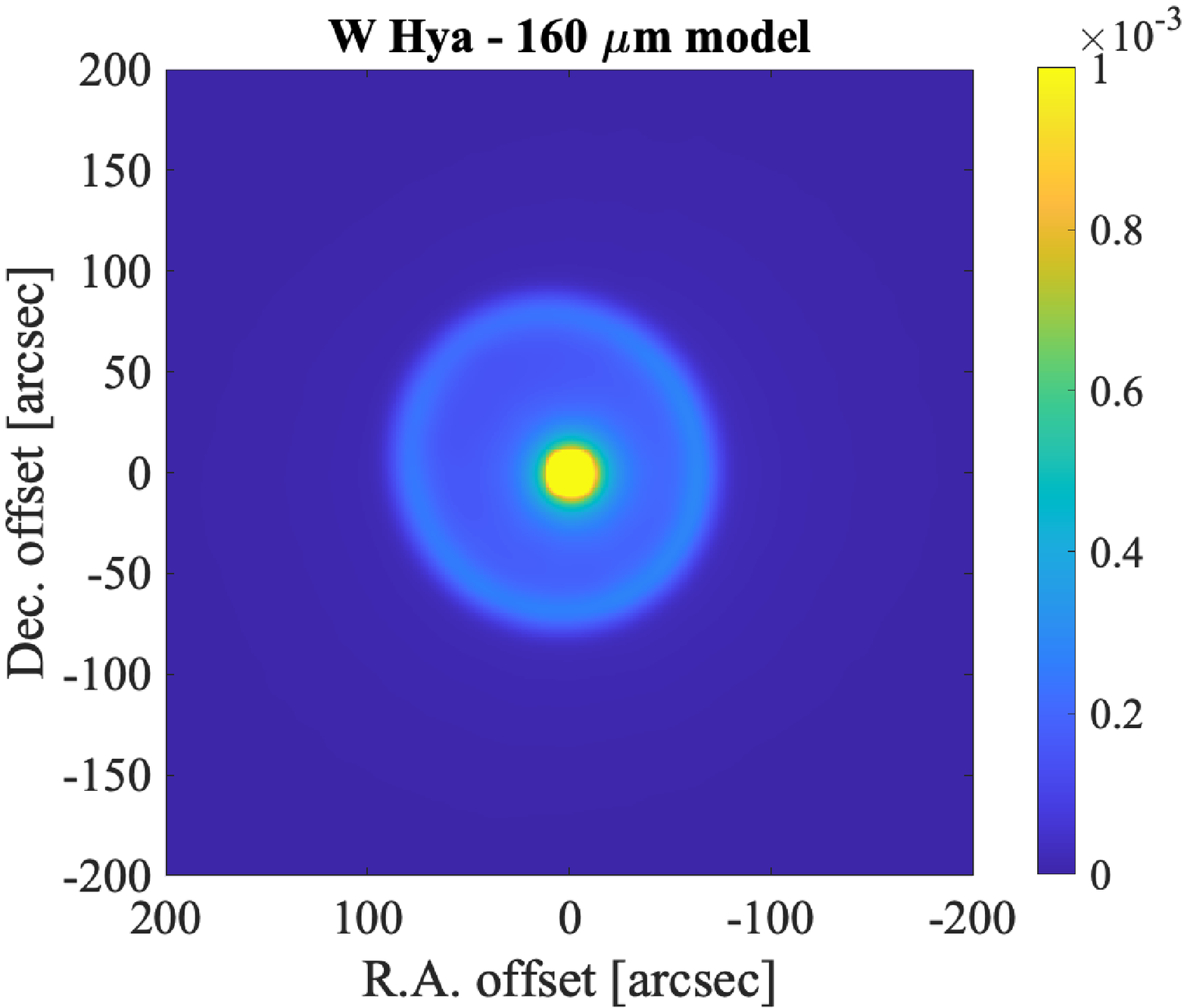}
\includegraphics[width=8cm]{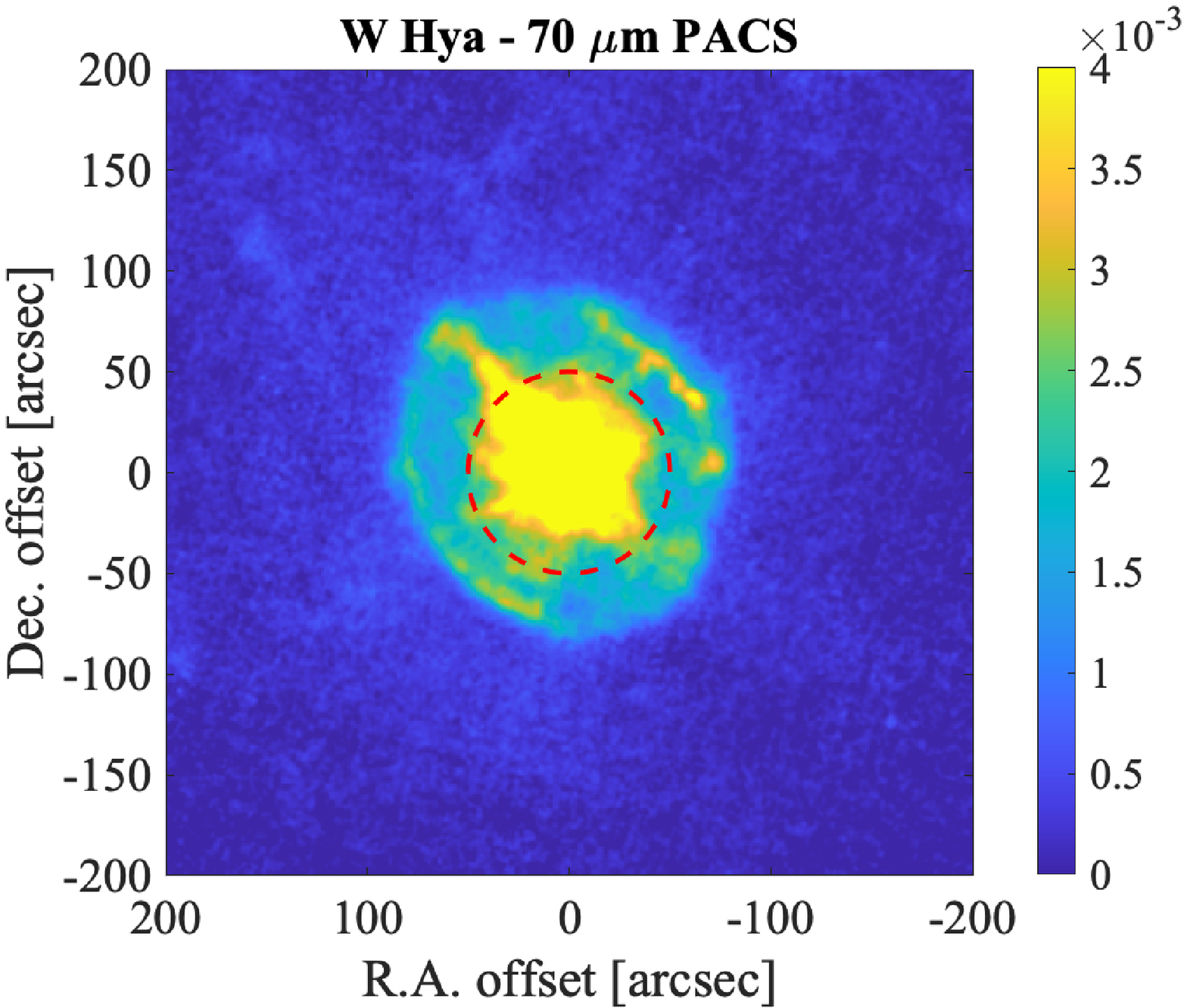}
\includegraphics[width=8cm]{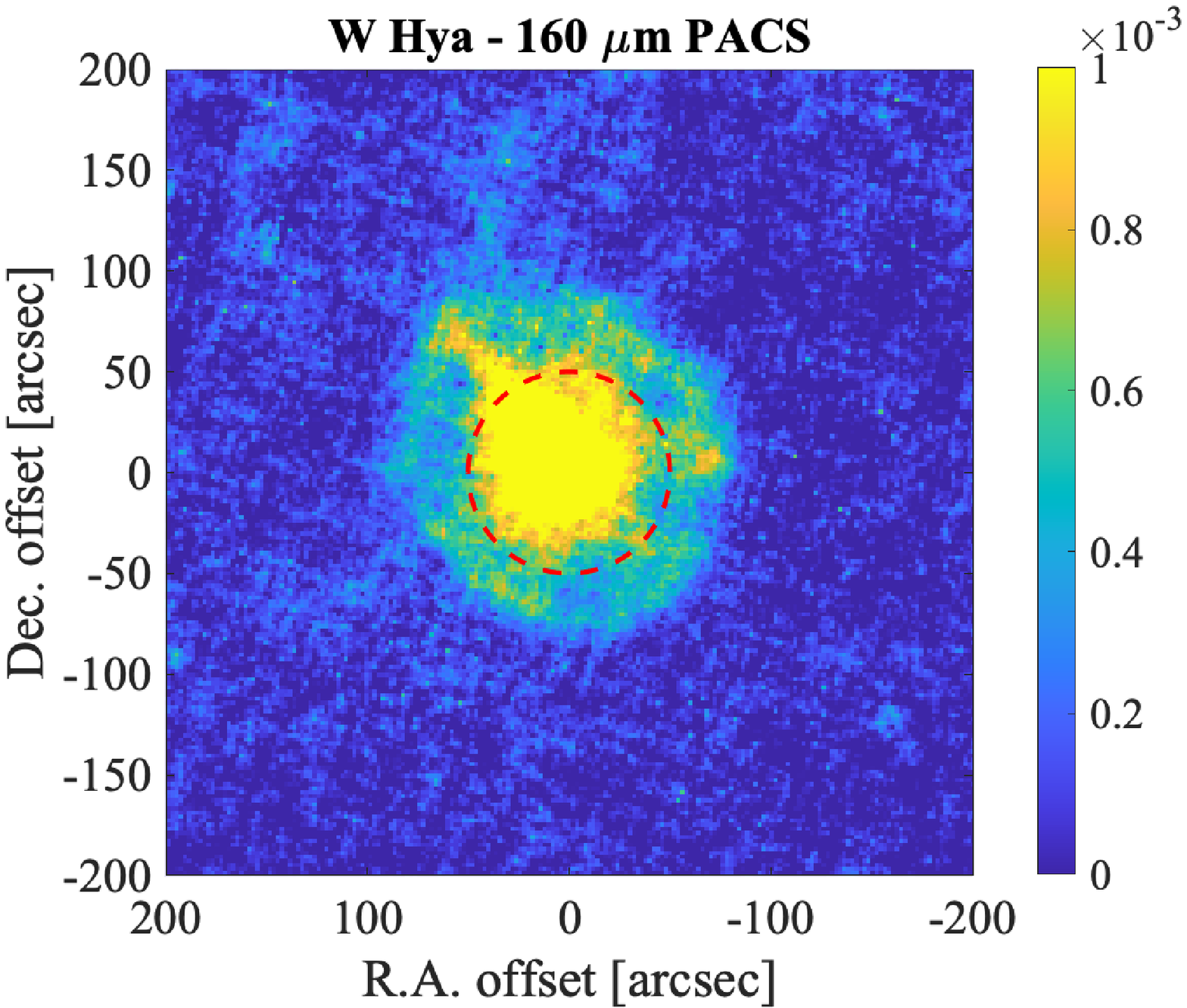}
\includegraphics[width=8cm]{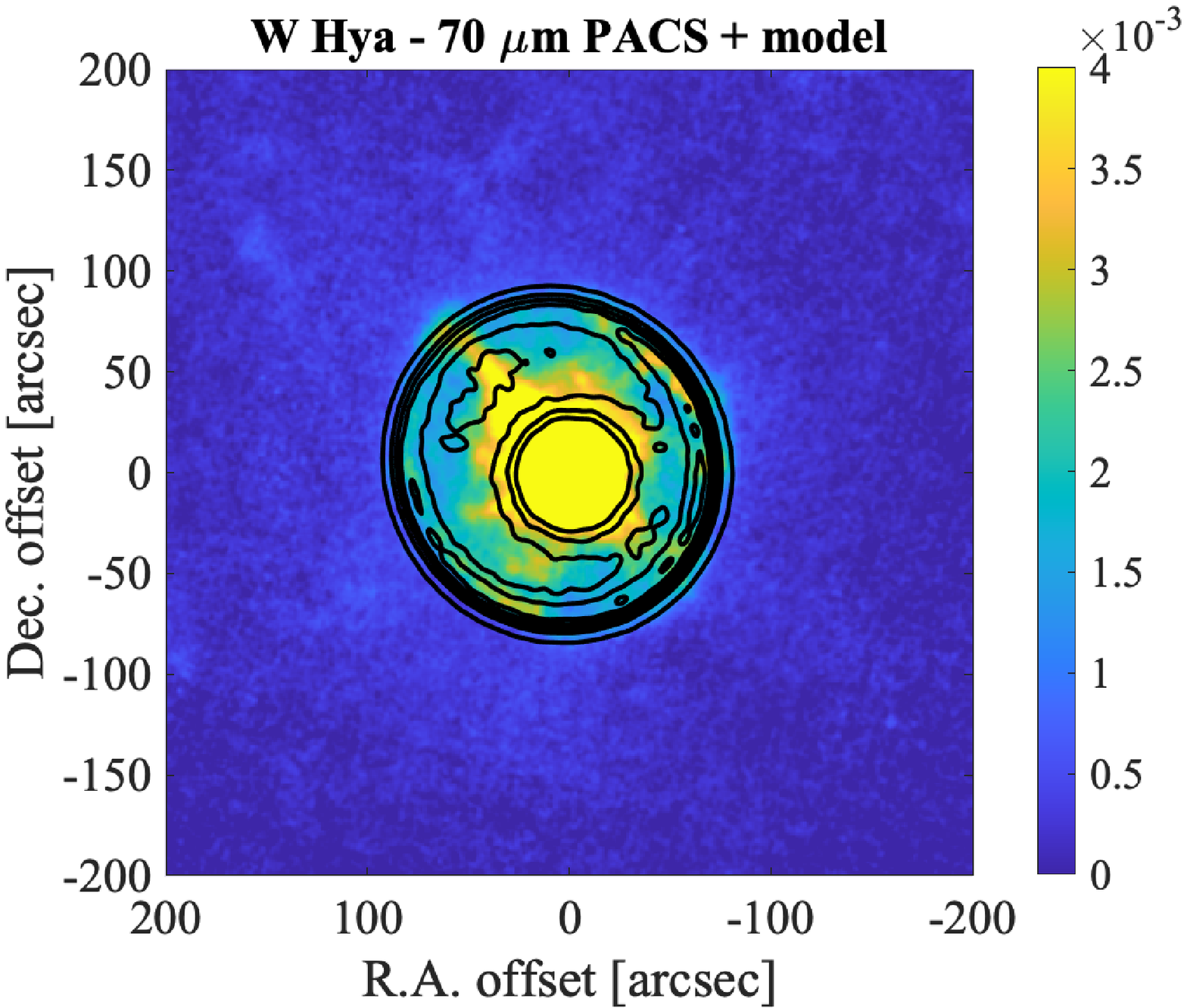}
\includegraphics[width=8cm]{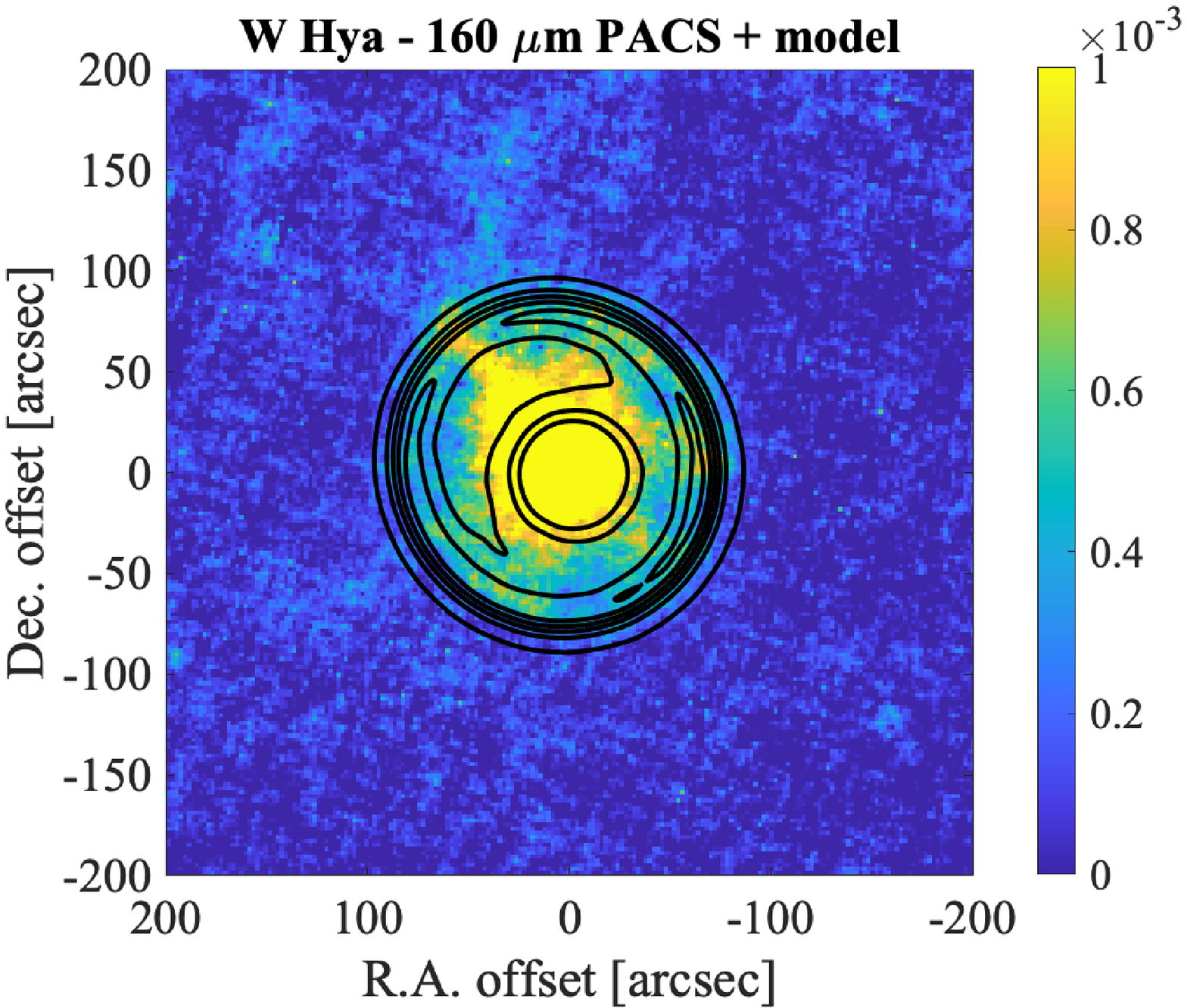}
\caption{W Hya: \emph{Top to bottom:} The Radmc3D model, the PACS image, and the PACS image with contours from the model. Images are for 70\,\micron~(left) and 160\,\micron~(right). Maximum contour levels are 3$\times10^{-3}$\,\Jyarcsec (70\,\micron) and 0.35$\times10^{-3}$\,\Jyarcsec (160\,\micron), respectively. Minimum contour levels are 10\% of maximum. The colour scale is in \Jyarcsec. The red dashed circle shows the mask used to measure the flux from the star and present-day mass-loss.}
\label{f:whya}
\end{figure*}

\begin{figure*}
\centering
\includegraphics[width=8cm]{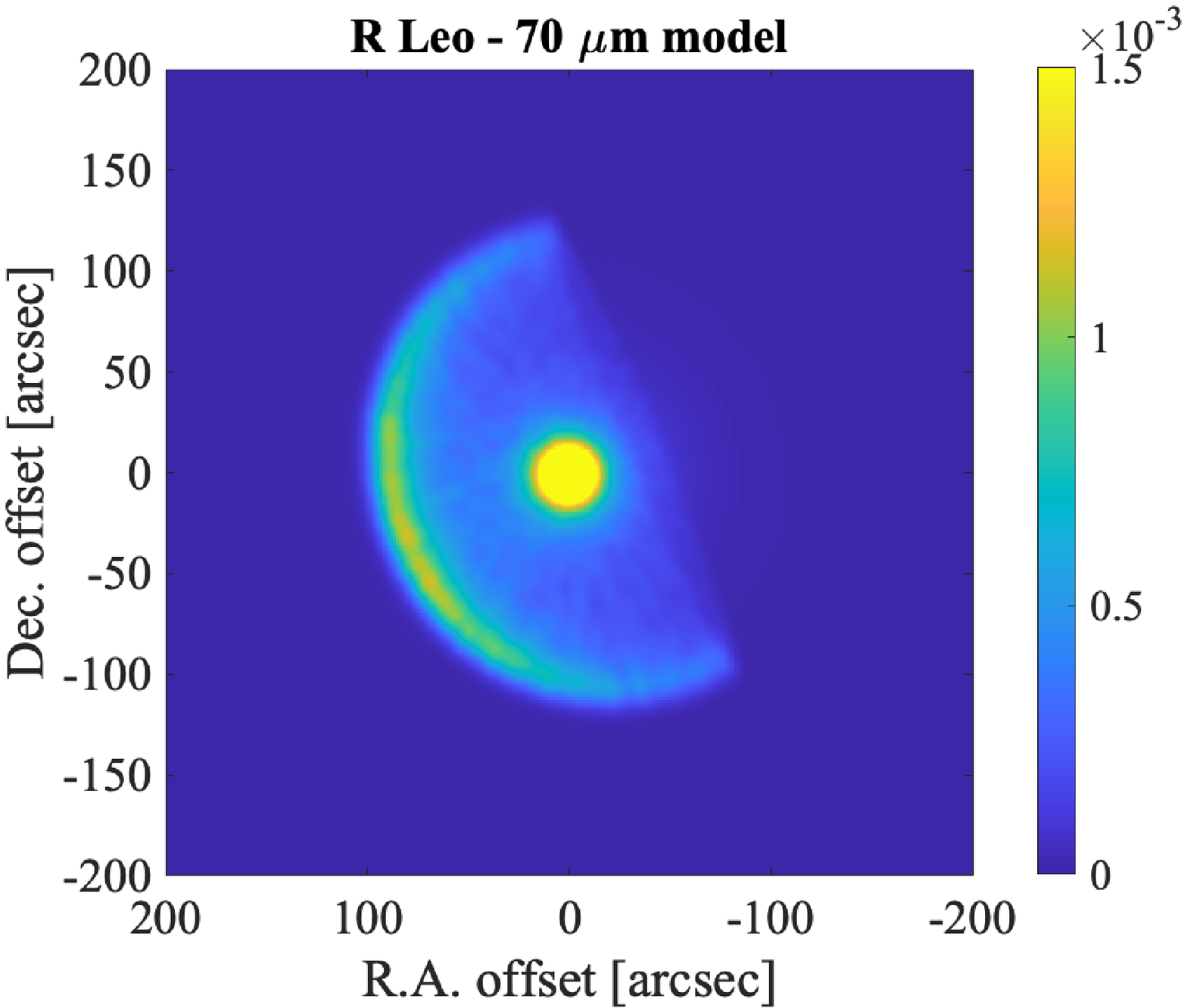}
\includegraphics[width=8cm]{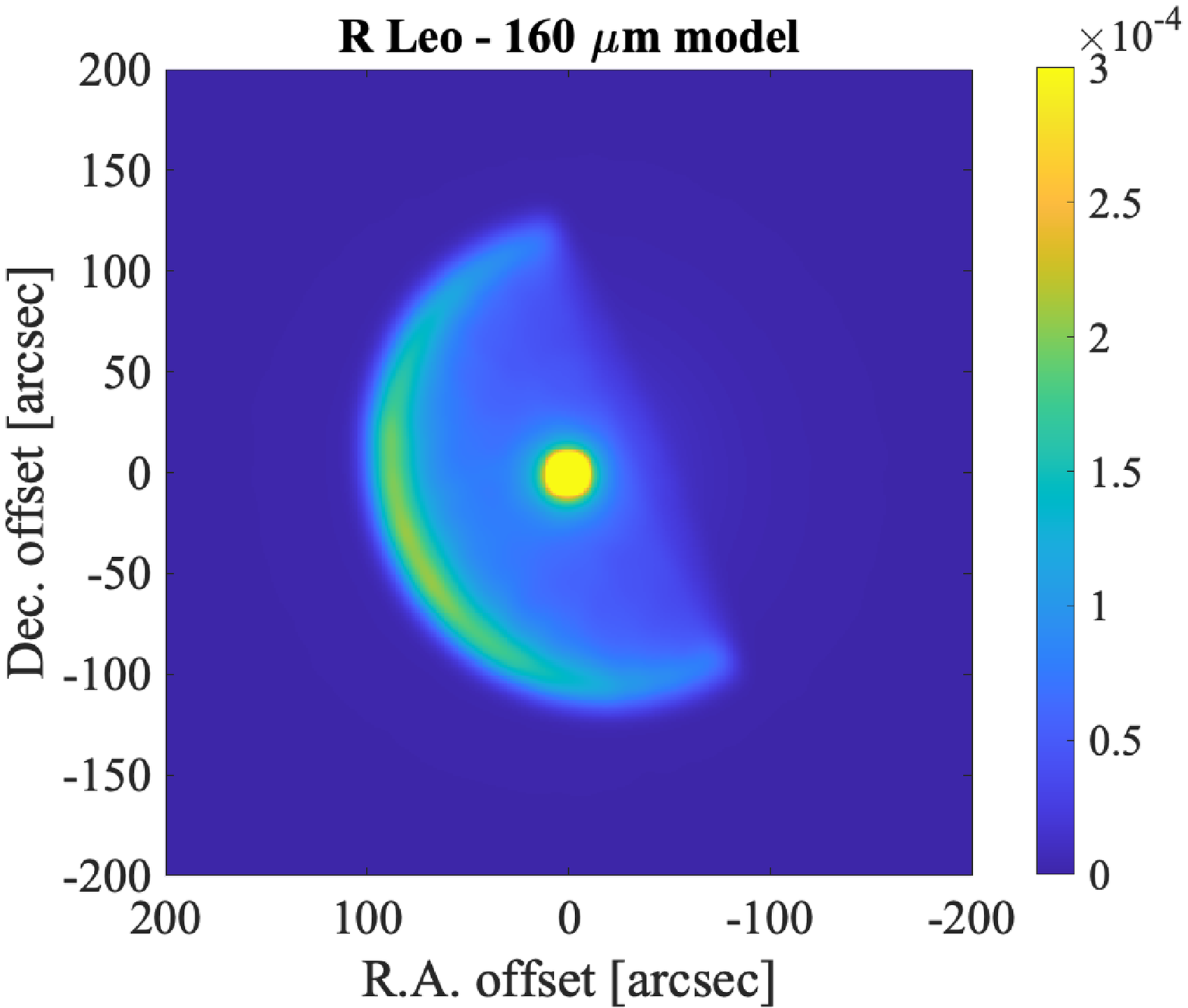}
\includegraphics[width=8cm]{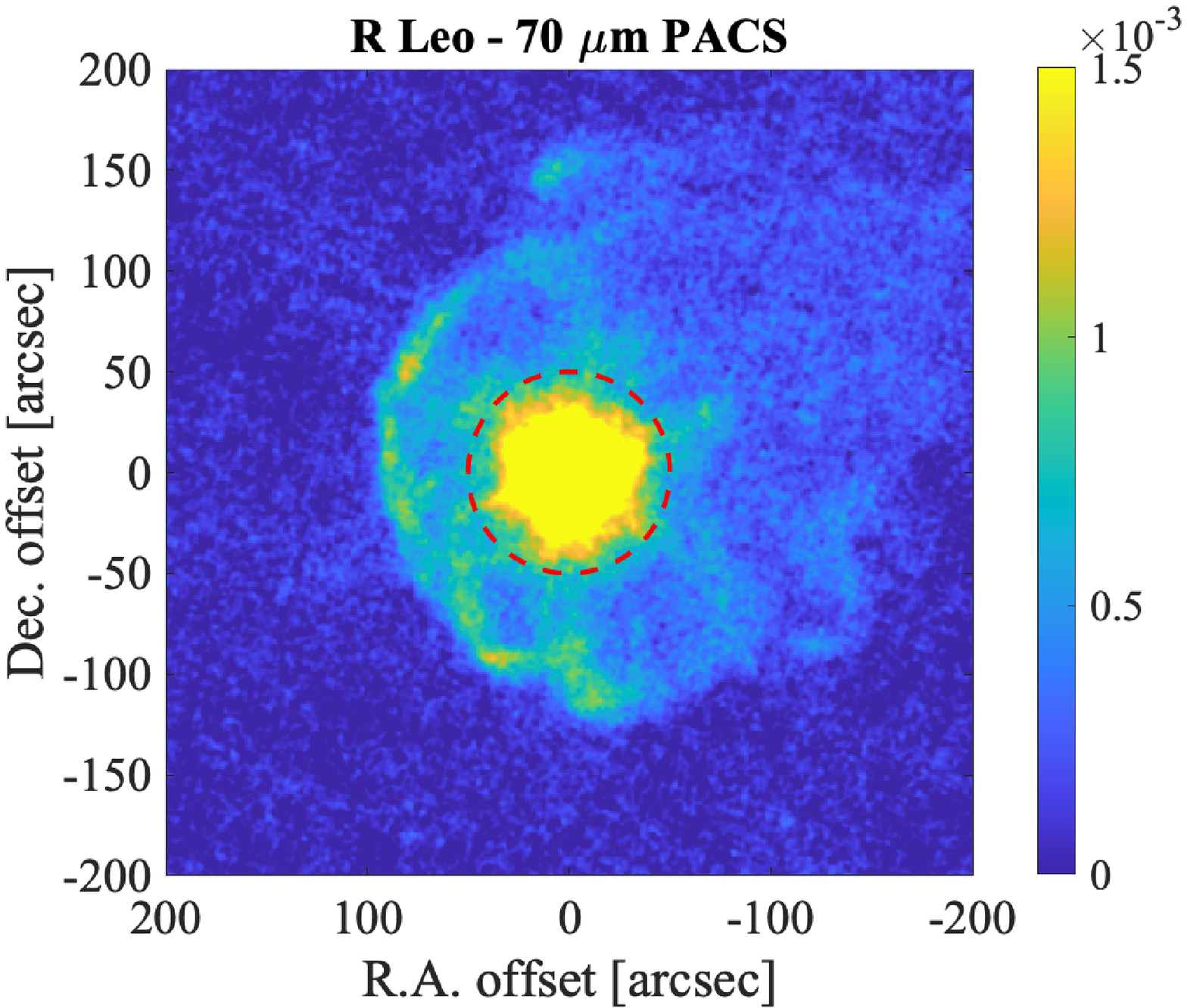}
\includegraphics[width=8cm]{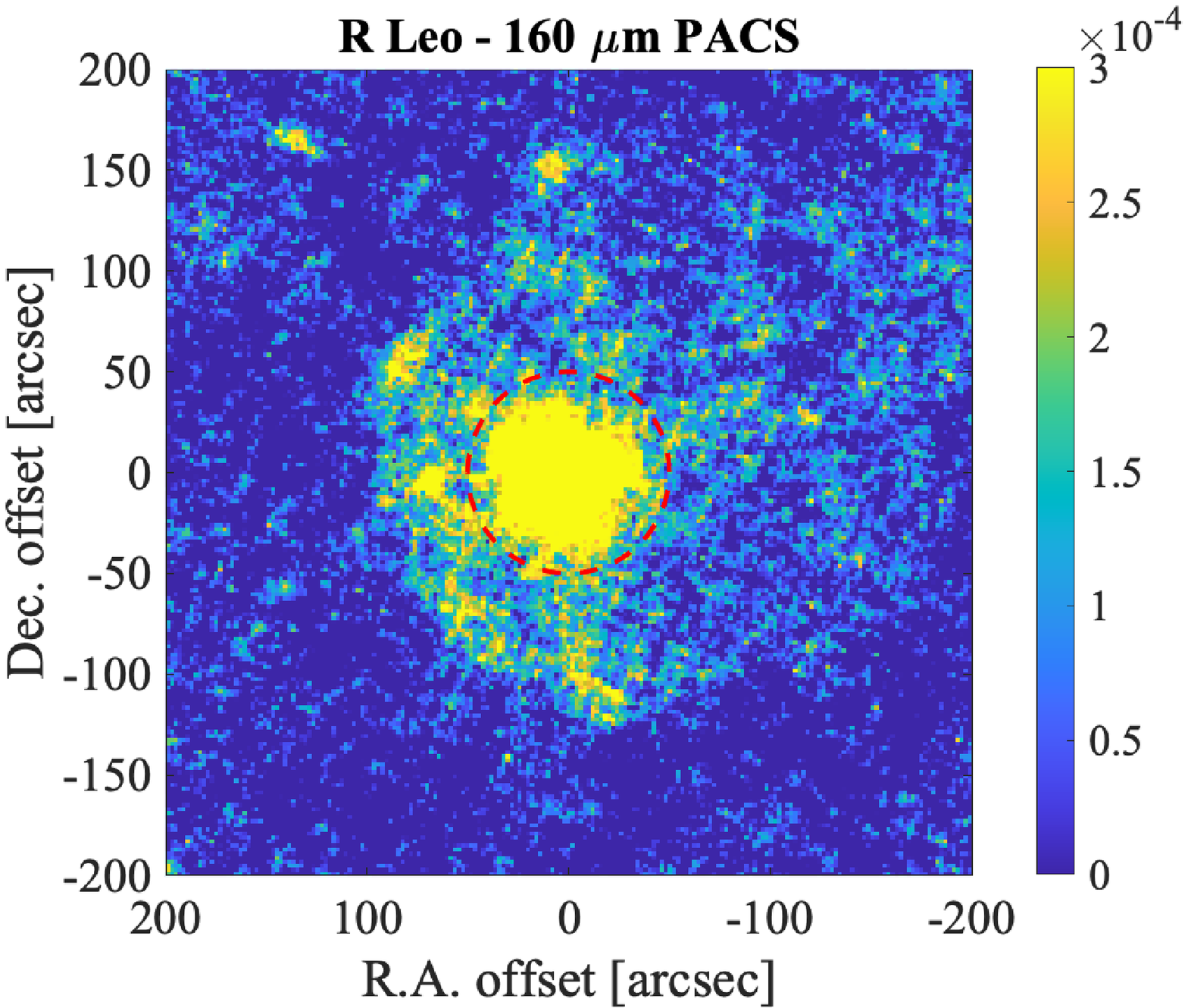}
\includegraphics[width=8cm]{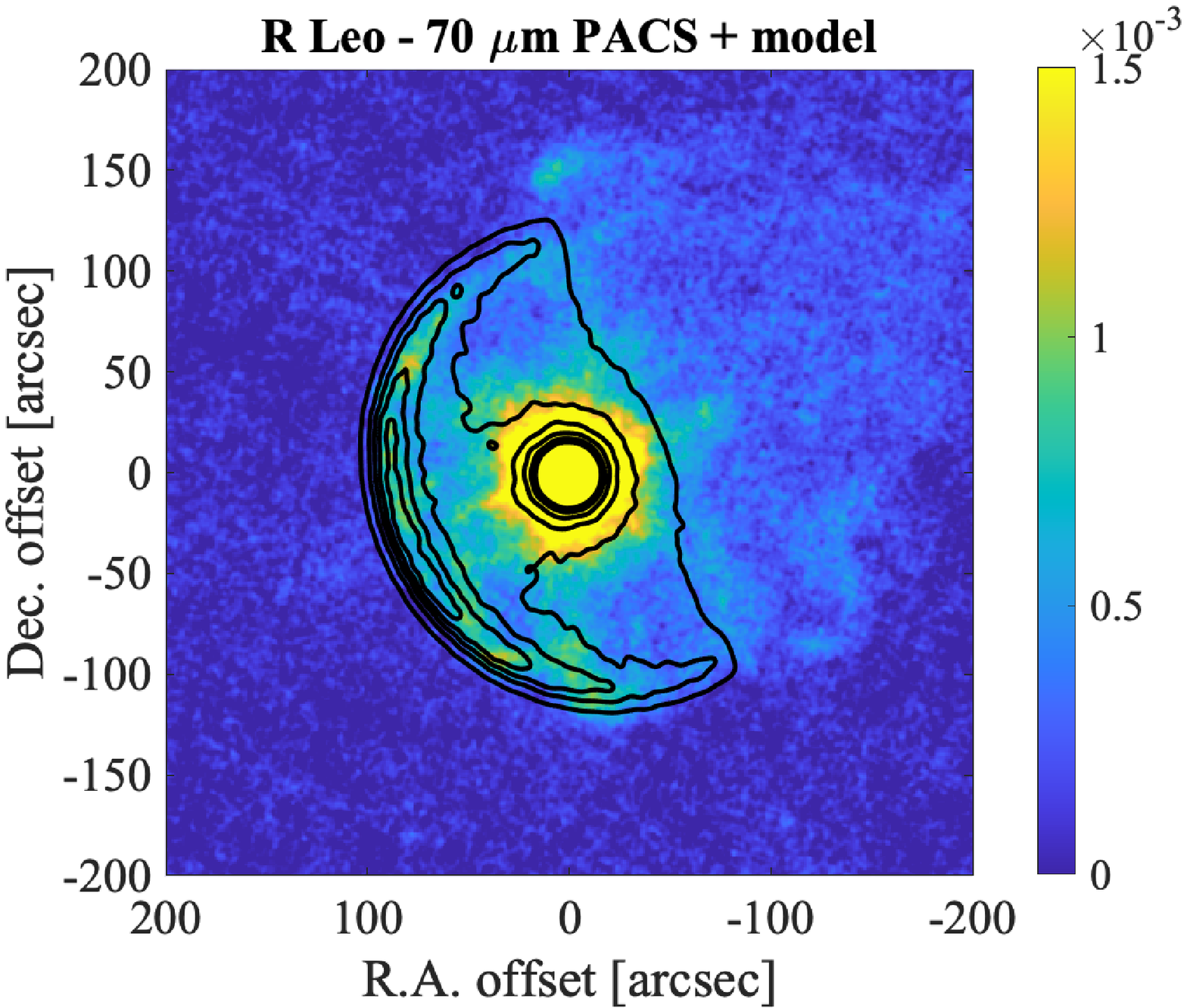}
\includegraphics[width=8cm]{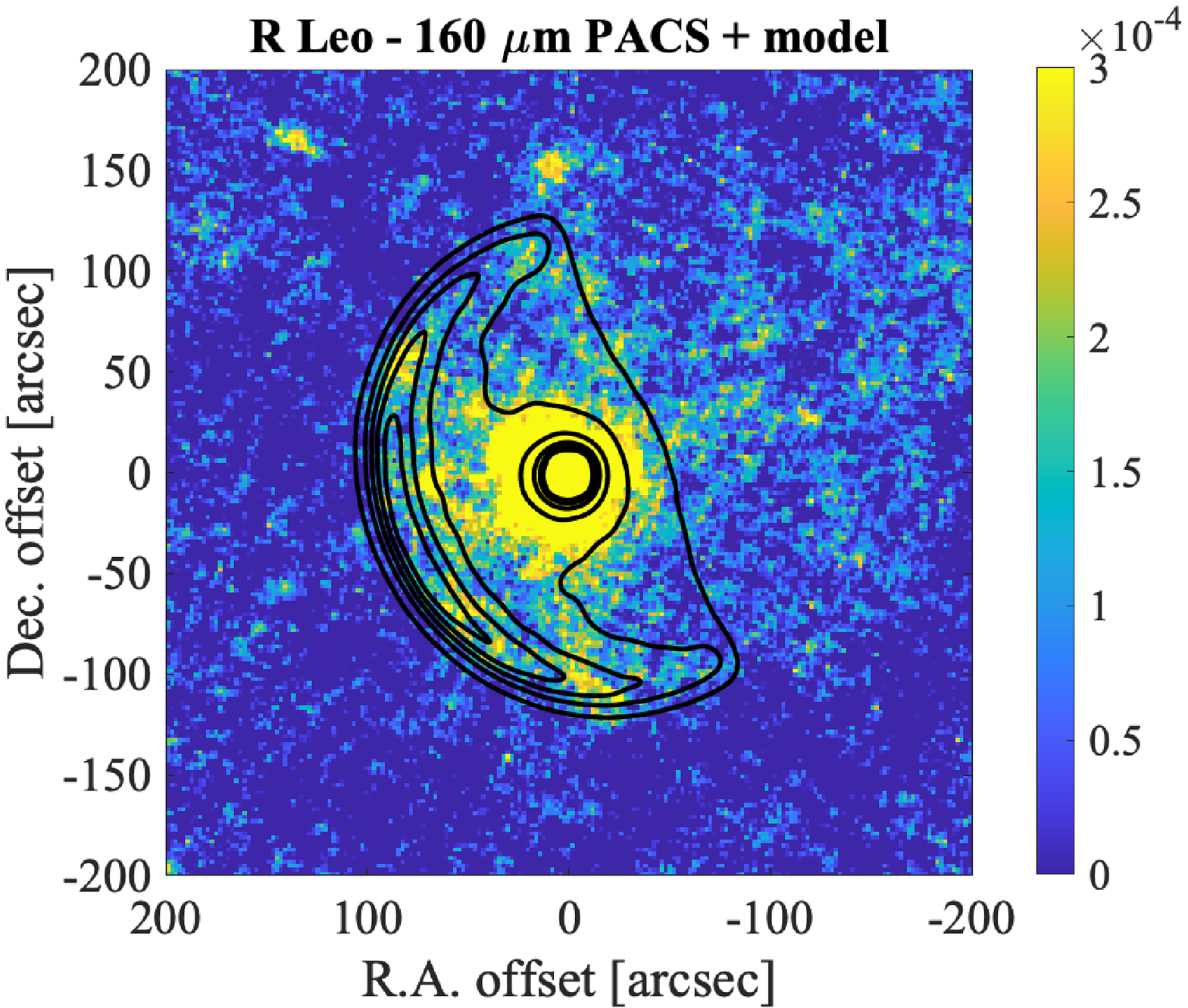}
\caption{R Leo: \emph{Top to bottom:} The Radmc3D model, the PACS image, and the PACS image with contours from the model. Images are for 70\,\micron~(left) and 160\,\micron~(right). Maximum contour levels are 1.1$\times10^{-3}$\,\Jyarcsec (70\,\micron) and 0.2$\times10^{-3}$\,\Jyarcsec (160\,\micron), respectively. Minimum contour levels are 10\% of maximum. The colour scale is in \Jyarcsec. The red dashed circle shows the mask used to measure the flux from the star and present-day mass-loss.}
\label{f:rleo}
\end{figure*}

\begin{figure*}
\centering
\includegraphics[width=8cm]{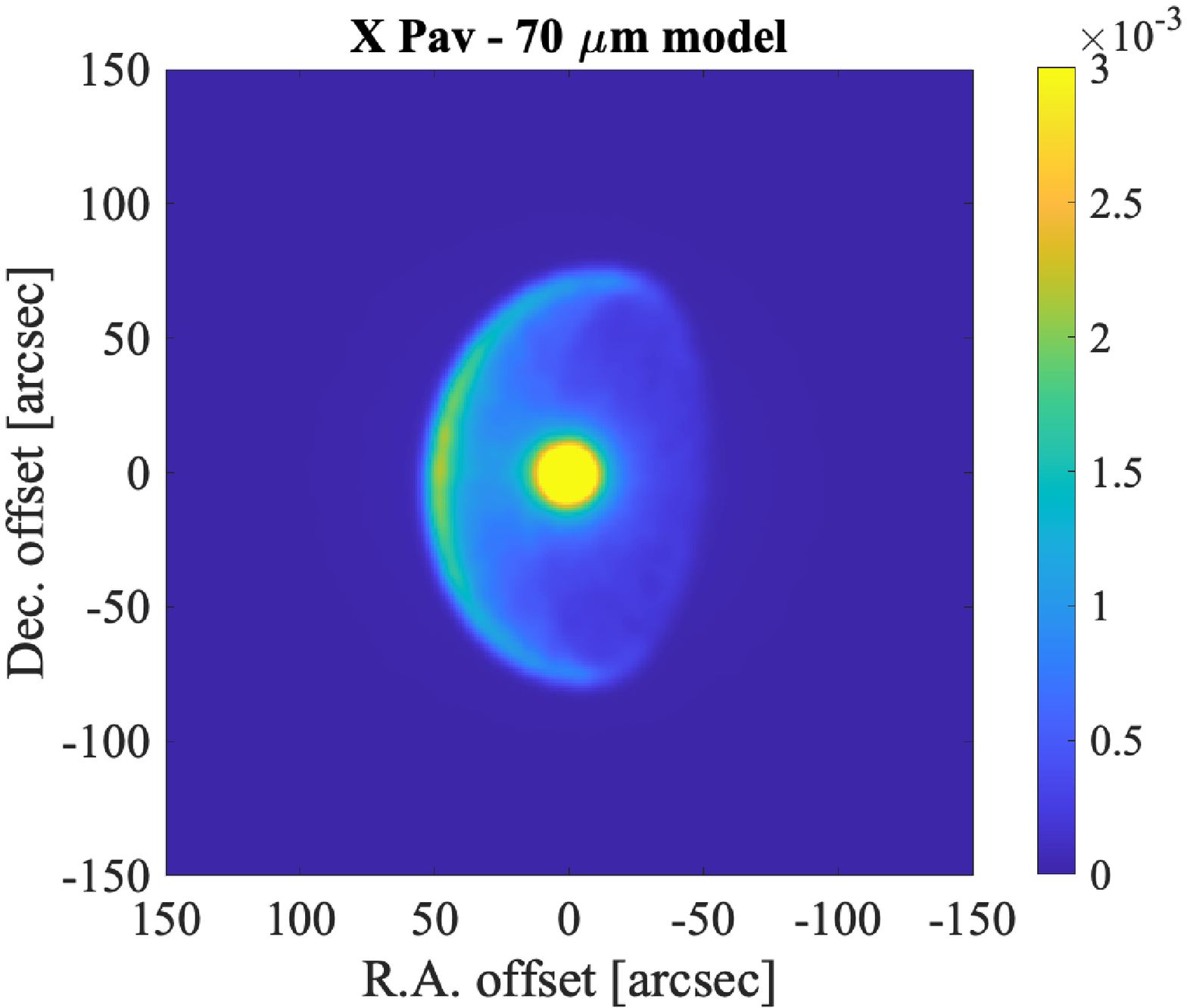}
\includegraphics[width=8cm]{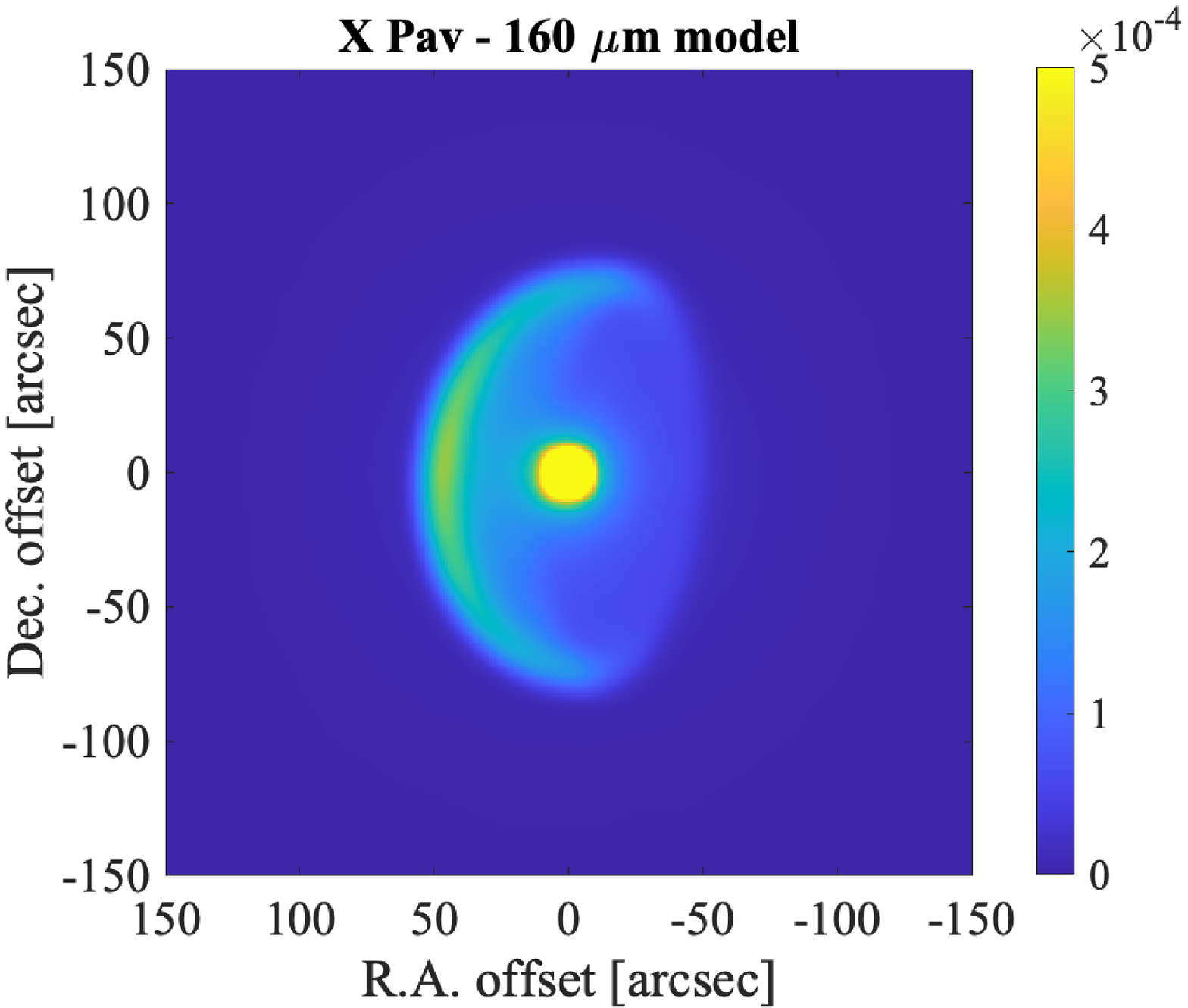}
\includegraphics[width=8cm]{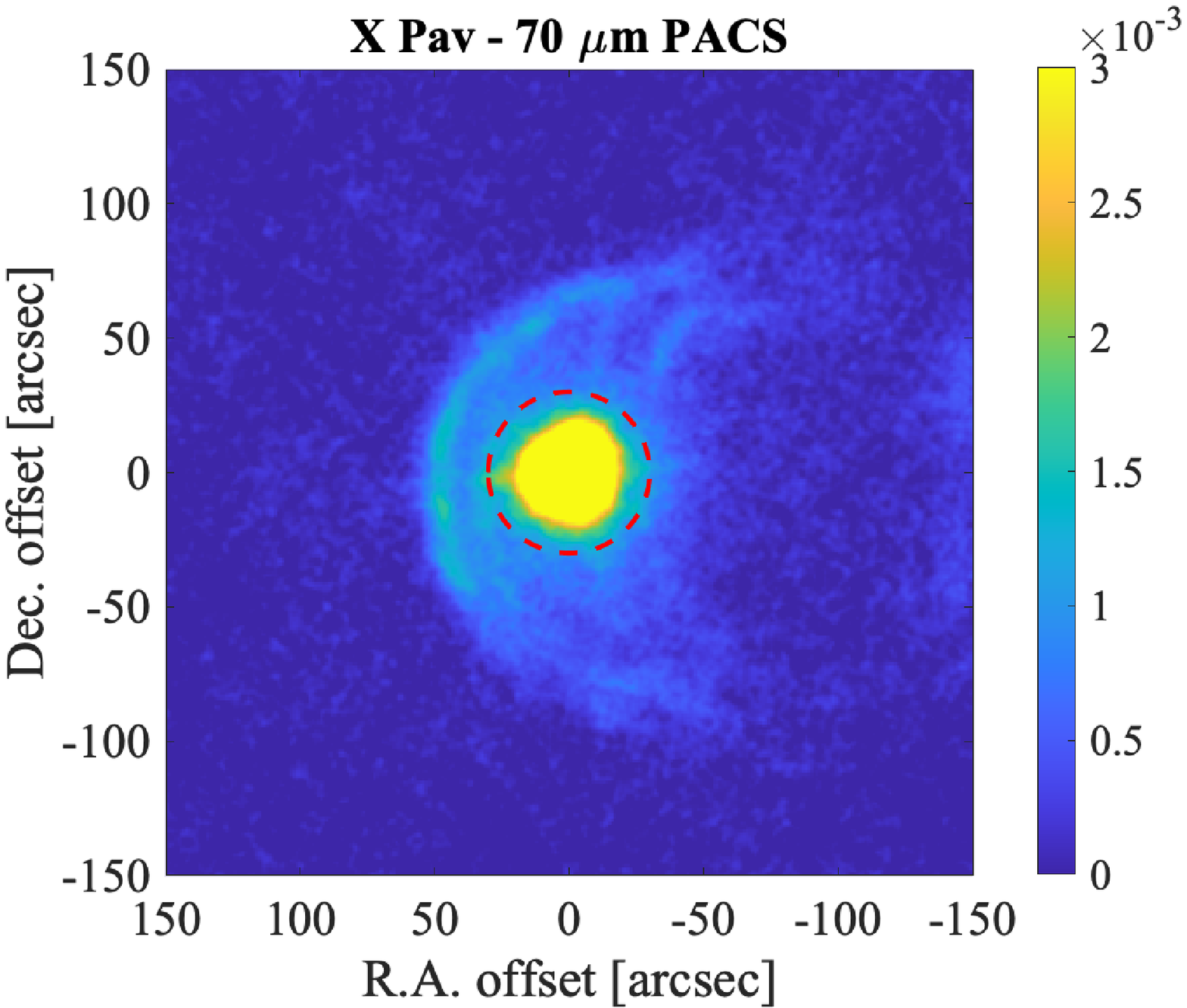}
\includegraphics[width=8cm]{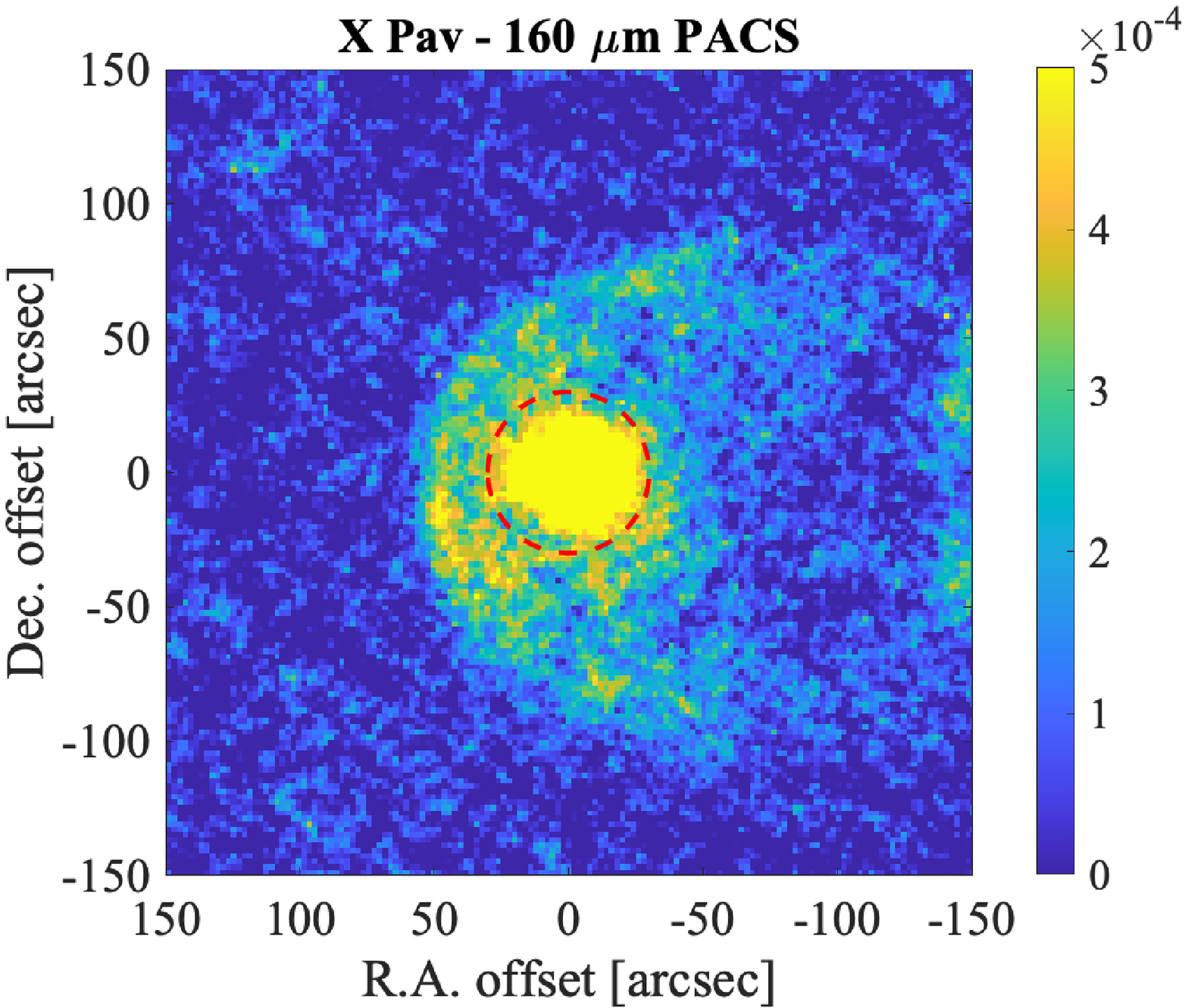}
\includegraphics[width=8cm]{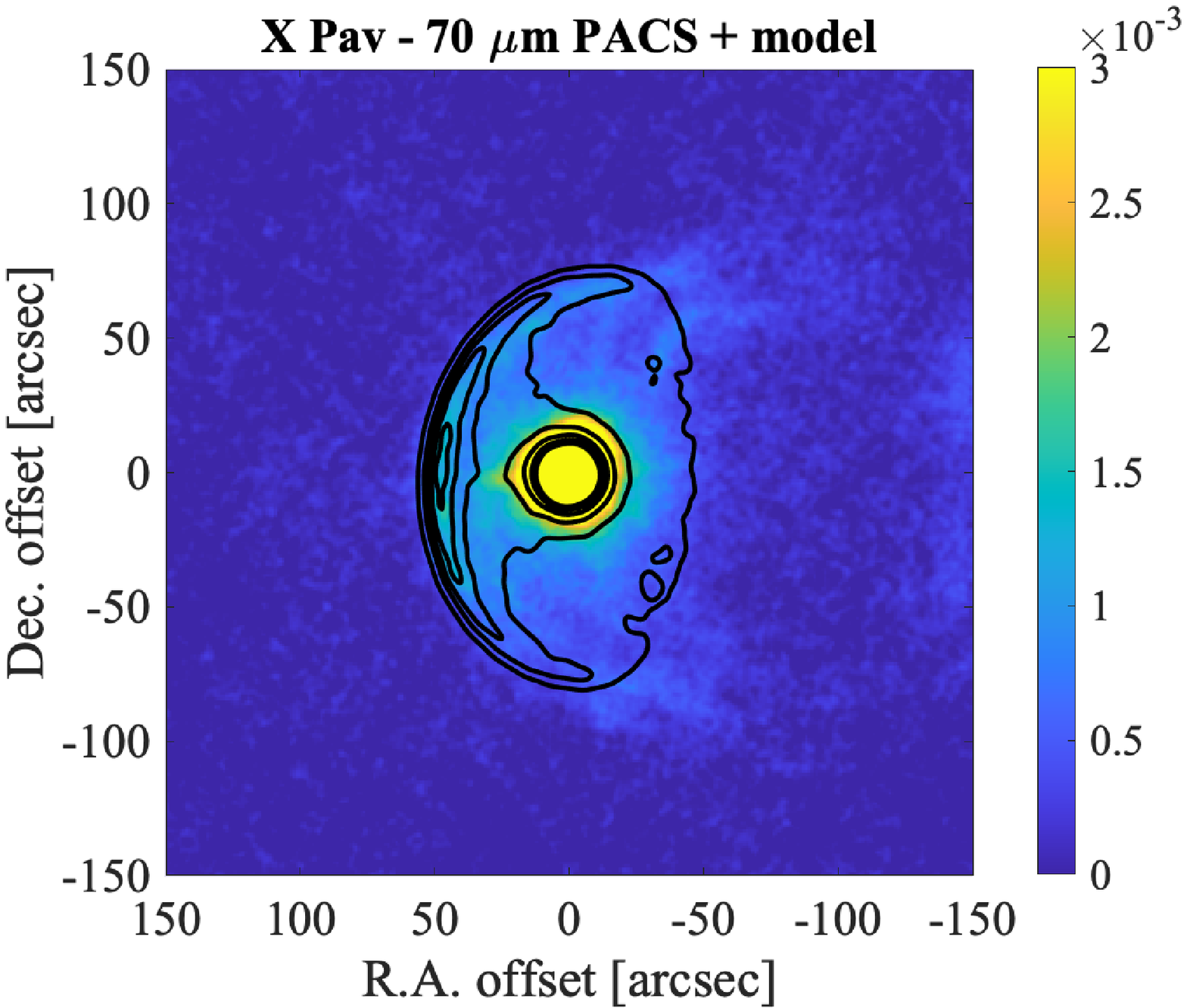}
\includegraphics[width=8cm]{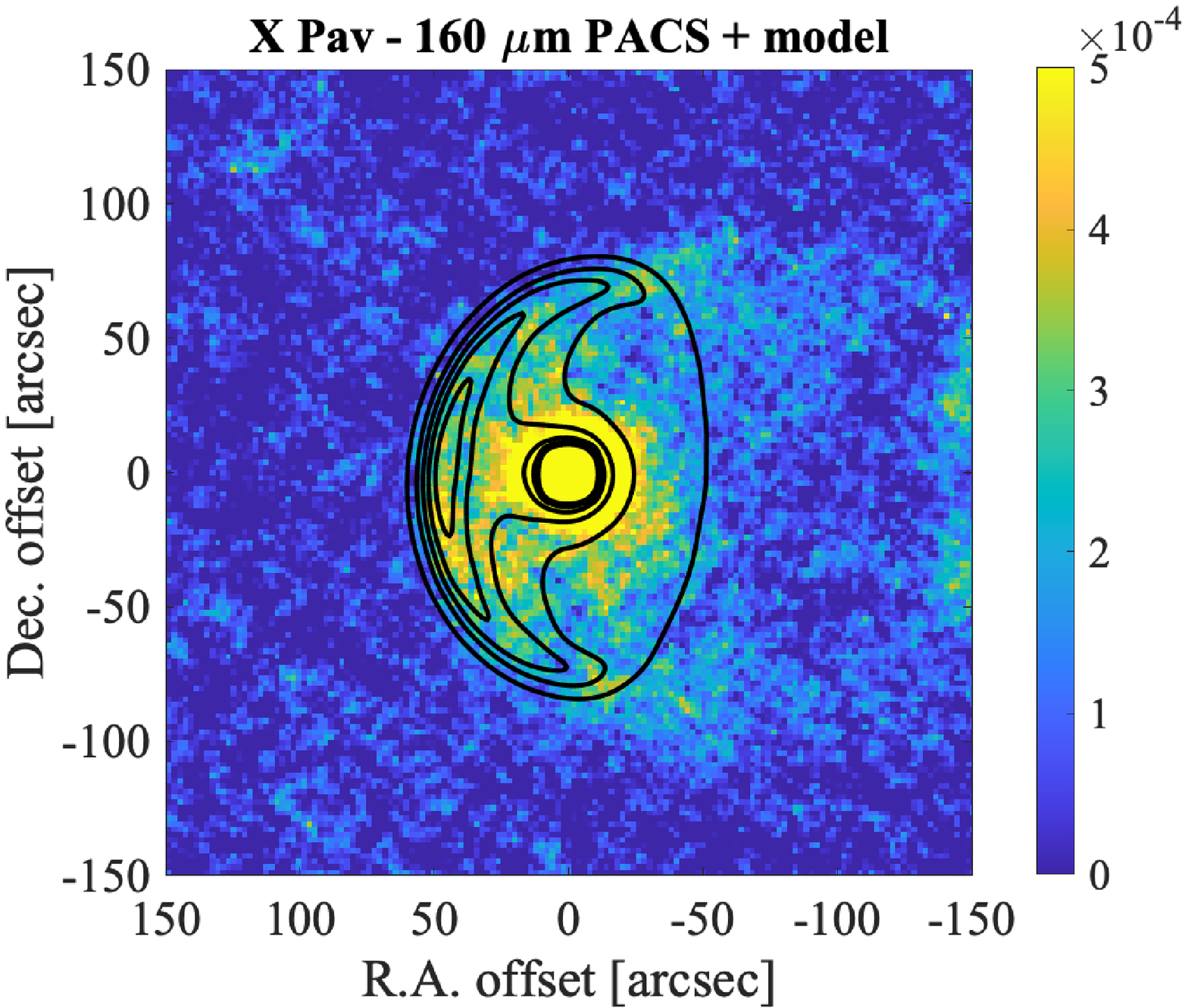}
\caption{X Pav: \emph{Top to bottom:} The Radmc3D model, the PACS image, and the PACS image with contours from the model. Images are for 70\,\micron~(left) and 160\,\micron~(right). Maximum contour levels are 2.2$\times10^{-3}$\,\Jyarcsec (70\,\micron) and 0.34$\times10^{-3}$\,\Jyarcsec (160\,\micron), respectively. Minimum contour levels are 10\% of maximum. The colour scale is in \Jyarcsec. The red dashed circle shows the mask used to measure the flux from the star and present-day mass-loss.}
\label{f:xpav}
\end{figure*}

\begin{figure*}
\centering
\includegraphics[width=8cm]{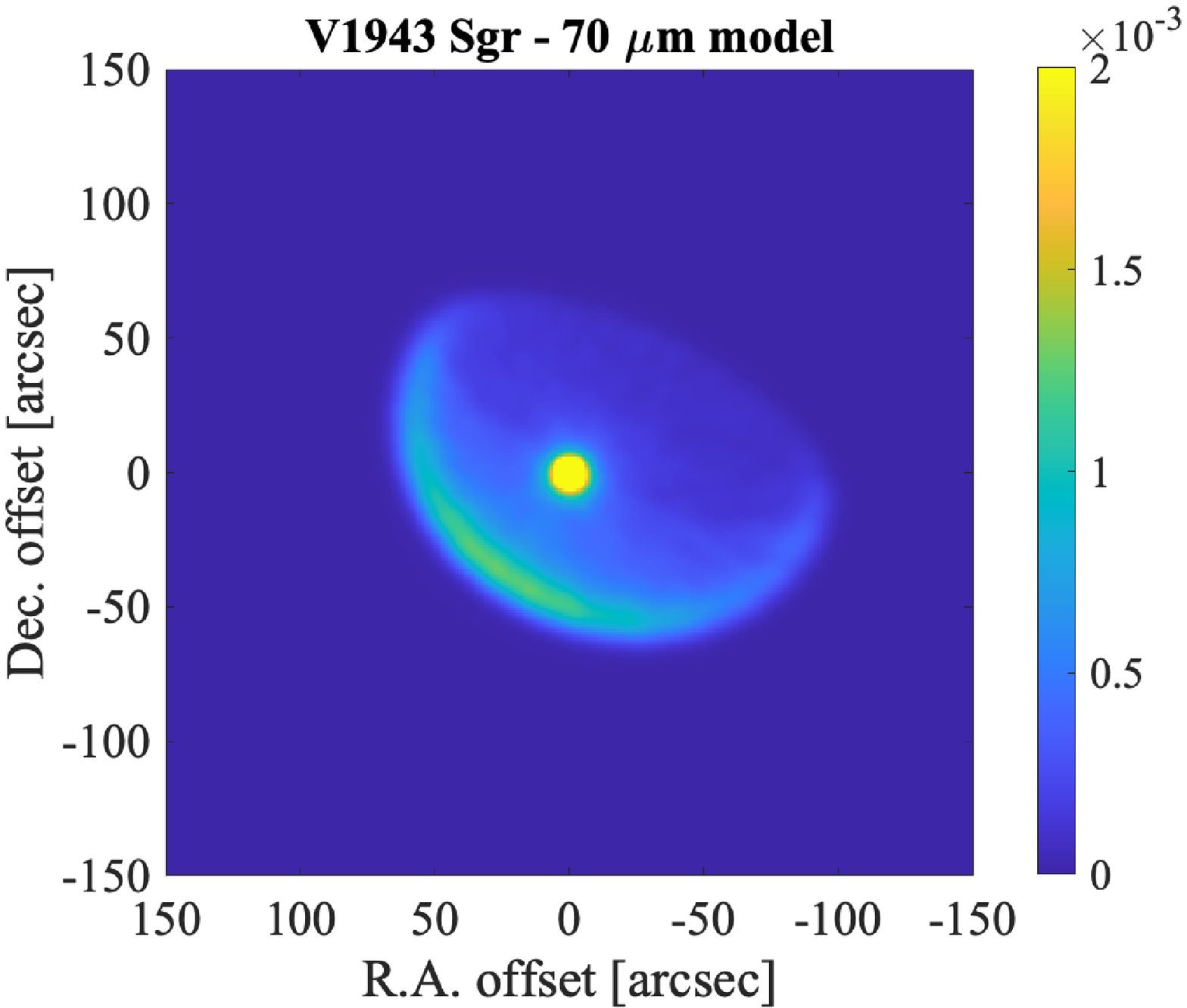}
\includegraphics[width=8cm]{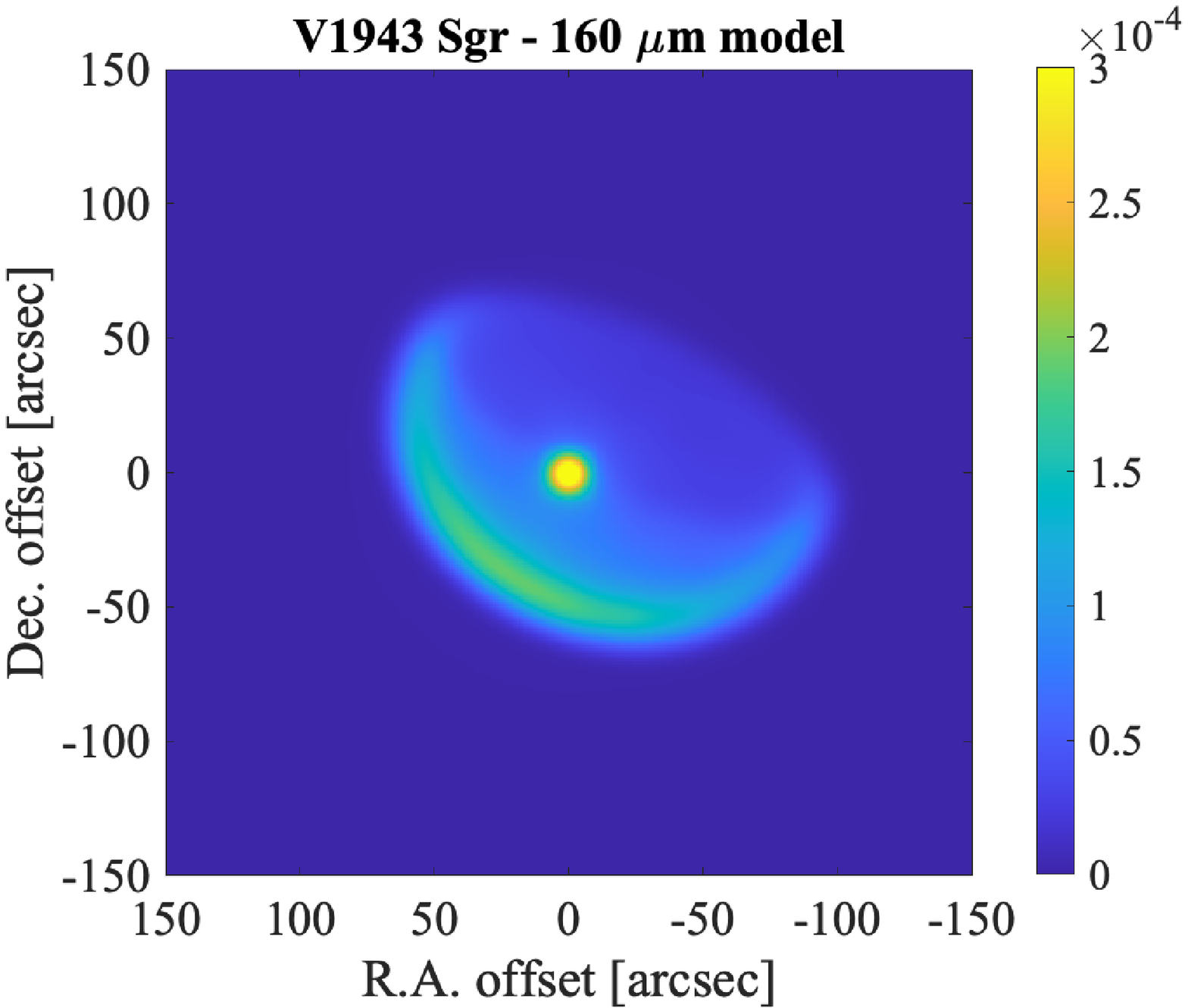}
\includegraphics[width=8cm]{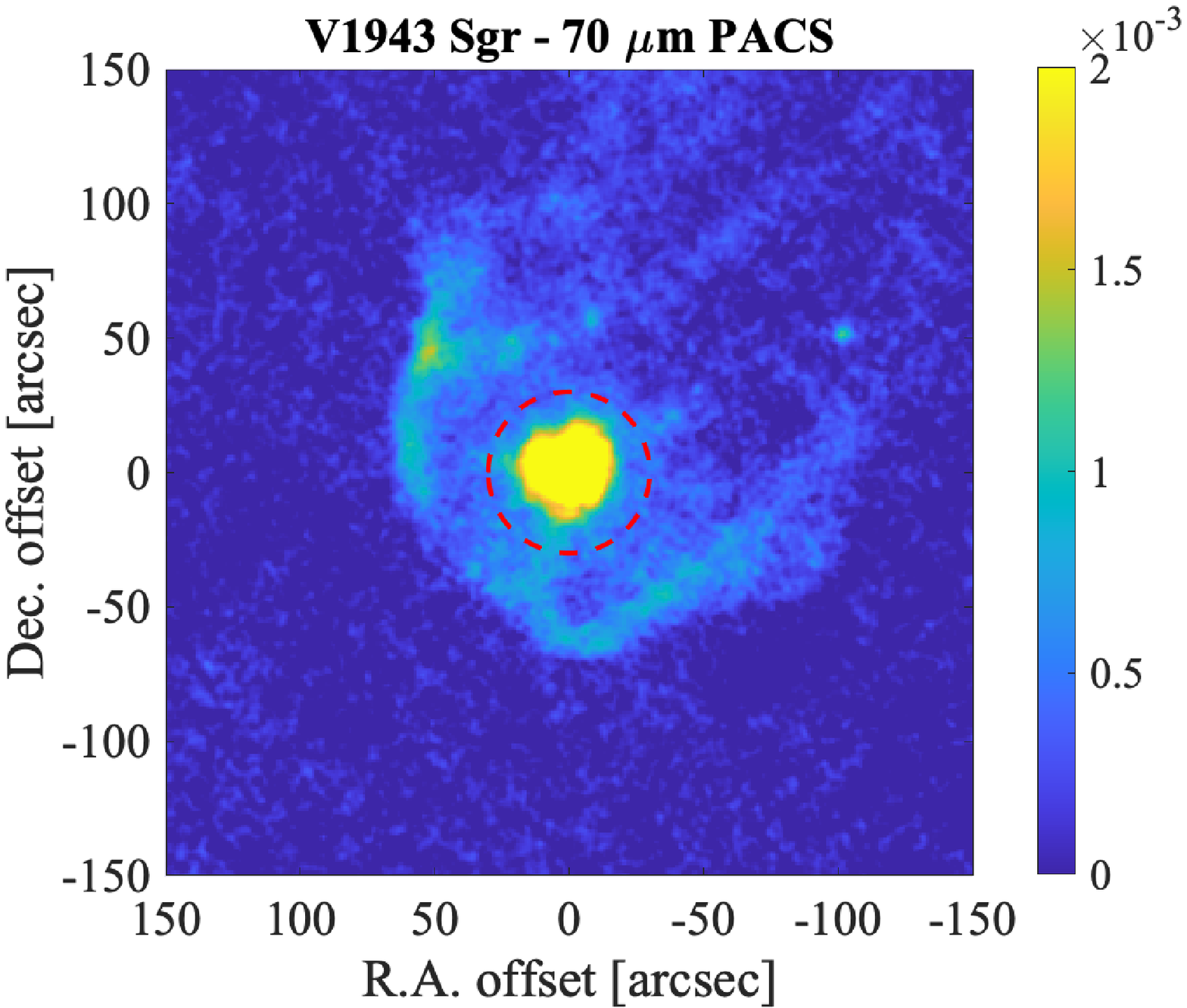}
\includegraphics[width=8cm]{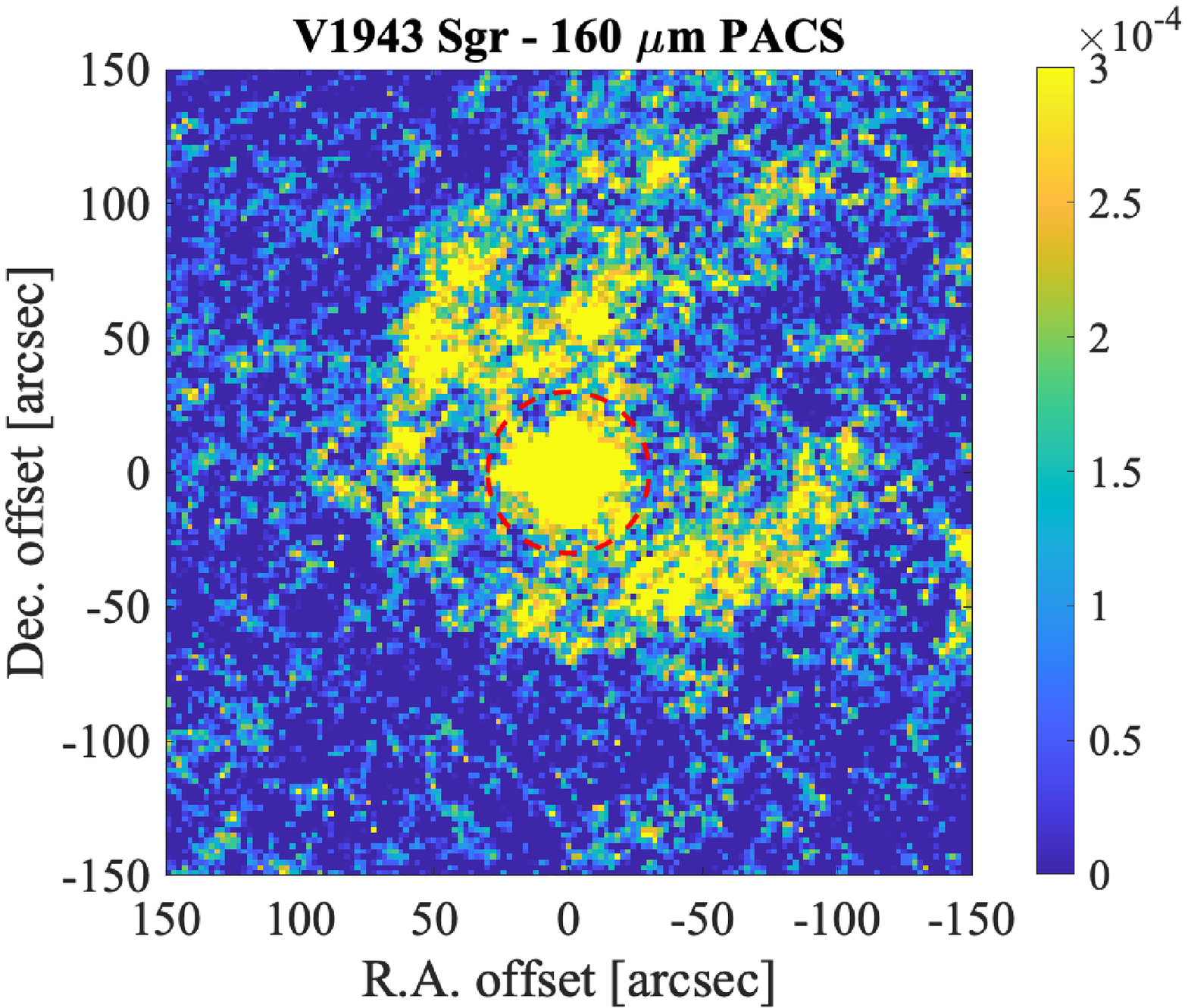}
\includegraphics[width=8cm]{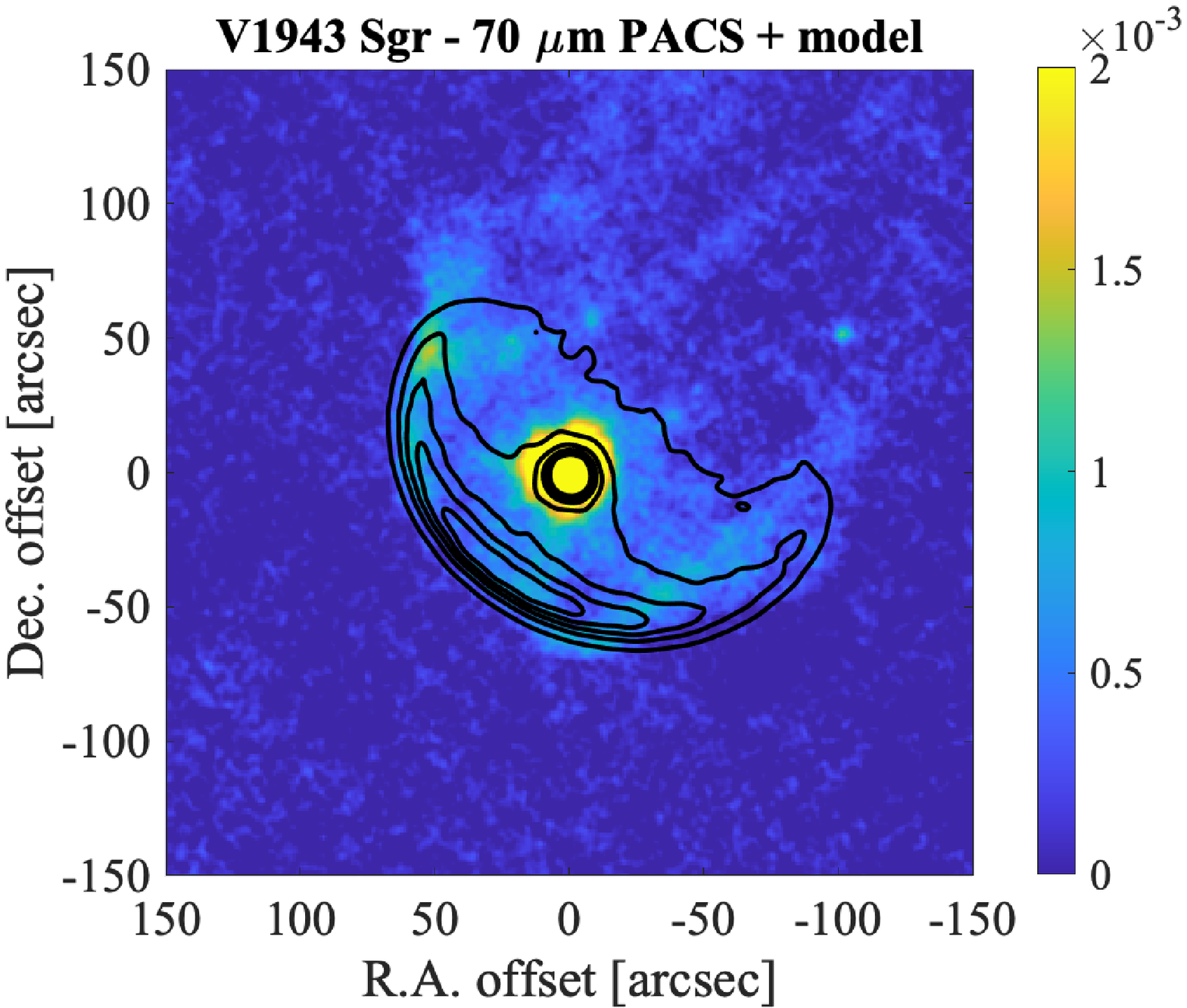}
\includegraphics[width=8cm]{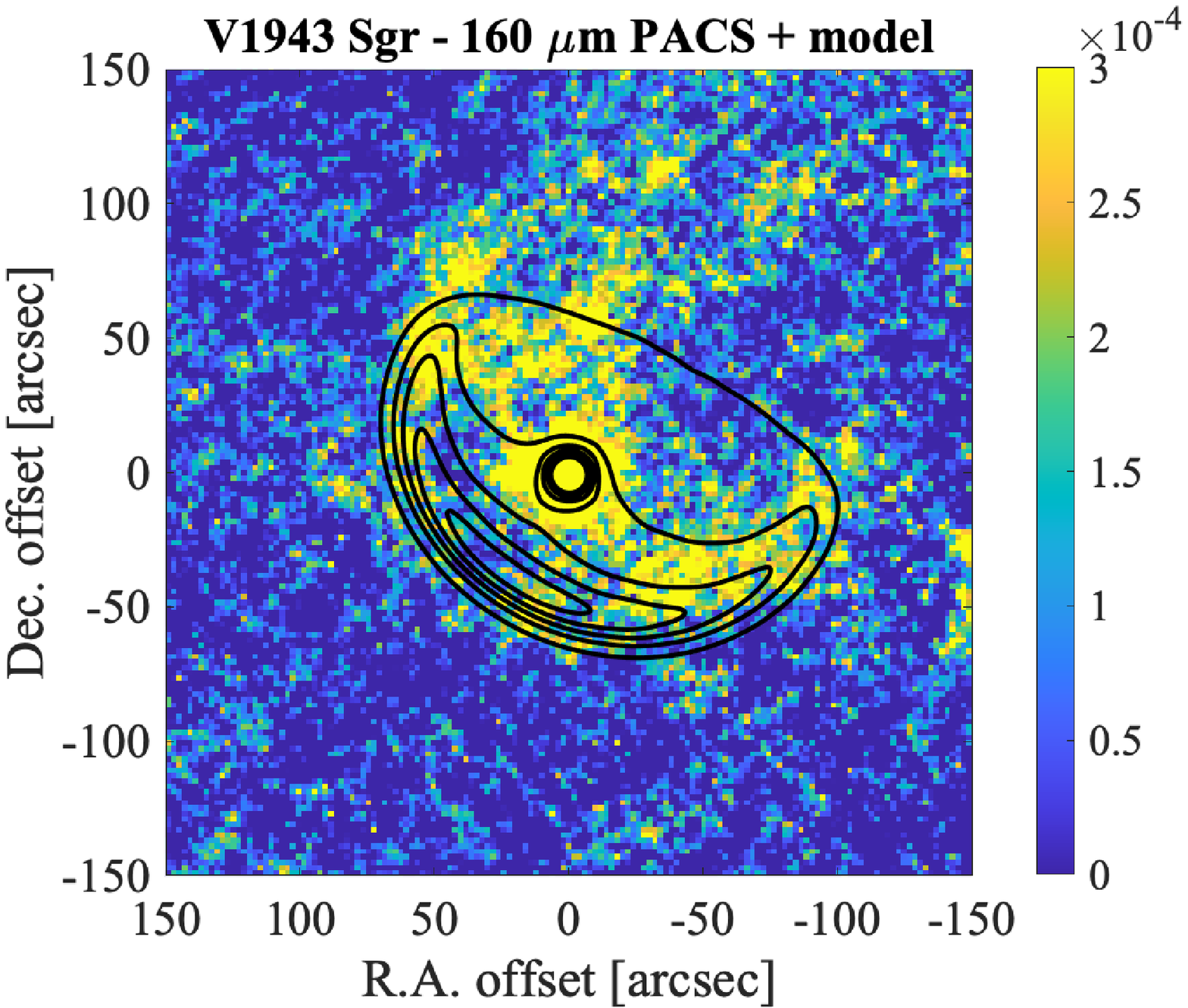}
\caption{V1943 Sgr: \emph{Top to bottom:} The Radmc3D model, the PACS image, and the PACS image with contours from the model. Images are for 70\,\micron~(left) and 160\,\micron~(right). Maximum contour levels are 1.2$\times10^{-3}$\,\Jyarcsec (70\,\micron) and 0.19$\times10^{-3}$\,\Jyarcsec (160\,\micron), respectively. Minimum contour levels are 10\% of maximum. The colour scale is in \Jyarcsec. The red dashed circle shows the mask used to measure the flux from the star and present-day mass-loss.}
\label{f:v1943sgr}
\end{figure*}

\begin{figure*}
\centering
\includegraphics[width=8cm]{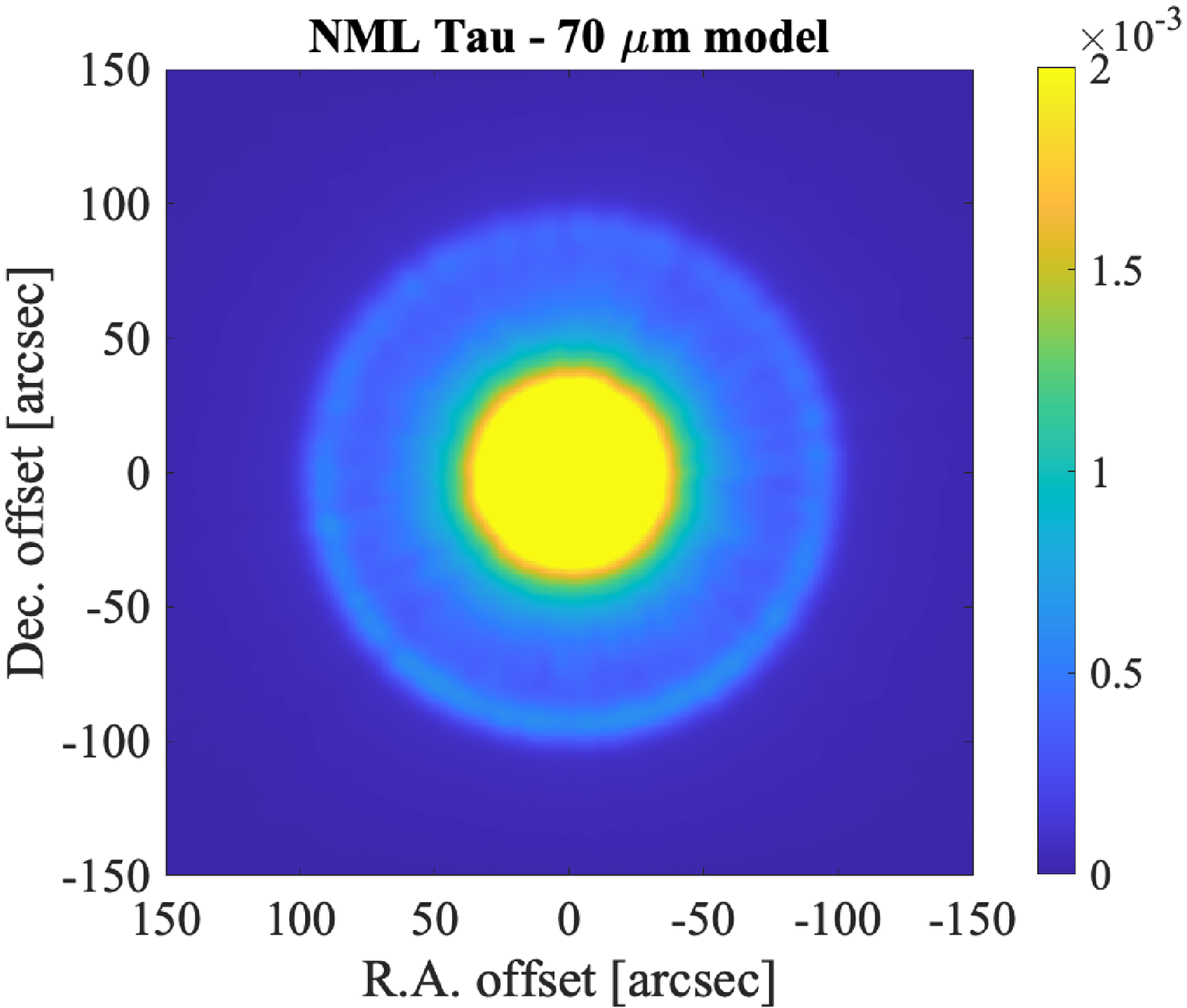}
\includegraphics[width=8cm]{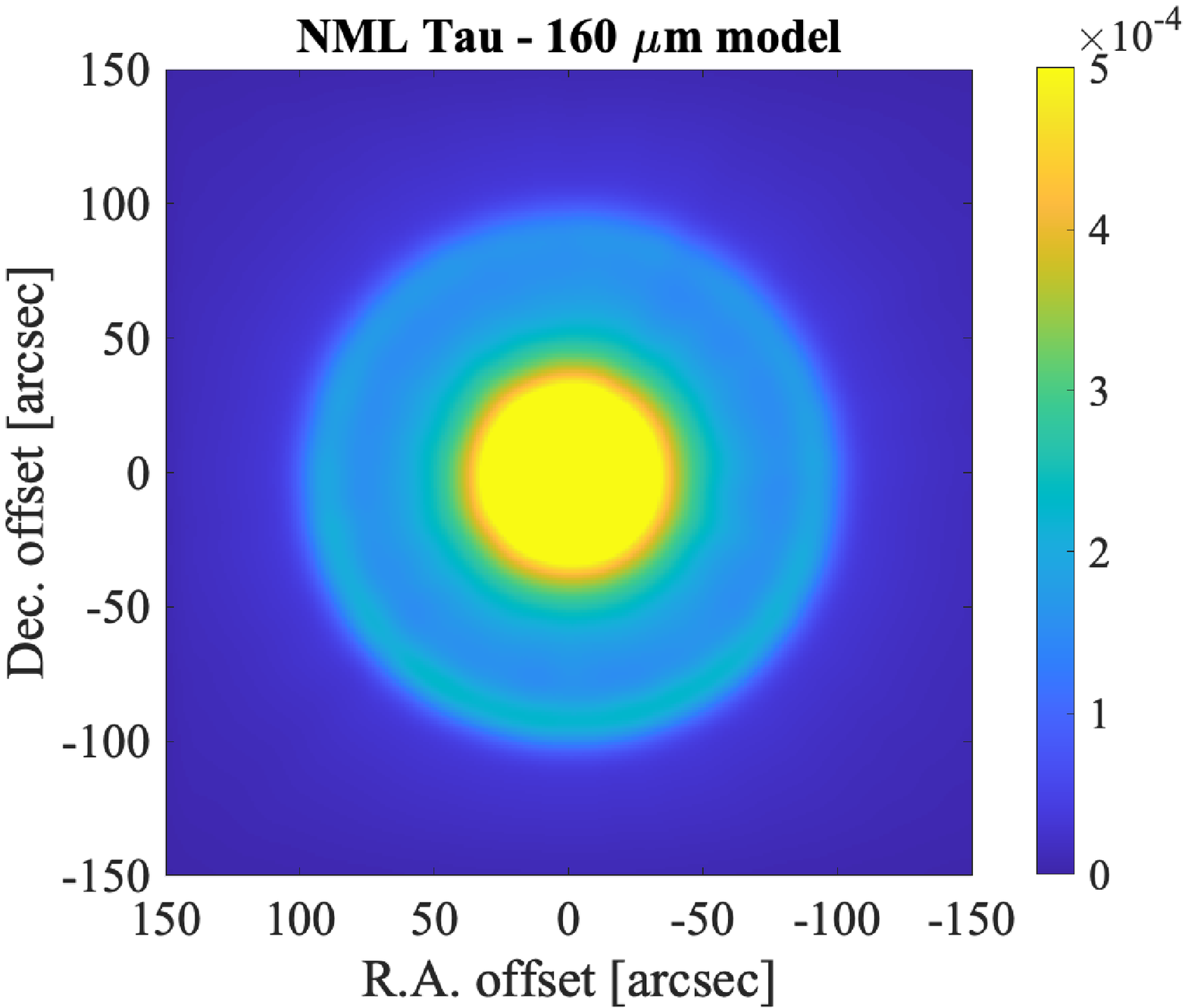}
\includegraphics[width=8cm]{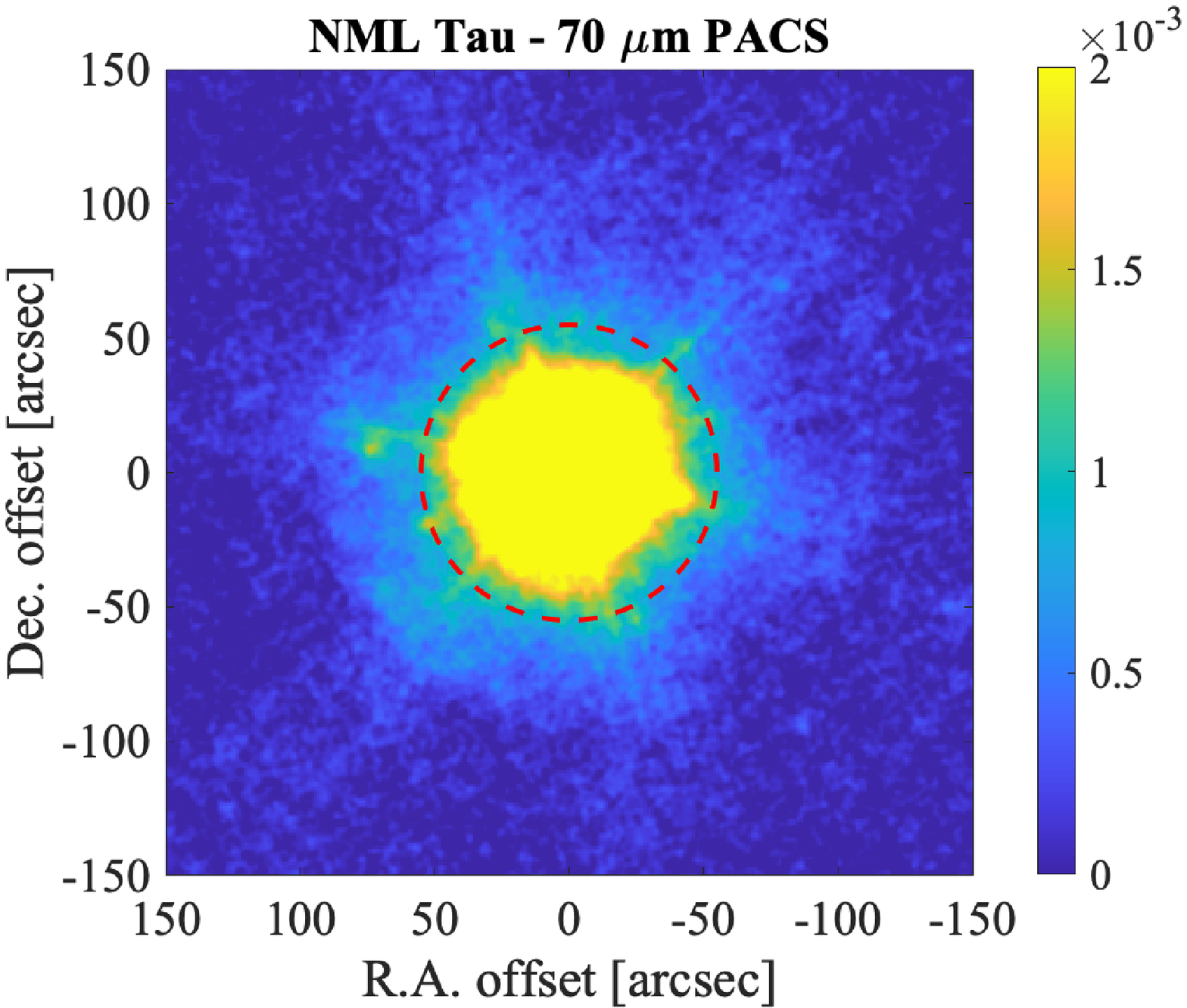}
\includegraphics[width=8cm]{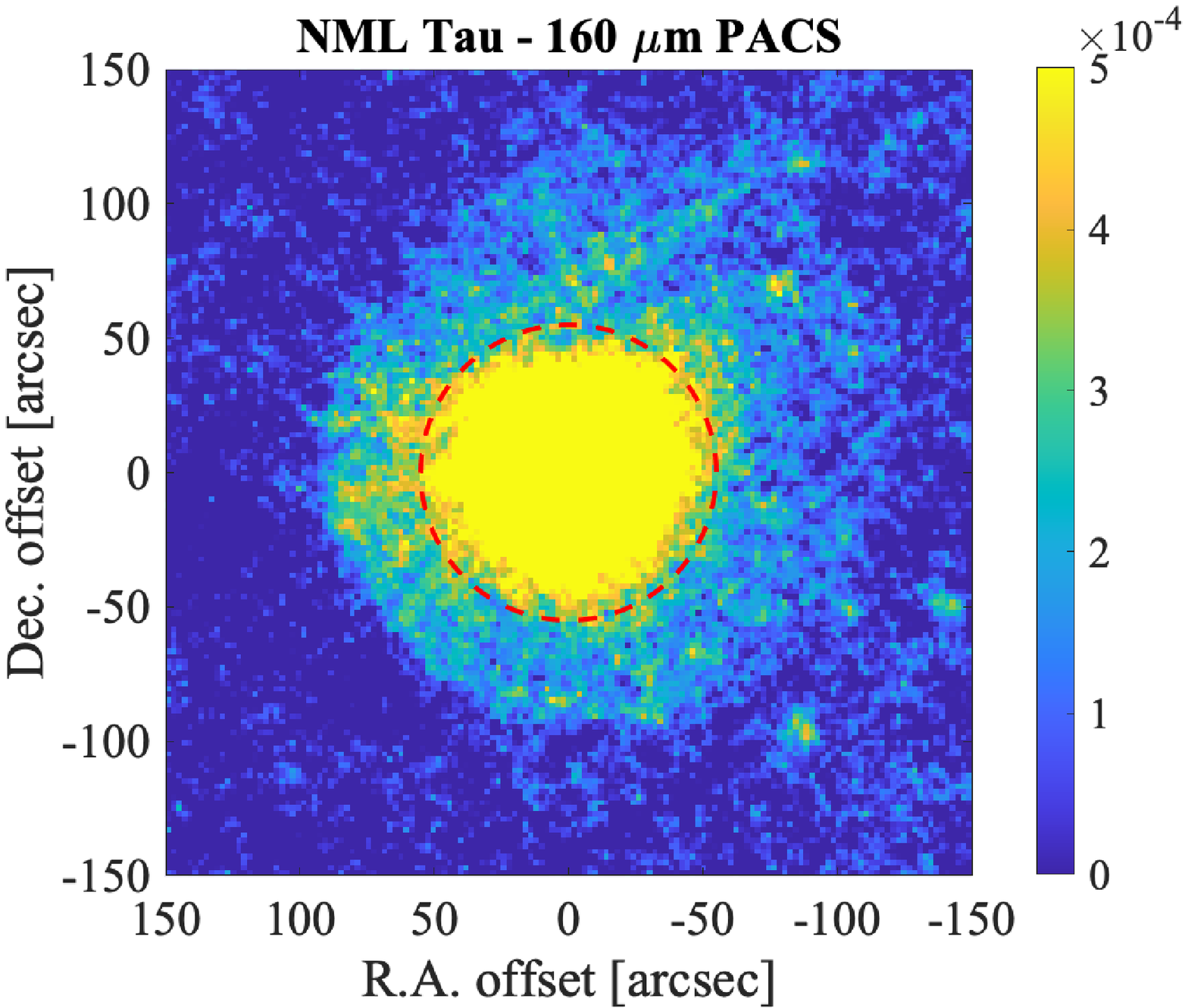}
\includegraphics[width=8cm]{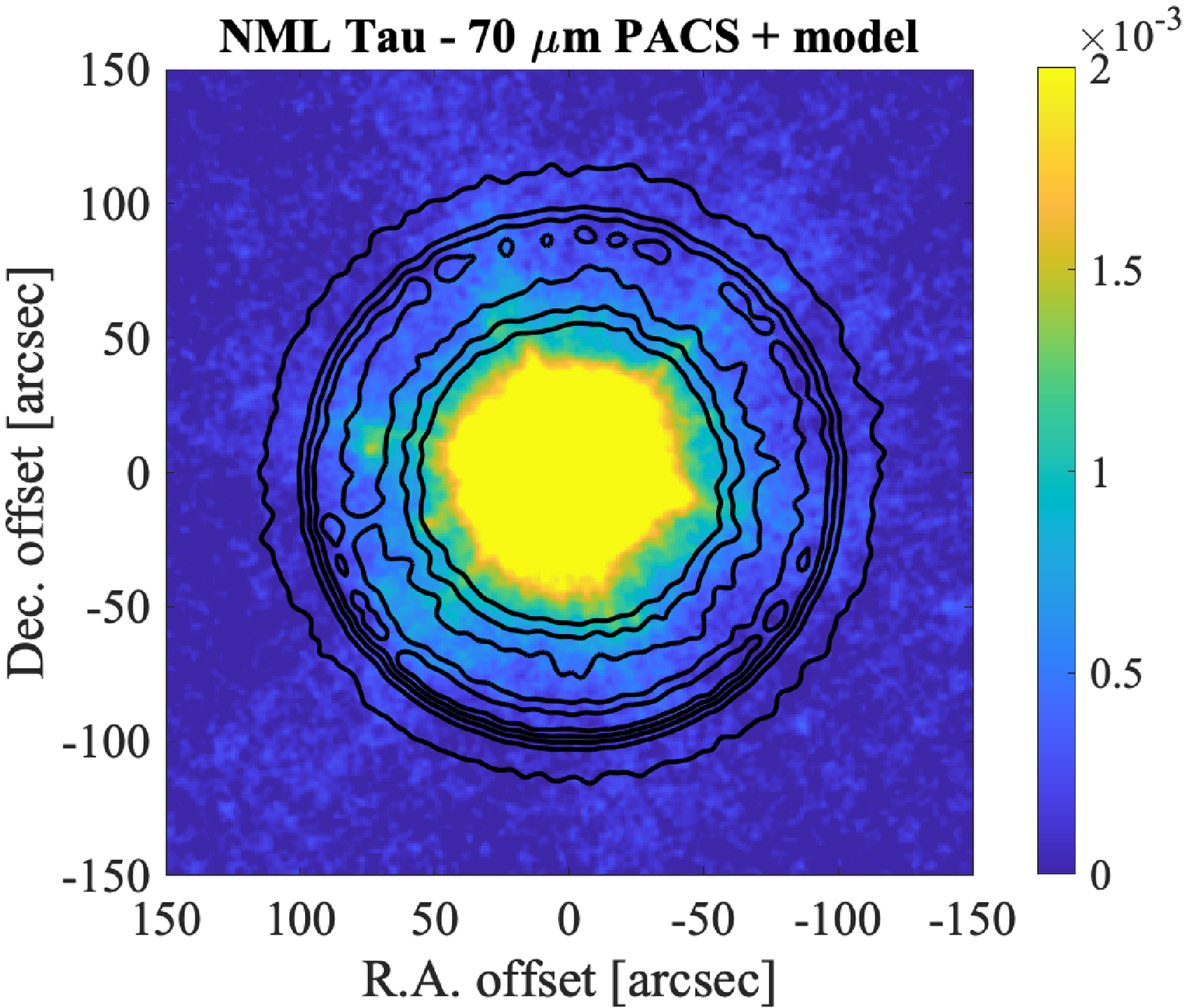}
\includegraphics[width=8cm]{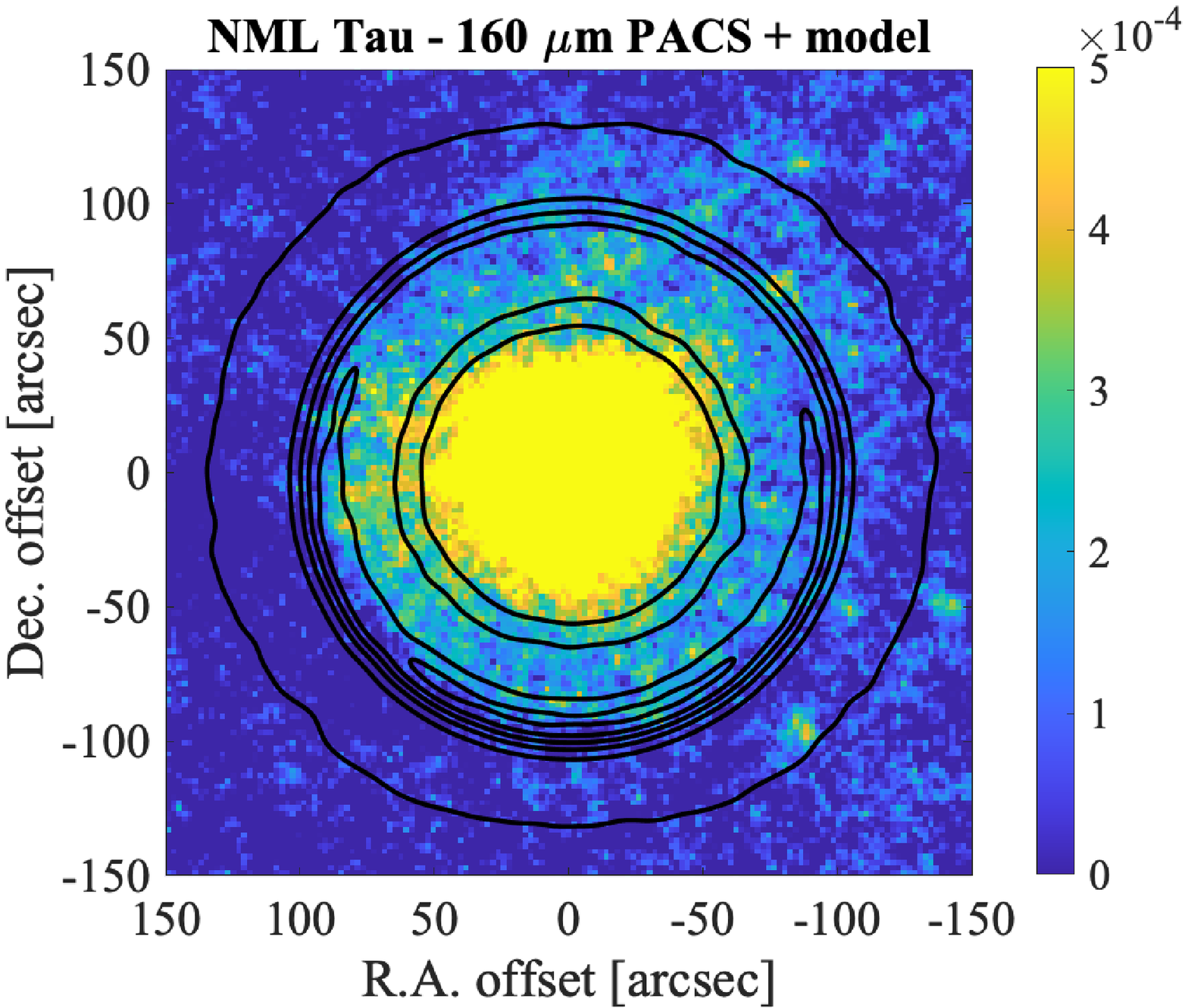}
\caption{NML Tau: \emph{Top to bottom:} The Radmc3D model, the PACS image, and the PACS image with contours from the model. Images are for 70\,\micron~(left) and 160\,\micron~(right). Maximum contour levels are 0.6$\times10^{-3}$\,\Jyarcsec (70\,\micron) and 0.2$\times10^{-3}$\,\Jyarcsec (160\,\micron), respectively. Minimum contour levels are 10\% of maximum. The colour scale is in \Jyarcsec. The red dashed circle shows the mask used to measure the flux from the star and present-day mass-loss.}
\label{f:nmltau}
\end{figure*}

\begin{figure*}
\centering
\includegraphics[width=8cm]{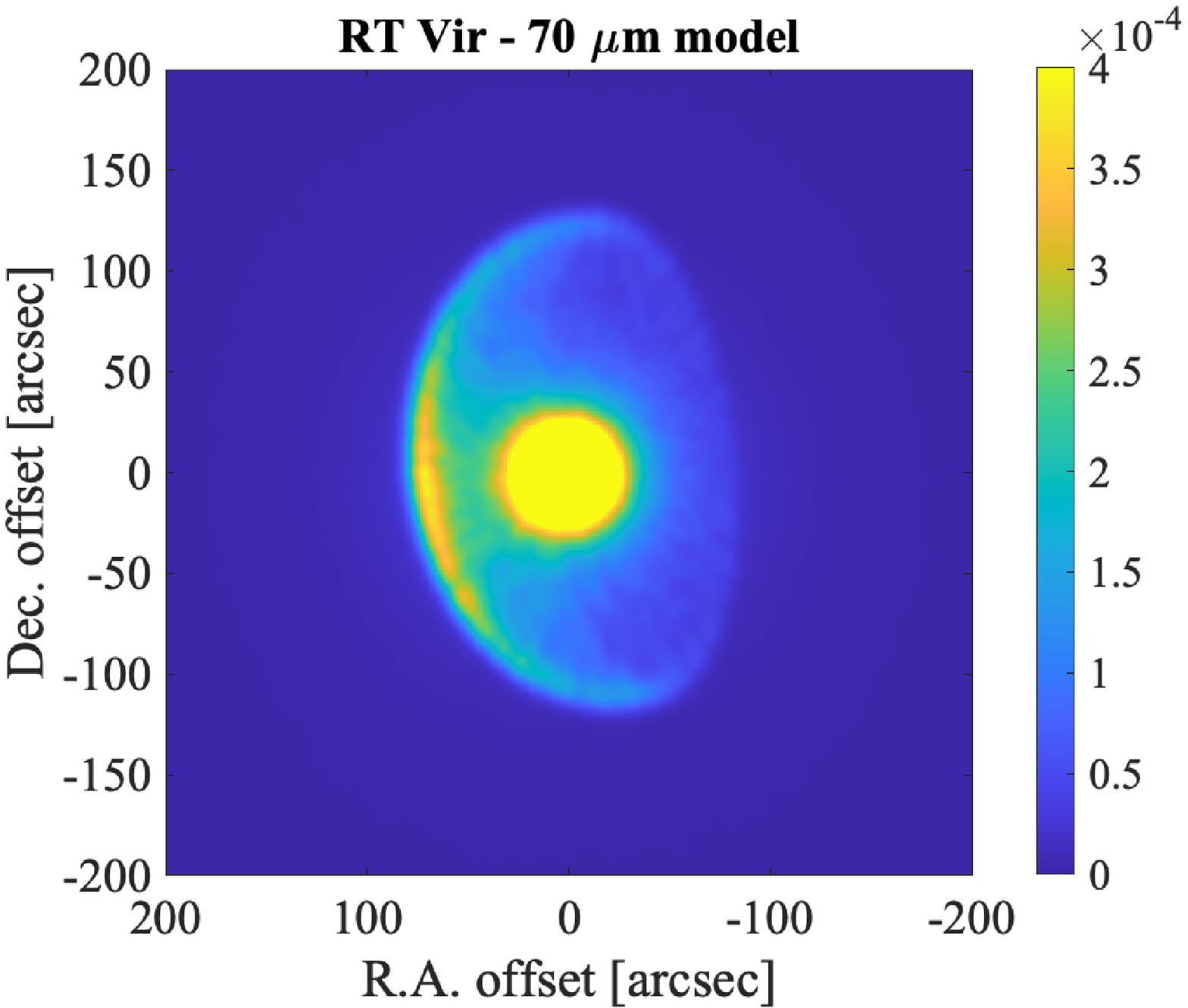}
\includegraphics[width=8cm]{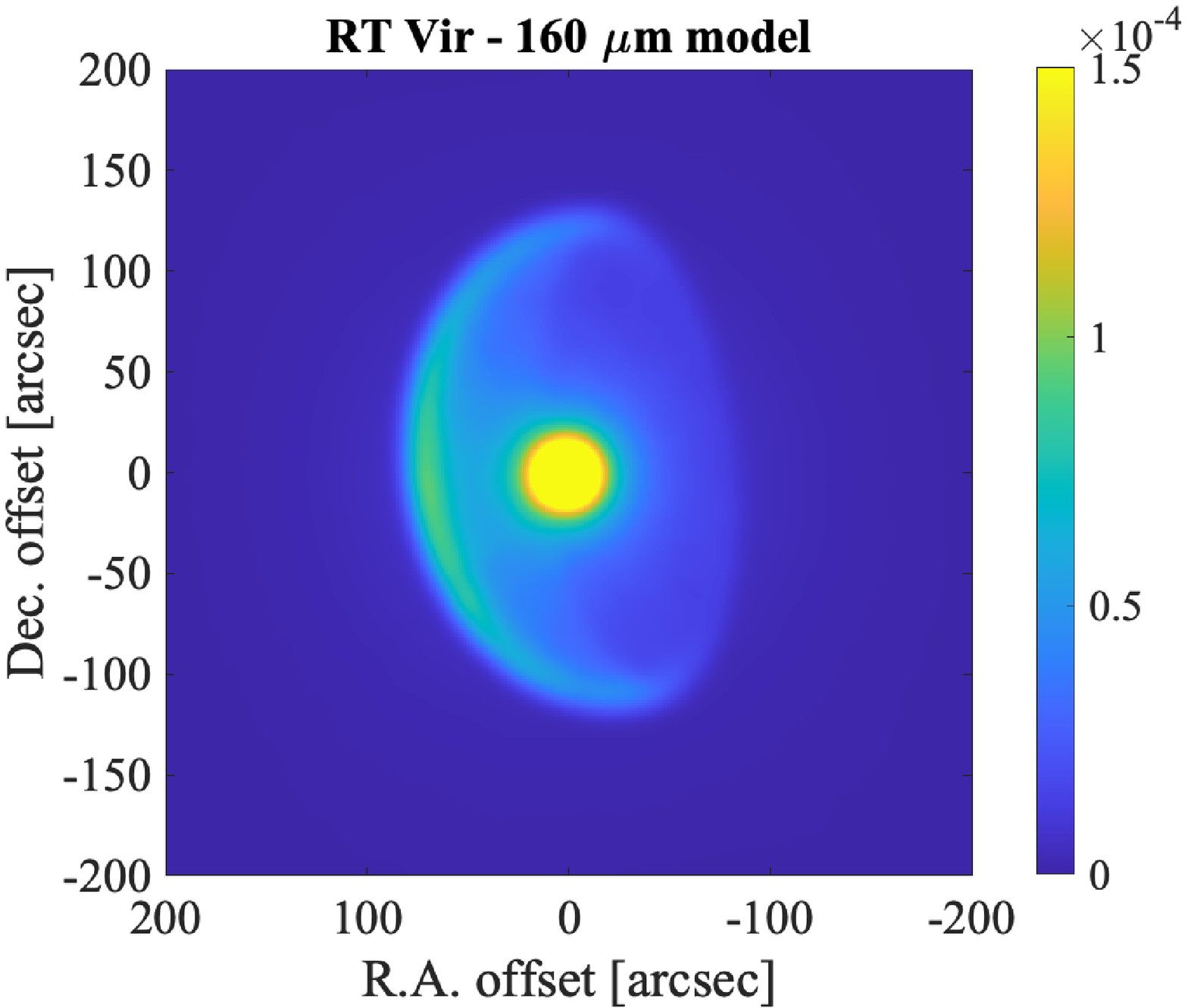}
\includegraphics[width=8cm]{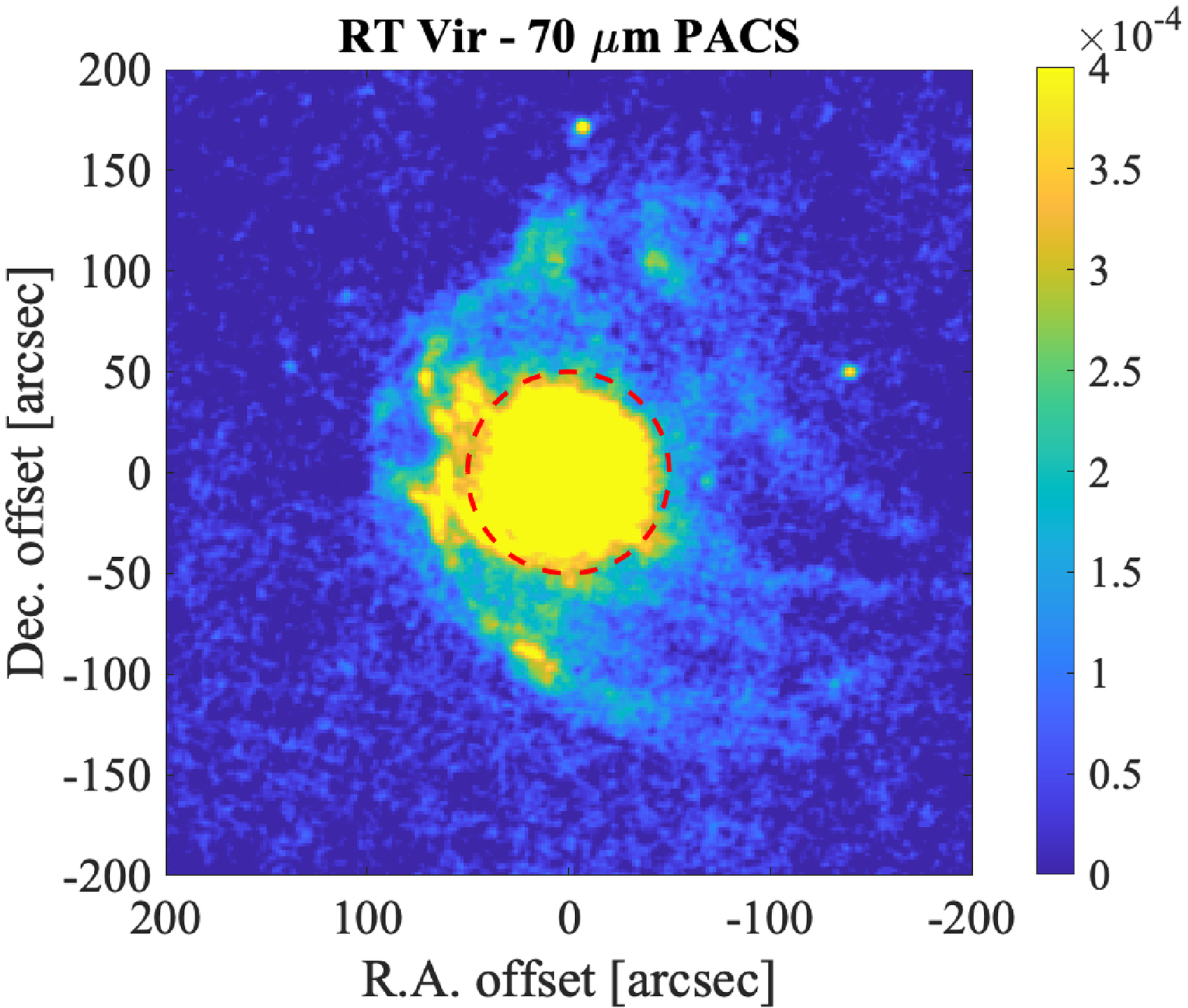}
\includegraphics[width=8cm]{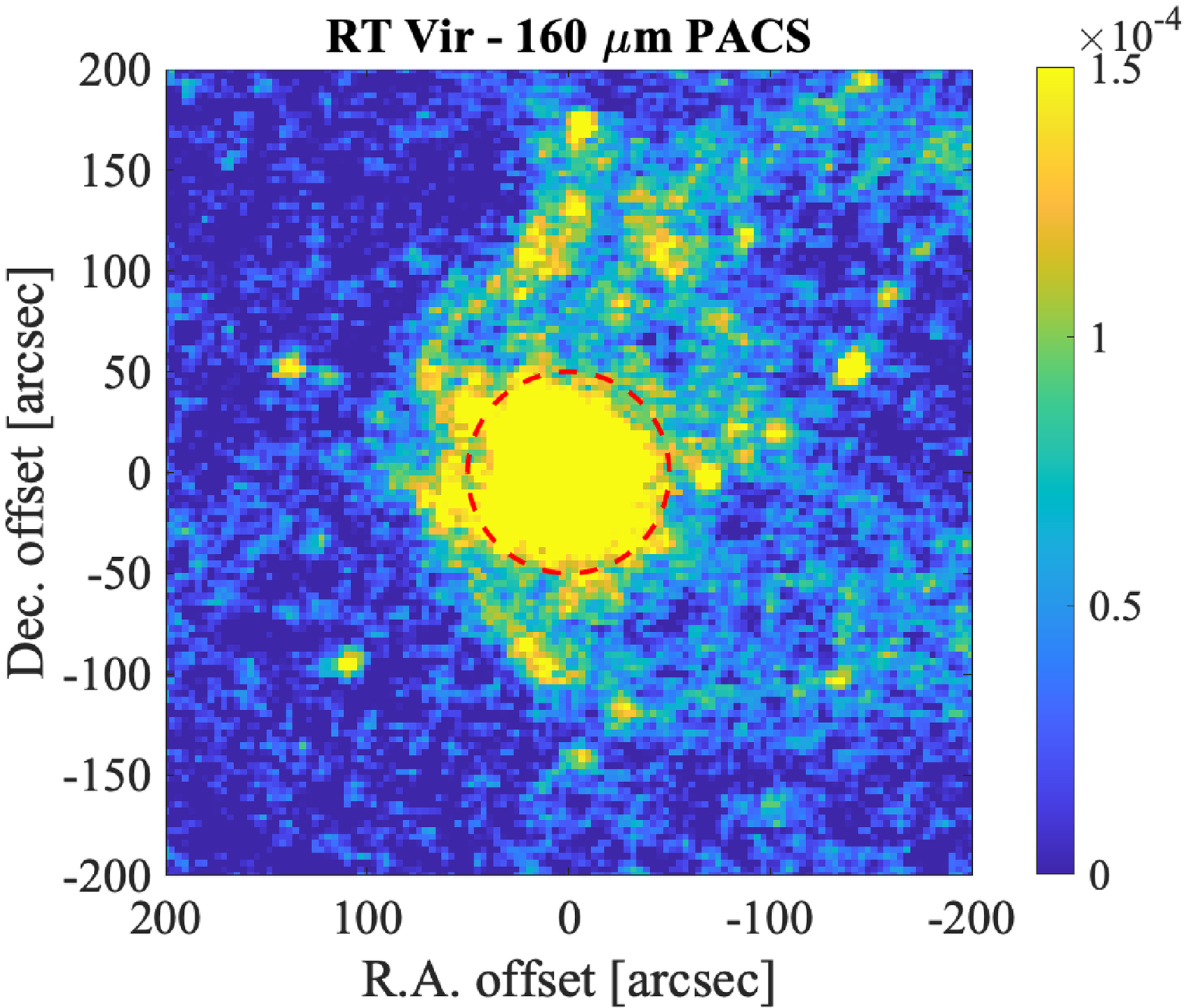}
\includegraphics[width=8cm]{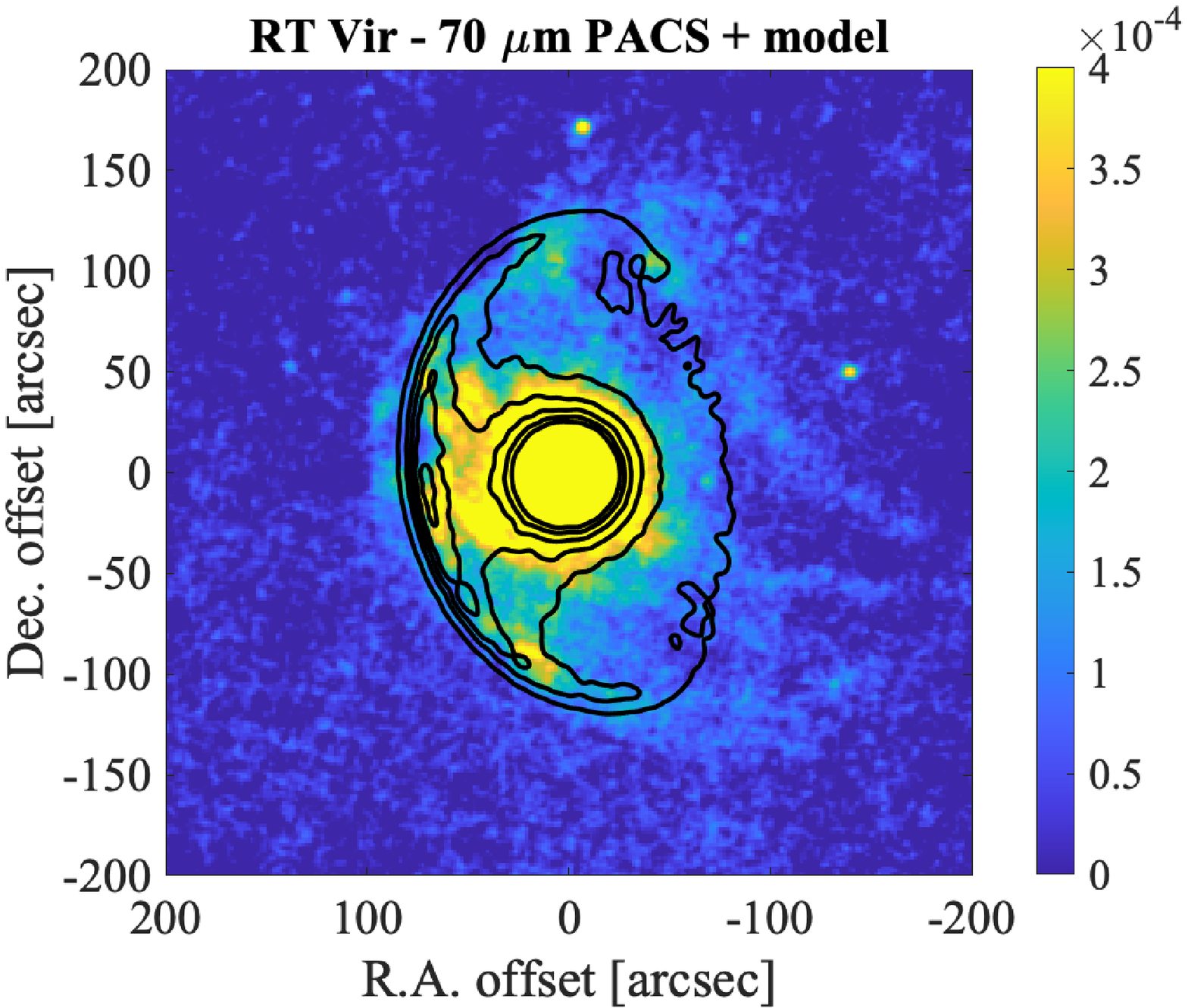}
\includegraphics[width=8cm]{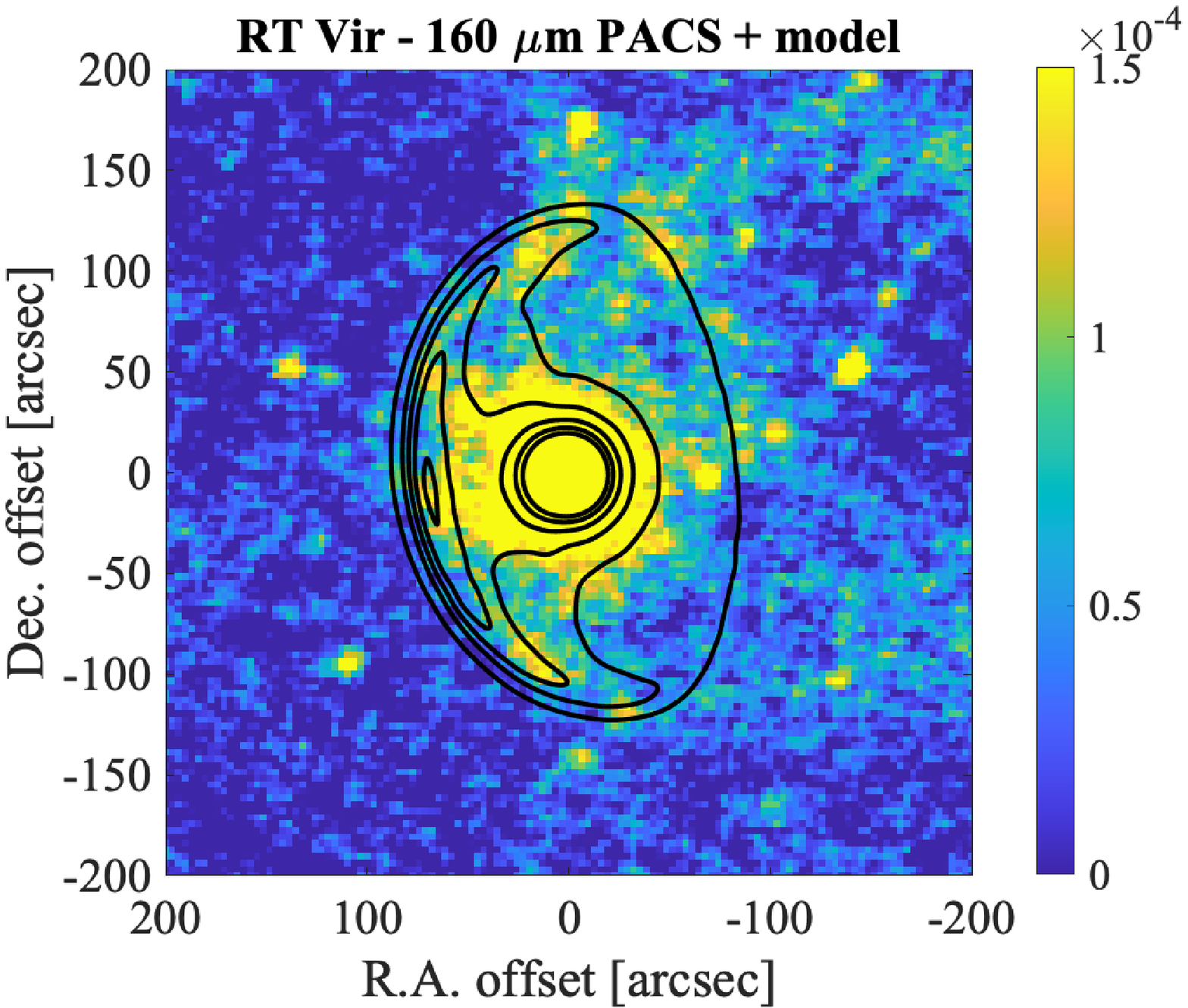}
\caption{RT Vir: \emph{Top to bottom:} The Radmc3D model, the PACS image, and the PACS image with contours from the model. Images are for 70\,\micron~(left) and 160\,\micron~(right). Maximum contour levels are 0.4$\times10^{-3}$\,\Jyarcsec (70\,\micron) and 0.1$\times10^{-3}$\,\Jyarcsec (160\,\micron), respectively. Minimum contour levels are 10\% of maximum. The colour scale is in \Jyarcsec. The red dashed circle shows the mask used to measure the flux from the star and present-day mass-loss.}
\label{f:rtvir}
\end{figure*}

\section{PACS images of sources that were not modelled}
\label{b:images}
\begin{figure*}
\centering
\includegraphics[width=8cm]{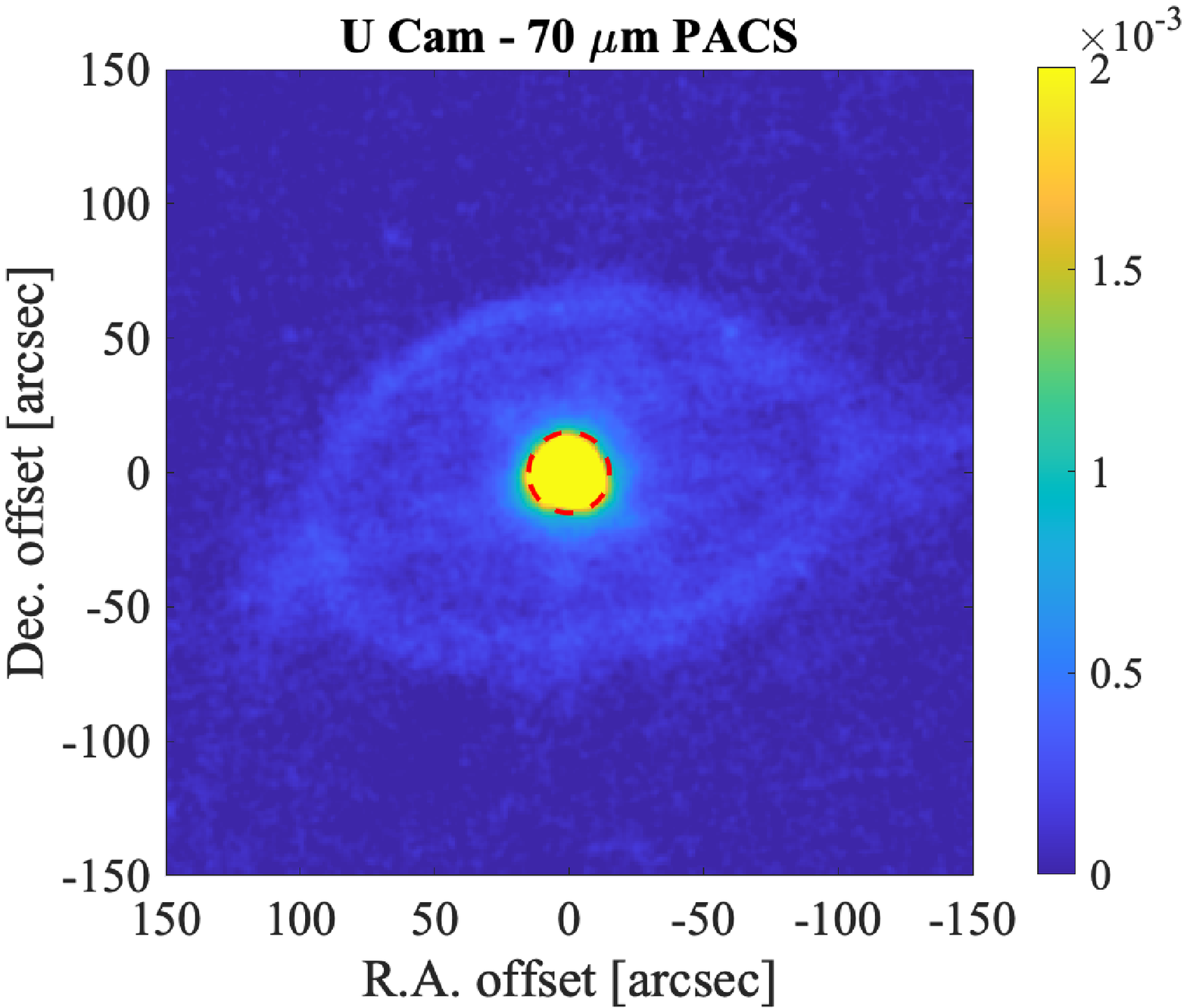}
\includegraphics[width=8cm]{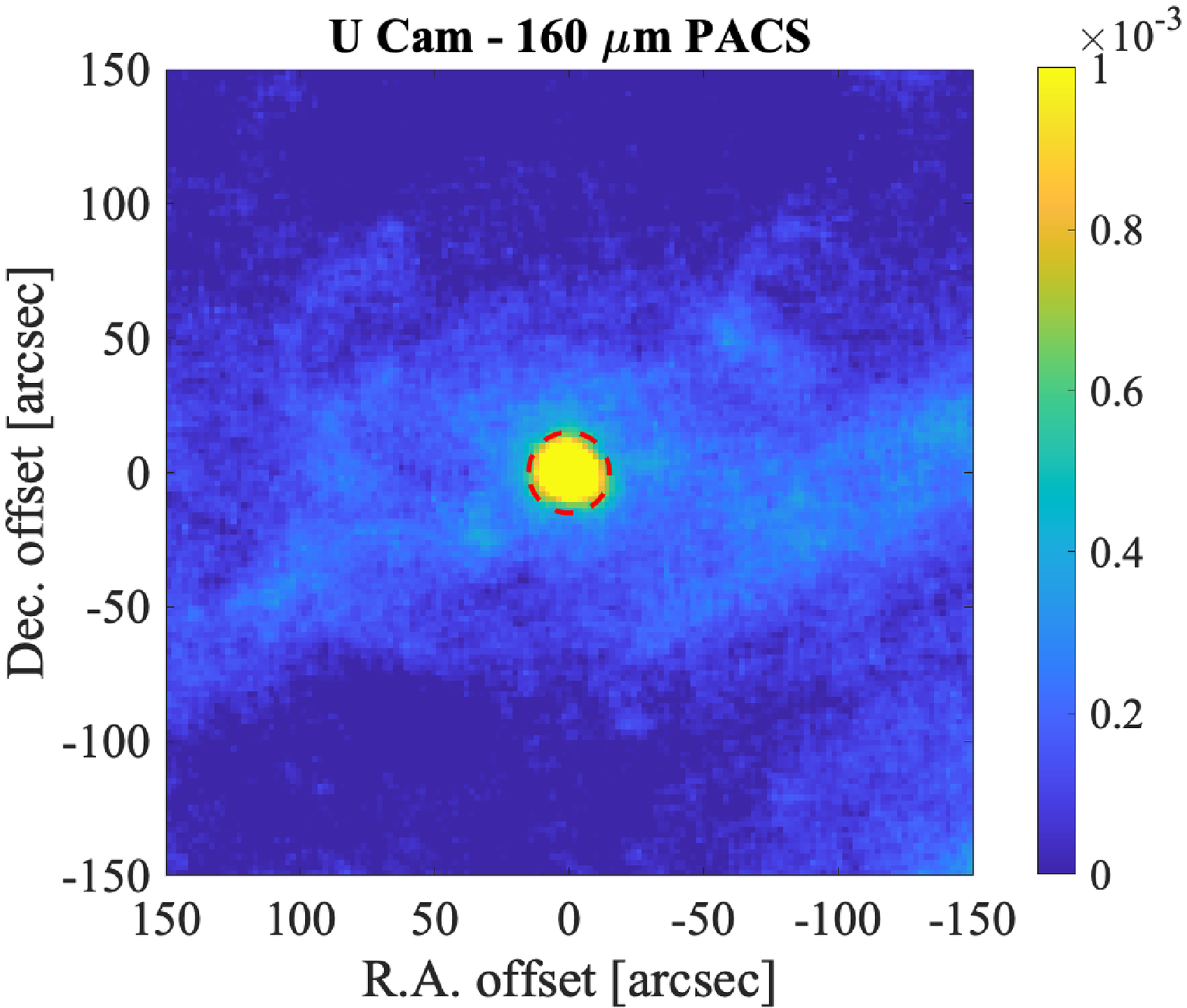}
\caption{U Cam: The PACS images at 70\,\micron~(left) and 160\,\micron~(right). The colour scale is in \Jyarcsec. The red dashed circle shows the mask used to measure the flux from the star and present-day mass-loss.}
\label{f:ucam}
\end{figure*}

\begin{figure*}
\centering
\includegraphics[width=8cm]{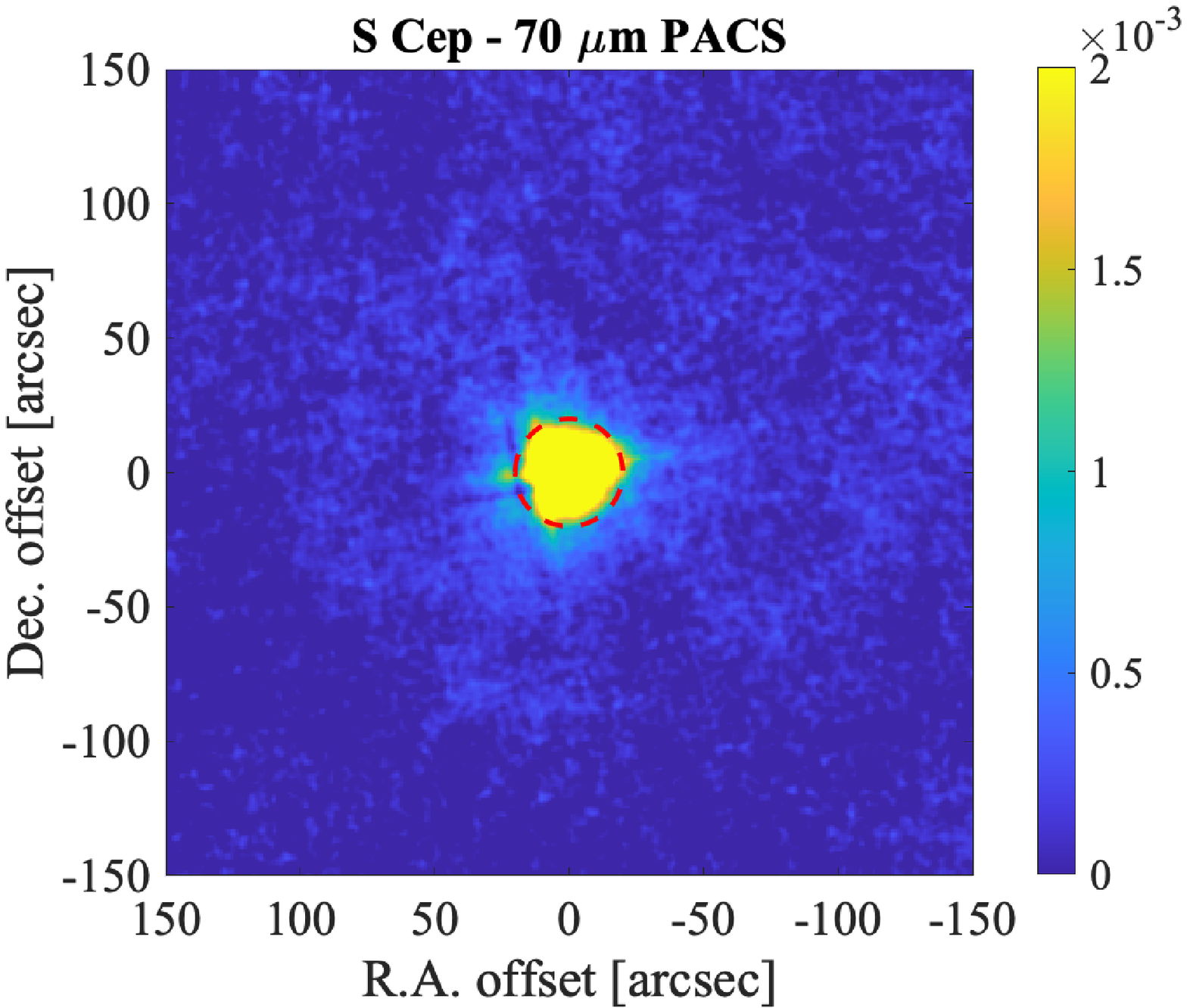}
\includegraphics[width=8cm]{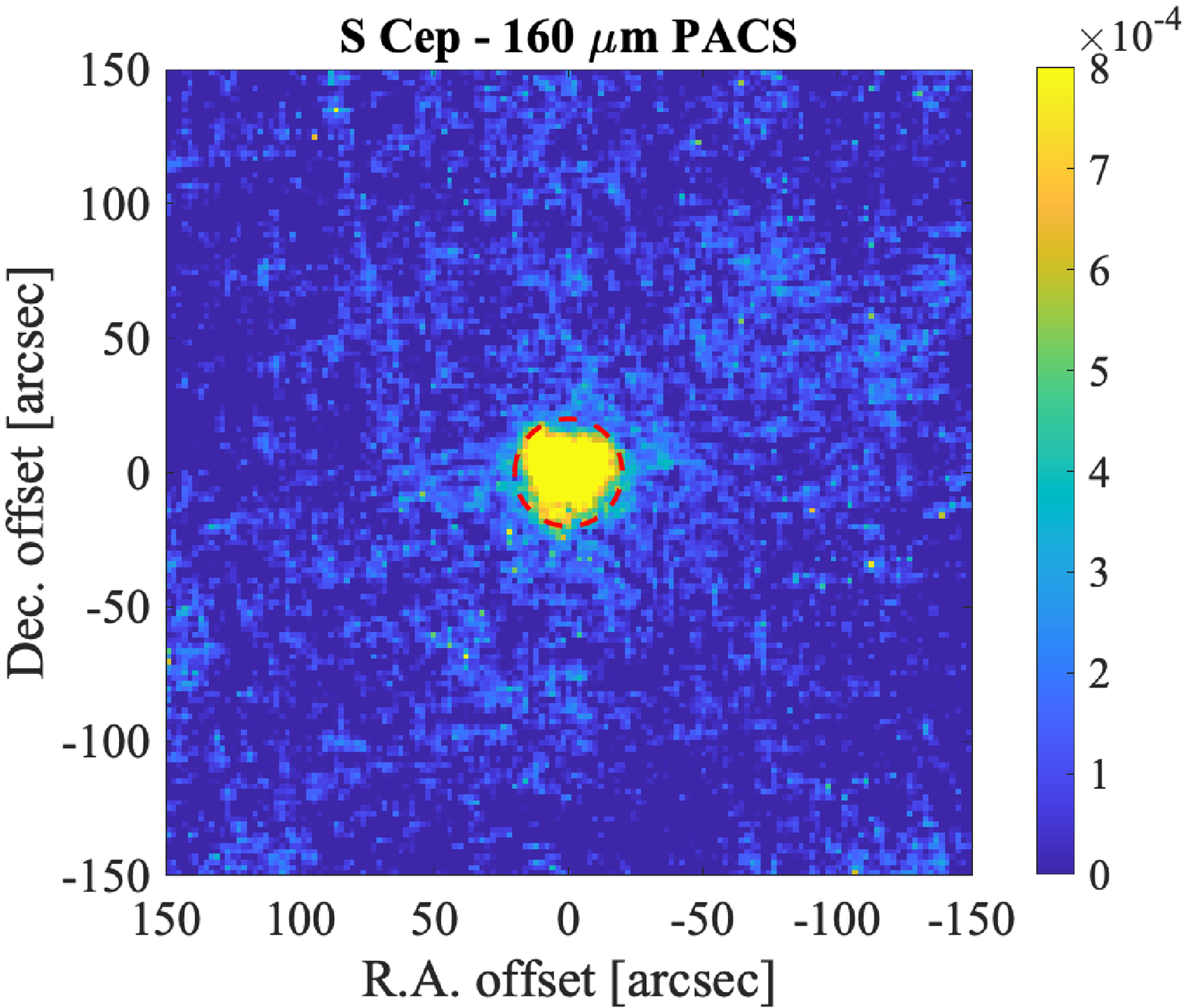}
\caption{S Cep: The PACS images at 70\,\micron~(left) and 160\,\micron~(right). The colour scale is in \Jyarcsec. The red dashed circle shows the mask used to measure the flux from the star and present-day mass-loss.}
\label{f:scep}
\end{figure*}

\begin{figure*}
\centering
\includegraphics[width=8cm]{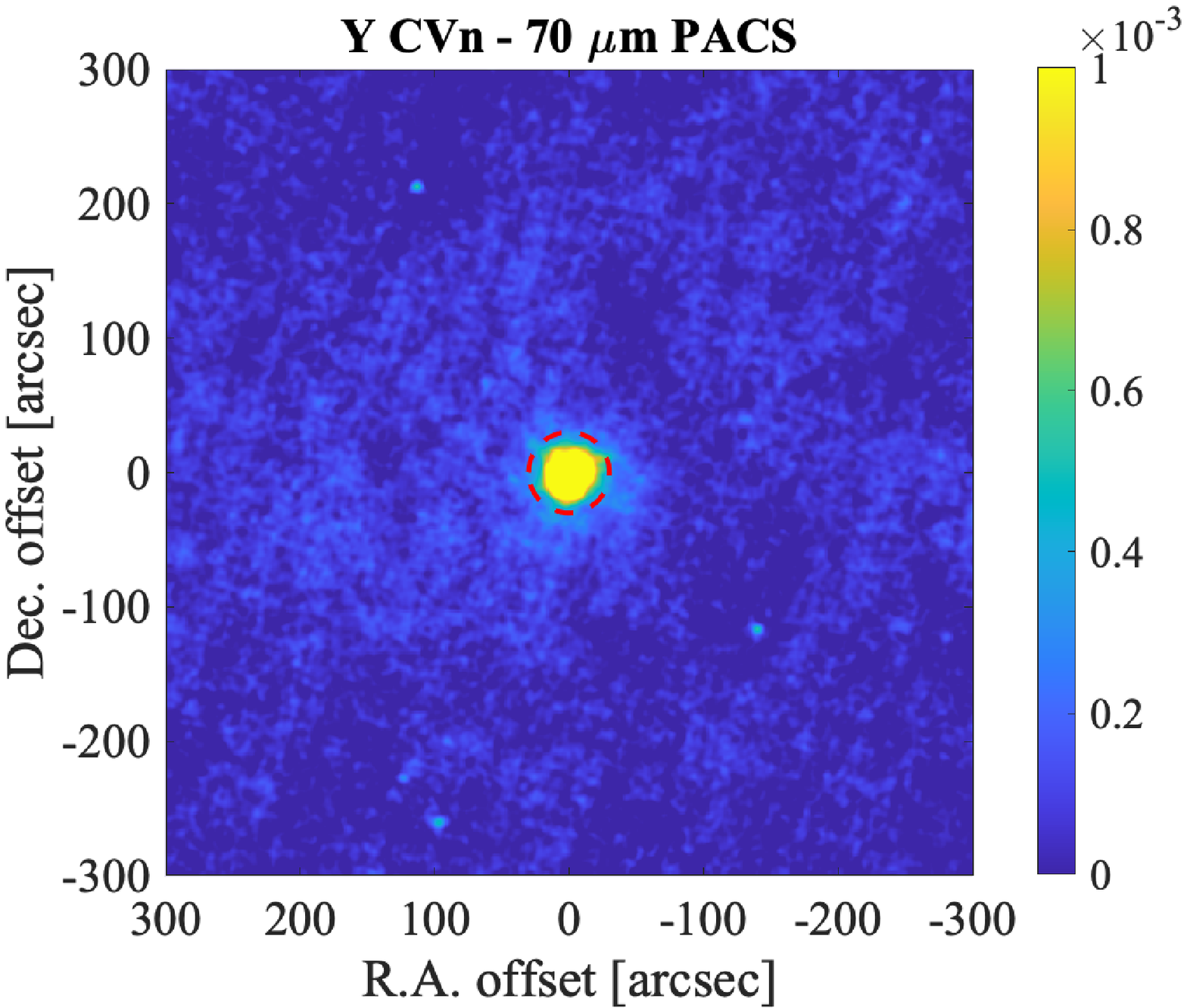}
\includegraphics[width=8cm]{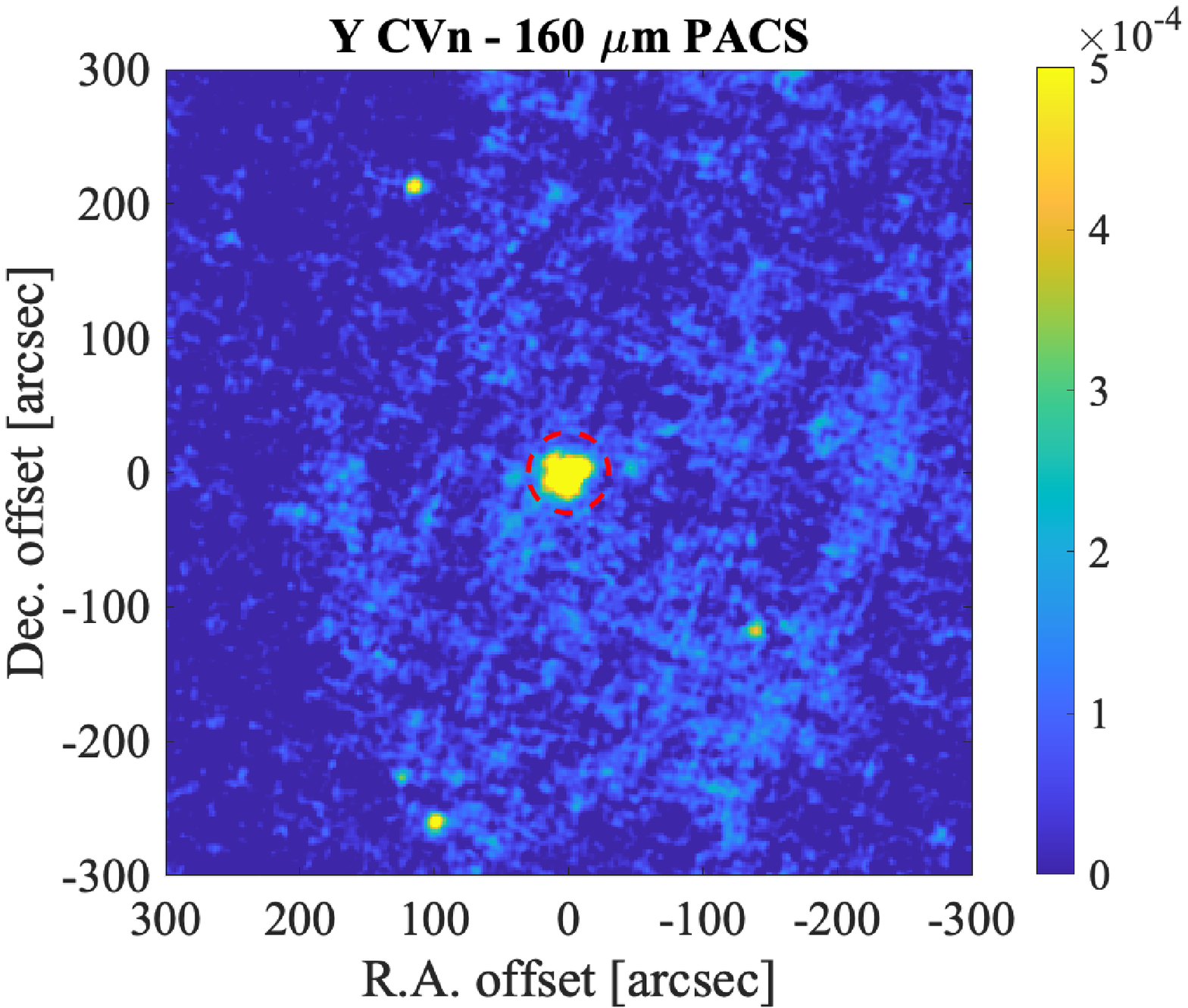}
\caption{Y CVn: The PACS images at 70\,\micron~(left) and 160\,\micron~(right). The colour scale is in \Jyarcsec. The red dashed circle shows the mask used to measure the flux from the star and present-day mass-loss.}
\label{f:ycvn}
\end{figure*}

\begin{figure*}
\centering
\includegraphics[width=8cm]{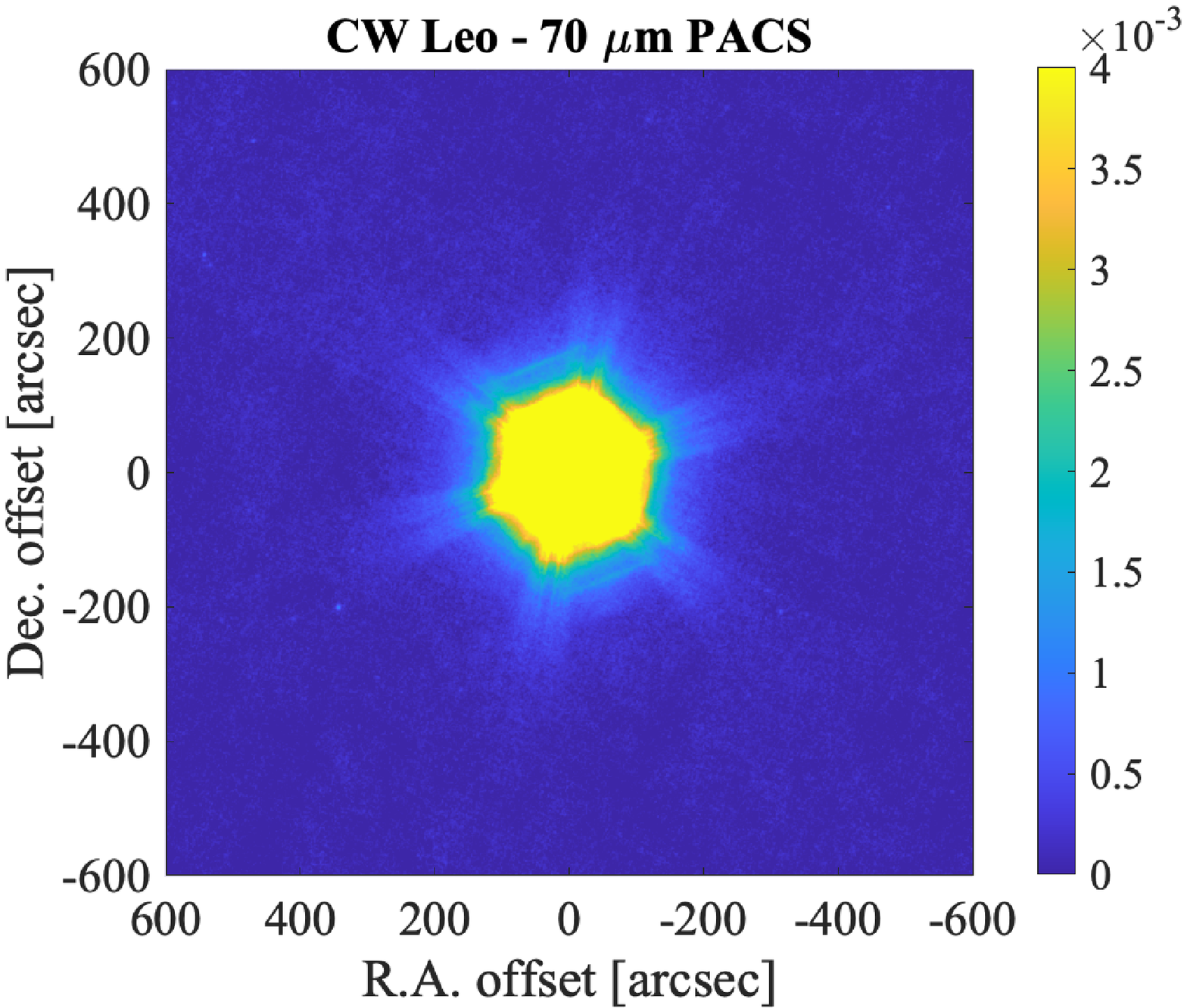}
\includegraphics[width=8cm]{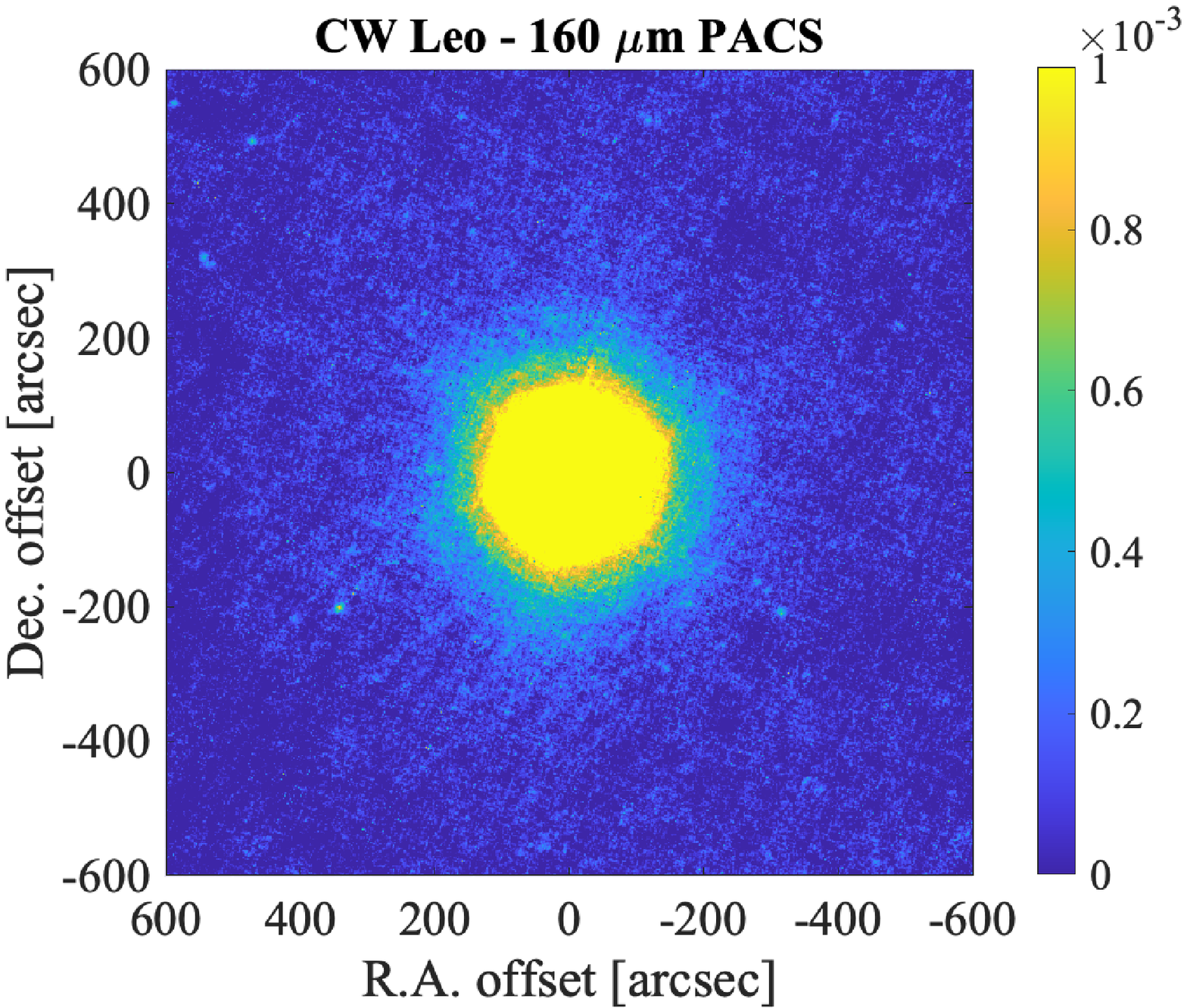}
\caption{CW Leo: The PACS images at 70\,\micron~(left) and 160\,\micron~(right). The colour scale is in \Jyarcsec. The red dashed circle shows the mask used to measure the flux from the star and present-day mass-loss.}
\label{f:cwleo}
\end{figure*}

\begin{figure*}
\centering
\includegraphics[width=8cm]{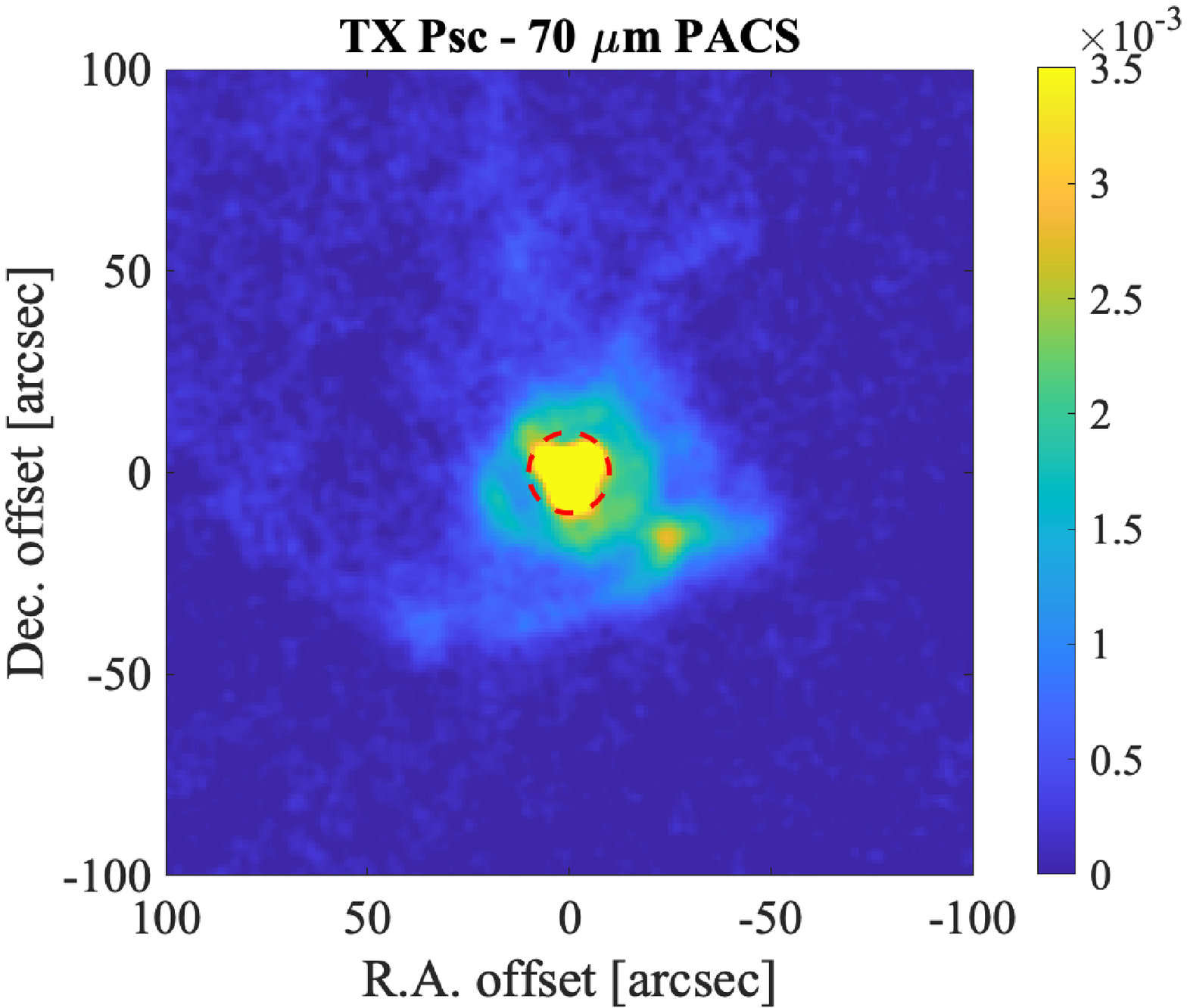}
\includegraphics[width=8cm]{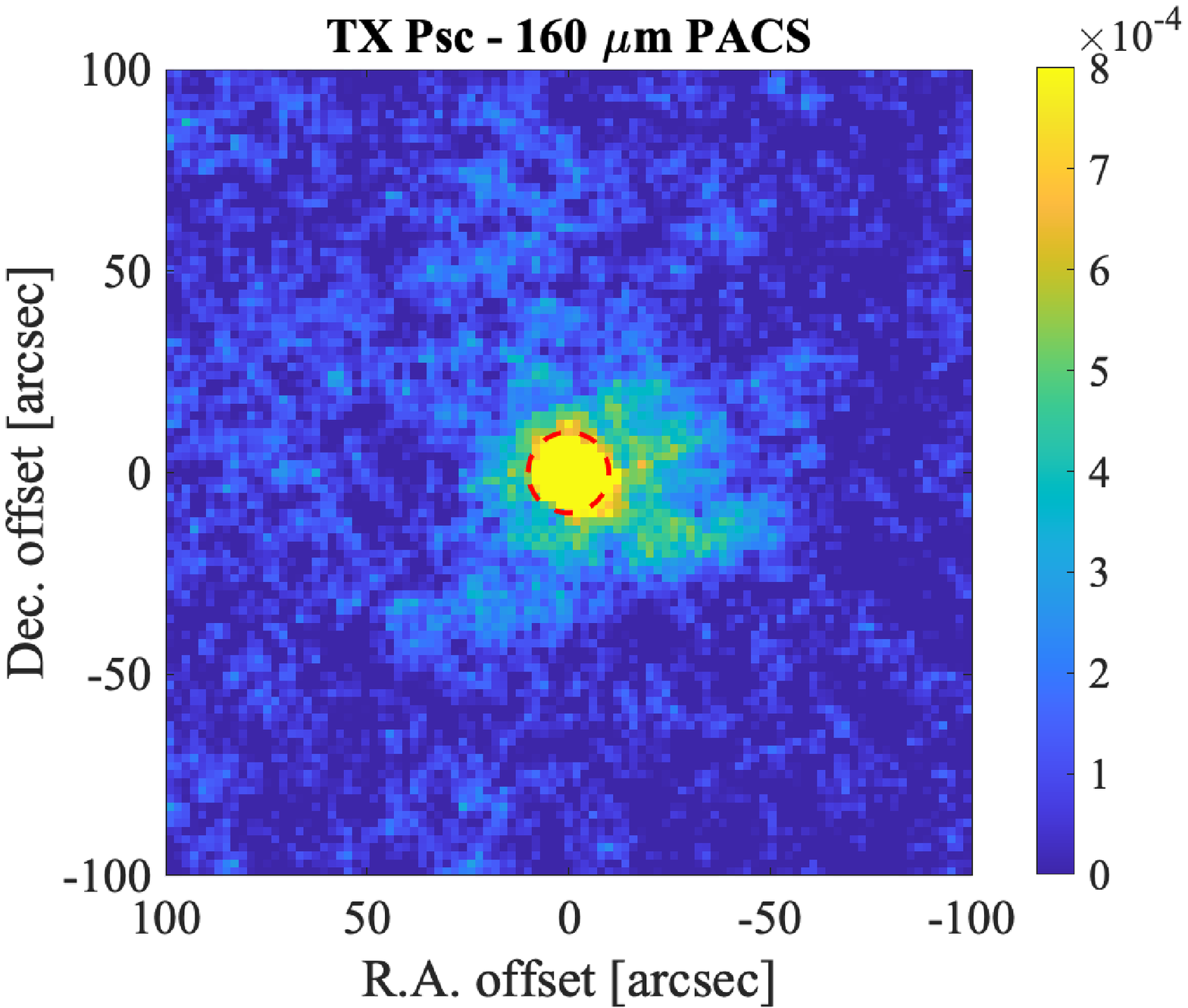}
\caption{TX Psc: The PACS images at 70\,\micron~(left) and 160\,\micron~(right). The colour scale is in \Jyarcsec. The red dashed circle shows the mask used to measure the flux from the star and present-day mass-loss.}
\label{f:txpsc}
\end{figure*}

\begin{figure*}
\centering
\includegraphics[width=8cm]{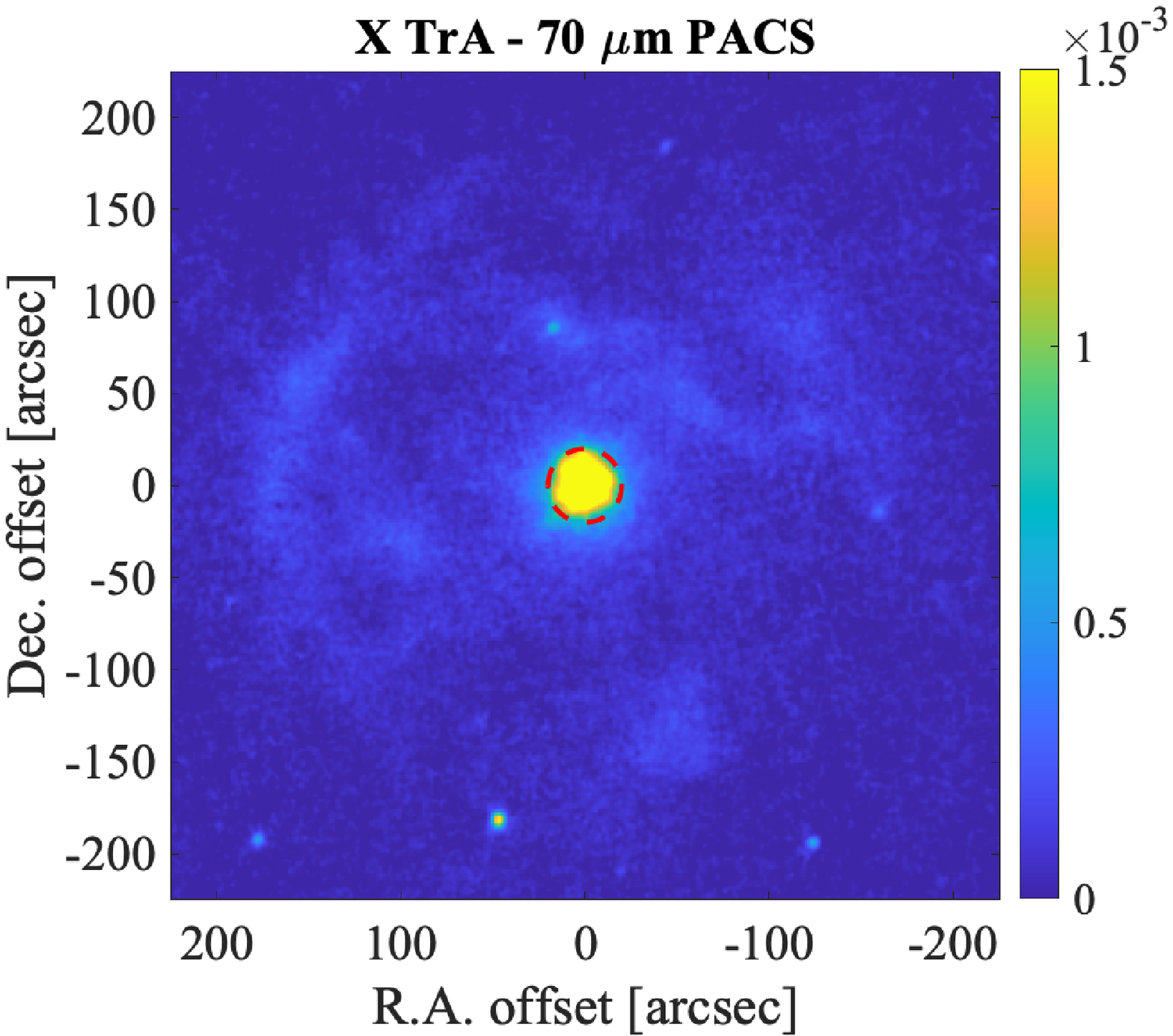}
\includegraphics[width=8cm]{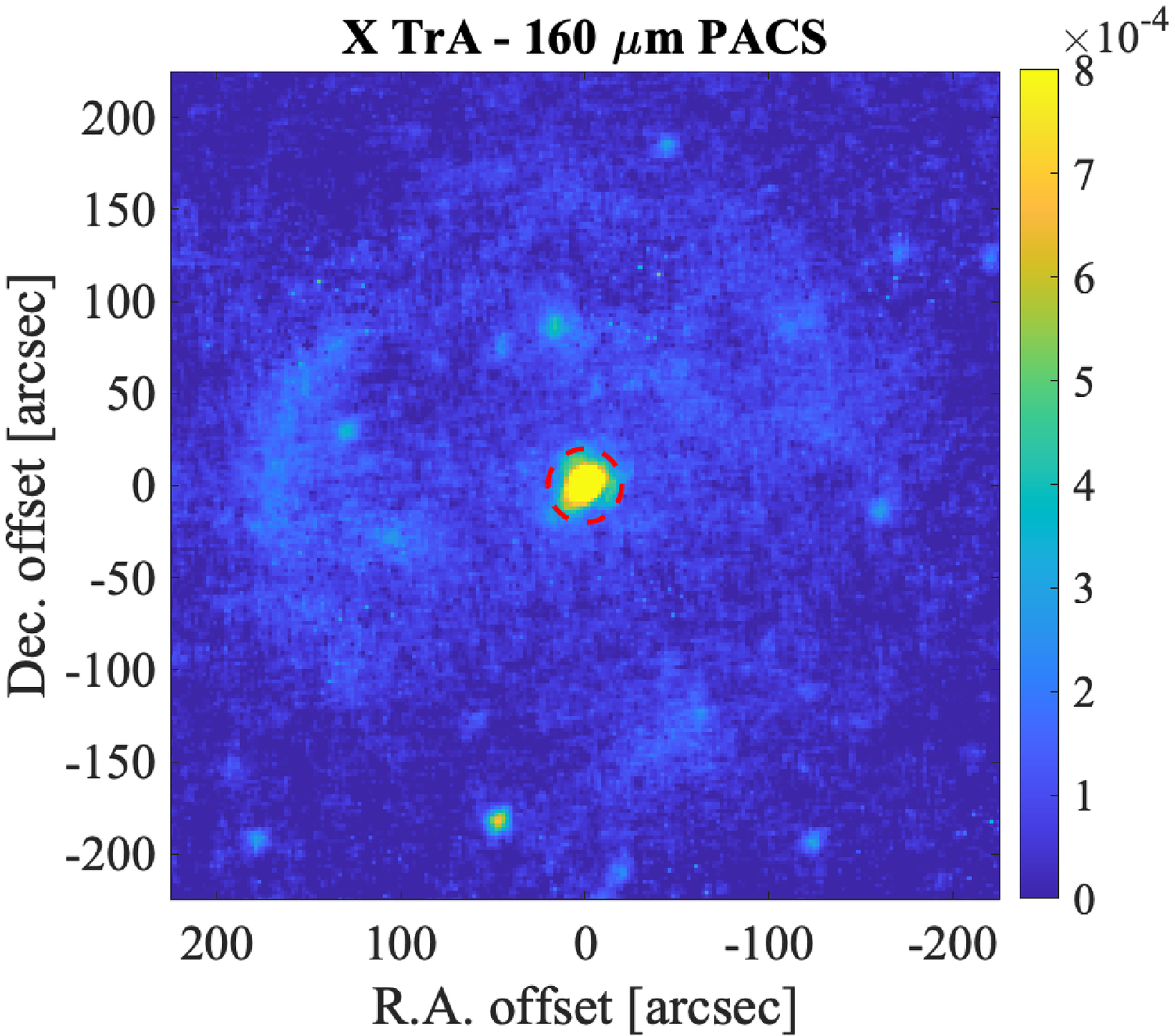}
\caption{X TrA: The PACS images at 70\,\micron~(left) and 160\,\micron~(right). The colour scale is in \Jyarcsec. The red dashed circle shows the mask used to measure the flux from the star and present-day mass-loss.}
\label{f:xtra}
\end{figure*}

\begin{figure*}
\centering
\includegraphics[width=8cm]{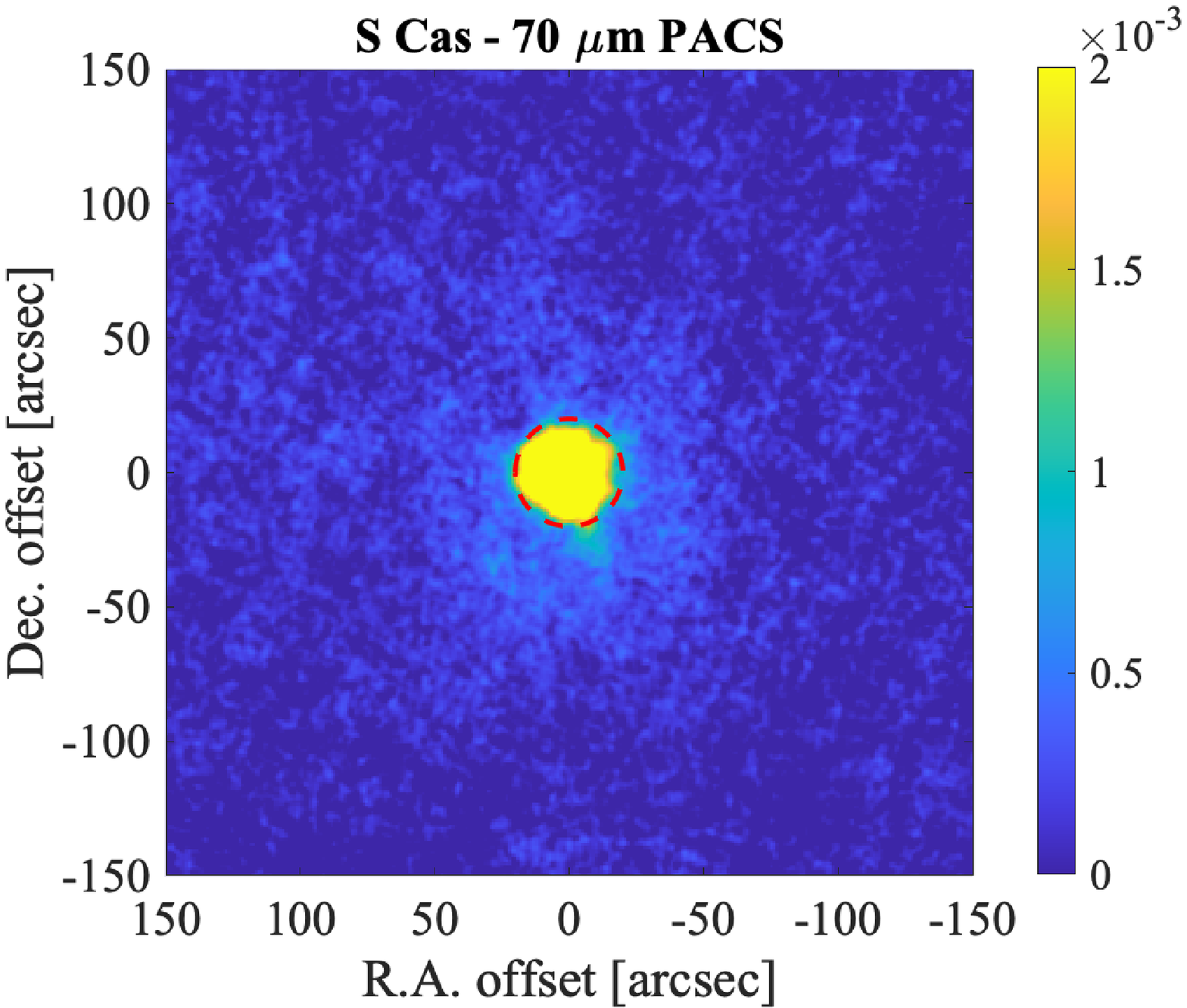}
\includegraphics[width=8cm]{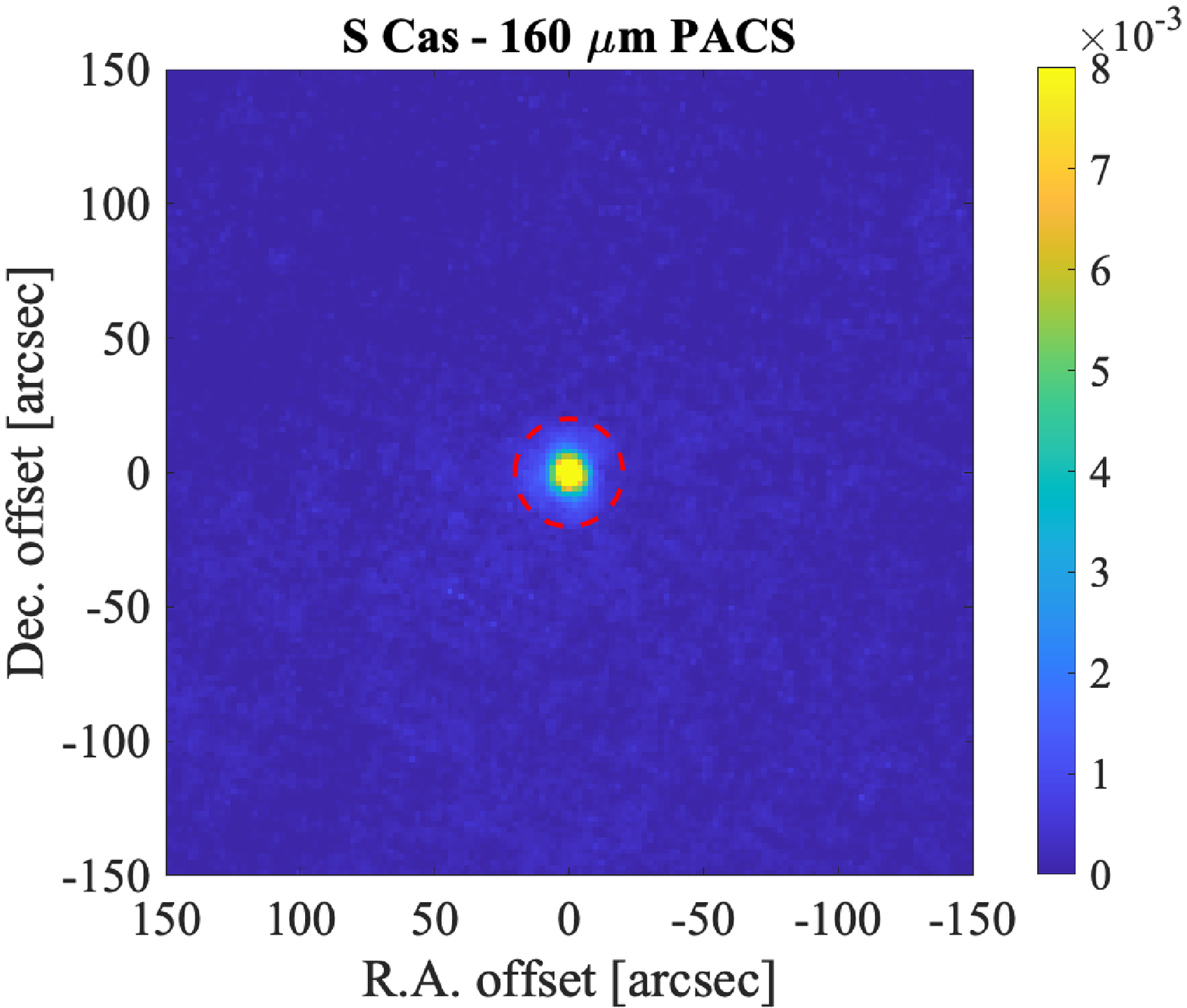}
\caption{S Cas: The PACS images at 70\,\micron~(left) and 160\,\micron~(right). The colour scale is in \Jyarcsec. The red dashed circle shows the mask used to measure the flux from the star and present-day mass-loss.}
\label{f:scas}
\end{figure*}

\begin{figure*}
\centering
\includegraphics[width=8cm]{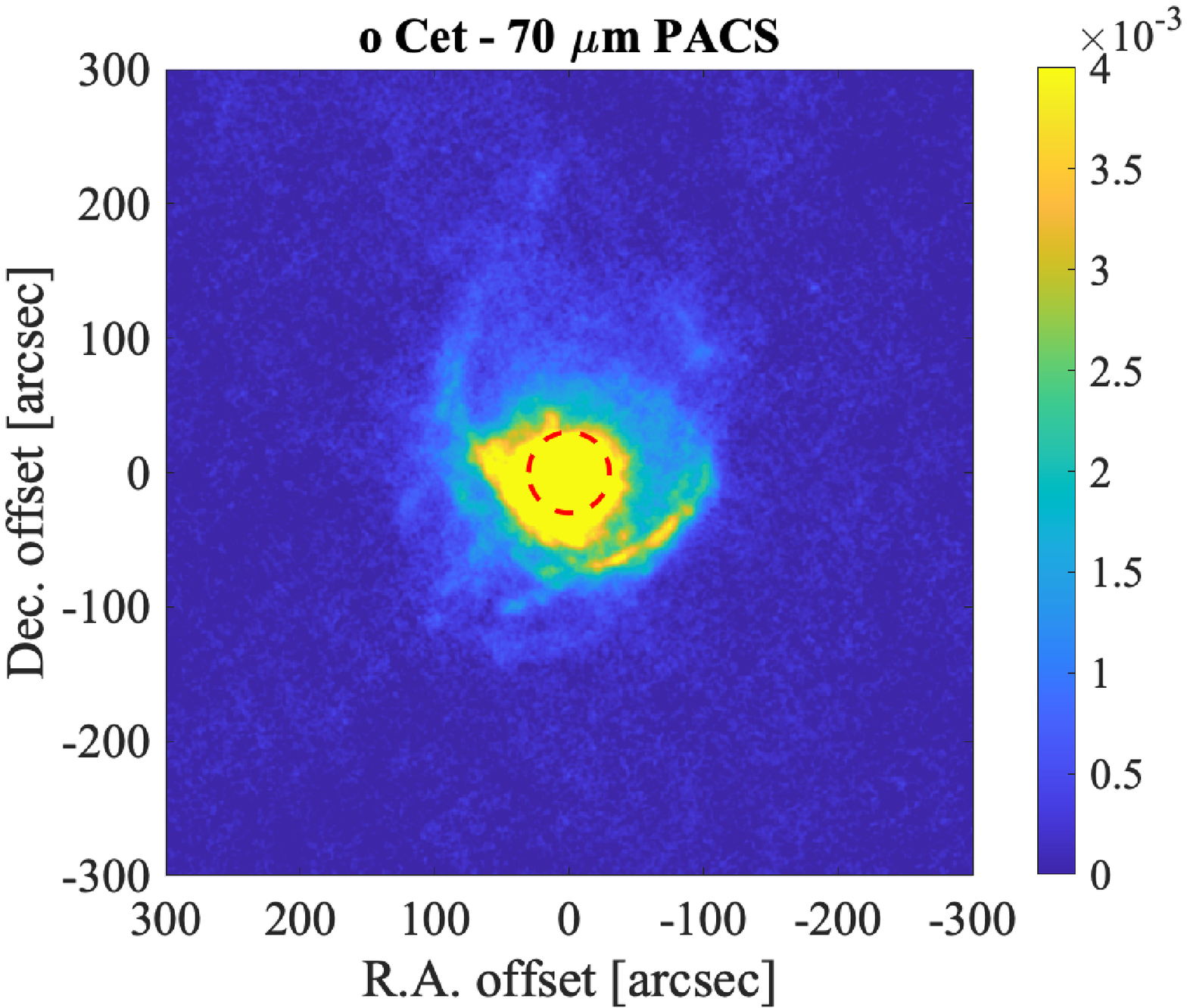}
\includegraphics[width=8cm]{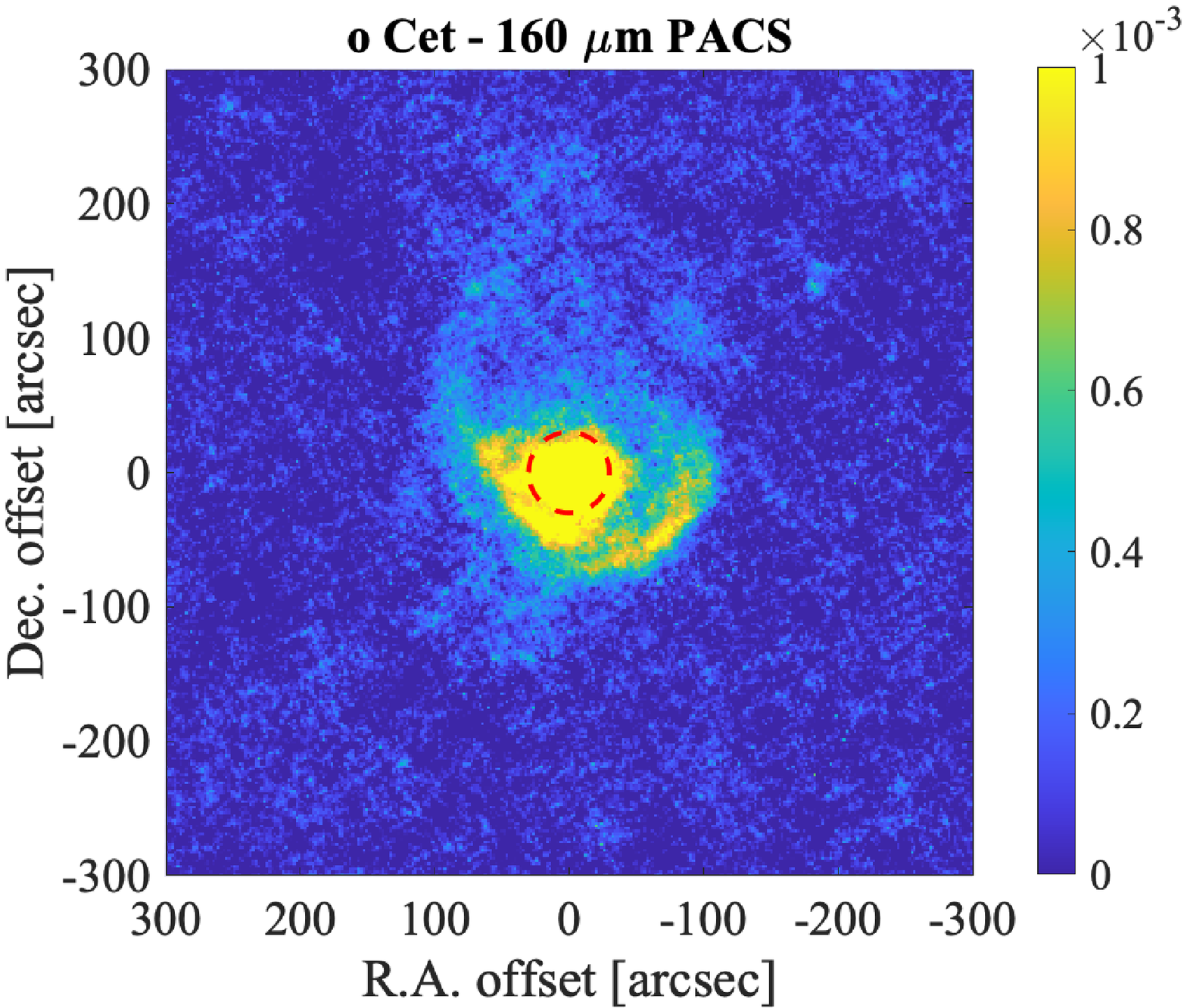}
\caption{$\omicron$ Cet: The PACS images at 70\,\micron~(left) and 160\,\micron~(right). The colour scale is in \Jyarcsec. The red dashed circle shows the mask used to measure the flux from the star and present-day mass-loss.}
\label{f:omicet}
\end{figure*}

\begin{figure*}
\centering
\includegraphics[width=8cm]{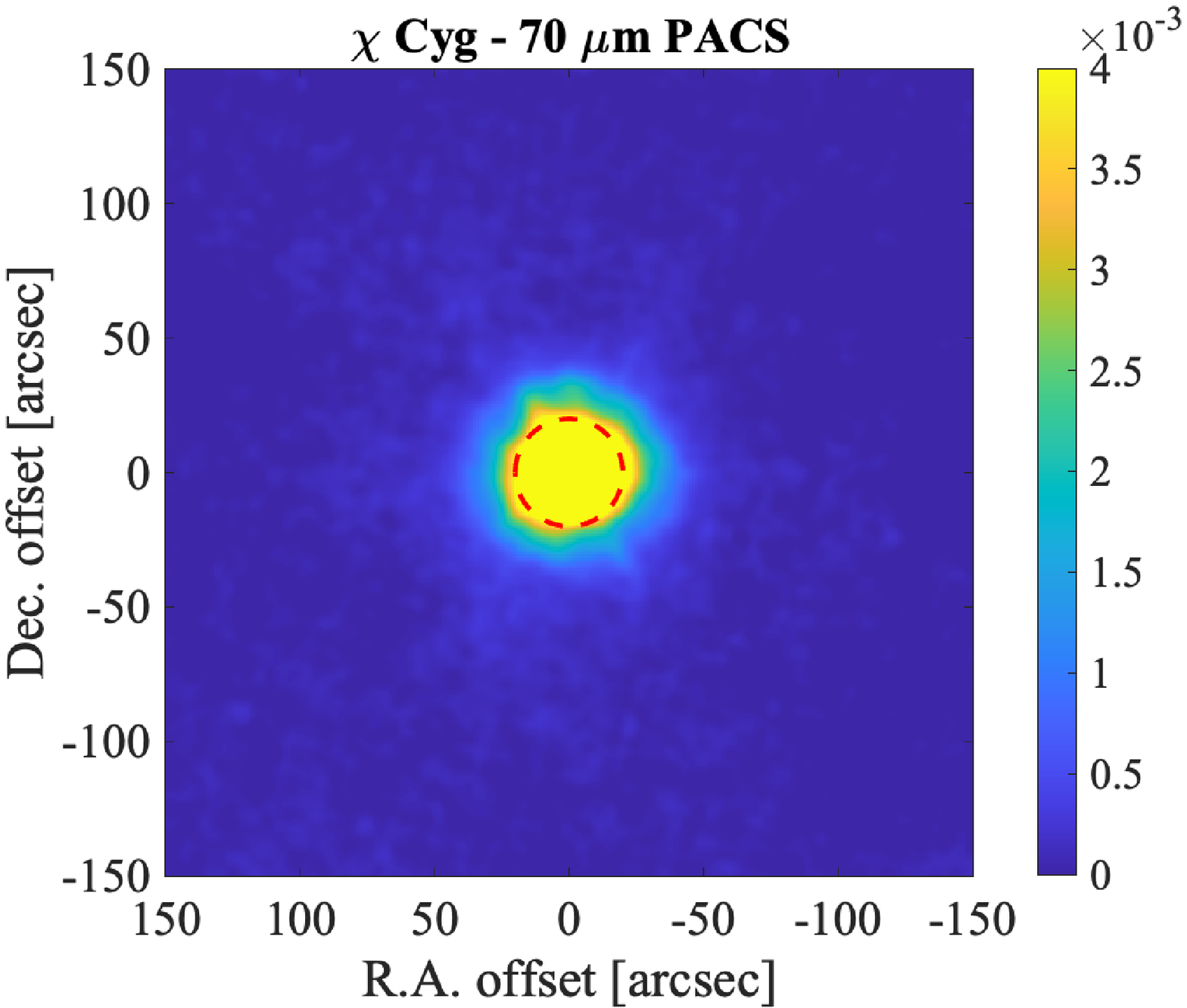}
\includegraphics[width=8cm]{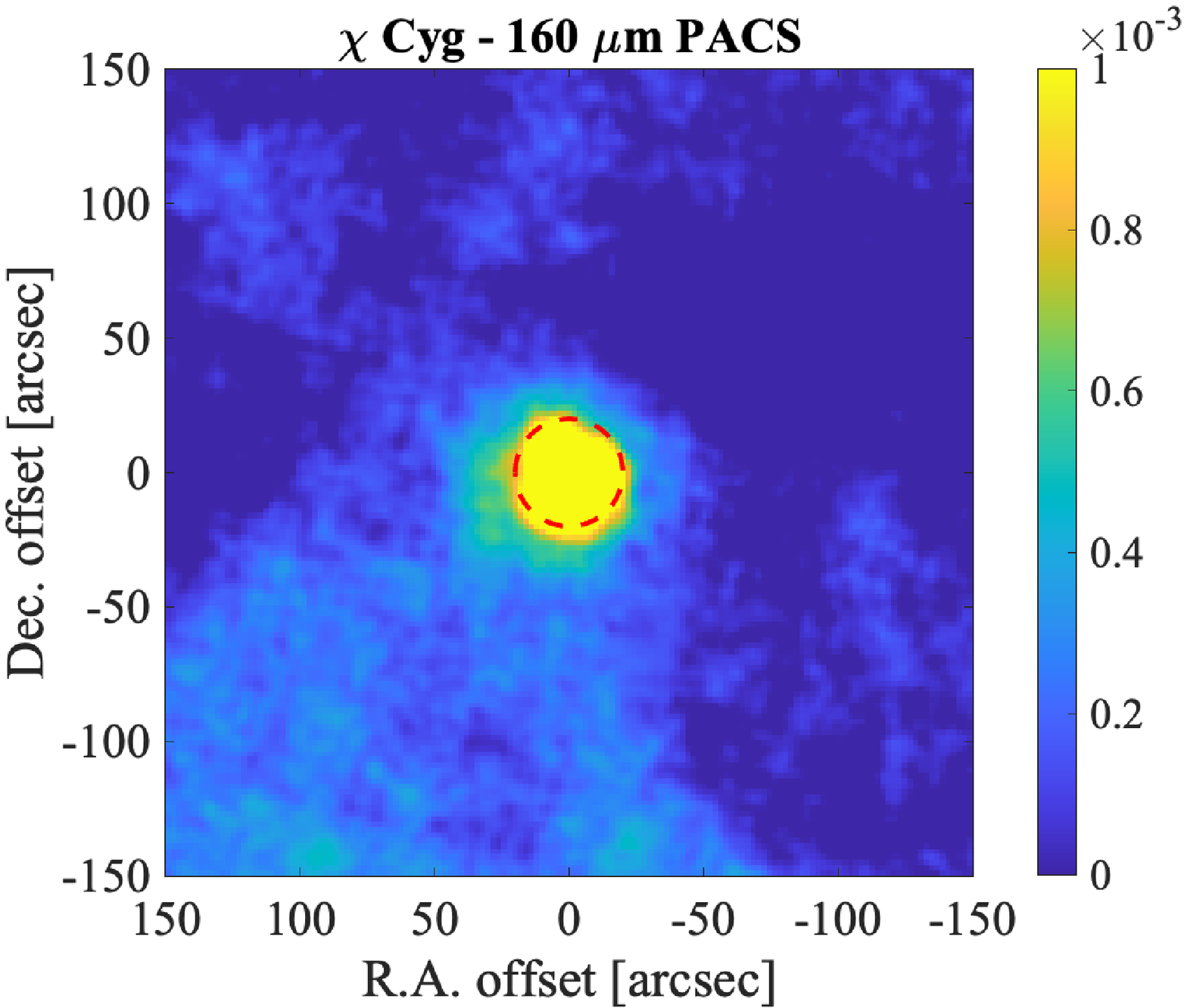}
\caption{$\chi$ Cyg: The PACS images at 70\,\micron~(left) and 160\,\micron~(right). The colour scale is in \Jyarcsec. The red dashed circle shows the mask used to measure the flux from the star and present-day mass-loss.}
\label{f:chicyg}
\end{figure*}

\begin{figure*}
\centering
\includegraphics[width=8cm]{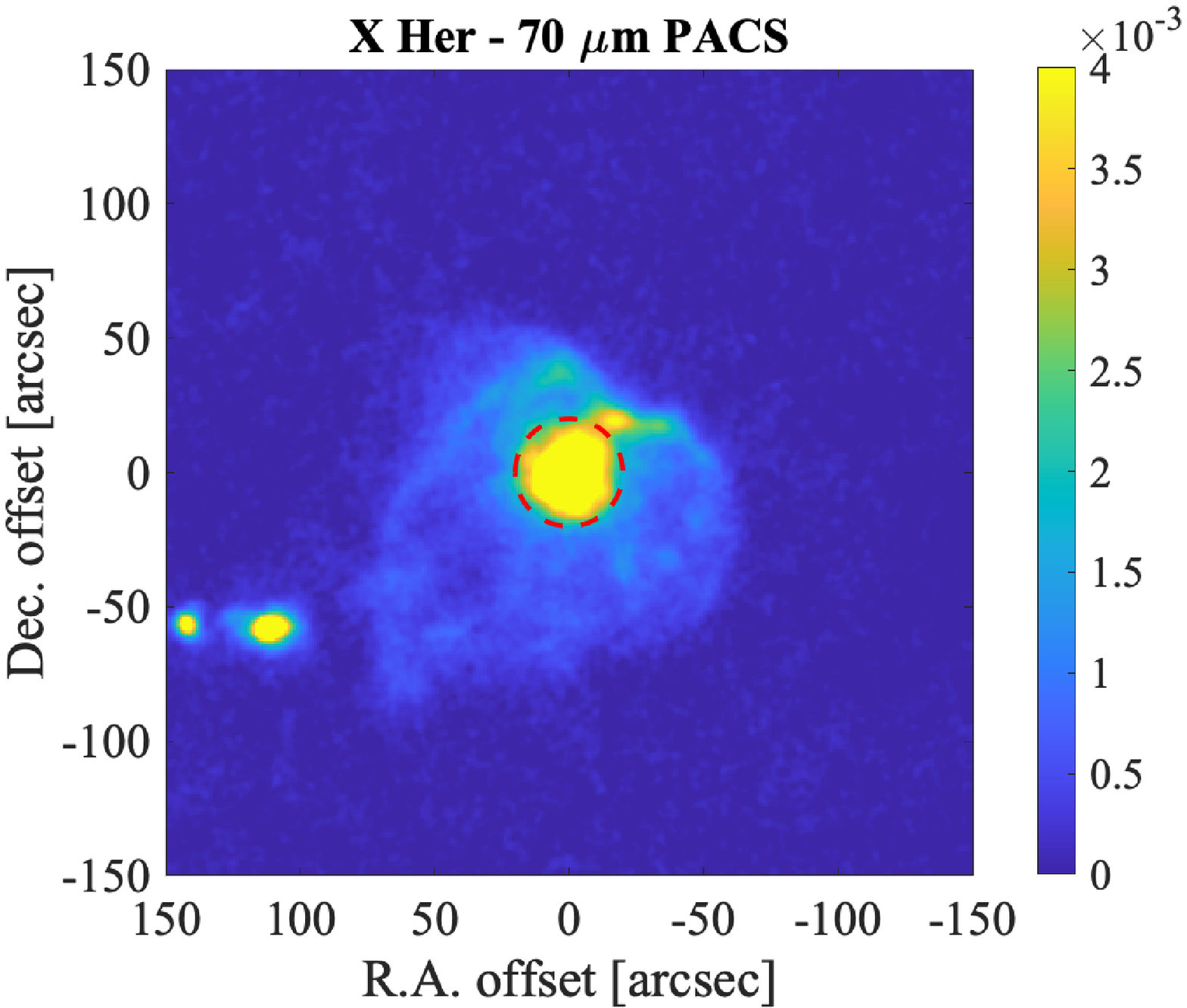}
\includegraphics[width=8cm]{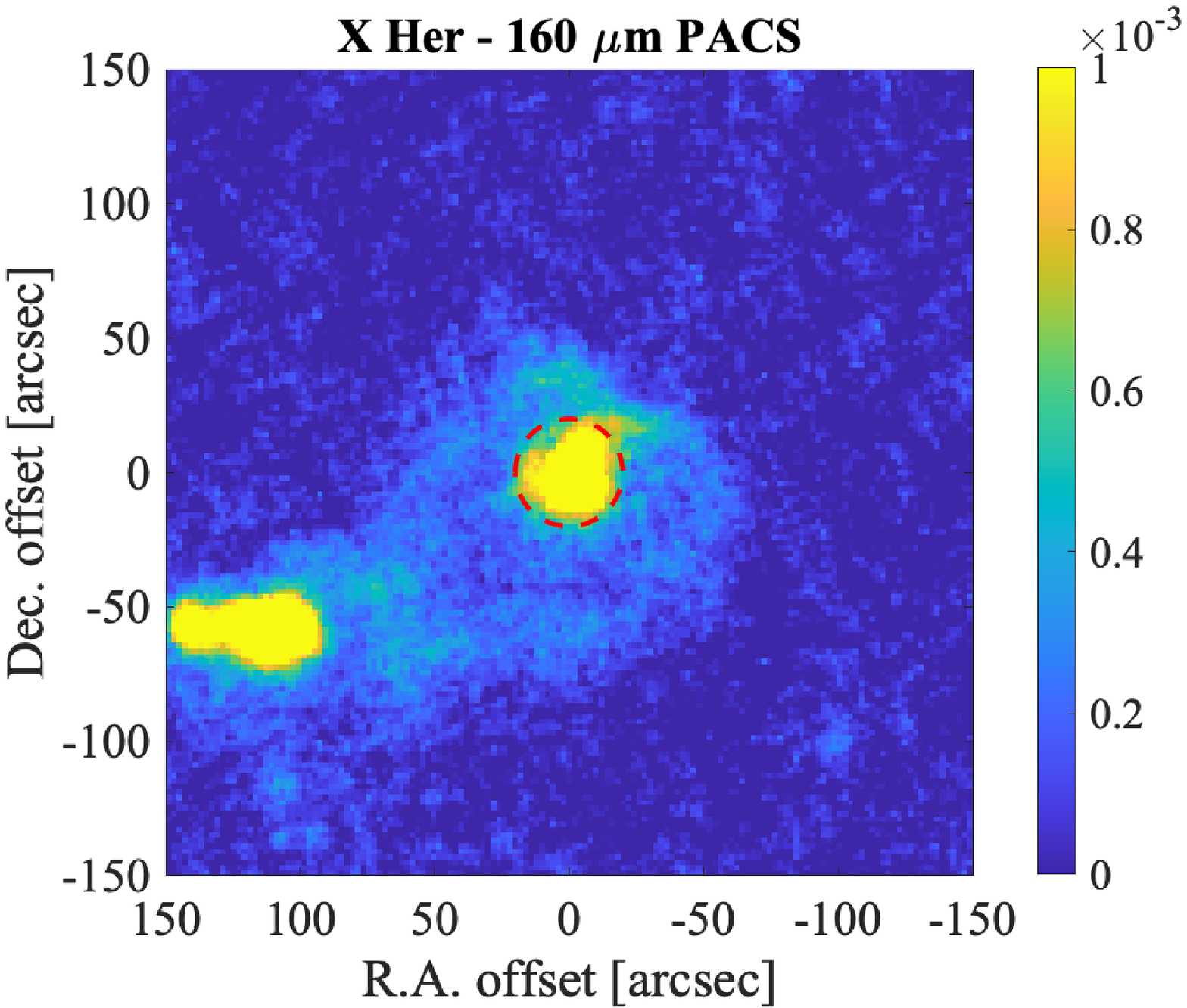}
\caption{X Her: The PACS images at 70\,\micron~(left) and 160\,\micron~(right). The colour scale is in \Jyarcsec. The red dashed circle shows the mask used to measure the flux from the star and present-day mass-loss.}
\label{f:xher}
\end{figure*}

\begin{figure*}
\centering
\includegraphics[width=8cm]{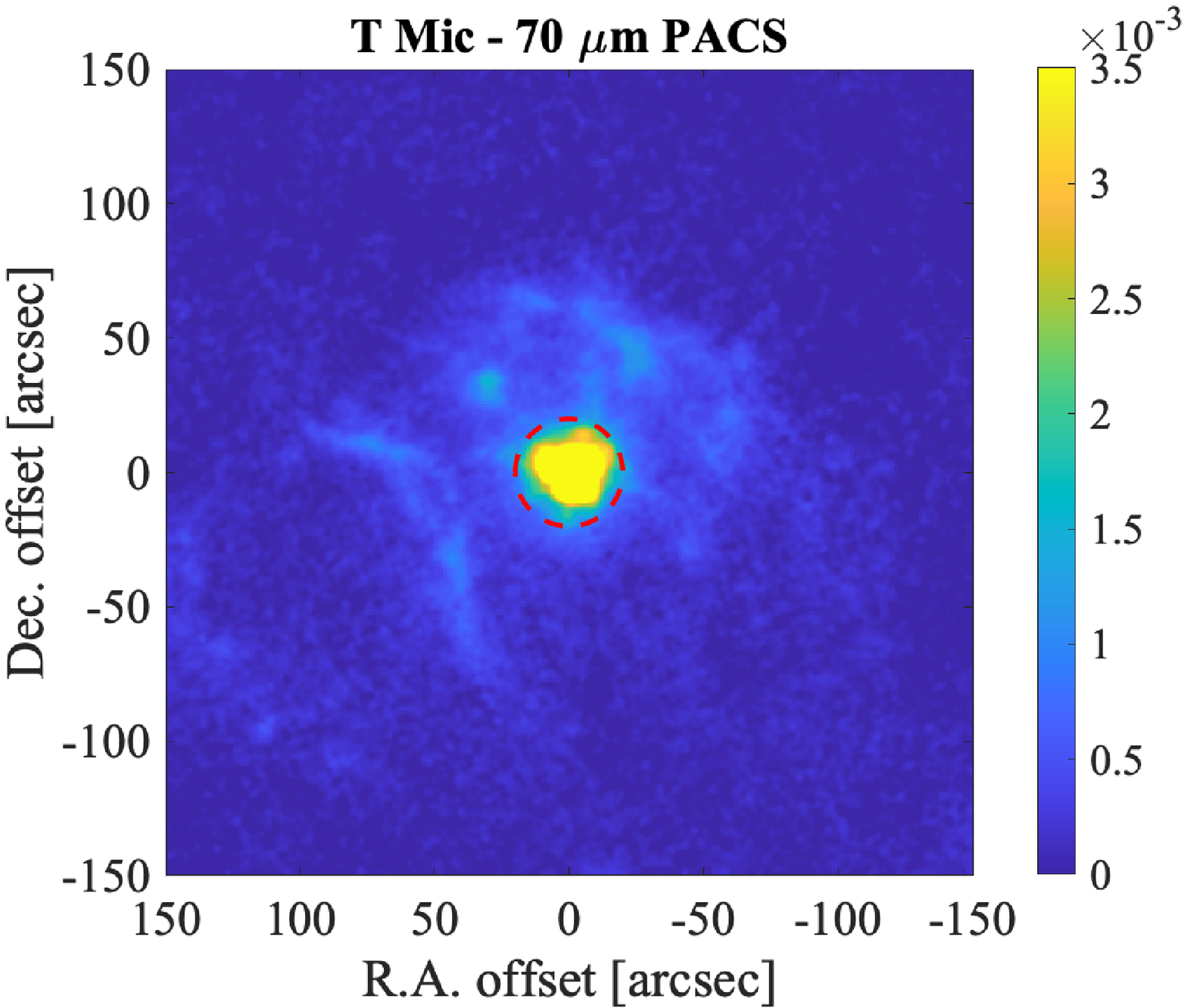}
\includegraphics[width=8cm]{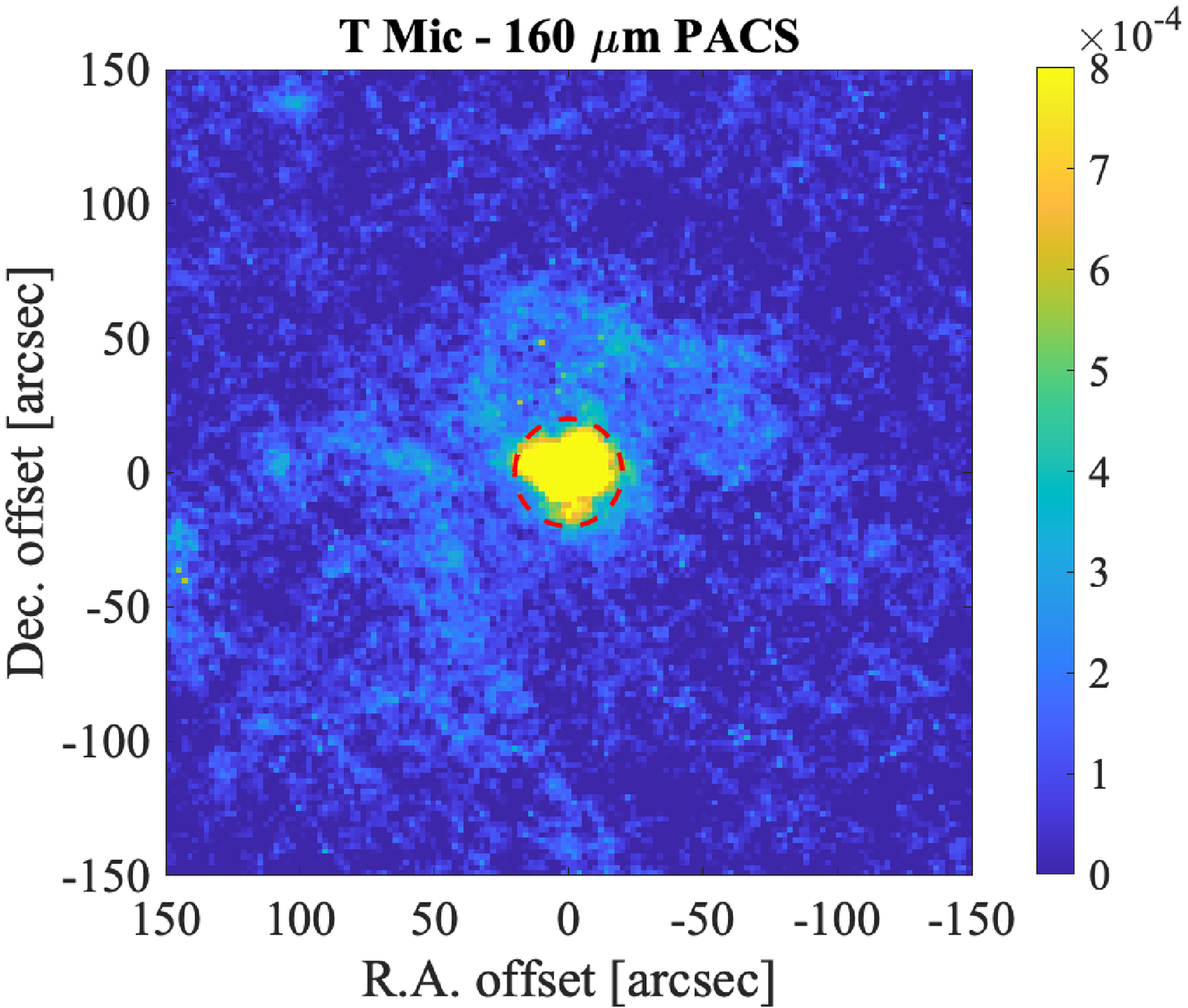}
\caption{T Mic: The PACS images at 70\,\micron~(left) and 160\,\micron~(right). The colour scale is in \Jyarcsec. The red dashed circle shows the mask used to measure the flux from the star and present-day mass-loss.}
\label{f:tmic}
\end{figure*}

\begin{figure*}
\centering
\includegraphics[width=8cm]{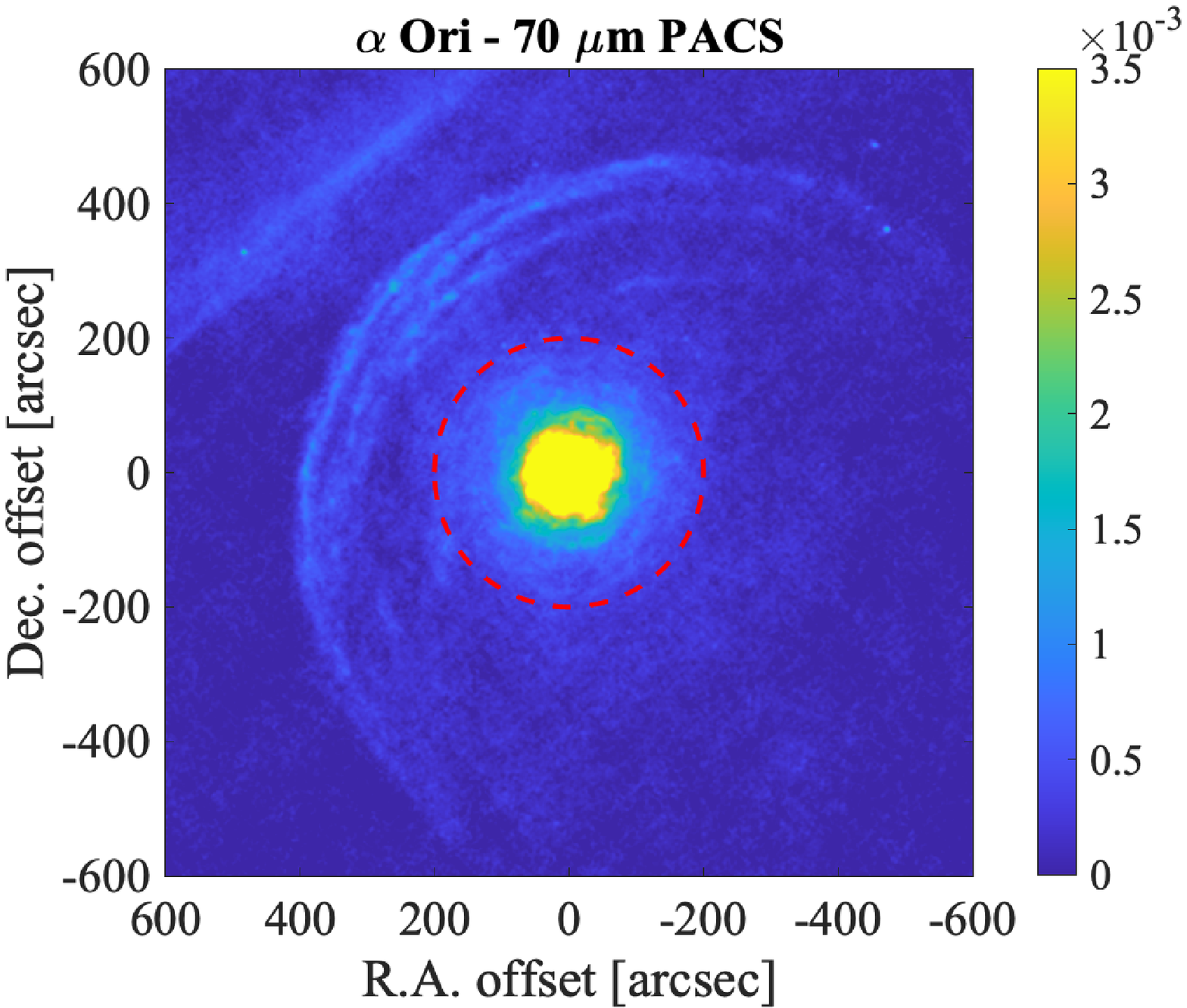}
\includegraphics[width=8cm]{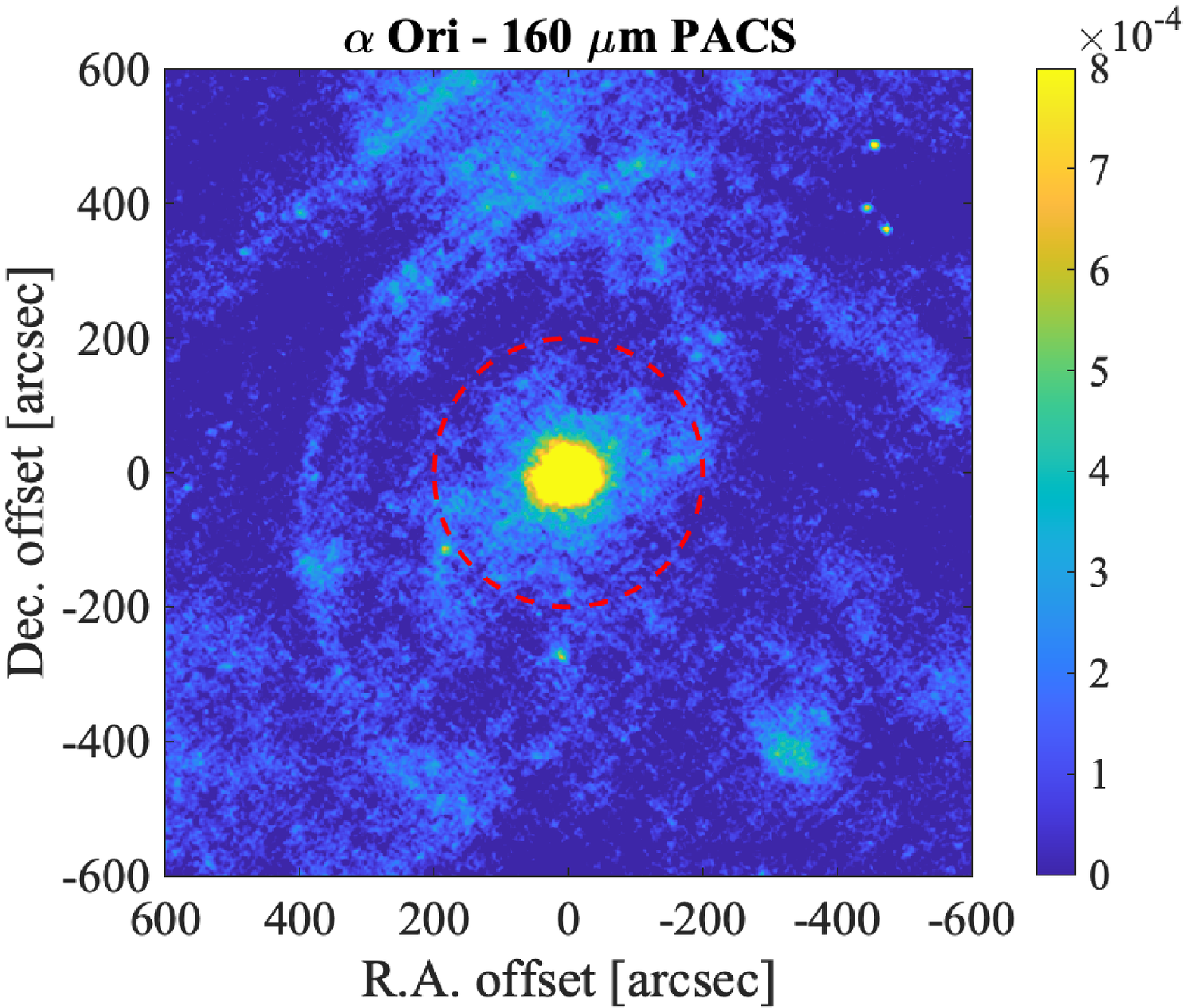}
\caption{$\alpha$ Ori: The PACS images at 70\,\micron~(left) and 160\,\micron~(right). The colour scale is in \Jyarcsec. The red dashed circle shows the mask used to measure the flux from the star and present-day mass-loss.}
\label{f:alphaori}
\end{figure*}

\end{document}